\documentclass[twocolumn,showpacs,preprintnumbers,amsmath,amssymb,prb,nofootinbib]{revtex4-1}

\usepackage{graphicx}
\usepackage{dcolumn}
\usepackage[tight]{subfigure}
\usepackage{amsmath}
\usepackage{simplewick}
\usepackage{amsfonts}
\usepackage{bbm} 
\usepackage{mathrsfs}
\usepackage{verbatim}
\usepackage{float}
\usepackage[usenames,dvipsnames]{color}
\usepackage{bm} 
\usepackage{cases}  
\usepackage{mathtools} 
\usepackage{dsfont}

\usepackage{enumitem} 

\usepackage{etoolbox} 

\usepackage{hyperref} 
\usepackage[normalem]{ulem} 
\usepackage{xspace}

\newrobustcmd{\mE}{master equation\xspace}
\newrobustcmd{\ME}{Master equation\xspace}
\newrobustcmd{\current}{current\xspace}
\newrobustcmd{\currents}{currents\xspace}
\newrobustcmd{\Current}{Current\xspace}
\newrobustcmd{\observable}{observable\xspace}
\newrobustcmd{\observables}{observables\xspace}
\newrobustcmd{\Observable}{Observable\xspace}

\usepackage{xparse}
\ExplSyntaxOn
\newrobustcmd{\inst}[1]
{
  #1 
  \peek_catcode_remove:NTF ^ 
   { \passerby_add_i:n }  
   { ^{\text{i}} }                  
 }
\newrobustcmd{\FCSinst}[1]
{
  #1 
  \peek_catcode_remove:NTF ^ 
   { \passerby_add_i:n }  
   { ^{\text{i}} }                  
 }
\newrobustcmd{\Landsberginst}[1]
{
  #1 
  \peek_catcode_remove:NTF ^ 
   { \passerby_add_i:n }  
   { ^{\text{i}} }                  
 }
\newrobustcmd{\adcor}[1]
{
  #1 
  \peek_catcode_remove:NTF ^ 
   { \passerby_add_a:n }  
   { ^{\text{a}} }                  
 }
\newrobustcmd{\Landsbergadcor}[1]
{
  #1 
  \peek_catcode_remove:NTF ^ 
   { \passerby_add_a:n }  
   { ^{\text{a}} }                  
 }

\cs_new_protected:Npn \passerby_add_i:n #1
 {
  ^{#1,\text{i}} 
 }
\cs_new_protected:Npn \passerby_add_a:n #1
 {
  ^{#1,\text{a}} 
 }
\ExplSyntaxOff

\newcommand{\currentKernel}[1]{W_{\hat{I}_{#1}}}

\newrobustcmd{\tot}[1]{{#1}^{\text{tot}}}
\newrobustcmd{\D}[1]{#1}
\newrobustcmd{\env}[1]{{#1}^{\text{res}}}

\newrobustcmd{\unit}{\mathbbm{1}}   
\newrobustcmd{\bra}[1]{\langle #1|}   
\newrobustcmd{\ket}[1]{|#1 \rangle}   
\newrobustcmd{\braket}[1]{\langle #1 \rangle}   
\newrobustcmd{\eval}[2]{\left . #1 \right | _{#2}}

\def\no#1{\mathop{:}\nolimits\!#1\!\mathop{:}\nolimits}

\newrobustcmd{\delR}{\vec{\nabla_R}}

\newrobustcmd{\Ldag}[1]{{{#1}^{\pmb{\dagger}}}}
\newrobustcmd{\Lket}[1]{{| {#1}  )}}
\newrobustcmd{\Lbra}[1]{{( {#1} |}}
\newrobustcmd{\Lbraket}[1]{{(  {#1}  )}}
\newrobustcmd{\expec}[1]{{\langle {#1}  \rangle }}
\newrobustcmd{\pureD}[1] {\ket{#1}\bra{#1}}

\newrobustcmd{\T}{\mathcal{T}} 

\newrobustcmd\scalemath[2]{\scalebox{#1}{\mbox{\ensuremath{\displaystyle #2}}}}

\newrobustcmd{\Tr}[1]{\underset{#1}{\mathsf{Tr}}}   
\newrobustcmd{\tr}{\mathsf{Tr}}   
\newrobustcmd{\nohat}[1]{#1}   
\newrobustcmd{\sdagger}{{\dagger}}   

\newrobustcmd{\vecg}[1]{{\bm #1}}   
\renewrobustcmd{\vec}[1]{{\mathbf{#1}}}   

\newrobustcmd{\timeint}[4]{ \underset{#1 \geq #2 \geq #3 \geq #4}{\int d #2 d #3 } }
\newrobustcmd{\timeintfour}[6]{ \underset{#1 \geq #2 \geq #3 \geq #4 \geq #5 \geq #6}{\int d #2 d #3 d #4 d #5} }

\newrobustcmd{\fcs}{FCS\xspace}   
\newrobustcmd{\ase}{ASE\xspace}
\newrobustcmd{\ar}{AR\xspace}

\newrobustcmd{\Cite}[1]{Ref.~\onlinecite{#1}}   
\newrobustcmd{\Cites}[1]{Refs.~\onlinecite{#1}}   
\newrobustcmd{\Ref}[1]{Ref.~\onlinecite{#1}}   
\newrobustcmd{\Refs}[1]{Refs.~\onlinecite{#1}}   
\newrobustcmd{\Tab}[1]{Table~\ref{#1}}   
\newrobustcmd{\tab}[1]{\ref{#1}}   
\newrobustcmd{\Fig}[1]{Fig.~\ref{#1}}   
\newrobustcmd{\Figs}[1]{Figs.~\ref{#1}}   
\newrobustcmd{\fig}[1]{\ref{#1}}   
\newrobustcmd{\Eq}[1]{Eq.~(\ref{#1})}   
\newrobustcmd{\Eqb}[1]{Equation~(\ref{#1})}   
\newrobustcmd{\Eqs}[1]{Eqs.~(\ref{#1})}   
\newrobustcmd{\eq}[1]{(\ref{#1})}   
\newrobustcmd{\App}[1]{Appendix~\ref{#1}}
\newrobustcmd{\Apps}[1]{Appendixes~\ref{#1}}
\newrobustcmd{\app}[1]{\ref{#1}}   
\newrobustcmd{\Sec}[1]{Sec.~\ref{#1}}   
\newrobustcmd{\Secs}[1]{Secs.~\ref{#1}}
\renewrobustcmd{\sec}[1]{\ref{#1}}   

\definecolor{myred}{rgb}{0.9,0,0}
\definecolor{myblue}{rgb}{0,0,0.5}
\definecolor{mygreen}{rgb}{0,0.6,0}

\robustify{\sum}
\robustify{\int}
\robustify{\nonumber}

\begin{document}

\title{
Gauge freedom in observables and Landsberg's nonadiabatic geometric phase:\\
Pumping spectroscopy  of interacting open quantum systems
}
\author{T. Pluecker$^{(1,3)}$}
\author{M. R. Wegewijs$^{(1,2,3)}$}
\author{J. Splettstoesser$^{(4)}$}

\affiliation{
	(1) Institute for Theory of Statistical Physics,
      RWTH Aachen University, 52056 Aachen,  Germany \\
  (2) Peter Gr{\"u}nberg Institut,
      Forschungszentrum J{\"u}lich, 52425 J{\"u}lich,  Germany \\
	(3) JARA-FIT, 52056 Aachen, Germany \\
  (4) Department of Microtechnology and Nanoscience (MC2),
      Chalmers University of Technology, SE-41296 G{\"o}teborg, Sweden \\
}
\date{\today}
\begin{abstract}
We set up a general density-operator approach to \emph{geometric} steady-state pumping through slowly driven open quantum systems.
This approach applies to strongly interacting systems that are weakly coupled to multiple reservoirs at high temperature, illustrated by an Anderson quantum dot.
Pumping gives rise to a \emph{nonadiabatic} geometric phase that can be described by a framework originally developed for classical dissipative systems by Landsberg.
This geometric phase is accumulated by the transported \emph{observable} (charge, spin, energy) and \emph{not} by the quantum state.
It thus differs radically from the adiabatic Berry-Simon phase, even when generalizing it to mixed states, following Sarandy and Lidar.

As a key feature, our geometric formulation of pumping stays close to a direct physical intuition (i) by tying gauge transformations to \emph{calibration of the meter} registering the transported observable and (ii) by deriving a \emph{geometric connection} from a driving-frequency expansion of the \emph{current}.
Furthermore, our approach provides a systematic and efficient way to compute the geometric pumping of various observables, including charge, spin, energy and heat.
These insights seem to be generalizable beyond the present paper's working assumptions (e.g., Born-Markov limit) to more general open-system evolutions involving memory and strong-coupling effects due to low temperature reservoirs as well.

Our geometric curvature formula reveals a general experimental scheme for performing \emph{geometric transport spectroscopy} that enhances standard nonlinear spectroscopies based on measurements for static parameters.
We indicate measurement strategies for separating the useful geometric pumping contribution to transport from nongeometric effects.

A large part of the paper is devoted to an explicit comparison with the Sinitsyn-Nemenmann full-counting statistics (\fcs) approach to geometric pumping,
restricting attention to the first moments of the pumped observable.
Covering all key aspects, gauge freedom, pumping connection, curvature, and gap condition, we argue that our approach is physically more transparent and, importantly, simpler for practical calculations.
In particular, this comparison allows us to clarify how in the FCS approach an ``adiabatic" approximation leads to a manifestly nonadiabatic result involving a finite retardation time of the response to parameter driving.

\end{abstract}

\maketitle

\newrobustcmd{\SINITSYN}{Sinitsyn07EPL}
\newrobustcmd{\REALTIME}{REALTIMEMACRO}
\newrobustcmd{\FCS}{Sinitsyn07EPL,Ren10,Yuge12,Liu13,Yuge13,Nakajima15}
\newrobustcmd{\GEOMETRICMAGNETISM}{Berry93}
\newrobustcmd{\GEOMETRICDENSITYOPERATORINTRO}{Uhlmann86,Dabrowski90,Sjoeqvist00,Zhou03,Ericsson03,Cohen03,Whitney03,Whitney04,Chu04,Tong04,Chaturvedi04,Sarandy05,Sarandy06,Fujikawa06,Sinitsyn07EPL,Goto07,Whitney10,Avron12}
\newrobustcmd{\GEOMETRICDENSITYOPERATOR}{Uhlmann86,Dabrowski90,Sjoeqvist00,Zhou03,Ericsson03,Cohen03,Whitney03,Whitney04,Chu04,Tong04,Chaturvedi04,Fujikawa06,Goto07,Whitney10}

\newrobustcmd{\ARNAIVE}{Splettstoesser06,Splettstoesser08a,Winkler09,Reckermann10a,Calvo12a,Avron12,Haupt13,Riwar13,Winkler13,Rojek14}
\newrobustcmd{\INTERACTIONINDUCED}{Sinitsyn07EPL,Sinitsyn07PRB,Sinitsyn09,Reckermann10a,Ren10,Yuge13,Riwar13,Yoshii13,Nakajima15}
\newrobustcmd{\CONTROLMECHANICS}{Marsden90,Marsden98,Cendra01,Bloch03,Andersson03thesis}

\section{Introduction\label{sec:intro}}

\subsection{Geometric effects in open quantum systems}
Currently there is a heightened  interest in geometric and topological properties of \emph{open} quantum systems where finite temperature, dissipation, and non-equilibrium transport play a key role.
For closed quantum systems topological properties are generally appreciated for their robustness against perturbations, assuming these perturbations keep the system closed.
This robustness, for example, underlies the successful topological classification of phases of closed quantum systems~\cite{Altland97,Read00,Snyder08,Nayak08,Kitaev09Conference,Ryu10,Hasan10,Qi11,Budich15a,Chiu15,Kennedy16}.
It is a pressing question as to how this scheme is affected when one opens up the system,
allowing for finite temperature, dissipation, and nonequilibrium~\cite{Diehl11a,Bardyn13,Iemini16}.
To address this issue, recently topological numbers for dissipative systems have been discussed~\cite{Huang14,Viyuela15,Budich15a,Budich15b} starting from  the Uhlmann connection for \emph{mixed} quantum states~\cite{Uhlmann86}.
One may also utilize topological robustness for \emph{controlling} quantum systems.
For example, in the area of quantum information processing this is exploited in geometric quantum computing~\cite{Zanardi99} and in topological error correction~\cite{DKP02,Nayak08rev}.
Ultimately,  topological properties stand in the foreground because of their superior robustness as compared to geometric properties.
However,  as for classical mechanics~\cite{\CONTROLMECHANICS},
also for quantum systems a deep understanding of the underlying \emph{geometric} properties is always a prerequisite for such control.
One of the reasons why the geometric properties of open quantum systems are more complex than for closed systems is that they are described by mixed states
rather than pure states~\cite{\GEOMETRICDENSITYOPERATORINTRO}.
However, as this paper will emphasize, this is not the only important difference.
There are also important geometric properties associated with \emph{observables}~\cite{Avron12},
including \emph{nonsystem} observables~\cite{Sinitsyn09}
defined (partly) on the system's environment
and their transport currents.

Indeed, geometric effects appear naturally in open quantum systems when considering \emph{pumping} of some observable quantity, e.g., charge, which has been studied extensively in electronic systems.
In the long-time limit the transport through a mesoscopic system exhibits steady-state pumping when its external parameters are periodically modulated in time~\cite{Grifoni98}.
For example, in highly-tunable quantum-dot systems  one can drive local system properties via electrical gates, modulate the coupling to external reservoirs\cite{Kaestner15} by tunnel barriers, or vary the external electro-chemical potentials~\cite{Brouwer01} or temperatures~\cite{Watanabe14}.
Such time-dependent control has been experimentally demonstrated even for atomic-scale junctions~\cite{Jehl03}
and is also of interest for realizing molecular motors~\cite{Chernyak09,Napitu15}.

When slowly driving an open quantum system, there are two effects to consider.
First, one generically obtains a nonequilibrium current at every instant in time if there is an instantaneous (possibly time-dependent) bias applied, either electrochemical or thermal or both.
This survives even in the adiabatic limit of vanishing driving frequency $\Omega$.
The net transported \observable after one cycle then
generically contains an average over these instantaneous currents (``sum of snapshots'').
Since this part of the current is invariant under inversion of the parameter driving,
its effect can always be canceled out experimentally or extracted theoretically.

However, there is also the \emph{pumping} current contribution which \emph{only} derives from this time-dependent driving~\cite{Buettiker85,Brouwer98}.
Physically, this contribution clearly differs from the instantaneous one:
it arises because the state of the open system cannot instantaneously follow the driving but lags behind.
It derives from the time-average of the nonadiabatic part of the current that is linear in driving frequency $\Omega$.
Nevertheless, this pumping effect is often referred to as ``adiabatic pumping'' ~\cite{Brouwer98,Switkes99,Splettstoesser06,Sinitsyn07EPL}
which we will emphatically avoid here.
In this paper, we focus on this \emph{adiabatic-response pumping} in open quantum systems
which has the hallmark of a \emph{geometric} quantity\cite{Resta00,Xiao10,Bohm,Marsden90,Nakahara03} in its simplest meaning:
the transported observable per cycle depends on the driving parameter curve alone and not on the driving frequency, in contrast to the nongeometric, instantaneous part.
Pumping thus arises as an adiabatic response to driving\cite{Avron12,Berry93}  (``lag'') and is inherently a
(first-order) \emph{non}adiabatic geometric effect,
in contrast to the more commonly considered geometric phases associated with adiabatic dynamics~\cite{Berry84,Bohm,Sarandy06}.

One reason for studying the geometric nature of pumping lies in robust control of transport of quantities like  charge, spin, or energy.
For example, for applications to charge-current standards\cite{Pekola13rev}
 the robustness of the geometric pumping with respect to frequency-fluctuations is relevant.
Therefore,  it is  of practical importance
 to be able to separate clearly this frequency-independent pumped observable,
responsible for ``clocked'' electron transfer~\cite{Roche13},
from the driving-frequency-dependent nongeometric contribution.
The present paper shows that a separate consideration of the geometric adiabatic-response part is also of theoretical importance
for identifying the physical origin of the gauge freedom underlying pumping.

Further motivation is provided by the interest in \emph{topological} pumping mentioned at the beginning:
namely, topological pumping arises when geometric quantities depend only on the ``type'' of the driving cycle (homotopy class). For example, one characteristic is the number of windings of the parameter cycle around a ``hole'' in parameter space (similar to the Aharonov-Bohm effect).  As long as this characteristic stays unchanged, the pumped quantity is even protected against continuous deformations of the geometric properties of the driving cycle.

In this work, we are particularly interested in identifying the geometric nature of the pumping contribution in systems where \emph{strong interactions} play a role.
This is largely motivated by experiments on quantum dot systems~\footnote{Pumping has also been studied intensively in superconducting systems.
There, the pumping of Cooper-pairs can be effectively expressed as a closed-system geometric phase~\cite{Fazio03,Governale05,Brosco08,Mottonen08,Gibertini13},
i.e., a Berry-Simon phase picked up by a superconducting state vector during the cyclic evolution. 
Here we study situations for which a pure-state description is not possible. }
 in which one can exploit strong Coulomb interactions to gain control over a \emph{single} electron
as already shown in early experiments on pumping~\cite{Pothier92}. 
More recently, accurately clocked sources of single charges~\cite{Feve07,Roche13,Pekola13rev} or spins~\cite{Xing14,Dittmann16} have been implemented. This illustrates the high degree of control over single electrons in a quantum-dot system \emph{in time}.

The final point of interest, going beyond the aspects of robustness and control, is the use of pumping as a \emph{``spectroscopic''} tool.
In this paper, we discuss how pumping effects can shed light on properties of an open quantum system that remain hidden when considering only non-driven, stationary transport.
This has use as an experimental tool, since one can infer, for example, the tunnel-coupling asymmetry and the spin-degeneracy of a quantum dot just by using the qualitative features of an \emph{interaction-induced} charge pumping~\cite{Reckermann10a,Calvo12a,Riwar13} effect.
However, it may also function as a theoretical tool similar to the usual linear response to a perturbation:
in  models that are theoretically hard to analyze, physical characteristics that are not revealed by stationary properties (e.g., due to renormalization effects) may well appear in an adiabatic-response calculation of pumping effects~\cite{Splettstoesser06}.


\begingroup
\squeezetable
\begin{table*}
\caption{\label{tab:compare}
Comparison of geometric density-operator approaches relevant to this paper.
}
\begin{ruledtabular}
\begin{tabular}{llll}
    \textbf{Approach}
  & \textbf{Prior works}
  & \textbf{Present work}
\\
\\
   (I) Adiabatic state
  & Adiabatic mixed-state geometric phase \cite{Sarandy05,Sarandy06}
  & Zero geometric phase for adiabatic steady-state [\Sec{sec:LackOfPhaseState}]
\\
    evolution (\ase)
  & Mixed-state adiabatic-\emph{response} correction \cite{Sarandy05,Sarandy06}
  & Zero geometric phase for nonadiabatic state [\Sec{sec:LackOfPhaseState}]
\\
  & Gauge freedom related to eigenvector rescaling \cite{Sarandy05,Sarandy06}
  & Restriction of gauge freedom by normalization and hermiticity [\Sec{sec:LackOfPhaseState}]
\\
\\
   (II) Full counting
  &  Geometric part of generating function \cite{Sinitsyn07EPL,Sinitsyn09,Chernyak12a,Chernyak12b,Chernyak10,Ivanov10}
  & Restriction of gauge freedom by real-valuedness observable [\Sec{sec:fcsGauge}]
\\
    statistics (FCS)                                                
  & All moments have geometric part  \cite{Sinitsyn07EPL,Sinitsyn09}
  & Clarification of ``adiabaticity'' in \fcs [\Sec{sec:fcsAdiabatic}]]
\\
	&
	& Conditions for applicability same as in \ar [\Sec{sec:fcsAdiabatic}]]
\\
\\
   (III) Adiabatic
  & Adiabatic-response of unique stationary state \cite{Avron12}
  & Adiabatic iteration for Born-Markov open system [\Sec{sec:LackOfPhaseState},\App{app:iteration}]
\\
    response (AR)
  & Geometric pumping of system observable \cite{Avron12}
  & Physical picture of gauge freedom in observable [\Sec{sec:GaugeFreedomObservable}]
\\
  &
  & Gauge freedom of \current memory kernels [\Sec{sec:GaugeFreedomObservable}]
\\
  & 
  & Geometric pumping of nonsystem observables [\Sec{sec:landsberg}]
\end{tabular}
\end{ruledtabular}
\end{table*}
\endgroup


\begingroup
\squeezetable
\begin{table}
\caption{\label{tab:crosslink}
Cross-links between geometric density-operator approaches discussed in this paper.
}
\begin{ruledtabular}
\begin{tabular}{lll}
\\ \textbf{Approaches}
& \textbf{Cross links in present work}
\\
\ase $\leftrightarrow$ \fcs
& \fcs is equivalent to the \ase approach   & \\
& with $\chi$ dependence & \\
\\
\ase $\leftrightarrow$ AR
& \emph{Nonadiabatic correction} to \ase [Eq. (23) of \Ref{Sarandy05}]  \\
& agrees with \ar result $\rho^\text{a}$ [\Eq{eq:adcorDensityOp}] \\
& but contributes \emph{zero} in unique steady-state [\Sec{sec:LackOfPhaseState}] \\
& \ar Geometric phase instead in observable.  \\
\\
AR $\leftrightarrow$ \fcs 
& Nonadiabatic Landsberg phase [\Sec{sec:fcsAdiabatic}]
\\
& equal to $\chi$-linear part of FCS-phase\cite{Nakajima15}
\end{tabular}
\end{ruledtabular}
\end{table}
\endgroup

Before we can formulate the open questions that our key results address,	we need to outline a number of existing theoretical approaches to pumping.
This also serves to keep the paper self-contained and makes it more accessible to readers with interest in either geometrical effects or open quantum systems or both.
This seems furthermore warranted since a number of quite different approaches,
designed to deal with different problems,
have been put forward.
We also point out a number of useful relations between cited references
that have received little attention so far.
A guide to our comparison of the geometric aspects of these approaches and different aspects put forward in this work is given in \Tab{tab:compare} and \tab{tab:crosslink}.

\subsection{Geometric density operator approaches}

For open systems \emph{without} interactions (beyond the mean-field level),
Brouwer's framework~\cite{Buettiker93,Buettiker94,Brouwer98} for pumping based on the Buttiker-Thomas-Pretre scattering theory for time-dependent setups is by now standard.
Within this approach, the geometric nature of charge pumping is
associated with unitary transformations of the scattering matrices~\cite{Altshuler99,Zhou03}.
This has played an important role, for example, in recent theoretical work on current-induced forces in nanoscale systems~\cite{Dundas09,Bode11,Thomas12b,Todorov14,Lue15}
and nanoscale motors~\cite{Haenggi09,Seldenthuis10,Bustos13,Napitu15,Arrachea16}.

However, when strong interactions become important one needs a different approach,
even though Brouwer-type formulas emerge also in this case~\cite{Calvo12a} (see \App{app:brouwer}).
Whereas Green's function approaches to pumping have been put forward~\cite{Splettstoesser05,Sela06,Fioretto08},
there is a well-established approach to strongly interacting systems based on the reduced density operator description.
However, within this approach the situation is less univocal regarding the geometric nature of pumping.
This is a primary topic of this paper.
Several geometric frameworks have been formulated based on the reduced density operator, including contexts unrelated to pumping.
We will tie together three of these formulations, found in \Refs{Sarandy05}, \onlinecite{Sarandy06} and \Ref{Avron12}, and \Refs{Sinitsyn07EPL}, \onlinecite{Sinitsyn09}, \onlinecite{Nakajima15}, respectively.
Before we outline the key results of our paper, we sketch these three geometric approaches, taking note of many other density-operator based works~\cite{\GEOMETRICDENSITYOPERATOR}.

(i) \emph{Adiabatic state-evolution (\ase) approach.}
A perhaps intuitive, but wrong expectation is that the geometric nature of pumping in open systems arises from the dynamics of the reduced quantum state.
However, in the following [cf. also \Ref{Pluecker16b}] it is still important to consider such geometric phases.
The geometric nature of this adiabatic \emph{mixed-state} evolution has been worked out by Sarandy and Lidar\cite{Sarandy05,Sarandy06}.
This closely follows the analogy to the adiabatic Berry-Simon phase for adiabatic evolution of a pure state of a closed quantum system.
In the \ase approach the mixed-state density operator $\rho(t)$ is considered as a ket vector $\Lket{\rho(t)}$ in Liouville (or Hilbert-Schmidt) space
evolving according to a time-local master equation
\begin{align}
\frac{d}{dt}\Lket{\rho(t)} = W[\vec{R}(t)]\, \Lket{\rho(t)}
  .
\label{eq:kineq}
\end{align}
Here, the kernel $W[\vec{R}(t)]$ takes over the role of the evolution generator played by the Hamiltonian $H[\vec{R}(t)]$ in the Berry-Simon case
based on the Schr\"odinger equation $\tfrac{d}{dt}\ket{\psi(t)} = -i H[\vec{R}(t)] \, \ket{\psi(t)}$ for natural units setting $\hbar  = \text{k}_\text{B} = \text{e} = 1$.
The time dependence enters entirely through the instantaneous values of the driving parameters $\vec{R}(t)$.
Similar to the Berry-Simon approach, in the \ase approach one expands the solution of the \mE in the eigenvectors $\Lket{v_n}$ to eigenvalues $\lambda_n$ of the kernel $W$, all with parametric time-dependence.
A gauge freedom emerges from the nonuniqueness of the normalization of these eigenvectors,
but in contrast to the Berry-Simon case, these changes in the normalization are nonzero complex numbers (nonunitary, noncompact gauge group), rather than phase factors (unitary, compact).
For slow driving, the solution of the master equation \eq{eq:kineq} follows (a sum of) these eigenvectors \emph{adiabatically} resulting in dynamical and geometric phases.
Several points discussed in this paper can  be understood as a formal application of this generalization of the Berry-Simon phase.
However, the \ase approach does not deal with \emph{steady-state} pumping
and the ASE phase essentially differs from the simple geometric pumping phase that
we work out here:
in our contexts the \ase phase for the steady state is identically zero, even when accounting for the first nonadiabatic correction (adiabatic-response) to the state.
This quenching of the Berry-Simon type phase of mixed states forms the starting point for the considerations of geometric steady-state pumping.

(ii) \emph{Full-counting statistics (\fcs) approach to pumping.}
Within the density operator framework, the geometric nature of \emph{pumping} of observables was first clarified
when Sinitsyn and Nemenmann\cite{Sinitsyn07EPL} applied the well-established FCS approach to pumping (``stochastic pumping"), introduced in more detail in \Sec{sec:fcs}. Interestingly, they found that pumping can be induced by interaction.
In the \fcs one uses an observable-specific generating function $Z^\chi$ depending on a ``counting field'' variable $\chi$ to obtain the statistics of a selected observable, i.e., all its moments and their dynamics.
From this generating function the change of the first moment of a reservoir observable $X^r$ can be obtained as
\begin{align}
  \expec{{X}^r}(t) - \expec{{X}^r}(0)=
  \partial_{i \chi} \left . Z^\chi(t) \right |_{\chi=0}
  .
  \label{eq:IntroPumpedObservable}
\end{align}
The generating function is obtained from a ``generating operator'' $\rho^\chi$, which is the ``adiabatic'' solution of a master-type equation similar to \Eq{eq:kineq} and exhibits a geometric phase similar to the one calculated in the \ase problem.
This elegant and powerful approach has been applied to various pumping problems and is reviewed in \Ref{Sinitsyn09}:
applications range from 
molecular reactions~\cite{Sinitsyn07EPL,Sinitsyn07PRL,Sinitsyn09},
to heat transport through strongly anharmonic molecules~\cite{Ren10,Liu13},
and strongly interacting quantum dots~\cite{Bagrets03,Utsumi07,Yuge12,Yoshii13,Nakajima15}.
It was also used to demonstrate that thermodynamic \emph{vector} potentials arise in slow but nonadiabatic transformations between non-equilibrium steady-states~\cite{Yuge13}
accounting for \emph{geometric} heat and excess entropy production.
In a recent paper~\cite{Nakajima15}, Nakajima et al.
addressed a possible point of confusion in the \fcs approach:
how can an ``adiabatic'' approach include physically nonadiabatic pumping?
In the last part of this paper we will further clarify this issue, extending their observations.

(iii) \emph{Adiabatic-response (\ar) approach.}
Finally, Avron \textit{et al.}  studied~\cite{Avron12} pumping in the density operator approach also starting from \Eq{eq:kineq}.
Interestingly, they considered pumping for both unique and  nonunique frozen-parameter stationary states.
The core idea of \ar is to first expand the \mE \eq{eq:kineq} in the driving frequency (smallest time scale) and solve for the density operator in zeroth ($\inst{\rho}$) and linear order ($\adcor{\rho}$) in this frequency:
\begin{align}
\rho(t) \approx \inst{\rho}(t) + \adcor{\rho}(t) + ... \;
.
\end{align}
Importantly,  the nonadiabatic part $\adcor{\rho}$ accounts for the ``laggy'' response and generates the pumping.
However, they restricted their analysis, using a Kato formulation, to pumping of \emph{system} observables
(i.e., with current operators related to particle transfer between parts \emph{within} the open subsystem)
and considered only a single reservoir.
For the case of a unique stationary state, the adiabatic-response pumping of system-observables calculated in \Ref{Avron12} was related to Berry-Robbins' ``geometric magnetism''~\cite{Robbins92,Berry93,Berry97,Cohen03} formulation of pumping.
In this paper, we instead study \emph{nonsystem} observables and their currents to \emph{multiple} reservoirs, which is crucial for describing the transport \emph{through} an open quantum system, enabling a geometric transport spectroscopy.
This requires account of an additional evolution equation,
namely, for the \emph{current}
$I_{X}^r$ of a nonsystem observable $X^r$ into reservoir $r$:
\begin{align}
	I_{X}^r = \tr \currentKernel{X^r} \rho(t)
	\label{eq:IntroCurrent}
	.
\end{align}
This brings in an additional observable-specific memory-kernel $\currentKernel{X^r}$ whose role in generating a geometric adiabatic-response has not been addressed so far.

\subsection{Summary of results}

The present paper was inspired by all three outlined approaches,
but in particular by a discussion  in \Ref{Avron12}  of the nonuniqueness of currents in relation to their observables,
reaching back to earlier works~\cite{Bellissard02,Gebauer04,Bodor06,Salmilehto12}.
Following up on an earlier remark in \Ref{Sinitsyn09} (p. 8) we combine this idea with Landsberg's approach~\cite{Landsberg92,Andersson03thesis,Andersson05} to dissipative systems with symmetries\cite{Kepler91,Ning92}.
The key point is to consider the physical role of the meter registering the pumping signal in the reduced density-operator formalism.
This results in an intuitive and clear physical picture that does not seem to have been worked out so far.

We outline the main steps and results of this paper:

(1) \emph{Landsberg geometric phase for pumping.}
In pumping a local gauge freedom emerges
 in the relation between a measurable pumped observable and its associated current operator [\Eq{eq:GaugeTransformationCurrentkernel}].
This is encoded in the simple adiabatic-response equations for the mixed quantum state,
\begin{gather}
  0= W \inst{\rho},
  \qquad
  \frac{d}{dt} \inst{\rho} = W \adcor{\rho}
	\label{eq:introAdsolution}
\end{gather}
and an ``enslaved'' equation for the current for a nonsystem observable:
\begin{align}
  \frac{d}{dt} \expec{  \hat{X}^r_g }
  =
  \tr \, \currentKernel{X^r_g} \left ( \inst{\rho} +\adcor{\rho} \right )
	.
	\label{eq:introExpecsolution}
\end{align}
Here $W$ and $\currentKernel{X^r_g}$ have only parametric time dependence.
Crucially, this observable  $\hat{X}^r_g(t) \coloneqq \hat{X}^r + g[\vec{R}(t)] \unit$ includes all possible parametrically time-dependent gauges $g$ relative to the ``bare'' time-constant observable $\hat{X}^r$.
Physically this gauge freedom corresponds to a calibration of the meter scale.
In the current kernel a gauge transformation $g \to g + f$, a recalibration, leads to
\begin{align}
  \currentKernel{X^r_g} \to
 \currentKernel{X^r_{g+f}}  = \currentKernel{X^r_g} + \delR f
  ,
\end{align}
 requiring an extension of the Heisenberg equation of motion [\Eq{eq:WIX}] to observables \emph{outside} the open system.
This makes the pumping contribution of the transported observable an instance of the geometric phase first considered by Landsberg
\begin{align}
  \Delta \adcor{X}^r = \oint_C d\vec{R} A_{X^r_g} [\vec{R}],
\end{align}
with the Landsberg \emph{connection} (gauge potential) [\Eq{eq:LandsbergPumpinConnection}]:
\begin{align}
 A_{X^r_g} = \Lbra{\mathds{1}}\currentKernel{X^r} \frac{1}{W} 
\Lket{ \delR \inst{\rho}}
  + \delR g
  ,
\end{align}
where the pseudo inverse $1/W$ is defined on the nonzero eigenspaces of $W$.

(2) \emph{Pumping determines a geometric effect.}
Geometrically, the observable (not the quantum-state) plays the role of a fiber (group) coordinate in a (principal) fiber bundle over the space of driving parameters with a tangible physical meaning.
The Landsberg geometric connection  on this space is essentially
 the \emph{adiabatic-response} part of the total \emph{current}
$I_{X^{r}_g}=\inst{I}_{X^{r}_g}+\adcor{I}_{X^{r}_g}$ of a gauge-dependent observable.
The geometric ``horizontal lift'' defined by this connection corresponds
to  maintaining the physical pumping current $\adcor{I_{X^{r}_g}}$  to be zero at each time instant by \emph{continuously adjusting the scale of the meter}.
The  geometric-phase ``jump,'' the holonomy of a horizontal lift curve, corresponds physically to the resulting cumulative adjustment of the meter's scale over a driving period: the  pumped \observable per period.

(3) \emph{Conditions for nonzero pumping curvature.}
The generic presence of gauge freedom implies that one can expect a geometric pumping contribution unless the connection $A_{X^r_g}$ is integrable for some special reason.
The gauge-invariant curvature (gauge field) 
\begin{align}
  B_{X^r} = \Lbra{\mathds{1}} \Big( \delR   \currentKernel{X^r} \frac{1}{W} \Big) \times \Big( \delR \Lket{ \inst{\rho} }\Big)
\end{align}
measuring this nonintegrability is just the pumped \observable per unit area of the driving parameter space.
This quantity is sensitive to crossings of lines in the parameter space where the open system is in resonance with the reservoirs.
This enables a geometric spectroscopy of open quantum systems.
Importantly, this formula also applies if there are no strict conservation laws, which is relevant, e.g., for spin- and heat transport.
It can however be easily simplified if such laws are present.
The application of our pumping formula to the explicit example of a single level Anderson quantum dot illustrates this spectroscopy,
showing that for a variety of driving protocols, \emph{interaction} is required to obtain a nonzero geometric pumping phase.

(4) \emph{Connections of different approaches.}
We find that the three approaches (\ase, \ar, \fcs) outlined in the previous section, are intimately related as summarized in \Tab{tab:crosslink}.
Our main line of comparison involves our \ar approach with the \fcs approach.
We show that when the \fcs is applied to the first moment of pumping, as done in many works, it is term-by-term equivalent to our much simpler \ar approach on all levels
(pumping formulas, gauge freedom, connection, curvature, and their limits of applicability).
We show that the \fcs not only unnecessarily complicates practical calculations, but is also less clear regarding the physical meaning of the geometric nature of pumping due to its ``mixing'' of effects of the quantum state and the observable [see discussion after \Eq{eq:complements}].

This comparison allows us to resolve the important issue regarding the ``adiabaticity'' of the \fcs  going beyond the scope of \Ref{Nakajima15}.
Also, we show how \emph{within} this physical nonadiabatic picture of the \ar approach the geometric nature of pumping can be fully understood,
independent of the \fcs formulation, thereby avoiding the nontrivial issue of its ``adiabaticity''.
In our comparison, the \ase approach turns out to be very relevant time and again.
We also connect our approach to the Kato formulation of the \ar approach of \Ref{Avron12}, shedding some new light on it.

\subsection{Adiabatic-response real-time approach
beyond the Markovian, weak-coupling limit}

An important implication of our work is that Landsberg's geometric framework is compatible with a more general \ar approach 
to pumping~\cite{Splettstoesser06} applicable to non-Markovian, strongly-coupled open quantum systems:
the gauge freedom we point out
derives from entirely general arguments.
Since this paper is written with this future extension\cite{Pluecker16b} in mind, it is important to briefly outline this more general \ar approach.

This general adiabatic-response approach to pumping in slowly driven open systems~\cite{Splettstoesser06} is based
on the exact time-nonlocal kinetic equation for the density operator,
\begin{align}
  \frac{d}{dt}\rho(t)  = -i[H[\vec{R}(t)],\rho(t)] 
  + \int_{-\infty}^t \hspace{-0.3cm}dt' W (t,t'; \{\vec{R}(\tau)\}) \rho(t')
  ,
  \label{eq:kineq-general}
\end{align}
here written for the time-dependent steady-state limit,
i.e., switching on the system-reservoir coupling at $t_0=-\infty$
and starting from an initially factorizing system-reservoir state.
This approach  is close in spirit to the \ar approach to pumping mentioned above under point (iii).
However, it goes beyond these by
incorporating the fact that the open-system evolution has a \emph{functional} dependence on the entire driving-parameter history, indicated by the dependence on $\{\vec{R}(\tau)\}$ of the kernel $W$.  This is accomplished by systematically accounting for processes of higher order in the coupling as well as the Laplace-frequency dependence of both the kernel and the density operator.
From this point of view, the superoperator $W$ in \Eq{eq:kineq} only accounts for the zero-frequency ($z=i0$) part of the Laplace-transform of $W(z;\vec{R}(t))$ of the kernel in \Eq{eq:kineq-general} after freezing its parameters at the latest time $\tau=t$.

Expectation values of nonsystem observables $X^r$, e.g., reservoir observables or reservoir-system currents,
are described by a similar time-nonlocal equation:
\begin{align}
  \braket{\hat{X}^r(t)} =  \tr \,  \int_{-\infty}^t dt' W_{\hat{X}^r} (t,t'; \{\vec{R}(\tau)\}) \rho(t')
  \label{eq:obs-general}
  ,
\end{align}
with an \emph{observable-specific memory-kernel} $W_{\hat{X}^r}$ that in general needs to be calculated separately in addition to $W$ in \Eq{eq:kineq-general}.
A key point of the paper is that this equation requires careful consideration in order to ensure explicit physical gauge covariance of the formalism.

For strongly interacting open systems at low temperature
the time-nonlocal kernels required in \Eqs{eq:kineq-general} and \eq{eq:obs-general}
can be systematically computed using the real-time diagrammatic technique~\cite{Koenig95,Schoeller09a}.
This provides a general framework for calculating kernels, including those required for noise~\cite{Aghassi06},
correlation functions~\cite{Schuricht09,BSchmidt10},
and~\footnote{Sinitsyn's master equation~\Eq{eq:GeneralizedMasterEquation} for the generating-operator of the \fcs can be derived  in both, the ``real-time'' and the ``Nakayima-Zwanzig' approach.
This underlines that neither label is a meaningful labels for distinguishing different approaches to pumping.}
for the full counting statistics~\cite{Braggio05}.
The flexibility of this approach is illustrated by the possibility of formulating a nonequilibrium renormalization group scheme for calculating $\rho(t)$~\cite{Schoeller09a,Eckel10,Pletyukhov10,Andergassen11a,Pletyukhov12a,Saptsov12a,Klochan13}.
For example, this enabled a nonperturbative adiabatic-response analysis~\cite{Kashuba12}
of interaction effects on the universal charge-relaxation resistance~\cite{Mora10,Hamamoto10,Lee11}
for  strong tunnel coupling and low temperature.

So far in this more general setting little attention has been paid to the geometric aspects of pumping.
This paper addresses two questions relevant to this:
First, where in the formalism does the gauge freedom responsible for geometric pumping arise?
What is its concrete physical meaning?
Second,  the general \ar approach to pumping
is based on real-time memory-kernels for nonsystem \emph{observables} [\Eq{eq:obs-general}].
The role of these kernels for the geometric nature of pumping has not been considered at all within the other \ar formulations outlined under point (iii) above.
What is this role?

These questions are intimately related and  lead to the insight that
\emph{observables}, rather than mixed quantum states, accumulate a geometric phase
that is responsible for steady-state pumping.
To see this, it is necessary, but not sufficient, to account for generically \emph{time-dependent} observables,
even when interested in the expectation values of \emph{time-constant} ones.
This is the fundamental difference to the \ar approaches listed under point (iii) and also turns out to provide  the link to the \fcs approach point (ii).
Fortunately, this can already be addressed in the much simpler setting of \Eq{eq:kineq}
instead of the general density-operator approach based on \Eq{eq:kineq-general}.
In this paper we thus start from this equation, i.e., the same kind of master equation as the approaches (i)-(iii)
reviewed above, allowing a useful three-way comparison.
The generalization starting from \Eqs{eq:kineq-general} and \eq{eq:obs-general} requires more care and will be discussed elsewhere~\cite{Pluecker17a}.

\subsection{Outline}
In summary, our aim is to set up a geometric framework for pumping
through strongly interacting open systems
 that
can deal with nonsystem observables,
that is more direct than the \fcs approach (when targeting only the first moment),
and that is a more suitable starting point for generalization to evolutions more complicated than \Eq{eq:kineq}.
The outline of the paper is as follows:

In  \Sec{sec:model}, we review how the kernels for the evolution of the state [\Eqs{eq:kineq} and \eq{eq:introAdsolution}] and for the observable expectation values [\Eq{eq:introExpecsolution}] can be derived.
We pay attention to issues related to inadvertent gauge fixing
 by the common procedure of normal-ordering expressions with respect to the reservoirs.
The key formula is the Heisenberg equation \eq{eq:WIX} for the current superoperator
\emph{after} the reservoirs have been integrated out.
At the end of this section, we formulate the guiding questions for the remainder of the paper.

In \Sec{sec:geometry}, we then show that in the pumping problem a gauge freedom emerges that is related to the physical calibration of the meter registering the transport of a nonsystem observable (reservoir charge, spin, heat, etc.).
The pumping problem precisely fits into the general geometric framework of Landsberg~\cite{Landsberg92} for driven dissipative systems with a continuous (gauge) symmetry.
The solution determines a geometric connection (gauge potential) on a simple fiber bundle of \emph{observables} over the manifold of driving parameters.
This connection is essentially the nonadiabatic current response and is closely related to a meter calibration.

In \Sec{sec:application}, we analyze the expression for the corresponding geometric curvature (gauge field), essentially the measurable pumped observable, and determine necessary conditions for a nonzero pumping effect.
We explain how under quite general circumstances pumping can be used to perform a \emph{geometric spectroscopy} of a weakly coupled open system.

Finally, in the extensive \Sec{sec:fcs} we compare the Landsberg-\ar approach in detail with the \fcs approach, when applied only to the first moment of the pumped \observable.
Despite the quite different formulation, we show that this approach is equivalent to the simpler and more direct Landsberg-\ar approach on all levels:
gauge freedom, connection (gauge potential), geometric pumping formula for the curvature (gauge field), as well as the limits of applicability.
Our formulation highlights the physical role of the meter
and allows us to further clarify the puzzling fact noted in~\Ref{Nakajima15}
that the ``adiabatic'' \fcs approach produces \emph{non}adiabatic contributions.

\section{Adiabatic-response approach
\newline
to pumping\label{sec:model}}

\subsection{Model, pumped \observables
\newline
and steady-state pumping\label{sec:model2}}

The adiabatic-response approach to pumping that we describe in this section
applies to very general open quantum systems.
We consider a quantum system with a discrete energy spectrum coupled to multiple noninteracting reservoirs indexed by $r$.
Whereas the reservoirs are assumed to be made up of either fermions or bosons, the system can be of mixed type as well.
We allow for possibly strong nonequilibrium conditions due to nonlinear biasing of the reservoirs' electrochemical potentials ($\mu^{r} \neq \mu^{r'}$).
Of central importance is that our findings also apply to a quantum system that is locally strongly interacting,
in contrast to several existing pumping approaches~\cite{Brouwer98,Buettiker94,Zhou03}.
For example, Coulomb interaction is crucial if one wants to describe driven transport through quantum-dot devices,
such as semi-conductor heterostructures~\cite{Vanevic16}, but also molecules~\cite{Napitu15} and single atoms~\cite{Jehl03}.
However, the approach applies equally well to bosonic models of pumping in chemical reactions between strongly interacting molecules~\cite{Sinitsyn07EPL}
and heat pumping using anharmonic~\cite{Ren10,Liu13} molecules.

The total system has the generic form of the Hamiltonian
\begin{align}
  H^\text{tot}(t) = H(t) + \sum_{r} H^r(t) + V(t)
  ,
  \label{eq:Htot}
\end{align}
with $H$ describing the system.
$H^\text{res}(t):=\sum_r H^r(t)$ accounts for the reservoirs including a driving term
$H^r(t)=H^r + \mathcal{V}^r(t)N^r$ for each reservoir $r$.
Finally, $V(t)=\sum_r V^r(t)$ is the coupling of the system to multiple reservoirs where
$V^r(t)$ describes the particle and energy exchange with reservoir $r$.
We denote the energy scale of the coupling by $\Gamma \propto V^2$, having in mind that for quantum-dot pumps this corresponds to the tunnel rate of particles.
In this case, $\Gamma^{-1}$ is the scale of the electron lifetime on the quantum dot.
To achieve pumping, we allow that all Hamiltonians in \Eq{eq:Htot} are driven time dependently through a set of parameters.
For example, for a quantum dot coupled to metallic electrodes,
 this means that aside from the reservoir electrochemical potentials and couplings,
any of the dot's parameters can be driven through applied voltages:
the single-particle energy levels, but also the two-particle interaction,\footnote{For experiments one should keep in mind that driving gate voltages defining a quantum dot changes the screening properties~\cite{Kaasbjerg08} and thus the effective interaction. This may well contribute to pumping and can be accounted for in our approach.} etc.

At the initial time where the driving and the coupling to the reservoirs are switched on
the initial equilibrium density operator of all reservoirs $r=\text{L},\text{R}, \ldots$, is:
\begin{subequations}\begin{align}
    \env{\rho}
    & := \prod_{r} e^{ - ( {H}^r  - \mu^r {N}^r )/T^r }
    \label{eq:rhoresa}
     \\
    & =  \prod_{r} e^{ - ( {H}^r(t) - \mu^r(t) {N}^r )/T^r }
    \label{eq:rhoresb}
		.
\end{align}\label{eq:rhores}
\end{subequations}
It is characterized by the constant temperatures $T^r$, the electrochemical\cite{Buettiker93,Pedersen98,Battista14a} potentials $\mu^r$
and the initial Hamiltonians $H^r$ without driving.
In the following we will use the form \eq{eq:rhoresb} in which we eliminated the undriven $H^r$ in favor of the driven Hamiltonian governing the dynamics, $H^r(t)$, by introducing a (canceling) time-dependence through driven electrochemical potentials $\mu^r(t)=\mu^r+\mathcal{V}^r(t)$.
The theory below can then be expressed entirely in terms of these \emph{parametrically} time-dependent quantities\footnote{The seemingly inconvenient cancellation of time-dependences in \Eq{eq:rhoresb} in the time-constant $\env{\rho}$ is actually advantageous.}.

We gather all driving parameters in one dimensionless vector $\vec{R}(t)$, i.e.,
each parameter is taken relative to a relevant scale, and all parametric dependences are denoted by ``$[\vec{R}]$''.
For example, in driven quantum dots, $\vec{R}(t)$ would include the applied voltages divided by temperature
[see the explicit example in \App{app:example}, \Eq{eq:Rexample}].
This ensures that $\dot{\vec{R}}$ has unit energy setting $\hbar=1$ [\Eq{eq:adcond}].
The parameters are cyclically driven in time at the frequency $\Omega$. We denote the period by $\T = 2\pi / \Omega$
and the traversed oriented closed curve in the parameter space by $\mathcal{C}$.

Other simplifying assumptions used in this work are that
the coupling is weak compared to temperatures,
i.e., 
\begin{align}
  \Gamma \ll T
  \label{eq:weak-coupling}
\end{align}
and that
 the driving velocity is slow on the scale of the system's inverse life-time, reading for $\vec{R}(t)=\bar{\vec{R}} + \delta \vec{ R} F(\Omega t)$
\begin{align}
  |\dot{\vec{R}}|
  \sim |\delta\vec{R}| \Omega
  \ll \Gamma 
  .
  \label{eq:adcond}
\end{align}
Note that this requires the product of amplitude $|\delta\vec{R}|$ and frequency $\Omega$ to remain small,
cf. \Eq{eq:condition-ar}.
Physically, this ensures that during one driving cycle many transport processes (due to the coupling $\Gamma \propto V^2$) occur, each process taking place for instantly frozen parameters to first approximation.

We are interested in the net change $\Delta X^r$ of a
physical \emph{reservoir} observable operator\footnote{We use a hat ($\hat{\, }$) only
when operators may be confused with their expectation values.} $\hat{X}^r$ after one driving period $\T=2\pi/\Omega$ in the \emph{time-dependent steady state}.
This state is established at any finite time as the time $t_0$ at which the system-reservoir coupling is switched on is sent to $t_0 \to -\infty$.
Aside from the slow-driving limit we always assume this steady-state limit, in which case
\begin{subequations}\begin{align}
  \Delta {X^r}
  & = \expec{\hat{X}^r}(\T)-  \expec{\hat{X}^r}(0)
    ,
 \label{eq:Xrpump1} \\
  & = \int_0^{\T}  dt \frac{d}{dt} \expec{ \hat{X}^r}  (t)
    .
  \label{eq:Xrpump2}
\end{align}\label{eq:Xrpump}\end{subequations}
Examples of such observables $\hat{X}^r$ are the charge, spin or energy of reservoir $r$.
We refer to $\Delta X^r$ as the net \emph{transported observable} per driving period to clearly distinguish it from the pumping contribution contained in it.
Note, that $\Delta {X^r}$ is \emph{not} the expectation value of an observable operator.
Instead, it is the result of a two-point measurement~\cite{Esposito09rev} at times $0$ and $\T$ in the steady-state limit $t_0 \to - \infty$.
In the \fcs approach discussed in \Sec{sec:fcs}  one calculates this quantity essentially using the first line \eq{eq:Xrpump1} via a moment generating function.
In the \ar approach, on which we focus instead, one calculates the second line \eq{eq:Xrpump2} by integrating the time-dependent \emph{current}-operator of the observable $\hat{X}^r$.

\subsection{Master equation \label{sec:masterEquation}}

In this section, we briefly review the derivation of the time-local \mE used to calculate the transported observables via the second equation \eq{eq:Xrpump2}.
Although much of this is standard, a number of important points related to the gauge freedom need to be highlighted.
Moreover, this prepares for a similar but less standard analysis for observables in \Sec{subsec:Currents}, in which a gauge freedom emerges.

In the simple limit of weak coupling and slow driving
we only need to consider the state evolution
 in the \emph{frozen parameter approximation}~\cite{Splettstoesser06}.
This amounts to calculating the evolution for fixed parameters $\vec{R}$ and in a second step inserting their instantaneous, time-dependent value $\vec{R}(t)$  [cf. \Eq{eq:MarkovMe}].
Thus, in the following
\begin{align}
  H^0 := H+H^\text{res}
\end{align}
as well as $V$ are all time-independent and the fixed parameter value $\vec{R}$ will not be written until it is needed again.
The master equation concerns the reduced density operator, the partial trace
\begin{align}
  \rho(t) := \tr_\text{res} \rho^\text{tot}(t)
  \label{eq:rhotdef}
  ,
\end{align}
of the density operator of system plus reservoir, as it evolves under the unitary time-evolution
\begin{subequations}\begin{align}
  \rho^{\text{tot}}(t) &= 
 U(t-t_0) \, \rho^{\text{tot}}(t_0) \, [U(t-t_0)]^\dagger
  ,
  \\
  U(t) &=  e^{-i (H^0+V)t} \label{eq:frozen}
  ,
\end{align}\label{eq:rhotot}\end{subequations}
starting from an initially factorizing state
\begin{align}
  \rho^\text{tot}(t_0)=\rho(t_0) \otimes \rho^\text{res}
  \label{eq:rhototfact}
	,
\end{align}
and letting $t_0 \to -\infty$ after taking the trace over the continuous reservoirs.
For the present purposes, an easy way of obtaining
the master equation for the reduced density operator $\rho(t)$ suffices.
We start from the Liouville equation for the density operator of the total system,
\begin{align}
  \frac{d}{dt} \rho^\text{tot}(t) =  -i [H^\text{tot},\rho^\text{tot}(t)]  \label{eq:VonNeumann}
  ,
\end{align}
which we integrate, then iterate once, and finally trace over the reservoirs.
Assuming that the coupling $V$ is partially normal ordered, i.e., $\Tr{\text{res}} V \rho^\text{res}=0$
(cf.  \App{app:kernels} and  \Sec{subsec:Currents}),
one obtains to leading order in the coupling $\Gamma$ [cf. \Eq{eq:AppModelME}]
\begin{align}
  \frac{d}{dt} \rho (t) =
  -i L \rho(t) +
  \int_{-\infty}^t d t^{\prime} W(t - t^{\prime}, \vec{R}) \rho(t^{\prime}) \label{eq:nonLocalMasterEquation}
\end{align}
where $L \bullet := [H, \bullet ]$ is the system Liouvillian superoperator
and the kernel $W(t,t')$ is the superoperator [cf. \Eq{eq:AppKernelFormula}]
\begin{align}
  & W(t - t^{\prime}, \vec{R}) \bullet =  \notag\\
  & -\Tr{\text{res}}
    \left[ V,  e^{-i \ H^0(t-t')}
    \left[ V, \rho^\text{res} \, \bullet
     \right] e^{ i H^0(t-t')}
    \right]
    \label{eq:DynamicsKernel}
    .
\end{align}
Here and below $\bullet$ denotes an arbitrary system operator
appearing as an argument of a superoperator.

Consistent with the weak coupling ($\Gamma$)
relative to the reservoir thermal fluctuations ($T$)
and the slow driving ($|\delta \vec{R}| \Omega\ll \Gamma$)
one should~\cite{Splettstoesser06} neglect the memory effects
by setting $\rho(t') \to \rho(t)$
in \Eq{eq:nonLocalMasterEquation}.
From 
$|\delta \vec{R}| \Omega
, \, 
                          \Gamma \ll T$
we thus obtain the Born-Markov master equation
\begin{align}
  \frac{d}{dt} \rho(t) = W[\vec{R}(t)] \rho(t)
  \label{eq:MarkovMe}
  ,
\end{align}
where we now again explicitly write the frozen parameter dependence.
Here we have conveniently defined the effective kernel $W$ as the sum of
the system Liouvillian
and
the zero-frequency Laplace transform of the kernel \eq{eq:DynamicsKernel} for fixed parameters $\vec{R}$,
\begin{align}
  W[\vec{R}]: = -i L[\vec{R}] + \lim_{z\rightarrow 0^+} \int_{0}^{\infty} dt e^{- z t} W(t; \vec{R})
  \label{eq:Wdef}
  .
\end{align}
In both terms the parameters are subsequently replaced by their time-dependent values, $\vec{R} \to \vec{R}(t)$.

We stress that the calculation of the required kernel raises no practical problems since it is based on the weak-coupling, high-temperature limit $\Gamma \ll T$ (see \App{app:example}).
It nevertheless accounts nonperturbatively for effects of \emph{strong interactions} on the system which enter the kernel through $H$ in \Eq{eq:DynamicsKernel}.
When going beyond this limit, the time-nonlocality of the kernel becomes important as discussed after \Eq{eq:nonLocalMasterEquation}.
However, this can be addressed transparently\footnote{ The “Wangsness-Bloch” approach used here to obtain \Eqs{eq:nonLocalMasterEquation} and \eq{eq:DynamicsKernel} runs into problems when going beyond the weak coupling approximation, see~\Ref{Koller10} for a discussion. The real-time approach allows for a systematic derivation of corrections \cite{Splettstoesser06} to \Eq{eq:nonLocalMasterEquation} and \eq{eq:DynamicsKernel} including higher-order coupling effects as well as non-Markovian effects. As a result of these corrections to the frozen-parameter approximation, the kernel's time dependence will in general not be mediated solely by the parameters as in \Eq{eq:MarkovMe}.}
by systematically extending the adiabatic expansion in the driving ($\Omega$) with the perturbative expansion in the coupling $\Gamma$ as established in \Ref{Splettstoesser06}.

\subsection{Observable and current kernels\label{subsec:Currents}}

We next review how in an analogous way the expectation value \Eq{eq:Xrpump} of a \emph{nonsystem observable}, i.e., also acting on the reservoir, can be obtained using the \emph{system} density operator, the solution of \Eq{eq:MarkovMe}.
In general, for a given density operator $\rho(t)$ the expectation value of a
system observable $\hat{X}(t) := \hat{X}[\vec{R}(t)]$ can be obtained from $\expec{\hat{X}}(t)=\tr \hat{X}(t) \rho(t)$
where $\tr$ is the trace over the system only.
However, this fails for nonsystem \observables
$\hat{X}^r(t) := \hat{X}^r[\vec{R}(t)]$
that have our interest here.
For this an additional piece of information, an \emph{observable kernel} or a related \emph{\current kernel}, is required.
Even though we are interested only in pumping of time-\emph{in}dependent observables,
it will be crucial to allow for parametric time-dependence of such observables
throughout the analysis and specialize only at the end, setting
$ \hat{X}^r[\vec{R}(t)] \to \hat{X}^r$.

\emph{Observable kernels and partial normal ordering.}
Analogous to the state evolution,
the expectation value of a nonsystem observable~\cite{Koenig96b,Schoeller09b,Saptsov12a} can be expressed as
[see \Eq{app:CurrentKernel}]
\begin{align}
  \expec{\hat{X}^r}(t) =
  \tr \int_{-\infty}^t dt' W_{\no{ \hat{X}^r} }(t,t') \rho(t')
  +
  \tr  \expec{\hat{X}^r}^\text{res}  \rho(t)
 .
  \label{eq:ObservableExpec}
\end{align}
Below it will be important that $\hat{X}^r(t)$ is allowed to be a hybrid system plus reservoir ($r$) operator.

We first discuss the second term, involving the partial average over the initial reservoir state
\begin{align}
  \expec{\hat{X}^r(t)}^\text{res}
   := \Tr{\text{res}} \hat{X}^r(t) \rho^\text{res}
  \label{eq:NotNormalOrderedObs} 
  .
\end{align}
Since we do not perform the system trace ($\tr$),
the resulting expression \eq{eq:NotNormalOrderedObs} is still an operator on the system Hilbert space.
Often, this second contribution to \Eq{eq:ObservableExpec} is not considered since
either by choice of observable or model  the partial trace \eq{eq:NotNormalOrderedObs} vanishes.
Such operators for which $\expec{\hat{X}^r(t)}^\text{res}=0$ we call \emph{partially}\footnote{``Partial'' distinguishes it from the usual operation of normal ordering that ensures that \emph{any single} Wick contraction of an operator expression is zero instead of just the average.}
\emph{normal ordered} with respect to the reservoirs.
The consideration of more general observables that are \emph{not} partially normal-ordered is important for the gauge freedom that underlies pumping.
Such a general observable can be split uniquely into two parts
\begin{align}
  \hat{X}^r(t) &= 
  \no{X^r(t)}
+
  \expec{\hat{X}^r(t)}^\text{res}  \unit^\text{res}
  \label{eq:nosplit}
  ,
\end{align}
thereby defining $ \no{X^r(t)}$.
The second, partially-averaged part of \Eq{eq:nosplit}
generates the second term in \Eq{eq:ObservableExpec}.

The first term of \Eq{eq:ObservableExpec} comes from
the first partially normal-ordered term in \Eq{eq:nosplit}.
To leading order in $\Gamma$,
a convenient explicit form of $W_{\no{\hat{X}^r}}(t,t')$ can be obtained formally from $W(t,t')$
by replacing in \Eq{eq:DynamicsKernel}
the leftmost $V \to i \tfrac{1}{2} {\hat{X}^r(t)}$ and the outer commutator by an anticommutator:
\begin{align}
  &
  W_{\no{\hat{X}^r}}(t,t')
  \bullet  = \label{eq:ObservableKernel}\\
  &- i \Tr{\text{res}} \tfrac{1}{2}
  \left [  \hat{X}^r(t)  , e^{-i H^0(t-t')}
  \left [  V, {\rho}^\text{{res}} \, \bullet
  \right ]    e^{+iH^0(t-t')}
  \right ]_+
  \notag
  .
\end{align}
This expression allows for a physically irrelevant redundancy (not to be confused with the gauge freedom) as one is free to add any term to it that vanishes under the trace [cf. discussion after \Eq{eq:redundancy-ar}].

We stress the importance of the decomposition \eq{eq:ObservableExpec}:
working with a partially normal-ordered observable, i.e., dropping the second term, removes the part of the observable in which the physical gauge freedom lies.
Such premature fixing of the gauge freedom is very common, motivated by valid practical reasons,  but obscures the simple geometric nature of the pumping from the very beginning.

\emph{Current kernels and Heisenberg equation.}
In the \ar approach, one follows the route \eq{eq:Xrpump2} and works with an observable \emph{\current} kernel to obtain the pumped nonsystem \observable.
The advantage is that the current becomes stationary for frozen parameters, in contrast to the observable $\hat{X}^r$ itself. 
As a result, for the slow parameter driving the current also evolves slowly in the steady-state limit, allowing for a Born-Markov adiabatic-response approximation very similar to the one made for the state evolution.

To this end, let $\hat{X}^r$ now be a reservoir-\emph{only} \observable. Its \current into reservoir $r$ reads as
\begin{align}
  I_{X^{r}}
  = \frac{d}{dt} \expec{\hat{X}^r}
  =\frac{d}{dt} \, \Tr{\text{tot}}{\hat{X}^r(t) \tot{\rho}(t)}
  .
  \label{eq:ExpectationCurrent}
\end{align}
The corresponding \current operator,
producing the time-derivative of the expectation value
\begin{align}
  I_{X^{r}} = \Big< {\widehat{\frac{d X^r}{d t}}} \Big>
 = \expec{\hat{I}_{X^r}} 
  ,
\end{align}
is given by the Heisenberg equation of motion
\begin{align}
  \hat{I}_{X^r}:= \widehat{\frac{d X^r}{d t}}
  = i [H^{\text{tot}},\hat{X}^{r}] + \frac{\partial \hat{X}^{r}}{\partial t} \label{eq:FluxOperator}
  .
\end{align}
This current is a ``hybrid'' nonsystem operator, i.e., acting on both system and reservoir.
Therefore, to integrate out the reservoirs by applying \Eq{eq:ObservableExpec} we need to decompose it according to \Eq{eq:nosplit} into two contributions.
First, for the partial average we obtain
\begin{align}
  \expec{\hat{I}_{X^r}}^\text{res}
    =
  \Big< \frac{\partial \hat{X}^{r}(t)}{\partial t} \Big>^\text{res}
  \label{eq:FluxOperatorav}
  .
\end{align}
Here we have assumed that
the nonsystem observable $\hat{X}^r$ is conserved inside each reservoir $r$
and conserves its particle number
for each value of the driving parameters, i.e.,
\begin{align}
  [ \hat{H}^r, \hat{X}^r ] = 0 ,
  \quad
  [ \hat{N}^r, \hat{X}^r ] = 0
  .
 \label{eq:HrXrcom}
\end{align}
This means that $\hat{I}_{X^r}$ is the operator for the net $\hat{X}^r$-current flowing out of the reservoir.
This is appropriate when the distribution of currents inside the reservoir is of no interest.
A consequence of \Eq{eq:HrXrcom} is that $[\hat{X}^r,\rho^\text{res}]=0$ [\Eq{eq:rhores}] and thus
\begin{align}
  \expec{ [V^r,\hat{X}^{r} ] }^\text{res}
= \tr_\text{res} [V^r,\hat{X}^{r}] \rho^\text{res} = 0
\end{align}
which we also used in writing \Eq{eq:FluxOperatorav}.
We stress that $\hat{X^r}$ is \emph{not} assumed to be conserved by the coupling $V$. This will be discussed separately, see \Eq{eq:Xconservation}.

Second, for the partially normal-ordered contribution of the current we obtain
\begin{align}
  \no{\hat{I}_{X^r}}
    = \no{ i [H^{\text{tot}},\hat{X}^{r}]}
  \label{eq:FluxOperatorno}
  .
\end{align}
Here we have assumed that the explicit time derivative of the observable has no partially normal-ordered part,
\begin{align}
  \no{
  \frac{\partial \hat{X}^{r}(t)}{\partial t}
  } = 0
  .
 \label{eq:Xrno}
\end{align}
To keep track of the gauge freedom of pumping it is sufficient to keep track of the limited class of observables whose time-dependent operators satisfy \Eq{eq:FluxOperatorav} 
(cf. \Sec{sec:geometry}).

Applying \Eq{eq:ObservableExpec} for $\hat{X}^r \to \hat{I}_{X^r}$ and using 
\Eq{eq:ObservableKernel}, \eq{eq:FluxOperatorav}, and \eq{eq:FluxOperatorno},
we obtain
\begin{align}
  \expec{\hat{I}_{X^r}}(t) =
  \tr \int_{-\infty}^t dt' W_{\no{ \hat{I}_{X^r}} }(t,t') \rho(t')
  +
  \tr  \expec{\hat{I}_{X^r}}^\text{res}  \rho(t)
 .
  \label{eq:CurrentObservableExpec}
\end{align}
We have thus traced out the reservoirs in the Heisenberg equation of motion.
We can now apply the Markov approximation to the first term in this equation in the same way as for the \mE \eq{eq:MarkovMe}, since the frozen-parameter \current becomes stationary.
We stress that the time dependence in the second term that we keep through \Eq{eq:FluxOperatorav}
can be arbitrary.\footnote{This implies that the gauge transformations $X^r \to X^r + g(t) \unit$ with \emph{arbitrary} time-dependent functions $g(t$), introduced in \Sec{sec:geometry}, do not break the validity of the Markov approximations.}
We then obtain the key formula for the current:
\begin{align}
	I_{X^{r}}(t) =  \tr \,  \currentKernel{X^r}[\vec{R}(t)] \, \rho(t) , \label{eq:MarkowFlux}
\end{align}
where we have defined 
 the effective \current kernel
\begin{align}
  \currentKernel{X^r}[\vec{R}]
   :=
  &
  W_{ \no{ i [V^r,\hat{X}^{r}] } } [\vec{R}]
   +
  \Big< \frac{\partial \hat{X}^{r}[\vec{R}]}{\partial t} \Big>^\text{res}
  .
\label{eq:WIX}
\end{align}
Equation \eq{eq:WIX} is of central importance:
it is the open-system equivalent of the Heisenberg equation \eq{eq:FluxOperator} for \emph{time-dependent nonsystem} observables
that obey \Eqs{eq:HrXrcom} and \eq{eq:Xrno}.
Here, the first term is the zero frequency Laplace transform of $W_{\no{\hat{X}^r}}(t,t')$
with $\no{\hat{X}^r} \to \no{ i [H^{\text{tot}},\hat{X}^{r}] }$
given explicitly by \Eq{eq:ObservableKernel}.
As mentioned before, often the last term in \Eq{eq:WIX} is not considered
because one assumes from the start that the observable is time independent.
This amounts to a premature fixing of the gauge similar to assuming partial normal ordering [see \Eq{eq:NotNormalOrderedObs} ff].

\subsection{Pumped observables -- ``Naive calculation''\label{sec:naive}}

With the master equation \eq{eq:MarkovMe} and the \current formula \eq{eq:MarkowFlux} carefully established,
it is now easy to calculate the transported \observable $\Delta X^r$ in adiabatic-response to the driving following the route via \Eq{eq:Xrpump2}.
We now discuss how this was done so far~\cite{\ARNAIVE}
and then formulate in \Sec{sec:questions} the questions that this calculation leaves open.

For slow driving
the density operator $\rho(t)$ can be expanded in powers of
the small driving velocity $| \dot{\vec{R}} | = |\delta \vec{R}| \Omega \ll \Gamma$ [\Eq{eq:adcond}]
\begin{align}
  \rho(t) \approx \inst{\rho}(t) + \adcor{\rho}(t) .
  \label{eq:adiabatic-response}
\end{align}
Here the first \emph{instantaneous} term is of order $O(1)$ and the second 
term is the \emph{adiabatic-response} $O(|\delta \vec{R}| {\Omega}/{\Gamma})$ accounting for the ``lag''.\footnote{Note the difference between ``lag'' (Markovian, non-adiabatic) that we keep and ``memory'' (non-Markovian) that we neglect:
Since thermal fluctuations are much faster than both coupling and driving,
$T \gg \Gamma, |\delta \vec{R} | \Omega$,
we can neglect the ``memory'' in the kernel.
This results in Markovian dynamics of $\rho(t)$ [\Eq{eq:MarkovMe}]
on time scale $\Gamma^{-1}$.
For driving velocities slower than this,
i.e., $|\delta \vec{R} | \Omega \ll \Gamma$, 
the solution $\rho(t)$ of \Eq{eq:MarkovMe} develops a small ``lag'' responsible for pumping that we do take into account.
}
Inserting this into \Eq{eq:MarkovMe} and collecting orders of $|\delta \vec{R}| \Omega/\Gamma$ one finds:
\begin{subequations} \label{eq:bothDensityOp} \begin{align}
  0 &= W(\vec{R}(t)) \inst{\rho}(t),
      \label{eq:instDensityOp} \\
  \frac{d}{dt} \inst{\rho}(t) &= W(\vec{R}(t)) \adcor{\rho}(t)
  .
  \label{eq:adcorDensityOp}
\end{align}\end{subequations}
These simple steps are equivalent to the asymptotic analysis / time-scale separation found in other works~\cite{Avron12a,Avron12,Landsberg92}.

Equation \eq{eq:instDensityOp} defines the instantaneous stationary state
$\inst{\rho}(t)=\inst{\rho}(\vec{R}(t))$,
i.e., the stationary state that would be reached if the parameters were frozen.
Throughout the paper we assume that this state is unique,
as is the case in many practical pumping problems,
see discussion in \Sec{sec:conclusion}.
This moreover helps to keep our discussion of the geometric phase effect accumulated by the \emph{observable} clearly separate from geometric phase effects related to quantum states (see \Sec{sec:geometry}).
Finally, most of the approaches we compare with rely on this assumption, see however \Ref{Avron12a}.

In contrast, \Eq{eq:adcorDensityOp} determines the {adiabatic response}, i.e., the first-order correction to the instantaneous evolution,
which depends on both the parameters $\vec{R}$ \emph{and} their velocities $\dot{\vec{R}}$ through $\frac{d}{dt} \inst{\rho}(\vec{R}(t))$. 
It can be expressed as~\cite{Calvo12a}
\begin{align}
  \adcor{\rho}(t) =
  \frac{1}{W(\vec{R})} \delR  \inst{\rho}(\vec{R}) \frac{d \vec{R}}{dt}(t)
  \label{eq:rhoasol}
	,
\end{align}
with the pseudo inverse $1/{W(\vec{R})}$, i.e., restricted to the subspace of nonzero eigenvalues of $W(\vec{R})$.

We now compute the observable as in most cited \ar works~\cite{\ARNAIVE}
 by assuming that $\hat{X}^r$ has no parametric time dependence to begin with:
setting $\partial \hat{X}^r/\partial t = 0$ in \Eq{eq:WIX},
\begin{align}
  \currentKernel{X^r}(\vec{R})
  = W_{ \no{ i [V^r,\hat{X}^{r}] }}(\vec{R})
  \label{eq:WIXnaive}
\end{align}
and inserting the expansion \eq{eq:adiabatic-response} into \Eq{eq:MarkowFlux}
we obtain an instantaneous part (``sum of snapshots'')
\begin{align}
  \Delta \inst{X}^{r} &
= \int_0^{{\T}} dt \tr \, \currentKernel{X^{r}}(\vec{R}(t)) \inst{\rho}(\vec{R}(t))
 \label{eq:Obs}
  ,
\end{align}
and an adiabatic-response part
\begin{align}
  \Delta \adcor{X}^{r} & =
 \int_0^{{\T}} dt \tr \, \currentKernel{X^{r}}(\vec{R}(t)) \adcor{\rho}\Big( \vec{R}(t),\dot{\vec{R}}(t) \Big)
 \label{eq:adCorrObs}
 .
\end{align}
The pumping current under the integral in \Eq{eq:adCorrObs} is clearly non-adiabatic, i.e., 
the system is ``lagging behind'', since \Eq{eq:rhoasol} $\propto \dot{\vec{R}}$.
Therefore, the \emph{pumped \observable} (cf. \Sec{sec:intro}) is geometric in the elementary sense that
it can be expressed as a line integral over the traversed driving parameter curve $\mathcal{C}$:
\begin{align}
  \Delta \adcor{X}^{r} =\oint_{\mathcal{C}} d\vec{R} \tr \currentKernel{X^{r}}(\vec{R}) \frac{1}{W(\vec{R})} \delR  \inst{\rho}(\vec{R})
  \label{eq:pumpedX}
	.
\end{align}

\emph{Scaling with parameters.}
The instantaneous part \eq{eq:Obs} and adiabatic-response part \eq{eq:adCorrObs}-\eq{eq:pumpedX} differ in their scaling with parameters, allowing them to be separately extracted from measurements,
both in principle and in practice. 
Since a physical meter that registers the (pumped) observable will be our key principle for understanding the geometric nature of pumping
we now discuss its scaling summarized as:
\begin{align}
  \Delta \inst{X}^{r}    \sim \frac{\Gamma}{\Omega}
   \qquad
  \Delta \adcor{X}^{r}  \sim 
                                             |\delta \vec{R}|^2 \text{ as $\delta \vec{R} \rightarrow 0$}
																						\label{eq:scaling}
																						.
\end{align}

First, the
 pumped observable $\Delta \adcor{X}^{r}$ does not depend on the parametrization of the driving cycle $\mathcal{C}$ and therefore is independent of the driving frequency $\Omega$.
However, its sign is reversed when inverting the orientation of the driving cycle $\mathcal{C}$.
In contrast to this, 
the instantaneous contribution $\Delta \inst{X}^{r} = \int_0^{{\T}} dt \inst{I_{X^{r}}} (\vec{R}(t))$ diverges $\propto \T \propto \Omega^{-1}$ at zero driving frequency
 because the instantaneous \emph{current} is frequency independent  (``infinite sum of snapshots'').

A second difference is that $\Delta \inst{X}^{r} \propto \Gamma$ since the currents scale up linearly with the strength $\Gamma$ of the coupling of the system to its environment.\footnote{In \Eq{eq:Obs} $\currentKernel{X^r} \propto V^2 \propto \Gamma$
by \Eq{eq:ObservableKernel} and \eq{eq:WIXnaive}.}
This effect is also present in \Eq{eq:adCorrObs}-\eq{eq:pumpedX} but there it is compensated by the downscaling of all relaxation times ($W^{-1}\propto \Gamma^{-1}$).
This makes the adiabatic-response pumping $\Delta \adcor{X}^{r}$  independent
 of the overall coupling scale $\Gamma$.
Physically speaking, for a more strongly coupled system the currents are larger
but the ``lag'' time is correspondingly shorter,
giving the same net pumping effect.
This difference between $\Delta \inst{X}^{r}$
and $\Delta \adcor{X}^r$ holds even when this scale is \emph{altered in time} and can be utilized experimentally.
In \App{app:pumping} we discuss how to use this scaling to extract the pumping contribution from measurements.

Finally, for fixed $\Omega$ but vanishing amplitude of driving $|\delta \vec{R}|$ around a working point $\bar{\vec{R}}$ 
the instantaneous part will saturate at a value
$\Delta \inst{X}^{r} \to \inst{I_{X^r}} (\bar{\vec{R}}) \T $
set by the stationary current which can be nonzero depending on  the parameter set $\bar{\vec{R}}$.
In contrast, the pumped observable always vanishes\footnote{\Sec{sec:application} shows that via Stoke's theorem the pumped charge can be expressed as an area integral which for small driving cycles scales as $|\delta \vec{R}|^2$.} $\Delta \adcor{X}^{r} \propto |\delta \vec{R}|^2$
as $|\delta \vec{R}| \to 0$,
see \Sec{sec:application}.

\emph{Limits of applicability.}
There are two restrictions that limit the applicability of the \ar approach (cf. also \Sec{sec:fcsAdiabatic}).
First, to be consistent, the sum of the instantaneous plus adiabatic-response correction to the \emph{state} must remain small relative to the neglected higher corrections, denoted by $\rho^{\text{rest}}$:
\begin{align}
   || \rho^{\text{rest}} ||   \ll    || \inst{\rho} + \adcor{\rho} || 
  ,
  \label{eq:condition-ar0}
\end{align}
where $||\bullet||=\Lbraket{\bullet|\bullet}$ denotes the operator norm.
As discussed in \App{app:ase} and \app{app:iteration}
this requires that for  all accessed values of the dimensionless driving parameters the velocity is sufficiently small compared to the  open-system's relaxation rates
\begin{align}
  | \dot{\vec{R}} | = \Omega \cdot |\delta \vec{R}|
  \ll  \Gamma(\vec{R})
  \label{eq:condition-ar}
  .
\end{align}
Here $\Gamma$ sets the magnitude of the nonzero eigenvalues of $W$ in \Eq{eq:bothDensityOp}.
Thus when driving with large dimensionless amplitude $|\delta \vec{R}|$ the restriction on the driving frequency $\Omega$ becomes more stringent\footnote{Often the quoted condition $\Omega \ll \Gamma$ for pumping implicitly assumes $|\delta \vec{R}| \sim 1$.}.
Also note that driving the coupling amplitude $\Gamma(\vec{R})$ 
plays a special role,
as compared to the other parameters:
the coupling amplitudes is additionally limited by \Eq{eq:weak-coupling}.
Using \Eq{eq:rhoasol} this implies that 
\begin{align}
  || \adcor{\rho} ||
  \sim 
\frac{\Omega \cdot |\delta \vec{R}|}{\Gamma(\vec{R})} 
  \ll
  1 \sim
|| \inst{\rho} || 
  .
  \label{eq:condition-ar-implied}
\end{align}

A second consistency condition is that the neglected higher nonadiabatic contribution,
$\Delta X^{r, \text{rest}}$,
to the transported observable is small relative to the first two contributions that are kept,
$\Delta \inst{X}^r$ and $\Delta \adcor{X}^r$ [\Eq{eq:Obs} and \eq{eq:adCorrObs}]:
\begin{align}
 |  \Delta X^{r, \text{rest}} |  \ll  |  \Delta \inst{X}^r +\Delta \adcor{X}^r |  
  .
 \label{eq:condition-ar2}
\end{align}
This was found to be of particular importance for pumping of energy and heat\cite{Juergens13}.
Although this is often not discussed, it may in fact impose tighter limits on the driving frequency than expected just from the first condition \eq{eq:condition-ar} for the expansion of the state.

Therefore we now briefly outline  how
 the expansion for the \emph{current} of some observable $\hat{X}^r$ may break down 
even if the expansion for the state $\rho(t)$ is good.
One can pictorially understand what may go wrong by considering operators $x$ as either vectors in Liouville space,
$\Lket{x}= x$ or covectors $\Lbra{x}=\tr x^\dag$.
The currents
$I^{k}_{X^r} = \tr \, \currentKernel{X^{r}} \rho^k 
=\Lbraket{V|\rho^k}$
for  $k=\{ \text{i}, \text{a},\text{rest} \}$
are Hilbert-Schmidt scalar products
of $\Lket{V}$ and
$\Lket{\rho^{k}}$,
i.e., the component of the latter along $\Lket{V}$.

One should now worry that if one chooses an arbitrary observable, i.e., the vector $\Lket{V}$, then its orientation may be such that 
the projection of the shorter $\Lket{\rho^\text{rest}}$  onto $\Lket{V}$ is larger than that of the longer
$\Lket{\inst{\rho}}+\Lket{\adcor{\rho}}$.
However, since
these two parts scale different with frequency $\Omega$ the importance of $I_{X^r}^{\text{rest}}$ relative to $\inst{I_{X^r}} + \adcor{I_{X^r}}$  can still be decreased by lowering the frequency and / or amplitude even further than required by condition \eq{eq:condition-ar}.

\subsection{Why is pumping geometric?\label{sec:questions}}

With \Eq{eq:pumpedX} the pumping problem is solved in great generality
under the assumptions stated in \Sec{sec:model2}.
This approach was formulated in \Ref{Splettstoesser06} and subsequently
 analyzed in detail in \Refs{Splettstoesser08a,Winkler09,Reckermann10a,Calvo12a,Haupt13,Riwar13,Winkler13,Rojek14}
and systematic higher-order corrections, beyond the Born-Markov approximation, were computed in \Refs{Splettstoesser06,Splettstoesser10}.

However, one should wonder about the geometric nature of the reported pumping effects in a more precise sense,
i.e., beyond ``the final answer can be written as a curve integral''.
It is clear from this that you can add a differential without changing the answer for mathematical reasons.
What this corresponds to physically is unclear.
Is the pumping effect, like so many other physical problems~\cite{\CONTROLMECHANICS}, related to some underlying gauge structure of the problem that is already \emph{physically} evident \emph{before} solving it?
If there is no gauge freedom, then a geometric effect can never arise.
Can the \ar-pumping problem be formulated in a manifestly gauge-covariant way?
We will show that
fully answering these questions will lead to a better physical understanding of
why and how pumping effects can appear at all.
This is not obvious in the \ar approach even though the calculations are simple.
Also, in more difficult situations involving strong coupling and memory effects\cite{Pluecker17a}
knowing about gauge structure in advance is helpful.

That there must be such a gauge structure in the \ar approach to pumping
was mentioned already in \Ref{Sinitsyn09} (p. 8) in relation to earlier works by Landsberg\cite{Landsberg92,Landsberg93}.
It was recently demonstrated\cite{Nakajima15}
that the, geometric, \fcs result coincides in general with the explicit \ar result \eq{eq:pumpedX}.
However, it is quite unsatisfactory that
the gauge structure must be inferred via the more complicated \fcs approach
instead of directly via the remarkably simple \ar-derivation:
above we found that one cannot really verify that the result \eq{eq:pumpedX} is gauge invariant, a crucial test for any geometric effect,
since it was obtained by (silently) fixing a gauge.
Therefore, in the remainder of the paper we address the following questions:

(i) \emph{What is the gauge freedom ``intrinsic'' to the \ar approach?}
In other words: through which physical quantity does a geometric phase enter the \ar pumping analysis?
From closed quantum systems~\cite{Thouless83,Cohen03}
 one might expect that the geometric phase of pumping resides in some freedom of the quantum state.
However, the open-system analog\cite{Sarandy05,Sarandy06} of the Berry-Simon geometric factor in the steady-state exhibits no change between start and end of the evolution.
A direct geometric origin of the pumping effect
is thus \emph{not} related to this Berry-Simon type geometric phase,
and has to be sought in the observable:
What then constitutes the physical gauge freedom for pumping of \emph{nonsystem observables}?
This remains unclear despite
 the elegant geometric formulations of the \ar approach to pumping of \emph{system} observables~\cite{Avron12,Avron12a,Avron11}.

(ii)  \emph{How does pumping generate a geometric effect?}
Given that the observable, instead of the quantum state, exhibits a geometric phase,
how is a geometric connection and curvature determined by the physics of \emph{pumping} leading up to \Eq{eq:pumpedX}?
The appearance of a geometric phase in such \ar-type calculations
is closely related to
Landsberg's~\cite{Landsberg92,Landsberg93} discussion of classical dissipative systems exhibiting a symmetry\footnote{See \cite{Andersson03thesis,Andersson05} for a detailed exposition and generalization to the non-Abelian case
and the review \Ref{Sinitsyn09} for related references.}.

(iii)  \emph{When is pumping nonzero?}
Under which conditions does a nonzero pumped observable, quantified by the geometric curvature,
actually arise?
In \Sec{sec:application} we exploit the simplicity of the geometric Landsberg-\ar pumping formula
[\Eq{eq:pumpedX}, \eq{eq:pumpedXlandsberg}]
to specify quite generally such necessary conditions,
and discuss simplifications that can be made when the pumped observable is conserved.

(iv) \emph{How are the \ar and \fcs geometric-pumping approaches related?}
Our key point is that the above questions can be answered entirely within the  simple \ar formulation:
the geometric nature of pumping does not require an \fcs formulation of the problem.
However, we believe that a detailed comparison with the established \fcs approach to pumping
is still warranted since it addresses important questions about this approach.
The large remainder of the paper, \Sec{sec:fcs},
is dedicated to this but can be skipped by readers mainly interested in the \ar approach put forward in this paper.
\section{Gauge freedom and \newline
geometry of pumping\label{sec:geometry}}
In this section, we will address questions (i) and (ii) regarding the geometric nature of pumping within the \ar approach.
The key idea is that
the gauge freedom responsible for pumping has the literal meaning of ``calibration'' of the \emph{meter} registering the measured value of the observable.
The differential-geometric notion of ``parallel transport,'' determining the connection and geometric phase in a relevant fiber bundle, corresponds physically to keeping the scale on the meter aligned with the needle during the pumping cycle.

\subsection{No gauge freedom in the quantum state\label{sec:LackOfPhaseState}}

To set the stage for answering question (i), we point out that pumping is not related to a Berry phase of the \emph{state}:
the parametrically driven time-dependent steady-state density operator $\rho(t) =\inst{\rho}(t)+\adcor{\rho}(t)$ [cf. \Eq{eq:adiabatic-response}] is continuous over a driving cycle within the mentioned approximations:
\begin{align}
  \rho(0) =\rho(\T)
  \label{eq:rhocont}
  .
\end{align}
Thus, a closed parameter curve produces a closed steady-state curve, \emph{without} any discontinuity.

For the instantaneous part $\inst{\rho}(t)$, one may derive the result \eq{eq:rhocont} using the \ase approach of Sarandy and Lidar~\cite{Sarandy05,Sarandy06}, mentioned in the Introduction.
At first, the continuity \eq{eq:rhocont} may seem at odds with their results in \Ref{Sarandy06},
where quite generally a Berry-Simon type geometric-phase discontinuity is predicted for the mixed quantum state $\inst{\rho}(t)$.
The crucial point is to consider the steady-state limit of their result,
which was not explicitly analyzed in \Ref{Sarandy06}.
In \App{app:ase}, we show that indeed their Berry-Simon-type phase vanishes in this limit, assuming only, as we do here, probability normalization and that a unique stationary state exists for frozen parameters [see \Eq{eq:instDensityOp} ff.].

To establish \eq{eq:rhocont} it remains to be shown that when including the adiabatic-response part, $\rho(t) \approx \inst{\rho}(t) + \adcor{\rho}(t)$,
the state is still continuous.
For this one can take the steady-state limit of the result reported in \Ref{Sarandy05} for the adiabatic-response correction $\adcor{\rho}(t)$
[\Eq{eq:adiabatic-correction}] to $\inst{\rho}(t)$.
This coincides with the result \eq{eq:rhoasol} of the \ar approach as we verify in \App{app:ase}.
The result is that $\rho(t)$ is continuous, again by trace normalization.

There is an elegant way of seeing that this continuity actually corresponds to  the vanishing of another geometric phase,
one that is associated with the \emph{nonadiabatic} part $\adcor{\rho}$.
This relies on a generalization of Berry's ``adiabatic iteration''~\cite{Berry87} to open quantum systems with a stationary state.
This we set up in \App{app:iteration}
where we again find that \Eq{eq:rhocont} is enforced by probability normalization,
not only when including the adiabatic-response $\adcor{\rho}$
but even when adding all higher nonadiabatic corrections.
Thus, the time-dependent steady-state exhibits no
discontinuity in \emph{any} order of the driving frequency when starting from the Born-Markov equation~\eq{eq:MarkovMe}.

Inquiring into question (i),
we must therefore conclude that within the reduced density operator approach steady-state pumping is associated with a geometric phase of an entirely different kind, unrelated to the quantum state.
In fact, as we will see in \Sec{sec:landsberg}, the quenching of the Berry-Simon type geometric phase of the quantum state allows the Landsberg geometric phase in the \emph{observable} to emerge.

\subsection{Gauge-freedom in pumped \observable
\label{sec:GaugeFreedomObservable}}

We now answer question (i) regarding the physical gauge freedom that underlies the geometric nature of pumping.
The key idea is that the \current is not uniquely defined \emph{in a pumping process}.
Nonsystem observables in such cyclic processes exhibit a gauge freedom that is not present in general for nonperiodic driving.

\begin{figure}[t]  
  \includegraphics[width=0.9\linewidth]{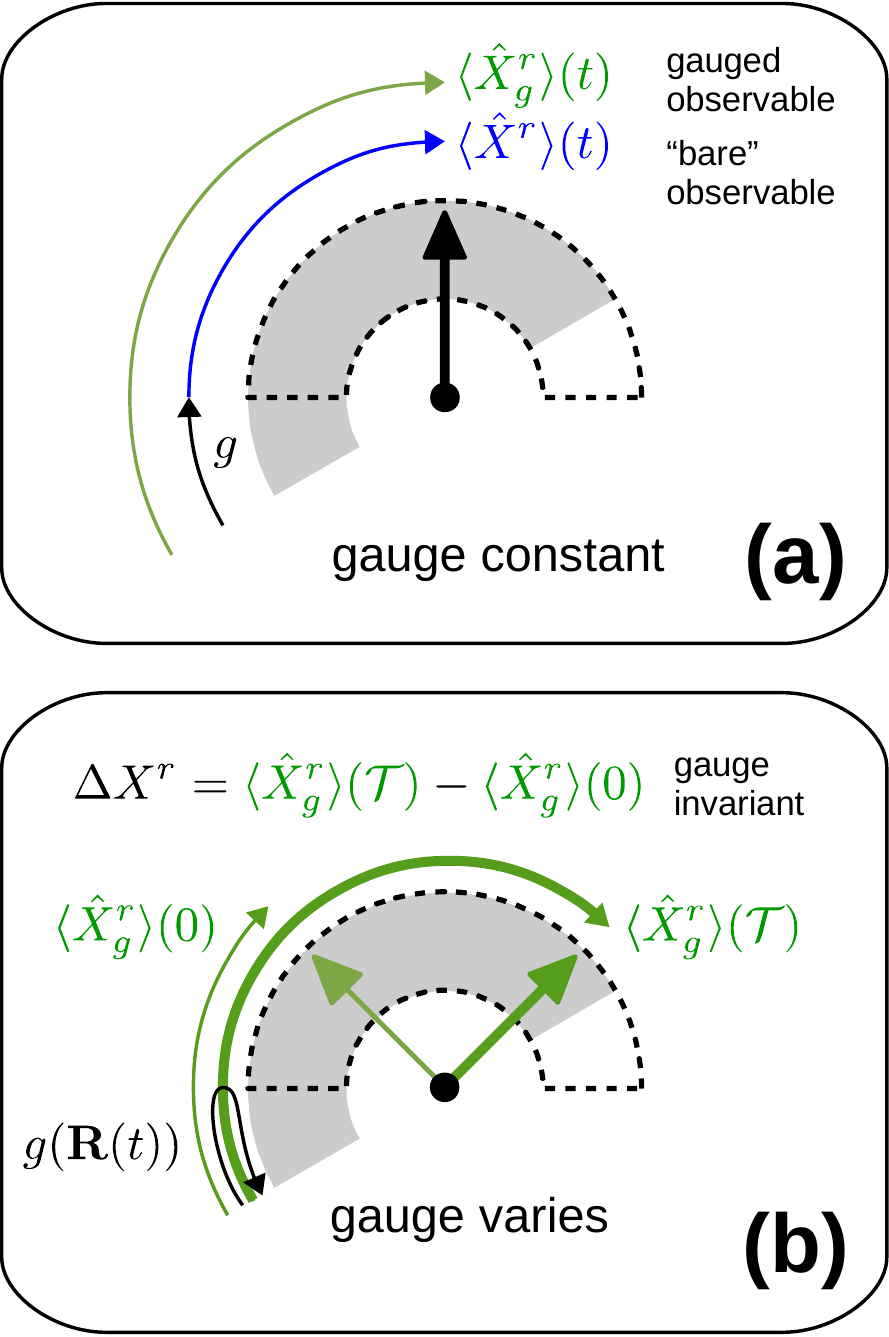}
  \caption{
    (a)
    Global gauge transformation $\hat{X}^r \to \hat{X}^r_g=\hat{X}^r+g \mathds{1}$
    leaving both currents $I_{X^r}$ and transported \observable $\Delta X^r$ invariant.
    The gray scalebar is translated by fixed amount $g$ relative to the ungauged one ($g=0$, dashed outline).
    The needle corresponds to the quantum state that from the observable $\hat{X}^r_g$
    the measurement expectation value $\expec{\hat{X}^r_g}=\expec{\hat{X}^r}+g$ indicated.
    (b)
    Local gauge transformation $\hat{X}^r \to \hat{X}^r_g=\hat{X}^r+g[\vec{R}] \mathds{1}$,
    changing the currents to $I_{X^r_g}$
    but leaving the transported \observable $\Delta X^r$ invariant.
    Since the meter scale is gauged in a \emph{continuous} way as function of the parameters $\vec{R}$,
    it returns to its original position every period: $g[\vec{R}(\T)]=g[\vec{R}(0)]$.
  }
  \label{fig:meter}  
\end{figure}

\emph{Total system description.}
On the level of the total system, the Heisenberg expression for the current operator \eq{eq:FluxOperator}, repeated here, reads
\begin{align}
  \hat{I}_{X^r}
  := \widehat{\frac{d X^r}{d t}}
  = i [H^{\text{tot}},\hat{X}^{r}] + \frac{\partial \hat{X}^{r}}{\partial t}
  .
  \label{eq:FluxOperator2}
\end{align}
An obvious transformation that leaves the \observable \current invariant~\cite{Sinitsyn09} is
\begin{align}
  \hat{X}^{r}(t) \rightarrow \hat{X}_{g}^{r}(t) = \hat{X}^{r}(t) + g \unit
  \label{eq:shift}
  ,
\end{align}
where $g$ is some fixed number independent of parameters and time.
Its physical meaning is clear when one accounts for the meter registering the measured value of the observable $\hat{X}^r$:
the number $g$ is simply a ``recalibration'' of that meter.
As illustrated in \Fig{fig:meter}(a), one can picture the observable $\hat{X}^r$ as the scale bar of a meter
whereas the meter's  needle corresponds to the quantum state producing the measured expectation value $\expec{\hat{X}^r}(t)$.
The recalibration \eq{eq:shift} is now a shift of the reference point of the scale bar
 behind the needle that indicates the measured value.
The Heisenberg equation \eq{eq:FluxOperator2} says that the current operator $\hat{I}_{X^r}$, and therefore also the transported \observable, remains unaltered.

However, these global, $\vec{R}$ independent, gauge transformations are a too narrow class for the present problem of \emph{pumping}:
here we require only that  the transported \observable, the integral of the current over a driving period [\Eq{eq:Xrpump2}], remains invariant.
This allows for a much larger group of \emph{local} gauge transformations:
for each parameter value $\vec{R}$ we can choose a different \emph{gauge for the observable},
determined by a continuous real function $g[\vec{R}]$:
\begin{align}
  \hat{X}^{r}[\vec{R}] \to \hat{X}_{g}^{r}[\vec{R}] = \hat{X}^{r}[\vec{R}] + g[\vec{R}] \unit
  .
  \label{eq:Xrgauge}
\end{align}
In the physical picture of \Fig{fig:meter}(b), this means that when driving $\vec{R}$ in time
the scale bar on the detector is allowed to \emph{vary in time} but only through the parameters.
Because $g[\vec{R}]$ is continuous,
this cannot affect the measurement of the pumped \observable since at the end of a driving cycle
the parameters, thus also the scale of the meter, has returned to its initial position:
one reads off the change correctly as
\begin{align}
  \Delta X^r_g = \expec{\hat{X}^r_g}(\T) - \expec{\hat{X}^r_g}(0)= \Delta X^r
\end{align}
for \emph{any} such calibration function $g$, continuous along the parameter driving curve $\mathcal{C}$.
We stress that during the driving cycle the currents \emph{have} changed due to the gauge transformation,
which is entirely physical\footnote{Whereas often gauge-invariance (dependence) is a test for ``(un)physicality'' of computed quantities,
this is not the case here:
Only a \emph{change} of an expectation value is gauge invariant.
One-point measurements, e.g., of the current $\expec{I_{X}}(\tau)$ in the interval $\tau \in [0,\T]$ are gauge dependent which is perfectly physical: changing the meter gauge changes the measured current.}.

A prime example of working with such a driving-parameter-dependent observable
is when one ignores (gauges away) the displacement charge currents when calculating the charge ($\hat{X}^r = \hat{N}^r$)  transported through a driven quantum dot from a capacitive model~\cite{Bruder94}.
In this case, the gauge function $g[\vec{R}]$ has the concrete physical meaning of minus the screening charge on electrode $r$ which depends on the time-dependent voltages ($\vec{R}$) applied to all the terminals of the circuit (see, e.g., \Ref{Calvo12a}, for a detailed discussion).
We stress, however, that our considerations hold equally well for other observables $\hat{X}^r$,
for example reservoir spin, energy, etc., 
for which there may be no obvious concept of displacement current or which may not be conserved.

In answer to question (i) we thus see that contained in every pumping problem
there is a simple local gauge group of meter recalibrations, which is much larger than the trivial global constant shifts \eq{eq:shift} of the observable.
It is nearly always hidden since one fixes the gauge to $g[\vec{R}]=0$
as soon as one decides to work only with the ``bare'' observable $\hat{X}^r$ which is time- and parameter independent, see \Eq{eq:NotNormalOrderedObs} ff.
This is one of the two ``naive'' things that we did in deriving the \ar pumping formula \eq{eq:pumpedX}.
However, we stress that our arguments so far did not invoke any ``open-system'' ideas
or related approximations (e.g., integrating out the reservoirs, Born-Markov or adiabatic approximation).
We also note that the gauge freedom \eq{eq:Xrgauge} related to the identity operator
is present for \emph{any} pumping problem:
it holds irrespective of the form of the parametrically time-dependent Hamiltonian $H^\text{tot}$.
It is thus truly a gauge freedom
  of the nonsystem observables that emerges for any periodic driving.
Thus, before having solved for, or even introduced, $\inst{\rho}$ or $\adcor{\rho}$
it is already clear that the geometric nature of pumping
is going to be associated with the freedom of calibrating the meter.

\emph{Open-system description.}
Now we show how this clear physical picture is reflected
in the reduced density operator description, i.e., \emph{after} integrating out the reservoirs.
This brings in open-system aspects.
For this we return to the \ar pumping equations \eq{eq:MarkovMe} and \eq{eq:MarkowFlux},
and our careful discussion of partial normal ordering and \current kernels in \Sec{subsec:Currents}.

The \current kernel equation \eq{eq:WIX} replaces the Heisenberg equation for the current in the total system description~\Eq{eq:FluxOperator2} in our above discussion.
Clearly, all observables differing by a constant lead to the same current kernel
because of the time-derivative in the second term of \Eq{eq:WIX}.
However, a time-local gauge transformation $\hat{X}^r[\vec{R}] \to \hat{X}^r[\vec{R}]+g[\vec{R}] \unit$
causes the current kernel \eq{eq:WIX} to transform as
\begin{align}
  \currentKernel{X^r} \to \currentKernel{X^r_g} =\currentKernel{X^r} + \frac{d}{dt} g[\vec{R}(t)] \mathcal{I}
  ,
   \label{eq:GaugeTransformationCurrentkernel}
\end{align}
where $\mathcal{I}$ denotes the identity superoperator.
For any gauge function $g[\vec{R}]$
this \current kernel produces the same transported observable
\begin{align}
  \Delta {X}^{r}_g 
  = \int_0^{\T} dt  \tr \currentKernel{X^r_g}[\vec{R}(t) ] \left[ \inst{\rho}(t) + \adcor{\rho}(t) \right]
  = \Delta X^{r}
  \label{eq:PumpingTrafo}
\end{align}
by virtue of the probability normalization of \Eq{eq:adiabatic-response} [implying $\tr \adcor{\rho}(t)=0$]
and the continuity $g[\vec{R}(\T)]=g[\vec{R}(0)]$.

Although the transported \observable is gauge invariant
the current kernel that produces it is not
(as we changed the meter gauge).
To relate this to the observable as in \eq{eq:Xrgauge}, or rather of its expectation values,
requires a little extra effort in the open system picture.
To this end we
separate the current in \Eq{eq:PumpingTrafo},
\begin{align}
 I_{X^r_g}(t) := \tr \, \currentKernel{X^r_g}[\vec{R}(t) ]  \left[  \inst{\rho}(t) + \adcor{\rho}(t) \right]
  \label{eq:Iidefg}
  ,
\end{align}
into an instantaneous, gauge independent part
\begin{align}
 \inst{I_{X^r}}(t) :=    \tr \, \currentKernel{X^r}[\vec{R}(t)] \inst{\rho}(t)
  \label{eq:Iidef}
  ,
\end{align}
and a remaining adiabatic-response part that is gauge dependent:
\begin{subequations}\begin{align}
\adcor{I_{X^r_g}}(t)
 & : =
I_{X^r_g}(t)- \inst{I_{X^r}}(t)
,
\\
 & = 
\tr \, \currentKernel{X^r}[\vec{R}(t)]  \adcor{\rho}(t) + \frac{d}{d t} g[\vec{R}(t)]
\label{eq:Iadef}
  .
\end{align}\end{subequations}
As before, the labels ``a'' or ``i'' indicate whether the current component depends on $\dot{\vec{R}}(t)$ or not.
Now we can identify the geometric part of the expectation value of the gauged nonsystem observable by splitting it up\footnote{This split-up is relative to $\expec{ \hat{X}^{r}}(0)$
since it can only be defined via the corresponding split-up of the current.
The latter exploits the gauge freedom \Eq{eq:Xrgauge} that emerges only for periodic driving.}
 at any time $t$ as
$  \expec{ \hat{X}^{r}_g }(t) = 
  \inst{\expec{ \hat{X}^{r}}}(t) +  \adcor{\expec{\hat{X}^{r}_g}}(t)
$ into an instantaneous, gauge independent part
\begin{align}
  \inst{\expec{ \hat{X}^{r}}}(t) - \expec{ \hat{X}^{r}}(0) & := \int_0^{t} d\tau \inst{I_{X^r}}(\tau)
  ,
\end{align}
and an adiabatic-response part that contains the gauge dependence:
\begin{align}
  \adcor{\expec{ \hat{X}^{r}_g}}(t) - \expec{ \hat{X}^{r}}(0) & = \int_0^t d\tau \adcor{I_{X^r_g}}(\tau)
	\label{eq:notGaugeInvariant}
  .
\end{align}
We stress that here we do not integrate over a driving period, but up to \emph{any} time $t$ within the driving period, $0 \leq t \leq \T$.
Thus, \emph{after} integrating out the reservoirs
the gauge dependence of the total system operator $\hat{X}^{r}_g= \hat{X}^{r} + g \unit$  resides in the \emph{adiabatic-response part of the observable}
\begin{align}
  \adcor{\expec{\hat{X}^{r}}}(t) \to 
  \adcor{\expec{\hat{X}^{r}_g}}(t) =
  \adcor{\expec{\hat{X}^{r}}}(t) + g[\vec{R}(t)]
  .
  \label{eq:Xrga}
\end{align}
and \emph{not} in the instantaneous one $\inst{\expec{\hat{X}^{r}}}$.
This is the open system equivalent of \Eq{eq:Xrgauge} that we sought.

\emph{Unphysical redundancy.}
At this point, it is important (cf. \Sec{sec:fcsApproach}) to note that the \current kernel has an additional, completely unrelated redundancy
that may obscure the above clear physical picture.
Even when fixing the gauge $g[\vec{R}]$ of the \observable $\hat{X}^r$,
the associated \current \emph{kernel} is still not unique: one can always add to it a time-dependent system superoperator $\Theta(t)$,
\begin{align}
  \currentKernel{X^r} \to \currentKernel{X^r} + \Theta(t)
  ,
  \label{eq:redundancy-ar}
\end{align}
for which $\tr \Theta(t) \bullet =0$, without changing any expectation value,
including the \ar part $\adcor{\expec{\hat{X}^{r}_g}}(t)$.
We actually made use of this when writing the \current kernel in the form \eq{eq:ObservableKernel}.
Importantly, this redundancy is independent of the physical gauge freedom\footnote{The rewriting of the result \Eq{eq:ObservableKernel} in \App{app:kernels} involves adding a commutator $[\hat{X}^r,\bullet]$ to the expression, which is invariant under the physical gauge transformations \Eq{eq:Xrgauge}.
This means one can do such rewriting at \emph{any} stage of the calculation.} and need not be considered further until we discuss the \fcs approach [cf.  \Eq{eq:redundancyrhochi}].

\emph{Geometric nature of pumping.}
Thus, the geometric nature of pumping
in open systems emerges naturally when one considers the \emph{current} of the transported observable, i.e., via the route \Eq{eq:Xrpump2}.
In the total (open) system description,
the gauge freedom lies in the nonunique association\footnote{The freedom in the assignment of a current operator to an observable was discussed in \Ref{Avron12} for geometric pumping of \emph{system} observables connected to a single reservoir,  motivated by other works~\cite{Bellissard02,Gebauer04,Bodor06,Salmilehto12}.
Here we consider more general \emph{nonsystem} observables and multiple reservoirs, requiring consideration of current kernels.
This allows steady-state transport \emph{through} the system to be discussed.
See further \App{app:pumpsys}.} 
of (the adiabatic-response part of) the transported observable
$\Delta X^r$ ($\Delta \adcor{X}^r$)
with a current  (kernel super-) operator $\hat{I}_{X^r_g}$ ($\currentKernel{X^r_g}$).

\begin{figure}[t]  
  \includegraphics[width=0.9\linewidth]{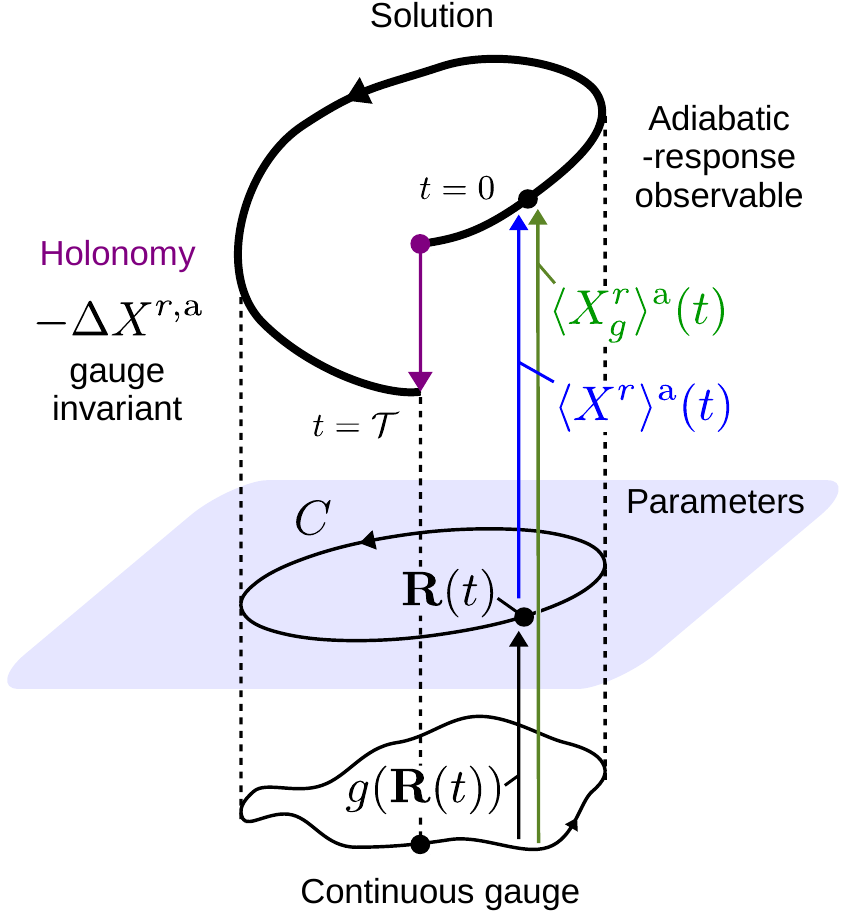}
  \caption{
    Fiber bundle space in which the pumping problem \Eq{eq:master}-\eq{eq:slave} is solved:
    the plane corresponds to the base space of driving parameters $\vec{R}$ containing the driving curve $\mathcal{C}$.
    The ``vertical'' space at each point $\vec{R}$ is formed by all adiabatic-response expectation values \eq{eq:Xrga}, $\expec{\hat{X}^{r}_g}^\text{a}$, each coordinatized by a real value of $g[\vec{R}]$ at that point $\vec{R}$.
  }
  \label{fig:bundle}
\end{figure}

Associated with the measurable transported \observable is thus a whole class of different, parametrically time-dependent observables $\hat{X}^r_g$.
We see that the space in which the physical pumping problem is solved is correspondingly much larger than thought initially based on our ``naive'' calculation in \Sec{sec:naive}.
More precisely, it has the structure of a simple fiber-bundle\cite{Nakahara03}, sketched in \Fig{fig:bundle}.
To each driving parameter $\vec{R}$ in the base space is attached a ``copy'' of the space of all possible gauge-equivalent,  adiabatic-response expectation values of the \observable,
i.e., \emph{all} possible gauge choices \eq{eq:Xrga} for fixed $\vec{R}$.
For the ``vertical'' coordinate in this space we can just take $g[\vec{R}]$,
i.e., our simple fiber is isomorphic to the real line.
This reflects the direct physical meaning of the real-valued $g[\vec{R}]$ as a calibration of the meter scale of \Fig{fig:meter}.

As in many other areas of physics~\cite{\CONTROLMECHANICS} where one solves a physical problem in such a fiber-bundle space,
a geometric phase is expected to emerge.
Viewed in this larger space it is now clear from the start
that there is  ``room'' for a geometric phase to develop along the ``vertical'' fiber direction of the \emph{observable},
 even though there is no ``Berry phase'' in the time-dependent steady-state evolution of the \emph{mixed-state} (\Sec{sec:LackOfPhaseState}).

\begin{figure}[t]  
  \includegraphics[width=0.9\linewidth]{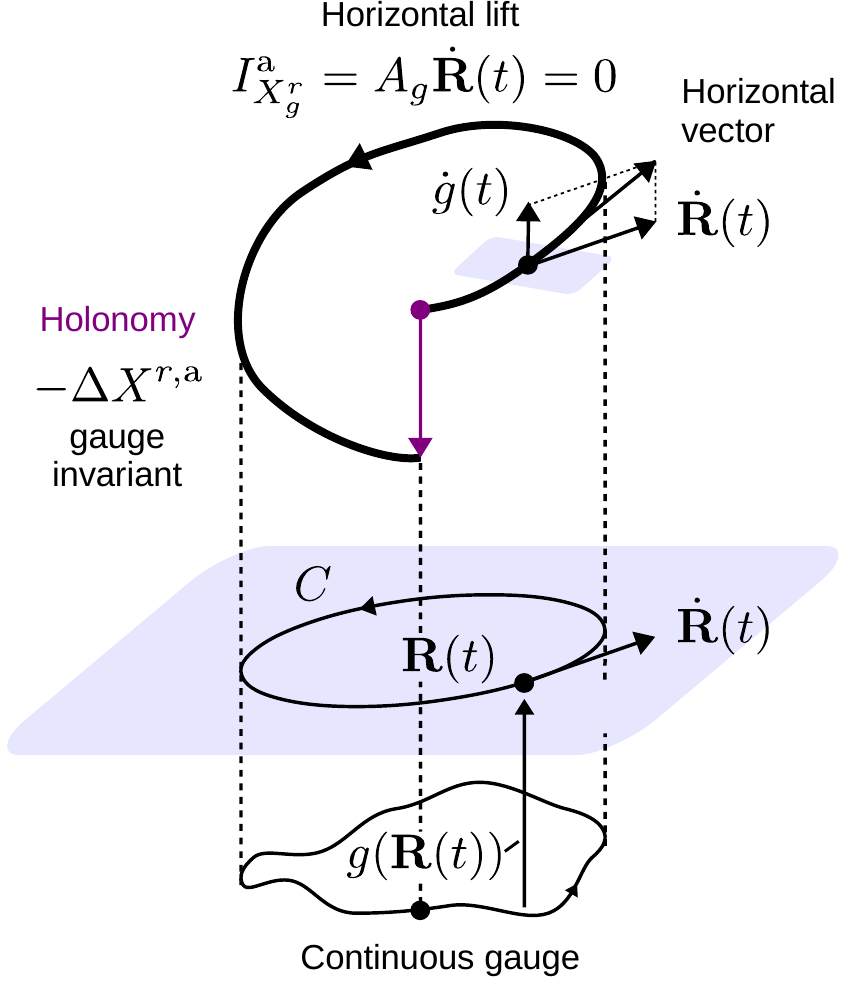}
  \caption{
    In a fiber-bundle space, the notion of what is ``horizontal'' is completely undefined, in contrast to ``vertical,''    which is naturally the direction along gauge coordinate $g$, the fiber space. 
    ``Horizontal'' cannot be defined as orthogonal to ``vertical'',
    since there is no physically motivated metric in this space.
    However, the function $A_g[\vec{R}]$ arising in the solution of the physical pumping problem of \Eqs{eq:master} and \eq{eq:slave} can be used to define what ``horizontal'' means at each point,
    thus defining a geometric ``connection.''
    The nonintegrability of the connection $A_g$ leads to the discontinuity in the horizontal lift of the curve $\mathcal{C}$ into the bundle space.
   This is the differential-geometric significance of Landsberg's connection for pumping.
  }
  \label{fig:horizontal}
\end{figure}

Returning to \Fig{fig:bundle}, we can visualize most clearly in what way the geometric origin of  pumping effects remains hidden if one starts from the ``bare'', time-independent \observable operator\footnote{Throughout the paper we assume that the ``bare'', ungauged observable $\hat{X}^r$ does not depend on the parameters: it is the ``probe'' used to detect a response of the driving ($\vec{R}$) and should be independent of the stimulus. However, when observable ${\hat{X}^r+\hat{X}}$ is conserved the corresponding system observable $\hat{X}$ may well dependent on $\vec{R}$ [see discussion after \Eq{eq:Xconservation}].}
and / or enforces partial normal-ordering of the current operator
(cf. \Sec{subsec:Currents}).
These technical assumptions physically amount to working in the fixed gauge $g=0$. Geometrically, this corresponds to using a special coordinate system relative to the plane in the sketches in \Fig{fig:bundle} and \Fig{fig:horizontal}.
However, \emph{all} smooth coordinate systems in this space are physically meaningful and equivalent for pumping.

\subsection{Landsberg's geometric pumping connection\label{sec:landsberg}}

Having answered question (i) by identifying the gauge freedom (the fiber bundle relevant for pumping) we will now answer question (ii):
we show how the solution of the pumping problem determines a \emph{geometric  connection} whose
geometric phase is just the pumped observable.
This determines the actual \emph{magnitude} of this geometric phase effect allowed by gauge freedom.
Following the \ar approach, the pumping problem is described by two equations (cf. \Secs{sec:masterEquation} and \sec{subsec:Currents}),
the state evolution
\begin{align}
  \partial_t \Lket{\rho(t)} =  W[\vec{R}(t)] \, \Lket{\rho(t)}
  \label{eq:master}
\end{align}
exhibiting a unique frozen-parameter stationary state, $W[\vec{R}]\Lket{\inst{\rho}}=0$,
and a second equation for a variable ``enslaved'' to this,
the gauge-dependent current
\begin{align}
  \frac{d}{dt} \expec{ \hat{X}^{r}_g}(t) 
  =  \Lbra{\unit} \currentKernel{X^r_g}[\vec{R}(t)] \Lket{\rho(t)} 
  \label{eq:slave}
	.
\end{align}
Such pumping equations fit~\cite{Sinitsyn09} into Landsberg's general framework of nonlinear, dissipative dynamics with a symmetry~\cite{Landsberg92,Landsberg93}, extended in~\cite{Andersson03thesis,Andersson05} to the non-Abelian case.
In our special case, the symmetry is an Abelian gauge freedom expressing changes of the coordinate system in the fiber bundle space as in \Fig{fig:bundle}.
Moreover, our dynamics is linear
as we have emphasized by introducing Liouville-space vector notation for operators and linear functions of operators, respectively:
\begin{align}
  \Lbra{\unit} := \tr , \qquad \Lket{\rho} = \rho .
  \label{eq:liouville}
\end{align}
Here, we highlight the two most relevant aspects of Landsberg's framework:

(a) Equation \eq{eq:master} alone does not exhibit a geometric phase,
i.e., in the time-dependent steady-state limit the solution should be continuous [see \Eq{eq:rhocont}].
For a unique stationary state that we consider here, we showed in \Sec{sec:LackOfPhaseState} that this is always the case  due to the general constraint of probability normalization.

(b) The variable $\expec{ \hat{X}^{r}_g}(t)$, enslaved to the dynamics of the state $\rho(t)$, is \emph{not} geometric as whole:
only its adiabatic-response part is geometric, as noted above.
However, even for this part to make sense in differential geometry,
the enslaved equation \eq{eq:slave} needs to transform in a specific way, as emphasized in \Ref{Andersson03thesis}.
If this were not the case, pumping could not be related to a connection and curvature, the basic concepts for relating physical results to geometric and topological properties of a fiber bundle.
In our case, this corresponds to the physical transformation law for the current kernel \Eq{eq:GaugeTransformationCurrentkernel},
which is essentially the Heisenberg equation of motion after integrating out the reservoirs.

Following \Refs{Andersson03thesis,Andersson05}, in Landsberg's approach one solves  \Eqs{eq:master} and \eq{eq:slave} for the time-dependent steady state
using the \ar procedure, now including the gauge dependence in contrast to \Eqs{eq:MarkovMe}-\eq{eq:MarkowFlux}, and extracts a geometric phase that is thus a leading-order \emph{nonadiabatic} effect (adiabatic response).
Proceeding as before in \Sec{sec:naive},
we compute $\inst{\rho}$ and $\adcor{\rho}$
and insert these into \Eqs{eq:Iidef} and \Eq{eq:Iadef}.
This gives for the instantaneous part of the transported \observable
\begin{align}
  \Delta \inst{X}^{r} = \int^{\T}_0 d t \inst{I}_{X^{r}}(t)
= \int_0^{\T} dt \Lbra{\unit}\currentKernel{X^r}[\vec{R}(t)] \Lket{\rho^i}
  ,
\end{align}
which is equal to \Eq{eq:Obs}.
However, the adiabatic-response correction now reads as
\begin{align}
  \Delta \adcor{X}^r   =  \int^{\T}_0 d t \adcor{I}_{X^{r}_g}(t) 
                      = \oint_{\mathcal{C}} d\vec{R} A_g[\vec{R}]
  \label{eq:pumpedXlandsberg}
	,
\end{align}
where using \Eq{eq:WIX} we introduced the expressions
\begin{subequations}\begin{align}
  A_g & := A_0[\vec{R}] + \delR g[\vec{R}]
  \label{eq:Ag}
  ,
  \\
  A_0 [\vec{R}] & = \Lbra{\unit} \currentKernel{X^r}[\vec{R}] \frac{1}{W[\vec{R}]} \delR \Lket{\rho[\vec{R}]}
  \label{eq:A0}
  .
\end{align}\label{eq:LandsbergPumpinConnection}\end{subequations}
The pumping is now explicitly seen to be geometric, in a more restricted sense,
since we have now formulated the problem without inadvertently fixing a gauge.
It is now clear why ``adding a differential under the integral'' must physically always be possible: it is a meter recalibration.
Answering question (ii), the pumped \observable
\eq{eq:pumpedXlandsberg} is indeed a geometric phase
determined by $A_g$, which, as we discuss below, plays the role of a geometric connection with a clear physical motivation.
By \Eq{eq:GaugeTransformationCurrentkernel} the gauge potential $A_g$  indeed shows the proper transformation to a new gauge
with the simple additive gauge group $\sim\mathds{R}$ of meter recalibrations:
\begin{align}
  A_f[\vec{R}] \rightarrow A_{f+g} = A_f[\vec{R}] + \delR g[\vec{R}]
  .
  \label{eq:ConnectionGaugeTransformation}
\end{align}
This ensures that the pumped observable is gauge invariant, even though it is computed from the gauge-dependent adiabatic-response part of the  current $I_{X^r_g}(t)\propto \dot{\vec{R}}(t)$.

When applying our considerations to the simpler case of pumping of \emph{system} observables and a single reservoir
our formulation recovers the geometric pumping result of Avron et al.\cite{Avron12} for the case of a unique stationary state.
This is worked out in \App{app:pumpsys}, further showing the complementarity to \Ref{Avron12}  which inspired the above.

\subsection{Physical meaning of horizontal lift, \newline parallel transport and holonomy\label{sec:HprizontalLift}}

The answer to question (ii), i.e., that pumping defines a geometric connection, 
can be further clarified by considering the physical meaning of key concepts of differential geometry of the fiber bundle of meter calibrations.
In this regard, the connection \eq{eq:Ag} has the advantage that it is a global connection~\cite{Nakahara03}, i.e., defined in the total space of the fiber bundle (in contrast to local connection forms independent of the gauge coordinate).
This global object has the most direct geometric significance, which is also why it is favored in differential geometry.
Also, physically it is most revealing as we now explain.

\begin{figure}[t]  
  \includegraphics[width=0.9\linewidth]{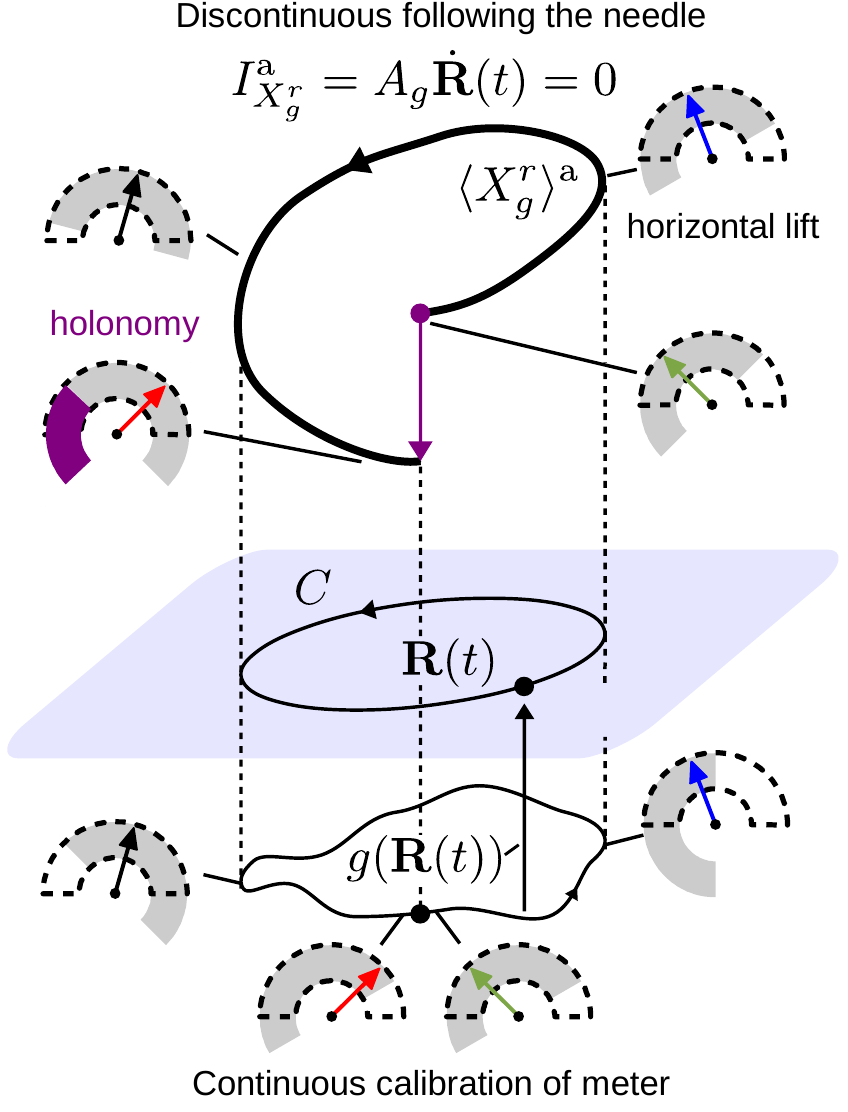}
  \caption{
    Physical meaning of differential geometric notions of \Fig{fig:horizontal} in the pumping problem.
    The linear space defined by the connection [\Eq{eq:zerocurrent}] locally determines a plane of ``horizontal'' vectors in \Fig{fig:horizontal}.
    Pumping corresponds to ``parallel transporting'' such a vector
    thereby producing a ``horizontal lift'' of the closed parameter drive curve $\mathcal{C}$.
    Physically, this signifies that one calibrates the meter for the observable, drawn in same way as in \Fig{fig:meter},
such that one maintains zero \emph{pumping-current} $I_{X^r_g}^\text{a}$ in this gauge.
    If there is pumping (the connection is nonintegrable) the lifted curve ``breaks'' as shown in \Fig{fig:horizontal}.
    Physically, the ``vertical'' discontinuity, the holonomy, is minus the pumped \observable,
    i.e., the cumulative calibration (purple) required to maintain zero current during one driving period.
Along the horizontal lift, the scale (grey) is made to follow
the pointer.
  }
  \label{fig:physical}  
\end{figure}

The central notion of ``horizontal space'' in a fiber bundle, can be defined directly by requiring this connection [a linear function of vectors in the total space, $(\dot{\vec{R}},\dot{g})$, transforming as \eq{eq:ConnectionGaugeTransformation}] to vanish. 
Since we avoid the use of differential forms, this should be written as
\begin{align}
  \adcor{I}_{X^{r}_g}(t)
  &= A_g[\vec{R}(t)]\dot{\vec{R}}(t) \notag\\
  & = A_0[\vec{R}(t)]\, \dot{\vec{R}}(t) + \dot{g}\, [\vec{R}(t)] = 0
    \label{eq:zerocurrent}
    .
\end{align}
This locally determines a linear relation between a direction in the base parameter space ($\dot{\vec{R}}$) and the ``vertical'' gauge direction ($\dot{g}$).
This is sketched in \Fig{fig:horizontal}.
From the point of view of differential geometry,
the pumping current can thus be used to \emph{define} the notion of a ``horizontal'' direction in the total space.
Moving tangent to this so-defined ``horizontal'' space is called ``parallel transport''
and by \Eq{eq:zerocurrent} physically corresponds to maintaining zero \emph{pumping} current as the measurement proceeds.
We stress that this zero-current condition \Eq{eq:zerocurrent} \emph{derives} ultimately from physics, i.e., from the state evolution [\Eq{eq:master}] plus the open-system Heisenberg equation of motion [\Eq{eq:slave}]: it is not imposed.
This is illustrated in \Fig{fig:physical} :  one calibrates the meter's scale in a
parameter-dependent way such that the needle always indicates a fixed
expectation value for $\hat{X}^r_g$
relative to this moving scale.

From the sketch of these calibrations in \Fig{fig:physical}
it is clear that, if there is a nonzero pumping effect,
this condition cannot be maintained in a continuous way along the
closed driving curve $\mathcal{C}$:
the ``vertical'' jump  at $\vec{R}(0)=\vec{R}(\T)$ is the cumulative
calibration required to maintain zero current during the driving period.
This calibration must equal minus the pumped value.
To see this, let $h(t)$ denote the curve
that solves the differential equation \eq{eq:zerocurrent} along the
closed base space curve $\mathcal{C}$.
This curve is known as the ``horizontal lift'' of $\mathcal{C}$ in
\Fig{fig:horizontal}.
Substituting $g[\vec{R}(t)] \to
h(t)$, we find for the discontinuity:
\begin{align}
  h(\T)-h(0) = - \oint_{\mathcal{C}} d\vec{R} A_0[\vec{R}] = - \Delta
  \adcor{X}^r
  .
  \label{eq:holonomy}
\end{align}
This discontinuity is the geometric phase
or the ``holonomy'' of the horizontal lift.
Indeed, in the present problem this is just minus the pumped \observable.

In summary,
by starting from the gauge freedom in  a system plus reservoir description
we arrived at the tangible physical meaning for all the relevant geometric notions
that arise in the reduced density operator description of steady-state pumping.
By simply formulating the \ar calculation of pumping
in the physically natural larger fiber-bundle space,
it becomes clear that there is always room for a geometric pumping phase to develop
for \emph{any} nonsystem observable.
Physically, this is just the space that includes all possible meter gauges.
Although in the present case the actual calculations are all easy,
this simple picture has received little attention so far.
Much attention has been given to the more general but also more complicated geometric \fcs approach.
The relative simplicity of Landsberg's approach is a crucial advantage when addressing more complicated models and dynamics~\cite{Pluecker17a}.
\section{Pumping curvature\label{sec:application}}

We now turn to question (iii) posed in \Sec{sec:questions}
by analyzing  necessary conditions 
for the pumped observable in the \ar approach  to be nonzero.

\subsection{Pumping curvature and
response covector\label{sec:covector}}

It is useful to rewrite the pumped observable
 \eq{eq:pumpedXlandsberg} as a surface integral over a curvature  $B_{X^r}[\vec{R}] = \delR \times A_{X^r}[\vec{R}]$:
\begin{align}
  \Delta \adcor{X}^r = \int_{\mathcal{S} } dS B_{X^r} .
  \label{eq:stokes}
\end{align}
By working with dimensionless parameters $\vec{R}$, this curvature has the direct physical meaning of pumped \observable per unit area of the driving parameter space.
In view of the following discussion, we now explicitly indicate  the considered observable $X^r$ in the curvature $B$.
Clearly, adiabatic-response pumping always requires at least two parameters to be driven. Otherwise, the driving curve $\mathcal{C}$ does not bound a surface $\mathcal{S}$ and \Eq{eq:stokes} gives a zero result. 
This rules out all driving schemes 
that are trivial one-dimensional curves in parameter space.

Given the driving cycle is two dimensional, we investigate when the integrand, i.e., the pumping curvature $B_{X^r}[\vec{R}]$, is nonzero
by writing it in the transparent form
\begin{align}
  B_{X^r}[\vec{R}] =   \Lbra{\delR \Phi_{X^r}[\vec{R}]} \times \Lket{ \delR   \inst{\rho}[\vec{R}]} 
  \label{eq:PseudoMagneticField}
\end{align}
deriving from the Landsberg connection \eq{eq:LandsbergPumpinConnection}.
We define $(a\times b)_{ij} := a_i b_j-a_jb_i$ component-wise,
which in two dimensions corresponds to the cross-product vector that is normal to the parameter plane.\footnote{For simplicity we consider the parameters $\vec{R}$ to form a vector space.
For general nontrivial parameter manifolds one should instead consider the exterior 2-form $B_{X^r}[\vec{R}] 
  =
\Lbra{\unit}  [ \text{d} ( \currentKernel{X^r} {W}^{-1} )
\wedge
   \Lket{ \text{d} \inst{\rho} }$
on the tangent space of the manifold.}
Extending ideas of \Ref{Calvo12a} [cf. also \App{app:brouwer}], here we introduced an \emph{adiabatic-response covector}
:
\begin{align}
  \Lbra{\Phi_{X^r}[\vec{R}]}
  := \Lbra{\mathds{1}} \currentKernel{X^r}[\vec{R}] \frac{1}{W[\vec{R}]} 
  .
  \label{eq:ResponseVectorDefinition}
 \end{align}

That the Landsberg curvature \eq{eq:PseudoMagneticField} is \emph{explicitly} gauge invariant [cf. \eq{eq:LandsbergPumpinConnection}],
indicates that it has a direct physical interpretation as a pumped observable:
by \Eq{eq:stokes} $B_{X^r}[\vec{R}]$ is the pumped observable per unit area of the parameter space, i.e., bounded by an infinitesimal pumping cycle at $\vec{R}$.
Thus, the pumping curvature can be experimentally obtained
 in the limit of small amplitude driving 
 as $ B_{X^r} = d (\Delta \adcor{X}^r)/d\mathcal{S}$.

The response covector \eq{eq:ResponseVectorDefinition}
has a physically transparent form:
a nonzero pumping curvature \eq{eq:PseudoMagneticField} necessarily requires that the combined effect
of the ``lag time'' of the retarded mixed quantum \emph{state}
(pseudo-inverse relaxation kernel $W^{-1}$)
together with
the \emph{observable} current (current kernel $\currentKernel{X^r}$)
is \emph{parameter dependent}.

The cross product in \Eq{eq:PseudoMagneticField} imposes a stringent condition:
the pumping curvature does not only vanish when  the gradient of the response-covector itself is zero
but also when it is parallel to the parametric gradient of the stationary-state,
schematically:
\begin{align}
  \delR  \Phi_{X^r}[\vec{R}] \quad \parallel \quad \delR  \inst{\rho}[\vec{R}]
  \label{eq:covector-para}
  .
\end{align}
Here, the gradients are vectors obtained from \emph{each} $\vec{R}$-dependent \emph{matrix element} of the two operators in the cross product in \Eq{eq:PseudoMagneticField}.
Expanding both operators in a basis one obtains a Brouwer-type formula (see \App{app:brouwer}),
which  shows that for a nonzero result
it is necessary that in this expansion at least some
components of the two operators
should pairwise have nonparallel gradients.
In general this is not yet sufficient since in the sum they may cancel. 
An explicit example of the curvature for an interacting quantum dot is is given in \Eq{eq:AppExampleCurvature}.

\subsection{Geometric spectroscopy\label{sec:spectroscopy}}

The schematic condition \eq{eq:covector-para}
 forms the basis for the \emph{adiabatic-response pumping spectroscopy}
proposed in \cite{Reckermann10a,Calvo12a,Haupt13} in the specific setting of quantum dots.
Here we outline how this spectroscopy works in more general terms,  indicating that it extends to a much broader class of open systems.

The physical idea is the following:
in \Eq{eq:covector-para} clearly both the kernel $W[\vec{R}]$, defining $\rho^{i}[\vec{R}]$  [\Eq{eq:instDensityOp}], and the current kernel $\currentKernel{X^r}$, defining $\Phi_{X^r}[\vec{R}] $  [\Eq{eq:ResponseVectorDefinition}],
depend on the spectral properties of the closed system, the Hamiltonian $H$,
and of the reservoirs, $H^\text{res}$.
Due to the weak coupling to the reservoirs
strong variations of these two quantities as function of the parameters $\vec{R}$
are expected when the driving curve $\mathcal{C}$ hits a parameter point
for which the system is in energetic resonance with the reservoirs.
For particle transport, for example, this happens when the electrochemical potential of one of the reservoirs lines up with one of the system's particle-addition energies.
When only such a single-resonance parameter value is traversed by the driving curve, the $\vec{R}$-dependence is thus effectively one dimensional
i.e., effectively we drive a single parameter.
We then have the situation \eq{eq:covector-para} of \emph{effective} one-parameter driving and the adiabatic-response is zero on general grounds.

However, the driving may also visit parameter values where the system simultaneously satisfies two (or more) resonance conditions.
For example, two or more system addition energies may line up with electrochemical potentials of two reservoirs.
Then the $\vec{R}$-dependence of the operators $\Phi_{X^r}[\vec{R}]$ and $\rho^{i}[\vec{R}]$ in \Eq{eq:covector-para} will in general be different,
giving the necessary lifting of condition \eq{eq:covector-para} for zero-pumping:
the magnitude of the pumping curvature is thus generically  expected to show a peak at crossings of resonances.

Indeed, in quantum dots this leads to a sharp pattern of ``spots''~\cite{Reckermann10a,Calvo12a,Yoshii13,Haupt13} at the crossings of  single-resonance parameter lines in the plane of applied voltages (``stability diagram'') when plotting $B[\vec{R}]$.
However, as the above argument indicates, the idea is more general:
a recording of $\Delta \adcor{X}^r$ or $B_{X^r}$  as function of the driving parameter working point $\vec{R}$
contains detailed information on the open system, both through its sign and magnitude.
Some of this information is not contained in the instantaneous transported observable $\Delta \inst{X}^r$.
For example,  in an interacting quantum dot the sign of the curvature reveals the spin-degeneracy through pumping
(without using a magnetic field)
 as well as the direction of the junction asymmetry~\cite{Reckermann10a}.
Also, different pumping observables have different resolving capabilities:
for example, some resonances that are hidden in charge pumping are revealed by spin pumping~\cite{Calvo12a,Haupt13}.

The basic principle of the \emph{geometric spectroscopy} is thus 
to probe the system's properties via its leading parameter-dependence in adiabatic-response to slow driving
 rather than to more drastic physical excitation.
The pumping formula~\Eq{eq:PseudoMagneticField} provides a straightforward approach for computing the detailed response for a variety of systems and observables~\cite{Pluecker16b},
illustrated in \App{app:example} for a driven quantum dot.
In general this formula can be written in a Brouwer-type~\cite{Brouwer98} form with additional terms (see \App{app:brouwer}).
Before we discuss simplifications of this formula in more special situations, we stress its generality:
It holds, within our approximations, for any weakly-coupled open quantum system, 
e.g., comprised of many orbitals with arbitrary interactions [as described by $H$ in \Eq{eq:Htot}].

\subsection{Pumping of conserved observables\label{sec:conserved-curvature}}

The Landsberg pumping curvature \eq{eq:PseudoMagneticField}
has the advantage that it can easily be simplified 
when the considered nonsystem observable $\hat{X}^r$
is conserved by the coupling [\Eq{eq:Htot}] at junction $r$ for each traversed parameter value:
\begin{align}
  [\hat{X}[\vec{R}]+\hat{X}^r , V^r[\vec{R}]]=0
  .
  \label{eq:Xconservation}
\end{align}
Here $\hat{X}[\vec{R}]$ is the corresponding system observable.
As for the reservoirs [cf. \Eq{eq:HrXrcom}],
we assume that this observable is also conserved inside the system (in the absence of coupling):
\begin{align}
  [ {H}[\vec{R}], \hat{X}[\vec{R}] ] = 0
  .
 \label{eq:HXcom}
\end{align}
Examples are
the charge ${N}^r+{N}$ for a tunnel junction to a normal-metal electrode
or the spin ${S}_z^r+{S}_z$ for a nonmagnetic tunnel junction to a ferromagnetic electrode with polarization in $z$ direction.
In these cases, both observables are $\vec{R}$ independent.

However, \Eq{eq:Xconservation} allows that the conservation law\footnote{For bilinearly coupled quantum dots, the energy current is conserved only to the first order in the coupling, which is considered here.
See \Ref{Gergs17a} for a detailed study of the corrections.}
yields a system observable that does depend on $\vec{R}$, even though $\hat{X}^r$ is $\vec{R}$-independent.
The prime example for this is the pumping of heat where in
${H}^r[\vec{R}] - \mu^r[\vec{R}] {N}^r$
the $\vec{R}$-dependence may cancel out between the two terms, as we assumed in \Eq{eq:rhores},
but the corresponding system observable
${H}[\vec{R}] - \mu^r[\vec{R}] {N}$
is still $\vec{R}$ dependent.
We stress that this parameter dependence \emph{cannot} be gauged away
since it is in general not of the form $g[\vec{R}]\unit$
and thus contributes to both the instantaneous and the adiabatic-response of the transported observable.
Our considerations apply to this case as well deserving a separate study~\cite{Pluecker16b}.

The conservation law \eq{eq:Xconservation} together with \Eq{eq:HXcom} implies
that the computation of $\currentKernel{X^r}$ can be avoided by using
\begin{align}
  \Lbra{\unit} \, \currentKernel{X^r}[\vec{R}]
  =
  -
  \Lbra{X [\vec{R}] } \, W^{r}[\vec{R}]
  .
  \label{eq:conservation}
\end{align}
Notably, on the right hand side, $\hat{X}$ is the \emph{system} observable
and $W^r$ is the part of the known \emph{time-evolution} kernel due to the coupling $V^r$ to reservoir $r$ alone.
The latter is easily obtained from $W$ as given by \Eq{eq:Wdef} by dropping $L$ and also all contributions from reservoirs $r' \neq r$.
We stress that in order to use \Eq{eq:conservation} only conservation at junction $r$ is required, not at all junctions.

The relation \eq{eq:conservation} is derived in \App{app:kernels} using the same assumptions and the same approach as in \Secs{sec:masterEquation} and \sec{subsec:Currents}.
Its use is illustrated in \Eq{eq:covectorexample}.
For weak-coupling and memoryless master equations this relation is well-known in a less convenient form
discussed in \App{app:ChargeKernel}.
The form \eq{eq:conservation}
has the advantage of having
 a very similar exact generalization~\cite{Saptsov12a} that relies only
 on the decomposition of the coupling as $V=\sum_r V^r$,
allowing the present considerations to be extended~\cite{Pluecker17a}.

Using \Eq{eq:conservation} the response vector \eq{eq:ResponseVectorDefinition} simplifies to
\begin{align}
  \Lbra{\Phi_{X^r}[\vec{R}]}
  = -
  \Lbra{X[\vec{R}]} W^{r}[\vec{R}] \frac{1}{W[\vec{R}]}
  \label{eq:ResponseVectorRoman}
  .
\end{align}
Physically, $W^r W^{-1}$ describes the ``fraction of $\hat{X}$'' in the system
that flows through the junction to reservoir $r$ where it `` turns into $\hat{X}^r$ ''.
Combined with \Eq{eq:PseudoMagneticField} we see that
the $\vec{R}$-dependence of this fraction
is the decisive factor for the pumping of a conserved observable:
\begin{subequations}\begin{align}
  &
  B_{X^r}[\vec{R}] = 
 \notag \\ &
  -
  \Lbra{X[\vec{R}]} \Big[ \delR W^{r}[\vec{R}] \frac{1}{W[\vec{R}]} \Big]
  \times \Lket{ \delR   \inst{\rho}[\vec{R}]} 
  \label{eq:PseudoMagneticFieldConserveda}
  \\ &
   -
  \Lbra{\delR X[\vec{R}]}  W^{r}[\vec{R}] \frac{1}{W[\vec{R}]} 
  \times \Lket{ \delR   \inst{\rho}[\vec{R}]} 
  .
  \label{eq:PseudoMagneticFieldConservedb}
\end{align}\label{eq:PseudoMagneticFieldConserved}\end{subequations}
The second term is entirely due to the above mentioned possible parameter dependence of the ``bare'' (ungauged) observable.
It is absent for constant $\hat{X}[\vec{R}]=\hat{X}$ such as charge or spin.
In this case the remaining term \Eq{eq:PseudoMagneticFieldConserveda}
nicely shows that nonzero pumping requires spatial symmetry breaking.
For example, for two reservoirs one expects that some electric or thermal bias or different coupling strengths at the two junctions 
is needed to break spatial symmetry for at least some of accessed parameters $\vec{R}$.
Otherwise, there is no net preferred direction of transporting charge
and the time-dependent charge current averages out over one pumping cycle.
This is clearly expressed by \Eqs{eq:ResponseVectorRoman} and \eq{eq:PseudoMagneticFieldConserveda}:
in such a case the kernels $W^{r}[\vec{R}]$, $W[\vec{R}]$  are proportional,
i.e., the fraction $W^{r}[\vec{R}] W[\vec{R}]^{-1}$ is constant.
The curvature then vanishes because the covector is \emph{zero} [not just parallel to $\delR \inst{\rho}$ as in \Eq{eq:covector-para}]:
\begin{align}
  \Lbra{\delR \Phi_{X^r}[\vec{R}]} = 0
  \label{eq:covector-zero}
  .
\end{align}
This then implies that there is no pumping of \emph{any} such conserved, constant observable $\hat{X}$.
For example, for a quantum dot with symmetric coupling to two normal metal electrodes $r=\text{L},\text{R}$ one finds
$W^{r}[\vec{R}] W[\vec{R}]^{-1} = \tfrac{1}{2} \mathcal{I}$ independent of $\vec{R}$ and $r$.
As a result,  
$\Lbra{\Phi_{X^r}}
 = - \tfrac{1}{2} \Lbra{X}
 =  - \tfrac{1}{2} \tr \, \hat{X}$
for the quantum dot observables
such as charge (${N}$) and spin (${S}_x$, ${S}_y,$ or ${S}_z$).

Finally, we note that the term \eq{eq:PseudoMagneticFieldConserveda}
can also be related to the expressions for the simple case of pumping of \emph{system} observables obtained using the \ar approach of \Ref{Avron12},
see \App{app:pumpsys}.

\section{Full counting statistics approach to geometric pumping\label{sec:fcs}}

We have completed our discussion of the Landsberg-\ar framework for geometric pumping.
Now we turn to the final question (iv) raised in \Sec{sec:questions} by comparing this approach with the established geometric \fcs density-operator approach~\cite{\SINITSYN},
which has been applied to various problems~\cite{\FCS}.
We first review its formulation for arbitrary moments of an observable $\hat{X}^r$,
making the same general assumptions\footnote{
We assume $[X^r,H^r]=0$ as in Eq. (15) of \Ref{Esposito09}.} as we made in \Sec{sec:model}.
Then we highlight some insights offered by the close analogy to Sarandy and Lidar's earlier geometric \ase approach to adiabatic mixed-state evolution \cite{Sarandy05}.
Then, focusing on the first moment only, we discuss how the \fcs formalism reduces exactly to the Landsberg-\ar approach:
first by giving a very simple \textit{a priori} argument
and then by simplifying the explicit \fcs expressions using standard perturbation theory in the counting field $\chi$.
Using the latter explicit relation, we show how the clear picture of the physical gauge freedom in pumping obtained in the \ar approach can be transferred to the \fcs approach.
Finally, we discuss how the ``adiabatic Berry phase'' of the \fcs can produce pumping effects of nonadiabatic origin and how the corresponding  ``adiabatic'' restriction on the driving relates to this.

\subsection{\fcs density-operator approach\label{sec:fcsApproach}}
The core idea of the full counting statistics density-operator approach~\cite{Bagrets03,Esposito09rev} is to describe transport processes by studying a single object, a generating function $Z^\chi(t)$.
This function incorporates the statistics for \emph{all} moments of a selected reservoir observable $\hat{X}^r$ (or several~\cite{Nakajima15}).
Its use is thus broader than the \ar approach
 based on the master equation \eq{eq:MarkovMe} complemented by an observable current kernel \eq{eq:MarkowFlux}.
Instead of taking the route via \Eq{eq:Xrpump2} by focusing on the current $d\hat{X}^r/dt$,
the \fcs approach follows the route via \Eq{eq:Xrpump1}, computing the cumulative \emph{change} of the expectation values of $\hat{X}^r$
between time $t$ and an initial time $t_0$.
In fact, the \fcs generalizes the transported observable \Eq{eq:Xrpump1} to  all moments $n=0,1,2,\ldots$
\begin{align}
  (\Delta {X}^r)^n(t) :=  \braket{ \hat{T} \left ( \hat{X}^r (t) - \hat{X}^r(t_0) \right )^n }
  ,
  \label{eq:deltaXndef}
\end{align}
where $\hat{T}$ denotes time ordering and $\hat{X}^r(t)$ is the Heisenberg-picture operator. 
It is obtained from the generating function $Z^\chi(t)$ by
\begin{align}
  (\Delta {X}^r)^n(t) = \partial^n_{i \chi} \left . Z^\chi(t) \right |_{\chi=0}
  ,
  \label{eq:PumpedObservable}
\end{align}
taking the steady-state limit $t_0 \to - \infty$ [cf. \Eq{eq:Xrpump} ff.].
Here the variable $\chi \in (-\pi,\pi )$ is the counting field
and we use the shorthand $\partial_{i\chi} := -i \partial/\partial \chi$.
Although we will not indicate this,
it is important to keep in mind that $Z^\chi$, and below any object depending on $\chi$, is specific to the selected observable $\hat{X}^r$.

The function $Z^\chi(t)$ is a transform of the probability density $P(t;\Delta X^r)$ to obtain a change $\Delta X^r$ in the discrete two-point measurement outcomes of observable $ \braket{\hat{X}^r}$ at the start and end of the time-interval $[t_0,t]$:
\begin{align}
  Z^\chi(t) := \int d\Delta X^r P(t; \Delta X^r) e^{i \chi \Delta X^r}
  .
\end{align}
In the following some general properties will be important:
(a) normalization of the probability distribution $P(t; \Delta X^r)$ implies at $\chi=0$
\begin{align}
  \left. Z^\chi(t) \right|_{\chi=0} = 1 \text{ for all $t \geq t_0$}
  .
  \label{eq:Znorm}
\end{align}
(b) Hermiticity of the observable $\hat{X}^r$ requires
\begin{align}
  [ Z^{\chi}(t) ]^{*} =  Z^{-\chi}(t)  \text{ for all $t \geq t_0$ and all $\chi$}
  \label{eq:Zherm}
  .
\end{align}
(c) At $t_0$ the changes in $\expec{X^r}$ are zero by definition \eq{eq:deltaXndef}:
\begin{align}
  Z^\chi(t_0) = 1 \text{ for all $\chi$}
  .
  \label{eq:Zinitial}
\end{align}

To obtain the required time-evolution of the generating function $Z^\chi(t)$, one studies an auxiliary operator $\rho^\chi(t)$ (here referred to as \emph{generating operator})  that produces this function upon tracing over the system:
\begin{align}
  Z^\chi(t) = \tr \, \rho^ \chi(t)
  .
  \label{eq:GeneratingFunction}
\end{align}
It is crucial for the following that in this step the \fcs approach introduces an additional redundancy, irrelevant for the gauge freedom, in the description: $\rho^ \chi(t)$ can be shifted by any time- and $\chi$-dependent traceless operator without altering $Z^\chi(t)$:
\begin{align}
  \rho^\chi(t) \to \rho^\chi(t) + \theta^\chi(t) .
  \label{eq:redundancyrhochi}
\end{align}
There is a corresponding redundancy also in the \ar approach [cf. \eq{eq:redundancy-ar}].

As reviewed in \Ref{Esposito09rev} the generating operator can in turn be expressed as
\begin{align}
  \rho^\chi(t) = \Tr{\text{res}} \, \rho^{\text{tot},\chi}(t)
  \label{eq:rhochi}
  ,
\end{align}
the partial trace over a generating operator $\rho^{\text{tot},\chi}(t)$ for system plus reservoir that evolves under a formal non-unitary time-evolution from the initially factorizing density operator \eq{eq:rhototfact}:
\begin{subequations}\begin{align}
  \rho^{\text{tot},\chi}(t)
   =& 
 U^{{\chi}/{2}}(t-t_0) \, \tot{\rho}(t_0) \, [U^{-{\chi}/{2}}(t-t_0)]^\dagger
  ,
  \\
  U^{\chi}(t) = & e^{i \chi X^r} U(t) e^{- i \chi X^r}
  .
\end{align}\label{eq:rhochitot}\end{subequations}
where $U(t)=e^{-i (H^0+V)t}$.
Here all driving parameters are frozen [cf. \Eq{eq:frozen}] and the limit $t_0 \to - \infty$ is taken after the reservoir trace.
This introduces a strong \emph{formal} analogy of the generating operator $ \rho^\chi(t) = \tr_\text{res} \rho_\text{tot}^\chi(t)$ to the reduced density operator $\rho(t)$ discussed in the \ase approach
will be exploited below.
An elegant aspect of the \fcs approach is that by the definitions \eq{eq:rhochi}-\eq{eq:rhochitot} the quantum state evolution is included in the $\chi=0$ part of the generating operator
\begin{align}
  \left. \rho^{\chi}(t) \right|_{\chi=0} & = {\rho(t)}  \text{ for all $t \geq t_0$}
  ,
  \label{eq:rhochizero}
\end{align}
ensuring condition \eq{eq:Znorm} holds,
\begin{align}
   \tr \, \rho^\chi(t)|_{\chi=0} =1
  .
   \label{eq:trrhochi}
\end{align}
Furthermore the dynamics \eq{eq:rhochitot} ensures condition \eq{eq:Zherm} by
\begin{gather}
  \rho^\chi(t)^\dag = \rho^{-\chi}(t)
    \label{eq:rho-chi} \qquad \text{ for all $\chi$}
  .
\end{gather}
The $\chi$-independent initial condition for the dynamics
\begin{align}
  \rho^{\chi}(t_0)  & = \rho(t_0) \text{ for all $\chi$}
  \label{eq:rhochiinitial}
\end{align}
guarantees condition \eq{eq:Zinitial}.
The flip side of this inclusion of $\rho(t)$ in $\rho^\chi(t)$ is that it becomes less clear what the ``adiabatic'' approximation for the formal time-evolution within the \fcs approach entails physically  [\Sec{sec:fcsAdiabatic}].
From $\rho^\chi(t)$ all moments of the transported observable can be obtained by \Eq{eq:PumpedObservable} and \eq{eq:GeneratingFunction}.

\subsection{\fcs approach to pumping\label{seq:fcsPumping}}

\emph{Born-Markov counting kernel.}
To calculate the generating operator $\rho^\chi(t)$ of the \fcs we can now exploit its analogy to the quantum state $\rho(t)$ in the \ase.
It allows, for example, to calculate the time-nonlocal kernels for the time-evolution of $\rho^\chi(t)$ using techniques developed for a quantum-state evolution kernel [cf. \Eq{eq:kineq-general}], e.g., using Nakayima-Zwanzig projections~\cite{Flindt08,Esposito09rev} or real-time diagrams~\cite{Braggio05},
with minimal modifications.
Using either technique one finds in the simple\footnote{Beyond the first moment, non-Markovian, higher order effects and initial correlations are important as shown in \Ref{Braggio05}.} Born-Markov limit that
the generating operator obeys a \fcs master equation~\cite{Bagrets03} with the counting field $\chi$ as a formal continuous parameter:
\begin{align}
  \frac{\partial}{\partial t} \Lket{\rho^\chi (t)} = W^\chi(t) \Lket{\rho^\chi(t)}
  .
  \label{eq:GeneralizedMasterEquation}
\end{align}
Here we have again introduced the Liouville-space notation \eq{eq:liouville} for operators, ${\rho^\chi(t)}=\Lket{\rho^\chi(t)}$.
The concrete expression for the \fcs kernel $W^\chi$ is not crucial for the following.
In \App{app:fcs} we give an explicit example for charge pumping through a single level quantum dot.
To preserve the general properties \eq{eq:rhochizero} and \eq{eq:rho-chi}
of the observable statistics
the \fcs kernel is restricted by, respectively,
\begin{align}
   W^\chi(t)|_{\chi=0} = W(t)
  ,
  \label{eq:generalizedtrace}
\\
  W^\chi(t) = K W^{-\chi}(t) K 
  ,
  \label{eq:generalizedHermiticity}
\end{align}
where
$K x := x^\dag$ denotes Hermitian conjugation.\footnote{$K$ is the antilinear superoperator that effects Hermitian conjugation of an operator $x$, see \Ref{Saptsov12a}, App. G.}

\emph{Pumping.}
We now outline how the \fcs approach applies to {pumping}.
In this case, the driving of the parameters $\vec{R}(t)$ is responsible for the time dependence, i.e., $W^\chi(t) \to W^\chi[\vec{R}(t)]$
in \Eq{eq:GeneralizedMasterEquation}.
Instead of following the original works~\Refs{Sinitsyn07EPL,Sinitsyn09}
we highlight the analogy to Sarandy and Lidar's earlier \ase approach~\cite{Sarandy05,Sarandy06} (summarized in \App{app:ase}) which can be applied to the \fcs master equation \eq{eq:GeneralizedMasterEquation}
with the sole modification of keeping track of the additional parameter $\chi$.

Thus, an approximate solution of the \fcs master equation \eq{eq:GeneralizedMasterEquation}
is obtained by first diagonalizing $W^\chi$ for fixed parameters $\vec{R}$ (instantaneous solution)
and then neglecting the couplings between different eigenspaces in the dynamics (see \App{app:ase}).
Formally similar to  adiabatic state dynamics, in the steady-state limit
only the left and right eigenvectors of $W^\chi(t)$ for the eigenvalue\footnote{As in the \ar approach, the kernel's eigenvalue with largest real part is assumed to be nondegenerate in the \fcs approach.} $\lambda^\chi_0(t)$ with largest real part
are required since the contributions of all other terms are exponentially smaller.
The condition for the validity of this approximation will be discussed at the end in \Sec{sec:fcsAdiabatic},
clarifying what the ``adiabaticity'' assumed in the \fcs physically entails.

Also analogous to the \ase approach is that the gauge freedom in the \fcs approach 
lies in the freedom of choosing the eigenvectors\footnote{See Eq. (19) ff. in \Ref{Sinitsyn07EPL} and p. 25. of \Ref{Sinitsyn09}.}:
for every $\vec{R}$ they can be multiplied by any nonvanishing complex function of $\chi$,
which we will discuss in more detail below [\Eq{eq:CountingFreedom}].
We will always choose\footnote{This is possible for driving cycles that can be covered by one single coordinate patch of the parameter manifold. For other cycles one can glue the solutions together in the standard way using the gauge invariance of the curvature, see, e.g., \Ref{Bohm}.}
the right eigenvector for eigenvalue $\lambda^\chi_0(t)$, denoted by $\Lket{v^\chi_0[\vec{R}]}$, to depend continuously on $\vec{R}$.
Normalization \eq{eq:generalizedtrace} further requires that
\begin{align}
  \Lbraket{\mathds{1}|v^\chi_0[\vec{R}]} |_{\chi=0}  & = 1
  .
  \label{eq:tracegauge}
\end{align}
The remaining gauge freedom amounts to specifying the trace of the operator $\Lket{v^\chi_0}$ as function of $\chi$.
In the following we choose
\begin{align}
  \Lbraket{\mathds{1}|v^\chi_0[\vec{R}]}   & = 1
                                             \text{ for all $\chi$}
  .
  \label{eq:choice}
\end{align}
It will turn out that by this we fix the physical gauge to the case $g=0$ of the \ar formulation.
We investigate other choices of the \fcs gauge later on.

In addition, it follows from \Eq{eq:generalizedHermiticity} that
$\lambda^{\chi*}_0=\lambda^{-\chi}_0$ and that the operators of the right eigenvectors for opposite counting fields  are related\footnote{
The general relation reads
$
 v^\chi_0[\vec{R}]   = \beta^\chi [v^{-\chi}_0[\vec{R}]]^\dag
$
where $\beta^\chi$ is a nonvanishing complex function
which is restricted as
$
  \beta^{\chi*} \beta^{-\chi} =1
  \label{eq:alpha}
$ by consistency when taking the adjoint of the relation and setting $\chi \to -\chi$.
By taking $\beta^\chi=1$ for all $\chi$ we obtain \Eq{eq:hermgauge}.
}:
one can always choose this eigenvector to additionally satisfy for any $\vec{R}$ and  $\chi$
\begin{align}
  v^\chi_0[\vec{R}]  & =[v^{-\chi}_0[\vec{R}]]^\dag
  .
  \label{eq:hermgauge}
\end{align}

With these choices, the ``adiabatic'' solution of \Eq{eq:GeneralizedMasterEquation} in the long time limit is given by
\begin{align}
  \Lket{\FCSinst{\rho}^\chi(t)}
  & = Z^\chi(t) \, \Lket{v^\chi_0[\vec{R}(t)]}
  \label{eq:AdRhoChit}
   ,
\end{align}
where we have set the initial condition $( \bar{v}_0^\chi [\mathbf{R}(0)] | \rho (0) ) = 1$  to ensure $Z^\chi(0) = 1$. 
This redefines $Z^\chi(t)$ to be the \emph{steady state} generating function relative to $t=0$ rather than $t=t_0$.
Only this quantity can be compared to the results of the AR approach, in which the steady state limit is taken from the start [Eq. (21) ff.].
Here, the label ``i'' is chosen in view of our later comparison with the \ar approach (cf. \Sec{sec:fcsAdiabatic}).
It should be noted that after the ``adiabatic'' approximation the $\chi=0$ part does \emph{not} keep track of the quantum state in the same way as the \ar does [$\adcor{\rho}(t)$ is missing].
Remarkably, as we will see, it does keep track of the pumping effects~\cite{Nakajima15} in the specific observable $\hat{X}^r$ that in \ar are caused by this missing term $\adcor{\rho}(t)$.

Inserting \Eq{eq:AdRhoChit} as an ansatz into \Eq{eq:GeneralizedMasterEquation} and solving along the closed  driving curve $\mathcal{C}$ traversed in period $\T$, one obtains
\begin{align}
  \Lket{\FCSinst{\rho}^\chi({\T})} & =e^{\Lambda^\chi({\T})} e^{\gamma^\chi({\T})} \Lket{v^\chi_0[\vec{R}(0)]} \label{eq:AdRhoChi}
   .
\end{align}
Here, the dynamical phase derives from the eigenvalue
\begin{align}
  \Lambda^\chi({\T}) & := \int_0^{\T} dt \lambda_0^\chi[\vec{R}(t)]
  \label{eq:CountingDynPhase}
  ,
\end{align}
whereas the geometric phase is obtained as
\begin{align}
  \gamma^\chi({\T}) & := -\oint_{\mathcal{C}} d\vec{R} A^\chi[\vec{R}]
  \label{eq:CountingGeoPhase}
\end{align}
from the corresponding left and right eigenvector through
\begin{align}
  A^\chi[\vec{R}] &= \Lbraket{\overline{v}^\chi_0[\vec{R}]|\delR|v_0^\chi[\vec{R}]}
  .
\label{eq:CountingConnection}                    
\end{align}

\emph{Geometric nature of the \fcs of pumping.}
In geometric terms (cf. \Sec{sec:HprizontalLift}),
the generating operator $\Lket{ \inst{\rho}^\chi ({\T}) }$ is, up to the dynamical factor $e^{\Lambda^\chi({\T})}$, a horizontal lift of the closed curve $\mathcal{C}$ in the driving-parameter space.
In the \fcs, the relevant ``vertical'' space attached to each parameter in the $\vec{R}$ plane
consists in the space of all possible instantaneous-eigenvector choices $\Lket{v_0^\chi[\vec{R}]}$ that one can make for the eigenvalue $\lambda_0^\chi[\vec{R}]$.
In this different space, a different notion of ``horizontal'' can be defined using
 the \fcs expression $A^\chi$, thus defining a geometric \fcs connection (gauge potential).
Therefore,
part of the generating operator has the properties of a geometric quantity:
it depends only on the driving cycle $\mathcal{C}$ and the geometric connection $A^\chi[\vec{R}]$
and is independent\footnote{This requires the transformation \eq{eq:Achitransform2} of the \fcs connection.} of the eigenvector gauge choice.
Finally, the total prefactor is just the moment generating function [cf. \Eq{eq:GeneratingFunction}],
\begin{align}
  Z^\chi({\T}) = \Lbraket{\mathds{1}|\rho^{\chi,i}({\T})} = e^{\Lambda^\chi({\T})} e^{\gamma^\chi({\T})}
  ,
  \label{eq:GeneratingFunction2}
\end{align}
and its exponent is by definition the cumulant\footnote{Since we focus on the first moment, cumulants need not be introduced here.}
 generating function.
Hence, the \fcs geometric ``phase'' is just the  geometric part of the cumulant generating function~\cite{Sinitsyn07EPL,Sinitsyn09}.
Its magnitude can be related by Stokes theorem to an \fcs curvature (gauge potential):
\begin{align}
  B^\chi[\vec{R}] :=  \Lbra{\delR \bar{v}^\chi_0[\vec{R}] } \times  \Lket{ \delR v^\chi_0[\vec{R}] } 
  \label{eq:Bchi}.
\end{align}
This comprises the \fcs approach to pumping.

\emph{First moment of the pumped observable.}
A merit of the \fcs  approach is that it provided the first density-operator formulation of geometric pumping applicable to strongly interacting systems.
Moreover, \fcs deals with the geometric nature of the \emph{entire} pumping process (all moments).
One extracts the first moment of the transported observable
$\Delta X^r = \Delta \inst{X}^r + \Delta \adcor{X}^r$ by
\begin{align}
  \Delta \inst{X}^{r} &= \partial_{i \chi}  \Lambda^\chi(T) \big|_{\chi=0}
 ,
 \label{eq:CountingStatisticsDynamical}\\
  \Delta \adcor{X}^{r} &= \partial_{i \chi}  \gamma^\chi(T) \big|_{\chi=0}
 .
 \label{eq:CountingStatisticsPumping}
\end{align}
The pumped \observable \eq{eq:CountingStatisticsPumping} is thus given by the \emph{$\chi$ derivative of} a geometric-phase \emph{function of $\chi$}
 obtained from the ``adiabatic'' solution of the counting master equation.
Although \Ref{Nakajima15} explicitly showed that after involved manipulations this coincides with the simple \ar result \eq{eq:pumpedX},
the pumped first moment does not emerge directly as geometric phase by itself:
due to the $\chi$-dependence it is not yet clear
what is the differential-geometric meaning of
$
\partial_{i \chi}  A^\chi[\vec{R}] \big|_{\chi=0}
$ and how it transforms under the \emph{physical} gauge transformations [\Eq{eq:Xrgauge}] that we found in the \ar approach.
To see this, we need to linearize \Eq{eq:AdRhoChi} with respect to the counting field $\chi$.

\subsection{Linearization in the counting field
\\ - Reduction to Landsberg-\ar approach\label{sec:fcsExpansion}}

\subsubsection{Equivalence of \fcs and \ar for first moment\label{sec:fcs-equivalence}}

Before specializing to pumping situations in the following sections,
we first show that the \fcs master equation, as regards the description of the first moment of an observable $\hat{X}^r$,
is exactly equivalent\footnote{Compare with a similar derivation given in \Ref{Nakajima15} see Eq. (22) there.} to the equations that form the \emph{starting point} of the \ar approach.
In the generating operator
\begin{align}
  \Lket{\rho^\chi(t)} &= \Lket{\rho(t)} + i \chi \Lket{\widehat{\Delta X}(t)}  + O(\chi^2)
  ,
  \label{eq:rhochi-linear}
\end{align}
the $\chi$-linear term is an operator that we denote here by
\begin{align}
  \widehat{\Delta X^r}(t) & := \partial_{i \chi} \left.  \rho^ \chi(t) \right|_{\chi=0}
	.
\end{align}
By definition \eq{eq:PumpedObservable} with $t_0 \rightarrow 0$ it produces the first moment by taking its trace over the system space (not: expectation value):
\begin{align}
  \Delta X^r(t) & = \tr \, \widehat{\Delta X^r}(t)
\label{eq:first-moment}
.
\end{align}
The operator $\widehat{\Delta X^r}(t)$ is thus \emph{not} an observable,
but just an auxiliary quantity to compute two-point measurement outcomes.
We now insert the $\chi$-linearization  of the generating operator \eq{eq:rhochi-linear}
and of the Born-Markov generator
\begin{align}
	W^\chi (t) & = W(t) + i\chi \partial_{i\chi} W^\chi(t)|_{\chi=0} + O(\chi^2)
        \label{eq:Wchi-linear}
\end{align}
into \Eq{eq:GeneralizedMasterEquation}
and compare the terms by powers of $\chi$.
The zeroth order of \Eq{eq:GeneralizedMasterEquation} accounts for the quantum state [\Eq{eq:rhochizero}] and the generator of its evolution,
\begin{align}
  \left. W^{\chi}(t) \right|_{\chi=0}  &= W(t)
	,
\label{eq:CountingExpansionDensityZero}
\end{align}
and gives the Born-Markov master equation \Eq{eq:MarkovMe},
$\frac{d}{dt} \Lket{\rho}=W\Lket{\rho}$.

The terms linear in $\chi$ give an equation of motion for
the operator $\widehat{\Delta X^r}(t)$ of \Eq{eq:first-moment}:
\begin{align}
  \frac{d}{dt} \Lket{\widehat{\Delta X^r}(t)} =
 \partial_{i\chi} W^\chi(t)|_{\chi=0} \Lket{\rho(t)} + W
\Lket{\widehat{\Delta X^r}(t)}
  \label{eq:linearorder}
	.
\end{align}
 The last term constitutes a redundant part because it is traceless by
probability normalization, $\tr W=0$.
We take the trace and comparing with the \ar result \eq{eq:MarkowFlux} expressed in the physical current kernel:
\begin{subequations}\begin{align}
  \frac{d}{dt}  \Delta X^r  (t)
  &= 
  \Lbra{\mathds{1}} \partial_{i\chi} W^\chi(t)|_{\chi=0} \Lket{\rho(t)}
    \\
   \stackrel{!}{=} \quad
  \frac{d}{dt}  \langle {\hat{X}^r} \rangle (t)
  & =
  \Lbra{\mathds{1}} \currentKernel{X^r} \Lket{\rho(t)}
  .
\end{align}\label{eq:compare}\end{subequations}
We conclude that the $\chi$-linear term of $W^\chi$ must be the \emph{current kernel} up to some time-dependent superoperator $\Theta(t)$ with $\Lbra{\unit}\Theta(t) =\tr \Theta(t) = 0$ [\Eq{eq:liouville}]:
\begin{align}
  \partial_{i\chi} W^\chi(t)|_{\chi=0} = \currentKernel{X^r}(t)+\Theta(t)
  \label{eq:Lambda}.
\end{align}
This $\Theta$ reflects that in the FCS and AR approaches one may choose the redundancy for the current kernel differently, see \App{app:fcs} for an example:
$\Theta$ is the difference between these conventions and can be dropped:
\begin{align}
  W^\chi (t) & = W(t) + i\chi \currentKernel{X^r}(t) + O(\chi^2) .
  \label{eq:CountingKernelExpansion}
\end{align}

With the $\Theta$-redundancy out of the way, it is now immediate from the linear expansions \eq{eq:rhochi-linear} and \eq{eq:CountingKernelExpansion} that the \emph{physically relevant, tracefull} part of the \fcs equation \eq{eq:GeneralizedMasterEquation} is exactly equivalent
to those of  the ``naive'' \ar approach ($g=0$) to \emph{pumping}. When consistently applied, these two approaches should thus
produce identical answers (\Sec{sec:fcsPumping}),
exhibit the same gauge freedom (\Sec{sec:fcsGauge}),
keep track of nonadiabatic ``lag'' and have the same limits of applicability (\Sec{sec:fcsAdiabatic}).

\subsubsection{Pumping formulas: connection and curvature\label{sec:fcsPumping}}
We now follow how the \fcs result for the pumped first moment
simplifies to the \ar result in practice.
Although this \emph{explicit} equivalence has been shown in \Ref{Nakajima15}, we here present an alternative derivation.
It employs  more standard operations and is easily extended to further important aspects discussed in the following.

The physically motivated form of the linearization of the counting kernel \Eq{eq:CountingKernelExpansion}  suggests how to proceed:
we calculate the eigenvectors perturbatively to first order in $i\chi$ utilizing the known unperturbed ($\chi=0$) eigenvectors of $W$ for the state evolution 
by treating the physical current kernel $\currentKernel{X^r}$ as perturbation.
Not writing the $\vec{R}$ dependence, the formulas for the $n$-th nondegenerate eigenvalue $\lambda^\chi_n$ of $W^\chi$ and its right ($\Lket{v_n }$) and left ($\Lbra{\overline{v}_n }$) eigenvectors to linear order in $i\chi$ are:
\begin{subequations}\begin{align}
  \lambda_n^\chi =& \lambda_n + i\chi \Lbraket{\overline{v}_n | \currentKernel{X^r} | v_n}, \\
  \Lket{v^\chi_n} =& \Lket{v_n }
- i\chi \sum_{m\neq n} \Lket{v_m} \frac{\Lbraket{\overline{v}_m|\currentKernel{X^r}|v_n}}{\lambda_m-\lambda_n}, \\
  \Lbra{\overline{v}^\chi_n} =& \Lbra{\overline{v}_n }
+ i\chi \sum_{m\neq n}                       \frac{\Lbraket{\overline{v}_n|\currentKernel{X^r}|v_m}}{\lambda_n-\lambda_m} 
                                \Lbra{\overline{v}_m}
  .
\end{align}\label{eq:perturb}\end{subequations}
Here $\lambda_n$, $\Lket{v_n }$, $\Lbra{\overline{v}_n }$ denote the corresponding quantities at $\chi=0$, i.e., those discussed for the quantum-state evolution studied in the \ase approach, see \Sec{sec:fcsAdiabatic} and \App{app:ase}.

The eigenvalue with largest real part, labeled by $n=0$, is nondegenerate by assumption.
As the eigenvectors and eigenvalues of  $\left. W^\chi \right |_{\chi=0}=W$ are known, we use $\lambda_0 =0$, $\Lket{v_0} = \Lket{\inst{\rho}}$ and $\Lbra{\overline{v}_0}=\Lbra{\mathds{1}} = \tr$ to obtain in leading order $i\chi$:
\begin{subequations}\begin{align}
  \lambda_0^\chi = &
    i\chi \Lbra{\mathds{1}} \currentKernel{X^r} \Lket{ \inst{\rho} },
\label{eq:CountingLeftEvExpandeda}
  \\
  \Lket{v^\chi_0} = &
  \Lket{\inst{\rho} }
    - i\chi  \frac{1}{W} \currentKernel{X^r}\Lket{\inst{\rho} }
		\label{eq:CountingRightEvExpandedc}
  \\   
\Lbra{\overline{v}^\chi_0} =&
 \Lbra{\mathds{1}}- i\chi \Lbra{\mathds{1}} \currentKernel{X^r}  \frac{1}{W} 
\label{eq:CountingLeftEvExpandedc}
     .
\end{align}\label{eq:CountingLeftEvExpanded}\end{subequations}
Here, as before: $W^{-1}$ denotes the pseudo inverse excluding the zero eigenvalue.
Inserting these expansions, the instantaneous part of the transported observable [\Eq{eq:CountingDynPhase}, \eq{eq:CountingStatisticsDynamical}] simplifies to
\begin{align}
  \Delta \inst{X}^{r}
  &=  \partial_{i \chi} \int_{0}^{\T} dt \ \left .  \lambda^\chi(t) \right |_{\chi=0}\notag \\
  &=  \int_{0}^{\T} dt \ \Lbra{\mathds{1}} \currentKernel{X^r}[\vec{R}(t)] \Lket{ \inst{\rho} [\vec{R}(t)]}
    \label{eq:SimplifiedCountingInst}
\end{align}
and the pumping part [\Eq{eq:CountingGeoPhase}, \eq{eq:CountingStatisticsPumping}] reduces to 
\begin{align}
  \Delta \adcor{X}^{r}
   =
  -
  \int_{C} d\vec{R} \partial_{i \chi} \left. A^\chi[\vec{R}] \right |_{\chi=0} 
   =
  \int_{C} d\vec{R} A_0[\vec{R}]
\label{eq:SimplifiedCountingPumping}
\end{align}
In the last step we used trace normalization for $\chi=0$, $\Lbraket{\overline{v}_0[\vec{R}]|\delR |v_0[\vec{R}]}=0$,
and we computed
\begin{align}
\partial_{i \chi} \left. A^\chi[\vec{R}] \right |_{\chi=0}
  &=  \left. \partial_{i \chi} \Lbraket{\overline{v}^\chi_0[\vec{R}]|\delR|v_0^\chi[\vec{R}]}  \right |_{\chi=0}  \notag\\
  &=  \left. \Big[ \partial_{i \chi} 
\Lbra{\overline{v}^\chi_0[\vec{R}]}
 \Big] 
\delR \Lket{  v^{\chi}_0[\vec{R}] }
\right|_{\chi=0}   \notag\\
  &=  -  \Lbra{\mathds{1}} \currentKernel{X^r}[\vec{R}] \frac{1}{W[\vec{R}]}  \delR \Lket{\inst{\rho}[\vec{R}]}  \notag\\
  &=  -  \Lbra{\mathds{1}} \currentKernel{X^r}[\vec{R}] \Lket{\adcor{\rho}[\vec{R}]}  \notag\\
  & = - A_0[\vec{R}]
\label{eq:dAchi}\end{align}
where $A_0$ is the gauge-independent part of the Landsberg connection \eq{eq:A0}.
Apart from the Liouville notation $\Lbra{\mathds{1}}=\tr$ and $\Lket{\inst{\rho}}=\inst{\rho}$, these are the expressions \eq{eq:Obs} and \eq{eq:adCorrObs}
obtained directly from the ``naive'' \ar approach to pumping, i.e., in the $g=0$ gauge.
When accounting for the conditions \eq{eq:tracegauge} and \eq{eq:hermgauge}, the $\chi$ linearization of the \fcs connection
in the \fcs gauge \eq{eq:choice} thus reduces exactly to the Landsberg connection of the \ar approach
\emph{in the $g=0$ gauge}:
\begin{align}
  A^\chi(t) = - i\chi A_0(t) + O(\chi^2)
  .
  \label{eq:Achilinear}
\end{align}
It follows that
the $\chi$-linear part of the \fcs curvature, $B^\chi := \delR \times A^\chi$,
which gives the first-moment pumping per unit parameter surface,
reduces exactly to the Landsberg \ar curvature $B = \delR \times A_0$:
\begin{align}
  B^\chi[\vec{R}] = -i\chi B[\vec{R}] + O(\chi^2)
  \label{eq:Bchilinear}
  .
\end{align}
The results for the pumped observable are identical
since both sides are gauge invariant.
However, the gauge transformations in the \fcs and \ar are two different, but related, constructions that will be discussed next.
As a practical matter, we note that before taking $\chi \to 0$ the explicit \fcs curvature, \Eq{eq:Bchi} does not seem to separate into physically distinct factors coming from the observable (response vector) and from the state,
as it does in the \ar curvature \eq{eq:PseudoMagneticField}.

\subsubsection{Gauge freedom and geometry of pumping\label{sec:fcsGauge}}

We now relate the \fcs and \ar approach to each other on the level of the gauge freedom.
This will allow us to clarify a few more points that were not addressed in \Ref{Nakajima15}.
In the \fcs approach, the choice of the (non-degenerate) eigenvector with largest real part for fixed $\chi$ and fixed parameters $\vec{R}$ is left free up to multiplication by a nonzero complex factor
\begin{subequations}\begin{alignat}{2}
  \Lket{v^\chi_0[\vec{R}]}
  &\rightarrow
  g^\chi[\vec{R}] \, \Lket{v^\chi_0[\vec{R}]} \label{eq:CountingFreedomRight}\\
  \Lbra{\overline{v}^\chi_0[\vec{R}]}
  &\rightarrow
  \frac{1}{g^\chi[\vec{R}]}
  \Lbra{\overline{v}^\chi_0[\vec{R}]} \label{eq:CountingFreedomLeft}. 
\end{alignat}\label{eq:CountingFreedom}\end{subequations}
This preserves the biorthonormality $\Lbraket{\overline{v}^\chi_0[\vec{R}]|v^\chi_0[\vec{R}]} = 1$
required of left and right eigenvectors
but changes the \fcs gauge \eq{eq:choice} to $\Lbraket{\unit | v^\chi[\vec{R}]}=g^\chi[\vec{R}]$.
To maintain the  conditions \Eq{eq:tracegauge} and \eq{eq:hermgauge}
we need to restrict the gauge transformations of the \fcs approach  by
(a) probability normalization
\begin{align}
   \left. g^{\chi}[\vec{R}] \right|_{\chi=0} = 1 \quad \text{for all $\vec{R}$}
  \label{eq:gnorm}
  ,
\end{align}
and (b) observable Hermiticity
\begin{align}
	[g^{\chi}[\vec{R}]]^{*} = g^{-\chi}[\vec{R}] \quad \text{for all $\vec{R}$ and $\chi$}
  \label{eq:gherm}
  .
\end{align}
Sinitsyn\footnote{See  Eq. (20) ff. in \Ref{Sinitsyn07EPL}.} emphasized that 
the generating operator is restricted by trace normalization \eq{eq:generalizedtrace} \emph{only} for $\chi = 0$, but not for $\chi \neq 0$.
This point has received little further attention,
 but turns out to provide the crucial link to the real-valued gauge freedom in the \ar approach, related to the physical calibration of the meter discussed in \Sec{sec:geometry}.
To connect this to the \fcs, we consider the $\chi$ linearization of the gauge transformation
\begin{align}
  g^\chi[\vec{R}] & = g^0[\vec{R}] + i\chi g^{1}[\vec{R}]  + \text{O}(\chi^2)
  ,
  \label{eq:gchilinear}                    
\end{align}
which is determined by the first two Taylor coefficients $g^0[\vec{R}]$ and $g^1[\vec{R}]$,
both being functions of the driving parameters.
The restrictions \eq{eq:gnorm} and \eq{eq:gherm} imply 
\begin{align}
  g^0[\vec{R}] =1, \qquad g^{1}[\vec{R}] \in \mathds{R} \text{ for all $\vec{R}$}
  .
  \label{eq:g0g1simple}
\end{align}

The identification of the gauge freedom now follows
by comparing the  gauge transformation appropriate to each connection.
Under the transformation \eq{eq:CountingFreedom} the \fcs connection $A^\chi[\vec{R}]$ given by \Eq{eq:CountingConnection} changes
to
\begin{subequations}\begin{align}
    A^{\chi}_{g^\chi}[\vec{R}]
   & =
    \Lbraket{\overline{v}_0^\chi[\vec{R}]|\frac{1}{g^\chi} \delR \Big[ g^\chi | v_0^\chi[\vec{R}]}  \Big] \\
   & =
     A^{\chi}[\vec{R}] + \frac{1}{g^\chi[\vec{R}]}\delR g^\chi[\vec{R}]
     \label{eq:Achitransform2}\\
   &=
     - i \chi \left[  A[\vec{R}] - \delR g^1[\vec{R}] \right] + \text{O}(\chi^2) \label{eq:Achitransform}
		.
\end{align}\end{subequations}
Comparing the $\chi$-linear part of \Eq{eq:Achitransform} with \Eq{eq:Achilinear} we see that
the Landsberg connection transforms as $A[\vec{R}]\to A_g[\vec{R}] = A[\vec{R}]+\delR g[\vec{R}]$ [\Eq{eq:ConnectionGaugeTransformation}] by
 the gauge transformation $\hat{X}^r \to \hat{X}^r + g[\vec{R}]\unit$ with a smooth, real function
\begin{align}
  g[\vec{R}]
  =
-  g^1[\vec{R}]
  .
  \label{eq:gauge-relation}
\end{align}
apart from an unimportant  constant.
We have thus located the physically relevant gauge freedom \eq{eq:Xrgauge}
in the first Taylor coefficient of the \emph{restricted}  \fcs-gauge function [\Eqs{eq:gnorm} and \eq{eq:gherm}].
The general relation between the \fcs connection and the Landsberg connection of the \ar approach,
both formulated in an arbitrary \emph{physical} gauge, reads as
\begin{subequations}\begin{align}
  A^\chi_{g^\chi}(t)  &= - i\chi A_g(t) + O(\chi^2)
   \label{eq:Achilinearg}, \\
  g^\chi[\vec{R}] & = 1- i\chi g[\vec{R}]+O(\chi^2)
  .
\end{align}\end{subequations}
This shows that the object $\partial_{i\chi} A^\chi_{g^\chi}(t)|_{\chi=0}$
obtained in the \fcs under the curve integral for the first moment [\Eqs{eq:CountingGeoPhase} and \eq{eq:CountingStatisticsPumping}] by itself is indeed a valid geometric connection:
it is just minus the Landsberg connection $A_g$ which has a clear and direct geometric meaning in a physically motivated fiber bundle, independent of the more complicated different geometric structure of the \fcs approach.

\subsubsection{``Adiabatic Berry phase'' of the \fcs approach\label{sec:fcsAdiabatic}}

We finally address the question how the generating operator $\Lket{\inst{\rho}^{\chi}}$,
obtained by an ``\emph{adiabatic}'' solution of the \fcs master equation \eq{eq:GeneralizedMasterEquation}, can produce the pumped observable
generated by the \emph{nonadiabatic} state correction $\Lket{\adcor{\rho}}$ in the \ar approach.
Does not the slow driving required for such ``adiabaticity'' in the \fcs imply that it should neglect such corrections?

\begin{figure}[t]  
  \includegraphics[width=0.9\linewidth]{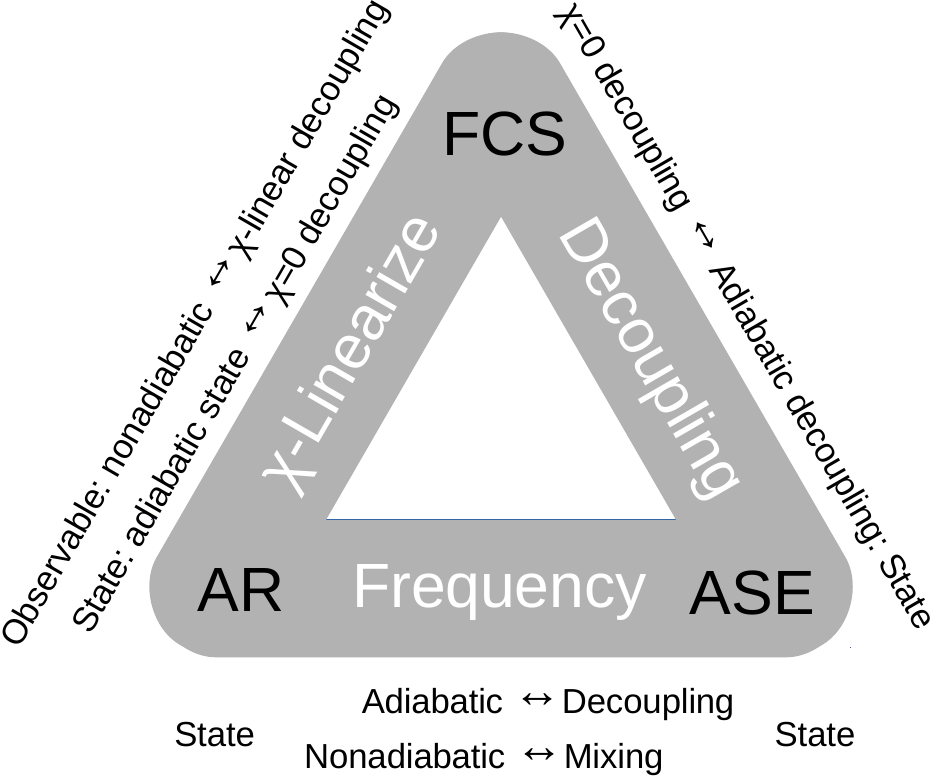}
  \caption{
    Relations between adiabatic-state evolution (\ase, \App{app:ase}),
    adiabatic-response (\ar, \Sec{sec:geometry})
    and full-counting statistics (\fcs, \Sec{sec:fcs}) approach discussed in the text.
  }
  \label{fig:compare}  
\end{figure}

To clarify this, we revisit key points of our comparison now that the details have been taken care of.
We make a three-way comparison of the \fcs, \ar, and \ase approach,
the key relations being illustrated in \Fig{fig:compare}.
The relevance of this issue was recognized in \Ref{Nakajima15},
where the \fcs was denoted as being ``$\chi$-adiabatic'' but without specifying which physical conditions on the driving limit the  applicability of the \fcs.

In the \fcs approach ``adiabatic'' is operationally understood in the same way as in the \ase approach, namely, 
as decoupling of the dynamics of different eigenspaces (cf. \App{app:ase});
this formally connects these approaches, cf. \Fig{fig:compare}.
In the case of the \ase approach we showed that in the steady-state limit the result of the (first correction to the) decoupling equals the \mbox{{(first-)}} zeroth-order term in the \ar frequency expansion of the \emph{quantum state}, denoted there by $\inst{\rho}$ ($\adcor{\rho}$).
Thus, the nonadiabatic term $\adcor{\rho}$ arises from coupling of the instantaneous stationary state to non-stationary decay modes.
This justifies our labeling of the \ase contributions with the corresponding labels ``i'' and ``a'' that were used in the \ar for instantaneous and adiabatic response, respectively.

In the case of the \fcs we have provisionally used the same labeling ``i'' for the ``adiabatic'' \fcs solution $\inst{\rho}^\chi$ to indicate the decoupling.
The first ``nonadiabatic'' correction to this decoupling, $\adcor{\rho}^\chi$, we correspondingly label by ``a''.
The crucial point is that the intimate connection of the decoupling and the frequency expansion, existing between \ase and \ar approaches, is \emph{not} present for the \fcs.
This means that our provisional labeling of the \fcs by ``adiabatic'' or equally by ``i''
is not uniquely related to a physical frequency expansion:
what it means depends on whether $\chi=0$ or $\chi \neq 0$.
In fact, this is unavoidable since precisely by the formal trick of including a counting field $\chi$ the \fcs is able to include the nonadiabatic effects into the framework of a formal ``adiabatic Berry-Simon phase''.

Guided by \Fig{fig:compare},
we now outline (a) how this works out for the terms that are kept in the three approaches
and (b) how the terms that are neglected limit the applicability of each approach.

\paragraph{``Adiabatic'' \fcs solution.}
The ``adiabatic'' solution of the \fcs master equation \eq{eq:GeneralizedMasterEquation}
can be written using \Eq{eq:AdRhoChi}, \eq{eq:CountingStatisticsDynamical} and \eq{eq:CountingStatisticsPumping} as
\begin{align}
  \Lket{ \inst{\rho}^{\chi}(t)} \label{eq:rhochiinst} 
  & = e^{i \chi \left[  \inst{\Delta X}^r(t) + \adcor{\Delta X}^r(t) \right] + \ldots} 
    \Big[ \Lket{\inst{\rho}[\vec{R}(t)]} + \ldots \Big] 
\end{align}
Here only the $\chi$-linear terms are indicated since we exclusively discuss the first moment.
A $\chi$-linear term in the eigenvector  [\Eq{eq:CountingRightEvExpandedc}] is also omitted since it is redundant [zero trace, cf. \Eq{eq:redundancyrhochi}].
From hereon we will not indicate such omissions ($\ldots$) for clarity.

\emph{$\chi=0$: \fcs is adiabatic.}
In this case, the ``adiabatic'' \fcs generating operator \eq{eq:rhochiinst}
only keeps track of the adiabatic, zeroth order in frequency,  result for the state:
\begin{align}
 \left . \Lket{ \inst{\rho}^{\chi}(t)} \vphantom{I^{I^I}}\right |_{\chi=0} = \Lket{ \inst{\rho}[\vec{R}(t)]}
\label{eq:fcsJustNeglected}
.
\end{align}
Thus, only that part of the \ar [\Eq{eq:rhochizero}] and \ase [\Eq{eq:quench}] result
is kept that describes the instantaneous dependence on the driving parameters.
Here the label ``i'' is thus appropriate.

\emph{$\chi\neq0$: \fcs is nonadiabatic.}
In contrast,
the generating operator $\Lket{ \inst{\rho}^{\chi}(t)}$ for nonzero $\chi$  is \emph{not} a function of the instantaneous parameters alone.
By introducing $\chi$, the FCS circumvents the normalization constraint
that prevents a Berry-Simon phase from appearing in the \ar and \ase approach for the steady-state (cf. \Sec{sec:LackOfPhaseState}).
The \fcs gauge freedom \eq{eq:CountingFreedomRight} for $\chi \neq 0$
allows the exponential term in \Eq{eq:rhochiinst} to accumulate a dependence on parameters at all previous times $t' \in [t,0]$.
We found [\Eq{eq:SimplifiedCountingInst}-\eq{eq:SimplifiedCountingPumping}] that at $t=\T$ this produces the sum
of the dynamical term $\inst{\Delta X}^r=\int_0^\T dt' \inst{I_{X^r}}[\vec{R}(t')]$ and the geometric term
$
\adcor{\Delta X}^r =\int_0^\T dt' \adcor{I_{X^r}}[\vec{R}(t')] = \oint_{\mathcal{C}}dR A_0[\vec{R}]
$ as given by the \ar expressions \eq{eq:Obs} and \eq{eq:pumpedX}, respectively  ($g=0$ gauge).
As in the original Berry-Simon situation, the geometric term ($\propto (\dot{\vec{R}})^1$) in the exponent is of one order higher in the driving frequency
 than the dynamical one ($\propto (\dot{\vec{R}})^0$, c.f. \Eq{eq:scaling} and \App{app:ase}).
In this way, the \fcs also includes the pumped observable of nonadiabatic origin by allowing the ``adiabatic'' solution \eq{eq:rhochiinst} to accumulate a phase.

Remarkably, this is an \emph{effect} of the nonadiabatic state correction $\adcor{\rho}$,
that we have just neglected in the $\chi=0$ part [\Eq{eq:fcsJustNeglected}].
How does the \fcs keep track of this effect without calculating $\adcor{\rho}$ explicitly?
A reconsideration of our perturbative treatment of the $\chi$-linearization $W^\chi=W+i\chi\currentKernel{X^r} $ sheds some light on \emph{how} this is achieved
 by combining the $\chi$-bookkeeping with ``adiabatic'' decoupling.
As in \Sec{sec:fcs-equivalence} this requires us to consider \emph{currents} and the \fcs master equation \eq{eq:GeneralizedMasterEquation}.
We now think for a moment of \Eq{eq:rhochiinst} as a solution \emph{ansatz},
$\Lket{ \inst{\rho}^{\chi}(t)}
= e^{i \chi \Delta X^r(t)}  \Lket{\inst{\rho}[\vec{R}(t)]}
$,
in which the transported observable $\Delta X^r(t)$ is to be determined.
One notes that the nonredundant part of right eigenvector, $\Lket{\inst{\rho}[\vec{R}(t)]}$,  from which it is built contains no nonadiabatic information whatsoever.
Neither does the corresponding eigenvalue:
it is just the instantaneous current,
$\lambda_0^\chi[\vec{R}(t)] = i\chi \inst{I_{X^r}}[\vec{R}(t)] $.
The nonadiabatic effect can thus only enter through the step of ``adiabatic'' decoupling,
i.e, when inserting this ansatz into the \fcs master equation $\frac{d}{dt}\Lket{\rho^\chi} = W^\chi \Lket{\rho^\chi}$ and projecting this onto the eigenspace
with the corresponding \emph{left} eigenvector
$\Lbra{\bar{v}^\chi_0} = \Lbra{\unit} - i \chi \Lbra{\Phi_{X^r}}$.
One immediately obtains an equation for the \emph{nonadiabatic} part of the total \emph{\fcs current} $\tfrac{d}{dt} \Delta X^r(t)$:
\begin{align}
  \adcor{I_{X^r}} := [ \tfrac{d}{dt} \Delta X^r(t) ] - \inst{I_{X^r}}
  =
  \Lbra{\Phi_{X^r}} \tfrac{d}{dt} \Lket{ \inst{\rho} }
  .
  \label{eq:complements}
\end{align}
This shows most directly that by the formal ``adiabatic'' decoupling,
 the instantaneous state $\Lket{ \inst{\rho} }$ is combined with a time-derivative and the nonadiabatic response covector $\Lbra{\Phi_{X^r}}$.
Only together they produce the nonadiabatic \emph{effect} for observable $\hat{X}^r$.
Even though the \ar expression $\adcor{\rho}= W^{-1}\tfrac{d}{dt} \Lket{ \inst{\rho} }$ does not explicitly appear, the \fcs thus keeps track of the three required pieces required on the right hand side of \Eq{eq:complements}  in three different places.

The response covector has precisely the right form  $\Lbra{\Phi_{X^r}} = \Lbra{\mathds{1}} \currentKernel{X^r} W^{-1}$ [\Eq{eq:CountingLeftEvExpandedc}] 
required to obtain the \ar result
by a perturbative mixing.
This is the point, where the physical ``lag''  enters the \fcs analysis [cf. the \ar discussion of \Eq{eq:ResponseVectorDefinition}].
The formal $\chi$-controlled mixing of $\Lbra{\unit}$ with $\Lbra{\Phi_{X^r}}$ into the left eigenvector thus substitutes for the nonadiabatic coupling induced by physical time evolution.
It is this mixing that allows the \fcs to circumvent the probability normalization obstructing geometric phase accumulation (\Sec{sec:LackOfPhaseState}).

Finally, we also note how the gauge transformation in the \fcs corresponds to the physical
meter-recalibration $X^r \to X^r_g = X^r + g\unit$,
discussed in the \ar approach [\Eq{eq:Iadef} ff.].
One obtains the same generating operator when changing the  gauge factor of the \emph{right} eigenvector while compensating for this by
using $X^r_g$ instead of $X^r$ in the ansatz
$\Lket{ \inst{\rho}^{\chi}(t)}
= e^{i \chi \Delta X^r_g(t)} \cdot
\big[
e^{-i \chi [g[\vec{R}(t)]-g[\vec{R}(0)]]}
\Lket{\inst{\rho}[\vec{R}(t)]}
\big] 
$.
The ``adiabatic'' decoupling then leads to a different \fcs current,
\begin{align}
   \adcor{I_{X^r_g}} := [ \tfrac{d}{dt} \Delta X^r_g(t) ] - \inst{I_{X^r}}
  =
  \Lbra{\Phi_{X^r}} \tfrac{d}{dt} \Lket{ \inst{\rho} } + \tfrac{d}{dt}g
\end{align}
which, however, integrates to the correct value $\Delta \adcor{X}^r$.

\paragraph{``Nonadiabatic'' corrections to \fcs.}

We now address the question  whether or not the slow driving required for the ``adiabaticity'' of the FCS implies that it should neglect nonadiabatic effects it produces.
It is therefore relevant to  compare the terms that the approaches in \Fig{fig:compare} neglect.
These conditions have received little attention in \fcs works that compute the first moment of pumping, see however~\Ref{Sinitsyn07EPL}.

For this we return to the general formulation of the \fcs of \Sec{sec:fcsApproach}
and decompose the generating operator as follows:
\begin{align}
  \rho^\chi = \inst{\rho}^\chi + \adcor{\rho}^\chi 
+ {\rho}^{\chi, \text{rest}}
.
\end{align}
As noted before, this labeling is tentative and we should distinguish zero and nonzero values of $\chi$.

\emph{$\chi=0$: \fcs neglects nonadiabatic effects.}
In this case each term reduces by \Eq{eq:rhochizero} to the corresponding contribution in the \ase and \ar expansion:
\begin{align}
  \rho = \inst{\rho} + \adcor{\rho} + {\rho}^{\text{rest}}
	.
\end{align}
For $\chi=0$ the labeling is thus appropriate.
For the \fcs to be consistent, the nonadiabatic correction
plus higher corrections
that are \emph{neglected}
 must be small relative to the ``adiabatic'' one that is kept:
\begin{align}
   ||  \rho^{\text{rest}} + \adcor{\rho} ||   \ll    || \inst{\rho} || 
  \label{eq:condition-fcs}
	.
\end{align}
This condition is satisfied if the condition \eq{eq:condition-ar} of slow driving,
 $|\dot{\vec{R}}| \ll \Gamma$, holds, since it guarantees
$||  \rho^{\text{rest}} || \ll || \adcor{\rho} ||   \ll    || \inst{\rho} ||$.
This can equivalently be expressed as a gap condition
commonly used to justify the decoupling of eigenspaces [\Eq{eq:StateGapCondition}].
For the \ase approach this leads to the same driving restriction as for the \ar approach [\Eq{eq:condition-ase}].
A possible source of confusion is that in the \ar approach the contribution $\adcor{\rho}$ is necessarily and correctly \emph{kept}, being aware that it is small
[\Eq{eq:condition-ar0}].

\emph{$\chi \neq 0$: \fcs keeps leading nonadiabatic effects, neglecting higher ones.}
In the \fcs approach consistency also requires  that the moments computed by \Eq{eq:PumpedObservable} with $t_0 \rightarrow 0$ from the ``adiabatic'' solution dominate the ones that are neglected, i.e.,
the first ``nonadiabatic'' correction and higher ones:
\begin{align}
  \partial_{i\chi}^n \Lbraket{\unit | \rho^{\chi, \text{rest}} + \adcor{\rho}^\chi}
  |_{\chi=0}
  \ll
  \partial_{i\chi}^n  \Lbraket{\unit | \inst{\rho}^\chi}
  |_{\chi=0}
	.
	\label{eq:NeglectFirstMomentFCS}
\end{align}
For the first moment, $n=1$, noting on the right hand side that the \fcs result reduces to the \ar expressions [\Eq{eq:SimplifiedCountingInst}-\eq{eq:SimplifiedCountingPumping}], we obtain
\begin{align}
 |  \Delta X^{r, \text{rest}} |  \ll  |  \Delta \inst{X}^r +\Delta \adcor{X}^r |  
  \label{eq:condition-fcs2}
  .
\end{align}
Both terms on the left in \Eq{eq:NeglectFirstMomentFCS} together form the term on the left in \Eq{eq:condition-fcs2},
while the term on the right in \Eq{eq:NeglectFirstMomentFCS} produces both terms on the right in \Eq{eq:condition-fcs2}. 
Here the tentative labeling is thus not appropriate but as mentioned above this is unavoidable.
Importantly,  the consistency condition \Eq{eq:condition-fcs2} is exactly that of the \ar approach [\Eq{eq:condition-ar2}], in which the small $\adcor{\rho}$ is kept to compute $\Delta \adcor{X}^r$.
Thus, within the \fcs this nonadiabatic effect is \emph{effectively kept},
despite the fact that one neglects ``nonadiabatic'' effects in the state based on a gap condition at $\chi=0$.

As summarized in \Fig{fig:compare},
 in the \fcs approach ``adiabaticity'' is merely a formal statement about the validity of the \emph{decoupling} similar to the \ase, but applied to a formal device, the {generating operator} $\rho^\chi$.
In contrast, in the \ar approach (non)adiabaticity is directly related to different terms in a frequency expansion,
corresponding to the (corrections to) decoupling of eigenspace in the \ase approach in the steady-state limit.
This underlines that the useful notions of ``adiabaticity'', ``Berry-Simon phase'' and ``decoupling'' should carefully be distinguished, in particular in open quantum systems with a unique stationary state.
Our discussion illustrates the usefulness of having a clear \emph{physical} picture of what all the geometric notions in a problem with gauge freedom stand for,
going beyond the level of ``the final result can be written as a curve integral''.

\subsection{Discussion of the \fcs approach\label{sec:fcs-discuss}}

Having answered question (iv) of \Sec{sec:questions} regarding the equivalence of the \fcs and \ar approach,
we emphasize the obvious advantage of the \fcs approach when one actually goes
 beyond the first moment of pumped quantities on which we focused here,
as, e.g., in~\Refs{Ren13,Li14}.
As discussed from its inception~\cite{Levitov93},
the \fcs can be considered a full description of an ideal meter detecting transport of an observable, see also~\Ref{Schaller09}.
The relation between meter recalibrations and gauge freedom in pumping formalisms, highlighted in our paper in the simplest setting,
deserves further consideration within the more general \fcs approach,
possibly, also for \emph{non-ideal} meters.

However, in applications where \fcs is used \emph{only} to calculate the first moment
it must be stressed that a definite overhead is introduced:
besides deriving a more complicated kernel $W^\chi$,
one needs to compute a specific instantaneous eigenvalue of $W^\chi$ with both its left and right eigenvectors in dependence of the continuous variable $\chi$.
Even when one \emph{afterwards} linearizes in $\chi$ to extract only the first moment,
 this is much more involved than solving the corresponding problem analytically or numerically in the \ar approach, even for very simple models.
For more complicated systems, the \fcs approach may become so involved that in order to make progress, one may be inclined to introduce further approximations, which are unnecessary when using the \ar approach.
As explicitly shown here and in \Ref{Nakajima15}, $\chi$-linearizing \fcs calculations \emph{before hand} amounts to using the \ar approach.
This overhead is even more relevant when considering strong coupling and memory effects [\Eq{eq:kineq-general}-\eq{eq:obs-general}] on geometric pumping that arise beyond the Born-Markov approximation~\cite{Splettstoesser06,Splettstoesser10}.

\section{Discussion and outlook\label{sec:conclusion}}

In this paper, we have discussed the geometric nature of pumping through open quantum systems using the reduced density-operator approach.
We focused explicitly on the memoryless, weak-coupling, high-temperature limit and the ubiquitous case of a unique parametric stationary state.
This allows treating geometric pumping phases of complicated discrete quantum systems with essentially arbitrary local interactions,
including in particular geometric phases that, in order to emerge, \emph{require interactions}.

We outlined how in our solution of the physical pumping problem (much simpler than equivalent alternative approaches) a local gauge freedom emerges in the relation between a measured pumped observable (two-point expectation value) and an observable operator [\Eq{eq:GaugeTransformationCurrentkernel}]:
the choice of gauge corresponds to a meter calibration.
In striking contrast to closed quantum systems, the gauge freedom is therefore a \emph{nonunitary} group, reflecting the very different physical setting in which geometry enters here:
the \emph{observable}, not the quantum-state, accumulates the geometric phase over one driving period, as physically expected in transport problems.

Our paper combines definite computational advantages with a clear view of the tangible physical meaning of all relevant geometric concepts.
We tie together the differential geometry defined by a connection $A_{X^r_g}$ with non-adiabaticity [$\dot{\vec{R}}(t)$] and the gauge freedom of recalibrations:
the Landsberg \emph{geometric connection} is essentially the \emph{nonadiabatic} part of the \emph{current}
\begin{align}
  A_{X^r_g}[\vec{R}(t)]
  = \frac{ \delta I_{X^{r}_g}(\vec{R},\dot{\vec{R}}(t)) }{\delta \dot{\vec{R}}(t)} 
  ,
\end{align}
with $X^r_g=X^r + g \mathds{1}$.
The corresponding curvature of this connection, its curl, is the pumped charge per unit area of the driving-parameter space.
We provided explicit formulas for the efficient computation of pumping phases applicable to very general interacting systems that are weakly coupled to reservoirs,
showing how computational overhead can be avoided
and conservation laws, present for specific observables, can be exploited.
We demonstrated these advantages for the example of a single-level Anderson quantum dot illustrating how a geometric pumping phase can be \emph{induced} by electron-electron interaction.

Moreover, we showed that the geometric ``horizontal lift'' defined by this connection corresponds to a (discontinuous) meter gauge
that maintains the physical pumping current $\adcor{I_{X^{r}_g}}$  to be zero at each time instant by \emph{continuously adjusting the scale} to follow the  \emph{meter's ``needle.''}
The  geometric-phase ``jump'' (the holonomy) exhibited by this horizontal lift is given by the resulting cumulative adjustment of the meter's scale over a driving period, i.e., the  pumped \observable per period.
We believe this is the most direct geometric significance one can attach to physical transport quantities in a pumping experiment. 
We demonstrated the importance of such a clear view by discussing the important issue how ``adiabatic'' approximations employed in the full-counting-statistics (\fcs) approach produces a (correct) geometric pumping phase that is due to \emph{nonadiabatic currents}.

It is an interesting question as to whether and how our approach relates to ideas for closed systems, in particular to the widely used Berry-Simon formulation of geometric phases appropriate for quantum \emph{state} evolution.
Such a relation can be established, but requires a construction different from --but relying on-- the one presented in this paper and will be discussed elsewhere~\cite{Pluecker17a}.
In fact, aside from the obvious consideration of \emph{topological} pumping effects in strongly interacting systems,
our work can be continued in several other directions:

(i) An important implication of our work is that Landsberg's geometric framework is compatible with the more general adiabatic-response approach to pumping based on real-time kernels~\cite{Splettstoesser06} [\Eq{eq:kineq-general}]:
the nonuniqueness in the choice of the observable operator during a driving cycle does not rely on the state dynamics. 
We thus expect that the present approach can be extended to non-Markovian, strongly-coupled open quantum systems with a unique parametric stationary state\footnote{This requires consideration of observable operators that are sufficiently general, i.e., explicitly time-dependent and not partially normal-ordered with respect to the reservoirs.}.
For this reason, we stressed that the gauge structure already requires attention when \emph{deriving} quantum master equations and expectation value of the current for a \emph{nonsystem} observable, an issue that has received little to no attention so far.
This is crucial since measurements that display geometric effects are performed in the \emph{environment}, that one would like to integrate out to keep the calculations simple.
These are situations which existing formulations, such as generalizations of Kato's method to open systems\cite{Avron12a}, have not dealt with.

(ii)
Our paper stressed the important distinction between the gauge freedom of the quantum state and that of the pumped \emph{observable}.
We showed that the \emph{state} gauge freedom is quenched in the ubiquitous case of a unique stationary state for fixed driving parameters, our well-motivated working assumption.
Clearly, an interesting extension of our work would be to consider nonunique stationary states~\cite{Avron12} leading to an interplay of the \emph{state} gauge freedom with the independent \emph{observable} gauge freedom in steady-state pumping.
Counter to standard intuition, a two-fold degenerate stationary state does not yet lead to a non-Abelian gauge freedom, but to an Abelian one due to physical restrictions imposed by trace and Hermiticity preservation, which are often not discussed.
For higher degeneracies of the stationary state, a non-Abelian gauge structure does seem to emerge, but this requires a careful account of the mentioned nontrivial restrictions.

(iii)
One may furthermore also consider a more complicated gauge group of non-Abelian character for the \emph{observable}. In fact, Landsberg's general geometric framework has already been extended to this case in \Ref{AnderssonThesis,Andersson05}.
However, its relation to the physical density operator formalism and to measurable pumping quantities (the starting point stressed throughout this  paper) needs to be clarified first.

In any case, to address these important problems, the present formulation of Landsberg's approach seems to provide the simplest and physically most transparent geometric picture: 
for the simple, but ubiquitous, class of problems addressed here, we demonstrated this by explicit comparison with several other approaches,
highlighting geometric, physical and computational advantages.
Our work illustrates that the standard intuition about geometric effects in quantum physics needs to be drastically reconsidered when turning to open quantum systems and motivates further work into this direction.

\acknowledgments
We acknowledge especially H. Calvo for discussions and for carefully reading the manuscript, M. Pletyukhov for pointing us to  \Ref{Sinitsyn09} and \Ref{Landsberg93},
and an anonymous referee for useful suggestions.
We further enjoyed useful discussions with M. Martin Delgado, M. Hell,  Y. Mokrousov, R.-P. Riwar, J. Schulenborg, D. Schuricht and B. Terhal.
T.P. was supported by the Deutsche Forschungsgemeinschaft (RTG 1995)
and J. S. by the Knut and Alice Wallenberg Foundation and the Swedish VR.

\appendix
\section{Born kernels and conservation law\label{app:kernels}}

In this section we derive the \emph{Born}-approximation\footnote{
The further \emph{Markov} approximation is  discussed in detail in the main text. 
Consistent with the Born-Markov  approximation the kernels \Eq{eq:AppKernelFormula} and \eq{eq:CurrentKernelFormula} can be calculated for \emph{frozen} parameters as discussed in the main text.
}
 to  the kernels governing the dynamics of the system state [\Eq{eq:DynamicsKernel}] and the expectation values of nonsystem observables [\eq{eq:ObservableKernel}].
We also derive the simplification \eq{eq:conservation} in case the observable is conserved at a junction.

\emph{Partial normal ordering.}
Although we follow the well-known Wangsness-Bloch approach~\cite{Koller10}, a key point, central to the issue of gauge freedom addressed in the main text, deserves to be highlighted:
it is crucial to split up both the coupling $V$
and the nonsystem observables $\hat{X}^r$ into a partially averaged term
and a partially normal-ordered term according to \Eq{eq:nosplit}.

Throughout the paper it is assumed that the coupling $V(t)$ is partially normal ordered, i.e., 
\begin{align}
 \expec{V(t)}^\text{res} = 0 . \label{eq:Vavzero}
\end{align}
This constitutes no approximation since for any Hamiltonian of the form of \Eq{eq:Htot} the coupling can be partially normal ordered by absorbing the partial average of the original coupling, which is a system operator, into the system Hamiltonian.\footnote{The system Hamiltonian may then depend on reservoir parameters (e.g., temperature, electrochemical potentials).}

Assuming initial decoupling of system and reservoir state, i.e., $\rho^{\text{tot}}(0) = \rho \otimes \rho^\text{res}$, and the interaction picture 
\begin{align} 
	A_I \coloneqq e^{i(H+H^\text{res}) t} A e^{-i(H+H^\text{res}) t}
\end{align}
the Liouville-von Neumann differential equation for the total system state reads as
\begin{subequations}\begin{align}
  \partial_t \tot{\rho}_I(t) &= -i \left [ V_I(t),\tot{\rho}_I(t) \right ]    \label{eq:liouvilleapp}
                      \\
  \tot{\rho}_I(t_0)  &= \rho_I(t_0) \otimes \env{\rho}
\label{eq:initfact}
  ,
\end{align}\end{subequations}
which is equivalent to the Dyson integral equation
\begin{align}
  \tot{\rho}_I(t)
  =
  \tot{\rho}_I(t_0) - i\int _{t_0}^t dt' \left [ V_I(t'), \tot{\rho}_I(t') \right ]
  .
  \label{eq:dyson}
\end{align}

\emph{Time-evolution kernel.}
To obtain the evolution equation for the reduced density operator $\rho_I(t) = \Tr{\text{res}} \tot{\rho}_I(t)$
to leading order $V^2\sim \Gamma$ in the coupling,
we insert \Eq{eq:dyson} into \Eq{eq:liouvilleapp}
and take the partial trace
\begin{align}
  \partial_t \rho_I(t) =& -i \Tr{\text{res}} \Big [V_I(t), \tot{\rho}_I(t_0) \Big] \notag \\
                &- \int_{t_0}^t dt' \Tr{\text{res}}  \Big[ V_I(t), \Big[ V_I(t'), \tot{\rho}_I(t') \Big] \Big] \label{eq:AppKernel1}
								.
\end{align}
The first term in \Eq{eq:AppKernel1} vanishes due to the partial normal-ordering \eq{eq:Vavzero}.
The second term in \Eq{eq:AppKernel1} can be simplified
consistent with the Born approximation by inserting
\begin{align}
  \tot{\rho}_I(t')
  \approx
  \rho_I(t') \otimes \env{\rho}
  \label{eq:born}
\end{align}
since the corrections, computed in \Ref{Koller10}, end up in higher orders of the kernel, which we neglect.
As a result, within the Born-approximation the kinetic equation reads in the interaction-picture
\begin{align}
  \partial_t \rho_I(t) =
 - \int_{t_0}^t dt' \Tr{\text{res}} \Big[ V_I(t), \Big[ V_I(t'), \rho_I(t') \env{\rho} \Big] \Big]
  \label{eq:kineqbornI}
\end{align}
and, correspondingly, in the Schr\"odinger picture
\begin{align}
  \partial_t \rho(t) = -i\left [ H, \rho(t) \right ] + \int_{t_0}^t W(t-t')  \rho(t') \label{eq:AppModelME}
\end{align}
with the time-nonlocal kernel \Eq{eq:DynamicsKernel} of the main text:
\begin{align}
  &W(t-t') \bullet 
  = \label{eq:AppKernelFormula} \\ 
  &-
  \Tr{\text{res}}
  \left [ V
  ,
  e^{-i\left[ H+ \env{H}\right](t-t')}
  \left [ V , \env{\rho} \bullet  \right ]
  e^{+i \left[ H + \env{H}\right](t-t')}
  \right] \notag
  .
\end{align}

\emph{Observable kernels.}
For an arbitrary time-dependent nonsystem observable we can proceed in close analogy.
The crucial step is \emph{not} to approximate the expectation value of $\hat{X}^r_I(t)$, but instead the expectation value of its partially normal-ordered part [\Eq{eq:nosplit}],
\begin{align}
  \expec{\no{\hat{X}^r(t)}}
  & = 
  \expec{\hat{X}^r(t)} - \tr \left ( \expec{\hat{X}^r(t)}^\text{res} \rho_I(t) \right )
    ,
    \label{eq:Xrnoav}
\end{align}
which does not exhibit fluctuations in the sense that
\begin{align}
  \expec{\no{\hat{X}^r(t_0)}}^\text{res} = 0
  .
  \label{eq:Xravzero}
\end{align}
Now, inserting the Dyson equation \eq{eq:dyson} into the expectation value we obtain two terms, similar to  \Eq{eq:AppKernel1}:
\begin{subequations}
\begin{align}
  \expec{\no{\hat{X}^r(t)}}
 =&\tr \Tr{\text{res}} \no{\hat{X}^r_I(t)} \tot{\rho}_I(t) \label{eq:totalSpaceCurrent0}
  \\ 
  = &\tr \Tr{\text{res}} \no{\hat{X}^r_I(t)} \tot{\rho}_I(t_0)
         \label{eq:totalSpaceCurrenta} \\
  &- i \tr \int_{t_0}^t dt' \Tr{\text{res}} \hat{X}^r_I(t) \left [  V_I(t'),\tot{\rho}_I(t')\right ] 
    \label{eq:totalSpaceCurrentb}
		.
\end{align}    \label{eq:totalSpaceCurrent}
\end{subequations}
Again, the first term \eq{eq:totalSpaceCurrenta} vanishes by virtue of partial normal ordering \emph{of the observable}:
inserting the initially factorizing state \eq{eq:initfact} and using \Eq{eq:Xravzero}
\begin{align}
  \tr \Tr{\text{res}} \no{\hat{X}^r_I(t_0)} \tot{\rho}_I(t_0)
  =\tr \expec{\no{\hat{X}^r_I(t_0)}}^\text{res} \rho_I(t_0) = 0
  .
\end{align} 
In the second term \eq{eq:totalSpaceCurrentb} we again insert the Born approximation
\eq{eq:born}.
In analogy to \Eq{eq:kineqbornI} we thus obtain
\begin{subequations}
  \begin{align}
&  
\expec{ \no{\hat{X}^r(t)} }
\\
  & =- i \tr \int_{t_0}^t dt' \Tr{\text{res}}
     \no{ \hat{X}^r_I(t) }
     \Big[ V_I(t'), \rho_I(t') \env{\rho} \Big] 
    \\
  & =- i \tr \int_{t_0}^t dt' \Tr{\text{res}}
     \tfrac{1}{2} \Big[
    \no{ \hat{X}^r_I(t) },
     \Big[ V_I(t'), \rho_I(t') \env{\rho} \Big]
     \Big]_{+}
    \label{eq:expectbornI}
		.
\end{align}
\end{subequations}
In the last step to \Eq{eq:expectbornI} we have used that for a product $A B = \tfrac{1}{2} [ A, B] + \tfrac{1}{2} [A,B]_+$
the commutator part does not contribute under the trace.
As mentioned after \Eq{eq:redundancy-ar}, this is equivalent to changing $W_{\no{X}} \to W_{\no{X}} + \Theta$ by a superoperator with $\tr \,  \Theta \, \bullet = 0$.
Using \Eq{eq:Xrnoav} we obtain the expectation value \eq{eq:ObservableExpec} of the main text:
\begin{align}
  \expec{\hat{X}^r(t)} 
  = 
   \tr \expec{\hat{X}^r(t)}^\text{res} \rho(t) +
   \tr \int_{t_0}^t dt' W_{\no{X^r}}(t,t') \rho(t')
    \label{app:CurrentKernel}
		,
\end{align}
with the time-nonlocal observable kernel \eq{eq:ObservableKernel}
\begin{align}
  &W_{\no{X^r}}(t,t') \bullet
  =   -i\tfrac{1}{2} \times \label{eq:CurrentKernelFormula} \\
	& \Tr{R}  
  \left [ \no{\hat{X}^r(t)} \,
     ,
  e^{-i\left[ H+ \env{H}\right](t-t')}
  \left [ V, \env{\rho} \bullet  \right ]
  e^{+i\left[ H+ \env{H}\right](t-t')}
  \right ]_+
 \notag
	,
\end{align}
which explicitly depends \emph{only} on the partially normal-ordered part of the observable $\no{\hat{X}^r}$.

\emph{Conservation law at a junction $r$.}
We now derive the identity \eq{eq:conservation}, which is used to simplify \Eq{eq:ObservableKernel},
assuming that the $\vec{R}$-independent  \emph{nonsystem} observable $\hat{X}^r$´
is conserved as in \Eq{eq:Xconservation}.
We follow \Ref{Saptsov12a} [Eq. (174) ff.]
adapting the simpler formulation used in the above derivations.
We allow the \emph{system} operator $\hat{X}[\vec{R}]$ to depend on $\vec{R}$ (cf. \Sec{sec:application}).

Accounting for our assumptions
that $\hat{X}^r$ is $\vec{R}$-independent,
$ [ \hat{H}^r, \hat{X}^r ] = 0$ [\Eq{eq:HrXrcom}]
and  $[ \hat{H}, \hat{X} ] = 0$ [\Eq{eq:HXcom}],
the current operators read as
\begin{align}
  \hat{I}_{X^r} = i [V^r,\hat{X}^{r} ]
  ,
  \qquad
  \hat{I}_X = i \sum_r [V^r,\hat{X}  ] + \frac{\partial \hat{X}}{\partial t}
	.
\end{align} 
The conservation law \eq{eq:Xconservation} implies that the part of the current \emph{due to the coupling} $V^r$, indicated by $|_r$, is conserved:
\begin{align}
       \expec{\hat{I}_{X^r} } 
& =   \expec{ i [V^r,\hat{X}^{r} ]  }
\\
& = - \expec{ i [V^r, \hat{X} ] }
:= - \expec{ \hat{I}_{X} - \tfrac{\partial \hat{X}}{\partial t} } |_{r} 
.
\end{align}
Going to the interaction picture, the right hand side is easily related to the state evolution:
\begin{subequations}\begin{align}
  &
\expec{ \hat{I}_{X} - \tfrac{\partial \hat{X}}{\partial t} } |_{r} 
\\
  & =
   -i \tr \Tr{\text{res}} \, [ \hat{X}_I(t),  V^r_I(t) ]  \tot{\rho}_I(t) 
  \label{eq:IXa}
\\
  & =
   \tr  \hat{X}_I(t) \Tr{\text{res}} \Big\{  -i [ V^r_I(t),   \tot{\rho}_I(t) ]  \Big\}
  \label{eq:IXb}
\\
 & =  - \tr  \hat{X}_I(t) 
  \int_{t_0}^t dt' \Tr{\text{res}} \left [ V_I^r(t),\left [ V_I(t'), \rho_I(t') \env{\rho} \right ]\right] 
   .
   \label{eq:IXc}
\end{align}
\end{subequations}
In step \eq{eq:IXa} we used
$\tr \tr_{\text{res}} [A,B] C = \tr \tr_{\text{res}} A [B,C]$
to pull the leftmost system operator out of the reservoir trace.
Under the partial trace in \Eq{eq:IXb}
we have precisely the right hand side
of the Liouville equation \eq{eq:liouvilleapp}
except for  $V_I(t)\to V^r_I(t)$ and we thus perform the same steps that led from \Eq{eq:AppKernel1} to \Eq{eq:kineqbornI}].

In the resulting \Eq{eq:IXc}, the time-integral is just the right hand side of the kinetic equation \eq{eq:kineqbornI} ``resolved with respect to junction-$r$'' at the latest time $t$.
Correspondingly, in the Schr\"odinger picture we obtain
\begin{align}
  \expec{ \hat{I}_{X^r} }(t)
 & =  - \tr  \hat{X}(t)  \int_{t_0}^t dt' W^r(t,t') \rho(t') 
   \label{eq:conservation2}
\end{align}
where $W^r(t,t')$ is obtained from \Eq{eq:AppKernelFormula} by simply replacing $V(t) \to V^r(t)$.
Finally, by comparing the Markov limit of \Eq{app:CurrentKernel} for $\hat{X}^r \to \hat{I}_{X^r}$ and the Markov limit of \Eq{eq:conservation2}
we obtain the result \eq{eq:conservation} of the main text 
\begin{align}
  \expec{ \hat{I}_{X^r} }(t)
   & = 
   \tr   \currentKernel{X^r} (t)  \rho(t)
	=
       - \tr  \hat{X}  W^r \rho(t) 
  \label{app:CurrentKernelMarkov}
\end{align}
with
$\currentKernel{X^r}
   :=    W_{\no{I_{X^r}}}(t) + \expec{\hat{I}_{X^r}}^\text{res}$
defined as in \Eq{eq:WIX}. 

\section{Conservation law - example
\label{app:ChargeKernel}}

Here we illustrate how \Eq{eq:conservation} in the main text
takes a more commonly known form
 when it is expressed in terms of ``current rates'' as in \Ref{Schoeller97hab}.
For the stationary-state charge current into the reservoir $r$:
\begin{align}
  I_{N^r}  = \Lbra{\unit} \currentKernel{N^r} \Lket{\inst{\rho}}
            = - \Lbra{N} W^r \Lket{\inst{\rho}}
						.
\end{align}
Let $\ket{k}$ be a particle-number eigenstate, i.e.,
$\hat{N}=\sum_k N_k \ket{k}\bra{k}$ with particle number $N_k$ in state $k$.
We now 
expand the density operator in basis operators $\Lket{kk'} : = \ket{k}\bra{k'}$
and
expand $\Lbra{N}\bullet =\tr \hat{N}\bullet = \sum_j N_j \Lbra{jj}$ in the dual basis $\Lbra{jj'}=\tr \ket{j'}\bra{j}$.
For simplicity we ignore contributions from the off-diagonal elements $k \neq k'$ of the stationary-state density operator in the $H$-basis.
Denoting the transition rates $\Lbra{jj} W^r \Lket{kk}:=W^r_{j,k}$
we obtain the well-known expression
\begin{align}
  I_{N^r} &
  =  - \sum_{k} \sum_{j}
  N_j W^r_{jk} \inst{\rho}_{kk} + \ldots
\\ &
  = - \sum_k \sum_{j \neq k}
 \left ( N_j - N_k \right) W^r_{j,k} \inst{\rho}_{kk} + \ldots
     ,
\end{align}
using only the probability conservation $W_{kk}^r=-\sum_{j \neq k} W_{j,k}^r$.
Physically, this counts minus the change in $N$ going from state $k \to j$,
which by conservation of $\hat{N}+\hat{N}^r$ should equal the change in $\hat{N}^r$.

\section{Pumping vs. instantaneous contribution\label{app:pumping}}

Here, we discuss how the different scalings with parameters, mentioned in \Sec{sec:naive}, allow the instantaneous part \eq{eq:Obs} and adiabatic-response part \eq{eq:adCorrObs} and \eq{eq:pumpedX} to be separately extracted from measurements,
both in principle and in practice.
How can one get rid of the nongeometric ``sum of snapshots'' of the instantaneous current, even in strong nonequilibrium situations?
In particular, one would like to do this directly in an experiment~\footnote{This can be done of course computationally if one has first measured the instantaneous currents at all parameter values accessed by the driving curve, but this seems less accurate as measurement errors for different times may accumulate.} in view of applications where pumping current is the main tool (\Sec{sec:intro}), but also in view of the pumping spectroscopy discussed in \Sec{sec:spectroscopy}.
We summarize three possible ways to extract the geometric pumped charge from the instantaneous background:

(a) \emph{Zero bias}: $\inst{I_{X^{r}}}(t)=\Delta \inst{X}^{r} = 0$.
The characteristic of this case of \emph{pure} pumping is that although at each instant the current for $\hat{X}^r$ is zero,
a cyclic parameter change can still transport this observable.

(b) \emph{Bias driving with cancellations}: $\inst{I_{X^{r}}}(t) \neq 0$ but $\Delta \inst{X}^r=0$.
When driving the applied bias the instantaneous current is generally nonzero.
The ``sum of snapshots'' may, however, still average to zero
if the bias-driving probes regions with opposing currents.
This happens, e.g., for charge transport through a symmetrically coupled quantum dot when driving gate and bias voltage at constant frequency $\Omega$.

If this cancellation is slightly incomplete for constant frequency, e.g., due to a nonsymmetric coupling,  then one may slowly modulate the frequency to shift weight between positive and negative contributions during the period in order to still achieve cancellation
and realize $\Delta \inst{X}^r=0$.
Importantly, this does not alter the adiabatic-response part because of its geometric nature:
it depends only on the traversed parameter curve.

Another possibility is to perform a global modulation of the coupling, i.e.,
a time-dependent, spatially uniform rescaling of all couplings by $V^r[\vec{R}] \to \sqrt{\alpha[\vec{R}]} V^r[\vec{R}]$ with a nonnegative function $\alpha[\vec{R}]$.
Such a rescaling 
modifies the instantaneous current by
$\inst{I_{X^r}} \to \alpha \inst{I_{X^r}}$
but drops out in the pumping current: 
$\adcor{I_{X^r}} \to \adcor{I_{X^r}}$.
This modification\footnote{Such a procedure is usually used to rectify the instantaneous current by suppressing it to values $\ll 1$ for parameter values $\vec{R}$ where the current flows in an undesired direction and setting it to $1$ otherwise.}
 of the instantaneous part can be used to cancel it to zero by equally weighting forward and backward instantaneous current contributions along the driving cycle. 
Then, after one driving cycle only a geometric pumped charge remains.

(c) \emph{General case with bias and/or bias driving}: $\inst{I_{X^{r}}}(t) \neq 0$ and $\Delta \inst{X}^r \neq 0$.
Finally, if the driving cycle probes only parameter values with a definite bias,
then one is ``pumping with / against the flow''
and the cancellation cannot be achieved in the way indicated above.
However, a simple modification of the measurement scheme as suggested in \Ref{Sinitsyn07PRB} still allows one to cancel out the instantaneous part:
one first drives the system for a large number of cycles $M$ and measures the total \observable during time $M\T$.
One then reverses the time-dependent driving protocol and repeats the measurement.
Subtracting the two measurements and dividing by $2M$ the instantaneous part cancels out (it does not change sign), leaving only the pumping part.

Situation (a) is at the focus of studies motivated by metrology~\cite{Jehl13}, where topological protection is desired.
Here, its geometric nature protects the pumping signal against various kinds of perturbations of the driving protocol.
For example, different parametrizations $\vec{R}(t)$ of the same curve,
 e.g., due to fluctuations in the driving speed, leave the pumping invariant.
Also, the deformation of the driving cycle has a smaller effect on the pumping contribution than on the stationary part
since the former scales with the area enclosed by the driving cycle for small amplitude driving.

Situations (b)-(c) are relevant to recent studies discussing pumping~\cite{Ren10,Nakajima15},
pumping-spectroscopy~\cite{Calvo12a,Reckermann10a,Haupt13,Riwar13}, and excess entropy production in nonequilibrium thermodynamics~\cite{Yuge13} in the presence of a nonlinear \emph{bias}.
The interesting point here is that the pumping contribution contains spectroscopic information~\cite{Splettstoesser06}
that is \emph{not} contained in the averaged stationary-state information of in the instantaneous part.

\section{Interaction-induced pumping through a quantum dot\label{app:example}}

Here we explicitly work out the \ar approach for the simple fermionic example of electron charge pumping through a single orbital quantum dot. 
This example was analyzed in \Ref{Calvo12a}. Here we extend it by allowing for driving of the tunnel barrier and in fact of any parameter. 
This serves as an illustration of the general approach presented in this paper.

The system Hamiltonian describing such an example reads as
\begin{align}
	H=\epsilon N + B S_z + \tfrac{1}{2}N_{\uparrow} N_{\downarrow}U
	.
\end{align}
with bilinear tunnel couplings $V^r$ specified through tunnel rates $\Gamma_r$ (treated in the wide-band limit)
and electron reservoirs $r= \text{L, R}$, characterized by $T$ and $\mu^r$.
The magnetic field $B$ is initially added for our comparison with the \fcs approach in \App{app:fcs}.  
We later focus on zero magnetic field results \Refs{Reckermann10a,Calvo12a,Yoshii13,Nakajima15},
and make use of the supplementary information to \Ref{Schulenborg16a} where more details can be found.

\emph{State dynamics.}
We first give the time-evolution kernel in \Eq{eq:MarkovMe}).
Due to spin and charge conservation the off-diagonal elements of the quantum-dot density operator $\rho$  in the $H$-eigenbasis, describing, e.g., the transverse spin $S_x$ and $S_y$, decouple from the diagonal ones describing the spin $S_z$ along $B$ and the charge ($N$) dynamics.
We can thus restrict the state dynamics to Liouville vectors spanned by pure states for an empty, spin-up, spin-down- and double occupied quantum dot, respectively:
\begin{align}
  \Lket{\rho} = p_0 \pureD{0} + \sum_\sigma p_\sigma \pureD{\sigma} +  p_2 \pureD{2}
	.
\end{align}
Acting on this subspace, the Liouvillian $L[\vec{R}]$ that we defined into the effective $W[\vec{R}]$ in \Eq{eq:Wdef} is zero and can thus be ignored.
In this basis the \mE \eq{eq:MarkovMe} for the example of this section reads
\begin{widetext}
\begin{align}
\partial_t
\begin{pmatrix} p_0 \\ p_\uparrow \\ p_\downarrow \\ p_2\end{pmatrix}
 = 
  \begin{pmatrix}  
    -W_{\uparrow,0}-W_{\downarrow,0}  & W_{0,\uparrow}                        & W_{0,\downarrow}                           & 0 \\
    W_{\uparrow,0}                             & -W_{0,\uparrow}-W_{2,\uparrow} & 0                                                 &  W_{\uparrow,2} \\
    W_{\downarrow,0}                          & 0                                          & -W_{0,\downarrow}-W_{2,\downarrow} & W_{\downarrow,2} \\
    0                                                & W_{2,\uparrow}                        &  W_{2,\downarrow}                           & -W_{\uparrow,2}-W_{\downarrow,2}
  \end{pmatrix}
  \begin{pmatrix}
    p_0 \\ p_\uparrow \\ p_\downarrow \\ p_2 
  \end{pmatrix}
  .
  \label{eq:AndersonMaster}
\end{align}
\end{widetext}
Letting $\bar{\sigma}$ denote the opposite of $\sigma = \uparrow,\downarrow$
and $\epsilon_\sigma=\epsilon+ \sigma B/2$, the rates $W_{x,y} =\sum_r W^r_{x,y}$ are given by
\begin{align}
  W^r_{\sigma,0} &= \Gamma_r f^+_r(\epsilon_\sigma) , &		W^r_{2,\overline{\sigma}} &= \Gamma_r f^+_r(\epsilon_\sigma + U) , \notag \\
  W^r_{0,\sigma} &= \Gamma_r f^-_r(\epsilon_\sigma) ,  &		W^r_{\overline{\sigma},2} &= \Gamma_r f^-_r(\epsilon_\sigma + U) ,
\end{align}
with the Fermi functions $f^\alpha_r(\omega) = 1 / (1 + \exp \alpha (\omega-\mu_r)/T)$ and temperature $T$.

Now focusing on the case of zero magnetic field, $B=0$, 
 the charge dynamics decouples from spin $S_z$ as well.
Changing the operator basis for $N=1$ to the mixed spin state $\tfrac{1}{2}\sum_\sigma \pureD{\sigma}$ and the spin-operator
$\tfrac{1}{2}\sum_\sigma \sigma \pureD{\sigma}=S_z$
and defining transition rates
\begin{align}
  W^r_{1,0} &=     \Gamma_r f^+_r(\epsilon) &
  W^r_{2,1} &=  2  \Gamma_r f^+_r(\epsilon+ U) \notag \\
  W^r_{0,1} &=  2  \Gamma_r f^-_r(\epsilon) & 		    
  W^r_{1,2} &=     \Gamma_r f^-_r(\epsilon + U) 
	\label{eq:appExampleRatesSpindegenerate}
\end{align}
 we obtain the \mE reduced to charge subspace:
\begin{align}
   \partial_t \begin{pmatrix} p_0 \\ p_1 \\  p_2\end{pmatrix} =
  \begin{pmatrix}
    -W_{1,0} & W_{0,1} &  0 \\ 
    W_{1,0} & -W_{0,1}-W_{2,1}  & W_{1,2} \\
    0 & W_{2,1}  & -W_{1,2}
  \end{pmatrix}
  \begin{pmatrix}
    p_0 \\ p_1 \\ p_2 
  \end{pmatrix}
   \label{eq:AndersonMaster2}
  .
\end{align}	
Here $p_N$ is the occupation probability of the charge $N=0,1,2$ state,
which is a mixed state for $N=1$ and a pure one for $N\neq 1$.

We now introduce the time-dependent driving through any of the parameters entering the \mE via \Eq{eq:appExampleRatesSpindegenerate}.
There is little need to explicitly express the rates $W_{N,N'}$ in terms of $\vec{R}$. 
We only illustrate for this example how to define these dimensionless parameters, required in the main text,
starting from the physical quantities $\epsilon, \mu^\text{L},\mu^\text{R},U,\Gamma^\text{L},\Gamma^\text{R}$.
Since $\epsilon, \mu^\text{L},\mu^\text{R},U$ enter the \mE through Fermi functions in \Eq{eq:appExampleRatesSpindegenerate},
temperature $T$ is a relevant energy scale.
The tunnel strengths $\Gamma^\text{L}, \Gamma^\text{R}$ play a special role since they set the scale of
the rates in \Eq{eq:appExampleRatesSpindegenerate},
in contrast to the other parameters.
The time-average of the total tunnel strengths $\bar{\Gamma}=\bar{\Gamma}^\text{L}+\bar{\Gamma}^\text{R}$
can be used to normalize them.
Note that both the scales $T$ and $\bar{\Gamma}$ are by our assumptions nonnegative.
Thus, the driving parameters\footnote{Choosing two tunnel rates $\Gamma^\text{L} / \bar{\Gamma}$ and $ \Gamma^\text{R} / \bar{\Gamma}$ as driving parameters does not lead to pumping. For nonzero rates the reason for this is that the pumping contribution depends only on the ratio of tunnel rates. Therefore driving of $\Gamma^\text{L} / \bar{\Gamma}$ and $ \Gamma^\text{R} / \bar{\Gamma}$ effectively amounts to single parameter driving.}
can be concretely chosen as
\begin{align}
  \vec{R} = \left(
  \frac{\epsilon }{ T } ,
  \frac{\mu^\text{L}-\mu^\text{R} }{ T } ,
  \frac{U}{T} ,
  \frac{ \Gamma^\text{L} }{ \bar{\Gamma} } ,
	\frac{ \Gamma^\text{R} }{ \bar{\Gamma} } 
  \right)
  .
  \label{eq:Rexample}
\end{align}

Using the notation of \Sec{sec:fcsPumping}
the real eigenvalues,  left and right eigenvectors of the superoperator $W$ in \Eq{eq:AndersonMaster2} are
\begin{widetext}
  \begin{subequations}  \label{eq:anderson-sindeg-eigenvectors}
    \begin{align}
      &\text{Eigenvalue} & &\text{Left eigenvector}  & &\text{Right eigenvector} \notag \\
      & \lambda_0 = 0
      & &{\overline{v}_0}=\scalemath{0.8}{ \begin{pmatrix} 1 & 1 & 1\end{pmatrix} }
      & &{v_0}= \scalemath{0.8}{
                          \frac{1}{\frac{W_{0,1}}{W_{1,0}}+\frac{W_{2,1}}{W_{1,2}}+1}
                          \begin{pmatrix} \frac{W_{0,1}}{W_{1,0}} \\ 1 \\ \frac{W_{2,1}}{W_{1,2}} \end{pmatrix}
                        } \label{eq:appExampleSteadyState} \\
     & \lambda_1 = -\frac{1}{2}(W_{1,0}+W_{1,2})
      & &{\overline{v}_1}=  \scalemath{0.8}{
             \begin{pmatrix} -\frac{W_{1,0}}{W_{0,1}-W_{2,1}} & 1 & \frac{W_{1,2}}{W_{0,1}-W_{2,1}} \end{pmatrix}
              }
      & &{v_1}= \scalemath{0.8}{
                        \scalemath{0.9}{
                        \frac{1}{\frac{W_{1,0}W_{0,1}+W_{1,2}W_{2,1}}{(W_{0,1}-W_{2,1})^2}+1}
                         }
                         \begin{pmatrix} -\frac{W_{0,1}}{W_{0,1}-W_{2,1}} \\ 1 \\ \frac{W_{2,1}}{W_{0,1}-W_{2,1}} \end{pmatrix}
                        }  \\
		 & \begin{minipage}{0.3\linewidth} 
		 			$\lambda_2 = -W_{1,0}-2W_{0,1} = -2 \Gamma$
		 	 \end{minipage}
      & &{\overline{v}_2}=  \scalemath{0.8}{\begin{pmatrix} \frac{W_{1,0}}{W_{0,1}} & -2 & \frac{W_{1,2}}{W_{2,1}} \end{pmatrix}}
      & &{v_2}= \scalemath{0.8}{\frac{1}{\frac{W_{1,0}}{W_{0,1}}+\frac{W_{1,2}}{W_{2,1}}+4}
                         \begin{pmatrix} 1 \\ -2 \\ 1 \end{pmatrix}}  
    \end{align}
  \end{subequations}
\end{widetext}
where  the parameter dependence has been suppressed.
The zero eigenvalue $\lambda_0$ is associated with preservation of the trace, represented by $\Lbra{\bar{v}_0}$,
and with the unique stationary state $\Lket{v_0}=\Lket{\inst{\rho}}$.
The negative eigenvalues $\propto \Gamma$
describe the decay mode of the charge $\hat{N}$, $\Lket{v_1}$, and fermion-parity~\cite{Contreras12,Saptsov12a,Saptsov14a} $(-\unit)^{\hat{N}}$, $\Lket{v_2}$,
 and their associated amplitude covectors.
These are additionally required to compute the response vector.

\emph{Observable response vector.}
We focus on charge pumping and choose for the observable operator
\begin{align}
	X^r = N^r \label{eq:appExampleObservable}.
\end{align}
As mentioned in the main text [\Eq{eq:Xrgauge}] and discussed in \Ref{Calvo12a}, at this point we already make use of the gauge freedom $N^r \to N^r + g[\vec{R}] \unit$:
to compute the physical time-dependent currents in a quantum dot one actually needs to account for screening currents~\cite{Bruder94}.
The resulting screening currents average out in one driving period. 
Thus, the gauge freedom in the observable is used from the start to discard these, \Eq{eq:appExampleObservable} giving the correct pumped charge.

Since charge is conserved by the tunnel coupling, i.e., $[V^r,N+N^r]=0$ [\Eq{eq:Xconservation}],
we can use the simplification \eq{eq:conservation} to obtain the connection \eq{eq:Ag}. 
With the pseudo inverse
$  W^{-1}
  =
  \sum_{k=1,2} \tfrac{1}{\lambda_k} \Lket{v_k} \Lbra{\overline{v}_k}
$
and using
\begin{align}
  2 W^r_{0,1}[\vec{R}]+W^r_{1,0}[\vec{R}]&=2 \Gamma^r[\vec{R}] \\
  2 W^r_{2,1}[\vec{R}]+W^r_{1,2}[\vec{R}]&=2 \Gamma^r[\vec{R}]
\end{align}
using notation borrowed from \Ref{Schulenborg16a}
the pumping response-covector \Eq{eq:ResponseVectorRoman} is obtained as:
\begin{subequations}
\begin{align}
  \Lbra{\Phi_{N^r}[\vec{R}]}
  &=
  -\Lbra{\hat{N}} W^r[\vec{R}] \frac{1}{W[\vec{R}]}
    \\
  &=
  \frac{\lambda_1^r[\vec{R}]}{\lambda_1[\vec{R}]}
  \, \left[ \Lbra{\hat{N} }  - N_0[\vec{R}] \, \Lbra{\unit} \right]
\label{eq:covectorexampleb}
  \\
	&\propto \Lbra{\overline{v}_1[\vec{R}]}
\label{eq:covectorexamplec}
  \, ,
  \end{align}\label{eq:covectorexample}\end{subequations}
where $N_0[\vec{R}] = \Lbraket{N | v_0[\vec{R}] }$ is the frozen-parameter stationary-state charge.
Since $\Lbraket{\mathds{1} | \delR \rho^{i}[\vec{R}] } = 0$
the $\vec{R}$-dependence responsible for charge pumping
enters only through the first nonzero eigenvalue
of $W^r$
\begin{align}
  \lambda_1^r[\vec{R}]=
  -\tfrac{1}{2} \big[ W^r_{1,0}[\vec{R}]+W^r_{1,2}[\vec{R}]  \big]
\end{align}
and the eigenvalue $\lambda_1= \sum_r \lambda_1^r$ of $W$.
The reason that the other eigenvalues $\lambda_2$ and $\lambda_2^r$ of $W$ and $W^r$, respectively, do not contribute
is that the eigenvector $\Lket{v_2} \propto \Lket{v_2^r} \propto (-1)^N$ is the fermion parity operator which plays a special role~\cite{Schulenborg16a}.

\emph{Pumping connection and curvature.}
Combining \Eq{eq:appExampleSteadyState} with \Eq{eq:covectorexample}
we obtain for the charge-pumping connection \eq{eq:Ag}
\begin{subequations}
	\begin{align}
&   A_0[\vec{R}] = \Lbraket{\Phi_{N^r} |  \delR \inst{\rho}} =
 \label{eq_AppExampleConnection} \\
 & \frac{\lambda_1^r[\vec{R}]}{2 \lambda_1^3[\vec{R}]}
 \Big[
 W_{1,0}[\vec{R}] \delR  W_{1,2}[\vec{R}] 
-
 W_{1,2}[\vec{R}] \delR   W_{1,0}[\vec{R}]
\Big] 
\label{eq_AppExampleConnection2}
	\end{align}
\end{subequations}
and for the pumping curvature \eq{eq:PseudoMagneticFieldConserved}
\begin{widetext}
\begin{align}
&   B[\vec{R}] =
\Lbra{\delR \Phi_{N^r}} \times \Lket{ \delR \inst{\rho}} =
  \label{eq:AppExampleCurvature1} \\
& = \frac{\lambda_1^r[\vec{R}]}{\lambda_1^3[\vec{R}]} \delR   W_{1,0}[\vec{R}] \times \delR   W_{1,2}[\vec{R}] 
& +  \delR \Big[\frac{\lambda_1^r[\vec{R}]}{2 \lambda_1^3[\vec{R}]} \Big] 
\times  \Big[ W_{1,0}[\vec{R}] \delR  W_{1,2}[\vec{R}] - W_{1,2}[\vec{R}] \delR   W_{1,0}[\vec{R}] \Big] \label{eq:AppExampleCurvature}
.
\end{align}
\end{widetext}

This result illustrates the geometric spectroscopy of \Sec{sec:application}:
when driving the system energies $H$ relative to the reservoirs $\mu^r$
for fixed coupling parameters $\Gamma^r$,
both $\Lbra{\delR \Phi_{N^r}}$ and $\Lket{ \delR \inst{\rho}}$ show sharp changes.
Due to the different functional dependence of covector and vector,
their gradients are nonparallel, i.e.,
\Eq{eq:covector-para} is violated and a pumping response 
can be expected at sharply defined parameter points.
Similar considerations hold for, e.g., the gradients $\delR W_{1,N}$ for $N=0,2$ in \Eq{eq:AppExampleCurvature1}.

In the present case, this pumping effect is \emph{induced} by the \emph{interaction}~\cite{Reckermann10a,Calvo12a,Haupt13} $U$ if the tunnel rates are not driven.
That the curvature vanishes in this case, can be seen from both formulas.
First, setting $U=0$ in the pumping curvature \Eq{eq:AppExampleCurvature} we find 
\begin{gather}
  \delR W_{1,0}[\vec{R}] + \delR W_{1,2}[\vec{R}] = \delR \Gamma[\vec{R}]
  \\
  \lambda_1^r[\vec{R}] = -\Gamma^r[\vec{R}],
  \quad
  \lambda_1[\vec{R}] = -\Gamma[\vec{R}]
\label{eq:AppExampleLambdaNonInteracting}
.
\end{gather}
where we used $f^{+}_r + f^{-}_r=1$ in \Eq{eq:appExampleRatesSpindegenerate}.
For constant tunnel strengths the gradients $\delR W_{1,0}$ and $\delR W_{1,2}$ are (anti)\emph{parallel} vectors, making the first term vanish,
even though the parameter dependence of rates for electron and hole processes \emph{is different}.
In the second term in \Eq{eq:AppExampleCurvature}, the first gradient is zero.

If one only wants to see that the curvature vanishes it is easier to use \Eq{eq:AppExampleCurvature1} and first set $U=0$. 
In this case, the form \eq{eq:covectorexampleb} of the response covector is simple
because the $N_0$ part drops out in the connection
\begin{align}
	A_0[\vec{R}] = \frac{\lambda_1^r[\vec{R}]}{2 \lambda_1^3[\vec{R}]} \Lbraket{N  |\delR \inst{\rho} [\vec{R}]}.
\end{align}
by probability conservation,
$\Lbraket{\mathds{1} | \delR \rho^{i}[\vec{R}] } = 0$.
Thus the only relevant part of the response covector is parameter independent
if the tunnel strengths are constant
and hence the pumping curvature vanishes [\Eq{eq:ResponseVectorDefinition} ff.].
This advantage of this derivation allows to avoid the evaluation of $\delR \inst{\rho}$ in \Eq{eq:covectorexamplec}.
We note that to compute nonzero curvature values for $U\neq 0$ the explicit form \eq{eq:AppExampleCurvature} is more convenient.

We therefore find that pumping without driving the tunnel rates is interaction \emph{induced}, i.e., there is no pumping effect in the absence of interaction, i.e., if $U=0$ during the whole driving cycle. 
This does not depend on the driving protocol or the choice of parameters, as long as the tunnel rates and interaction stay constant.
We note that when \emph{driving} the interaction $U$ itself, a pumping contribution can still arise as parameter regions with nonzero interaction are visited during the driving cycle.

The second curvature contribution in \Eq{eq:AppExampleCurvature} was not given in \Ref{Reckermann10a,Calvo12a,Haupt13} since the tunnel rates were assumed to be constant from the start. If this is not the case, driving of tunnel rates may result in pumping even without interaction.
Similar interaction-induced pumping effects have been discussed in different contexts\cite{Sinitsyn07EPL,Ren10}.

\section{Pumping of system observables -
\newline
\ar approach of Avron, et. al~\cite{Avron12}
\label{app:pumpsys}}

In this appendix we show how the \ar approach of Avron, Fraas, and Graf, in \Ref{Avron12} fits into our  considerations of gauge freedom of observables in \Sec{sec:geometry}
when we replace nonsystem observable $\hat{X}^r$ by a system observable $\hat{X}[\vec{R}]$.
We extend their considerations by allowing for possible parametric time-dependence of the observable including that introduced by a gauge transformation.

The Heisenberg current superoperator for a gauged \emph{system} observable $\hat{X}_g(t) := \hat{X}[\vec{R}(t)]+ g[\vec{R}(t)] \unit$ reads when written under the trace $\Lbra{\unit}=\tr$
\begin{align}
  \Lbra{\unit}  \currentKernel{X_g} [\vec{R}(t)]
  = \Lbra{X} W[\vec{R}(t)]  + \frac{d}{dt}g[\vec{R}(t)]  \Lbra{\unit}
  \label{eq:FluxOperatorAvron}
  .
\end{align}
In this simpler situation of a system observable, this can be directly seen:
following \Ref{Avron12}, the right hand side of \Eq{eq:FluxOperatorAvron} is calculated  by inserting
$\frac{d}{dt}\rho(t) = W(t) \rho(t)$ into
the expectation value
$ \frac{d}{dt}  \expec{ \hat{X}_g }(t)
= \frac{d}{dt} \Lbraket{ \hat{X}_g  | \rho(t) }
= \Lbra{\unit} \currentKernel{ X_g } \Lket{ \rho(t) }
$.
This is the \emph{system}-observable analog to the Heisenberg equation of motion \eq{eq:WIX} for nonsystem observables where $W[\vec{R}]$ is effectively replaced by $W^r[\vec{R}]$.
Here, it is not necessary to start from the total system current \eq{eq:FluxOperator2}
as we did in deriving the current kernel \eq{eq:WIX}.
Note, in particular, that in the above we did not assume that $\left [ H,X \right ]=0$, as we did in \Eq{eq:HXcom}.

As in \Ref{Avron12}, we now solve the master equation for the adiabatic response,
$\adcor{\rho}=\frac{1}{W} \delR \inst{\rho}$,
and obtain
for the transported observable $\Delta X = \Delta \inst{X} +\Delta \adcor{X}$:
\begin{align}
  \Delta \inst{X}  & = \int_0^{\T} dt \Lbra{\unit} \currentKernel{X}[\vec{R}(t)] \Lket{\inst{\rho}[\vec{R}(t)]}
\\
  \Delta \adcor{X} &= \int_{\mathcal{C}} d \vec{R} A_g[\vec{R}]
                   = \int_{\mathcal{S}} d S B[\vec{R}]
\end{align}
where we recognize the Landsberg connection $A_g[\vec{R}] = A_0[\vec{R}] + \delR g[\vec{R}]$ with
\begin{align}
  A_0[\vec{R}] & = \Lbra{X[\vec{R}]}
                 Q[\vec{R}]
                 \Lket{\delR \inst{\rho}[\vec{R}]}
  \label{eq:A0Avron}
  .
\end{align}
As in \Ref{Avron12} the result is expressed in the non-Hermitian projector on the Liouville subspace
complementary to the unique stationary-state zero eigenvalue
on which the pseudo-inverse $W^{-1}$ is computed:
\begin{align}
  Q[\vec{R}]
  := W[\vec{R}] \frac{1}{W[\vec{R}]}
  =\mathcal{I}-\Lket{\inst{\rho}[\vec{R}]}\Lbra{\unit} 
\end{align}
Our derivation emphasizes that the projector $Q$ has two factors:
$W$ comes from the system observable [\Eq{eq:FluxOperatorAvron}],
 whereas $W^{-1}$ accounts of the ``lag'' time of the evolution [\Eq{eq:ResponseVectorDefinition} ff.].
The pumping connection \eq{eq:A0Avron} corresponds to Theorem 9 of \Ref{Avron12}.
We refer to this work for a detailed discussion, as well as additional interesting results which find no parallel in the present paper, e.g., the case of nonunique stationary states.
The corresponding curvature
\begin{subequations}\begin{align}
  B[\vec{R}] & = 
               \Lbra{X[\vec{R}]} \Big[ \delR  Q[\vec{R}] \Big]
               \times
               \Lket{\delR \inst{\rho} [\vec{R}]}
               \label{eq:Avron}
               \\ &
             +
             \Lbra{\delR X[\vec{R}]}
             \,  Q[\vec{R}]
            \times
               \Lket{\delR \inst{\rho} [\vec{R}]}
               \label{eq:notAvron}
                .
\end{align}\end{subequations}
shows that the $\vec{R}$-dependence of projector $Q$ in the term \eq{eq:Avron} generates pumping as shown in \Ref{Avron12}
[cf. our discussion in \Sec{sec:application} of conserved nonsystem observables \Eq{eq:ResponseVectorRoman}].
The term \eq{eq:notAvron} does not appear in \Ref{Avron12} since there $\hat{X}$ is explicitly assumed to be $\vec{R}$ independent,
but it is related to our discussion following \Eq{eq:HXcom}.

The geometric pumping response of \Ref{Avron12}, there formulated elegantly in terms Liouville-space Kato projections, can thus also be considered as
another instance of Landsberg's nonadiabatic geometric phase.
The gauge freedom responsible for the geometric effect
lies in the observable and the relevant fiber bundle is the same as that discussed in \Figs{fig:meter} - \fig{fig:physical} with $\hat{X}^r \to \hat{X}$.
Our discussion of \Ref{Avron12} thus underscores a number of key points of of the main text  in a simpler setting:

(i) 
Pumping involves a  geometric \emph{non}adiabatic response
of the quantum state relative to the instantaneous solution
of the Berry-Robbins type (``geometric magnetism'')~\cite{\GEOMETRICMAGNETISM}.

(ii)
Although geometric, the pumping effect is \emph{not} a ``Berry-Simon'' phase,
i.e., associated with adiabatic \emph{state} evolution:
the gauge freedom responsible for the result \Eq{eq:A0Avron} lies in the observable, not in the state.

(iii)
The connection \Eq{eq:A0Avron} relies critically on the decay to a unique frozen-parameter \emph{stationary} state, a key feature distinguishing  Landsberg's phase in dissipative systems from the Berry-Simon phase.

(iv)
From \Sec{sec:fcs} it is now also clear that the results of an \fcs approach for pumping of the first moment of \emph{system} observables will reduce to those obtained by the simpler \ar approach of \Ref{Avron12}.

\section{Response covector -- Generalized Brouwer formula.\label{app:brouwer}}
In \Ref{Calvo12a}, pumping-response \emph{coefficients} were introduced that are closely related to our response covector \eq{eq:ResponseVectorDefinition} (see also \Ref{Avron12}).
These allow to bring the Landsberg curvature \eq{eq:PseudoMagneticField} into a form similar to the well-known Brouwer formula~\cite{Brouwer98} valid in the noninteracting / mean-field picture of transport. Additional terms arise for conserved parameter-dependent system observables.
In the presence of interactions, model-specific calculations~\cite{Splettstoesser06,Splettstoesser08a,Reckermann10a,Calvo12a,Riwar13} have been shown to take such a form.
Here we briefly describe how the considerations of \Ref{Calvo12a} extend to our more general setting.

To obtain the Brouwer-type form, one expands the instantaneous stationary state $\inst{\rho}$
in \emph{any} complete orthonormal basis of observables $\hat{Y}[\vec{R}]=\hat{Y}^\dag[\vec{R}]$,
i.e., for which $\sum_Y \Lket{Y[\vec{R}]}\Lbra{Y[\vec{R}]}=\mathcal{I}$
with $\Lket{Y[\vec{R}]}=\hat{Y}[\vec{R}]$, $\Lbra{Y[\vec{R}]}\bullet =\tr \, \hat{Y}[\vec{R}] \bullet$.
For example, for an interacting, single-orbital  quantum dot coupled to normal metal electrodes,
one possible set of such observables is
$\{\hat{Y}\} = \{ \unit,\hat{N},\hat{S}_z,(-\unit)^{\hat{N}} \}$.
In the expansion
\begin{align}
  \Lket{\inst{\rho}[\vec{R}] } = \sum_Y \Lket{Y[\vec{R}]} \,  \inst{\expec{Y}}[\vec{R}] \label{eq:appBrouwerStateExpansion}
  .
\end{align}
the coefficients
are the stationary-state averages of these observables,
$\inst{\expec{\hat{Y}}}[\vec{R}]
= \Lbraket{Y[\vec{R}]|\inst{\rho}[\vec{R}]}
= \tr \, \hat{Y}[\vec{R}] \inst{\rho}[\vec{R}]$.
In Brouwers formula, the pumping response of \emph{reservoir}-observable $\hat{X}^r$ is expressed in terms of the responses of the averages of these \emph{system} observables. In our case, using \Eq{eq:PseudoMagneticField} the pumping curvature with the expansion \eq{eq:appBrouwerStateExpansion} is
\begin{align}
  B_{X^r}[\vec{R}] = 
	\sum_Y &\Lbraket{ \delR  \Phi_{X^r}[\vec{R}]|Y[\vec{R}]} 
  \times \delR  \inst{\expec{\hat{Y}}}[\vec{R}] \notag \\
	+& \Lbra{ \delR  \Phi_{X^r}[\vec{R}]} \times \Lket{\delR Y[\vec{R}]} 
  \ \inst{\expec{\hat{Y}}}[\vec{R}]
  \label{eq:brouwer}.
\end{align}

The first term in \Eq{eq:brouwer} is similar to Brouwer's formula. It contains the parameter gradient of the average system observables, whereas in the second term these averages appear by themselves.
Moreover in the first term the role of response vector \eq{eq:ResponseVectorDefinition} is to assign to each system observable $\hat{Y}[\vec{R}]$ its response-coefficient $\Lbraket{\Phi_{X^r}[\vec{R}]|Y[\vec{R}]}= \tr \, \Phi_{X^r}^\dag[\vec{R}] \hat{Y}[\vec{R}] $ under the $\delR$ in \Eq{eq:brouwer}.
Although similar to Brouwer's formula, \Eq{eq:brouwer} holds for strongly interacting, weakly coupled open quantum systems,
with (possibly nonconserved) nonsystem observable $\hat{X}^r$.

Without further assumptions, there is no natural choice for the $\{\hat{Y}[\vec{R}]\}$:
unitary transformations of this basis of operators leave \eq{eq:brouwer} invariant.
This nonuniqueness may seem a disadvantage that \Eq{eq:brouwer} as compared to the basis-independent response-covector \Eq{eq:ResponseVectorDefinition}.
However,
when $\hat{X}^r+\hat{X}[\vec{R}]$ is conserved at junction $r$ [\Eq{eq:Xconservation}]
we can exploit the simplification \eq{eq:ResponseVectorRoman}
by including the corresponding system observable $\hat{X}[\vec{R}]$ in the expansion set $\{\hat{Y}[\vec{R}]\}$.
The pumping-response of the conserved observable $\hat{X}^r$ can then be
written
as a ``self-response'' plus cross-responses of other, linearly independent observables $\hat{Y}[\vec{R}]$:
\begin{widetext}
\begin{align}
  & B_{X^r}[\vec{R}] = \label{eq:brouwer2}
\\
& 
  -
  \Big[ \delR \Lbra{X [\vec{R}] }W^r[\vec{R}]\frac{1}{W[\vec{R}]} \Big] \Lket{X}
  \times \delR  \inst{\expec{\hat{X}[\vec{R}]}}
  \notag 
 -
\Big[ \delR \Lbra{X [\vec{R}] }W^r[\vec{R}]\frac{1}{W[\vec{R}]} \Big] \times \Lket{\delR X[\vec{R}]} \
  \inst{\expec{\hat{X}[\vec{R}]}} \notag \\
& -
  \sum_{Y\neq X}
	\left \{
   \Big[ \delR \Lbra{X [\vec{R}] } W^r[\vec{R}]\frac{1}{W[\vec{R}]}  \Big] \Lket{Y}
  \times \delR  \inst{\expec{\hat{Y}[\vec{R}]}} 
 - 
	\Big[ \delR \Lbra{X [\vec{R}] } W^r[\vec{R}]\frac{1}{W[\vec{R}]}  \Big]  \times \Lket{\delR Y[\vec{R}]} \
  \inst{\expec{\hat{Y}[\vec{R}]}}
	\right \} \notag
\end{align}
\end{widetext}
For example, for charge pumping, with parameter independent $X$ and $Y$,
in \Ref{Splettstoesser06} only the first term played a role  ($\hat{X}=\hat{N}$) due to spin-symmetry,
whereas in \Ref{Reckermann10a} the cross-talk with the spin ($\hat{Y}=\hat{S}_z$) in the third term is relevant as well.
In \Refs{Calvo12a,Haupt13} spin pumping also involved two terms ($\hat{X}=\hat{S}_z$ and $\hat{Y}=\hat{N}$).\footnote{The explicit pumping curvature result \Eq{eq:brouwer2}
corresponds to expressing in the intermediate calculation,
for example, the \ar charge current as
$\adcor{I}_{N^r}(t) = \Lbraket{\Phi_{N^r}|N} \cdot \frac{d}{dt}\inst{\expec{\hat{N}}}$
[\Ref{Splettstoesser06}] and
$\adcor{I}_{N^r}(t) = \Lbraket{\Phi_{N^r}|N} \cdot \frac{d}{dt}\inst{\expec{\hat{N}}}
+ \Lbraket{\Phi_{N^r}|S_z} \cdot \frac{d}{dt}\inst{\expec{\hat{S}_z}}$
[\Refs{Reckermann10a}].}
Importantly, \Eq{eq:brouwer2} also applies to discrete many-particle systems whose stationary state is specified by a larger set of observables.
\section{Adiabatic geometric phase for mixed states:
\newline
\ase approach of Sarandy and Lidar  \label{app:ase}}

In the main text we make repeated use of insights offered by the \ase approach of \Refs{Sarandy05,Sarandy06}, an approach not related to pumping.
Here we summarize the required key points,
but also discuss some relevant issues that can {not} be found in \Refs{Sarandy05,Sarandy06}.

\emph{Adiabatic approximation.}
Sarandy and Lidar consider a time-local \mE for the density operator
\begin{align}
  \frac{d}{dt} \Lket{\rho(t)} = W(t) \Lket{\rho(t)}
  \label{eq:MarkovMeapp}
	.
\end{align}
Since this approach can deal with non-steady-state evolution it is convenient to start the evolution from state $\rho(0)$ at $t_0=0$ (instead of $t_0 \to -\infty$ as in the main text).
In our setting, we assume the kernel decomposes as
\begin{align}
  W(t)=\sum_n \lambda_n(t) \Lket{v_n(t)} \Lbra{\overline{v}_n(t)}
  \label{eq:Wspectral}
\end{align}
with the right- and left eigenvectors $\Lket{v_n(t)}$ and $\Lbra{\overline{v}_n(t)}$ to the possibly complex eigenvalues
 $\lambda_n(t)$. The eigenvalues are assumed to be nondegenerate for simplicity
and to have non-positive real parts.
The density operator is expanded in the right eigenvectors
\begin{align}
  \Lket{\rho}(t)= \sum_n e^{\Lambda_n(t)}  \rho_n(t)  \Lket{v_n(t)} \label{eq:rhoExpansion}
  ,
\end{align}
where the anticipated dynamical ``phase'' $\Lambda_n(t) = \int_0^t dt' \lambda_n(t')$
is split off right away.
Inserting this ansatz into \Eq{eq:MarkovMeapp}
yields coupled equations for the coefficients:
\begin{align}
  \partial_t \rho_n(t) =& -\Lbraket{\overline{v}_n(t)|\partial_t v_n(t)} \rho_n(t)
  \label{eq:exactCoefficientsForDensity}\\
  &- \sum_{m \neq n} \Lbraket{\overline{v}_n(t)|\partial_t v_m(t)} \rho_m(t) e^{\Lambda_m(t) - \Lambda_n(t)} 
    \notag
		.
\end{align}
At this point we apply the adiabatic approximation for open quantum systems, discussed carefully by Sarandy and Lidar~\cite{Sarandy05},
which amounts to neglecting the second line in \Eq{eq:exactCoefficientsForDensity}.
The condition under which this is valid is described in \Eq{eq:StateGapCondition}.
The adiabatic  approximation, here labeled by ``i'' for ``instanteneous''\footnote{The adiabatic solution is the instantaneous solution with the right ``phase'' factors.}
as in the main text,
for the non-steady-state solution of the master-equation \eq{eq:MarkovMeapp} is given by:
\begin{align}
  \Lket{\inst{\rho}(t) } = \sum_{n}
e^{ \Lambda_n(t)}
e^{\gamma_{n}(t)}
\rho_n(0) \Lket{v_n(t)} \label{eq:adiabaticSolution}
.
\end{align}

Thus, in the adiabatic evolution the \emph{components} of the mixed state $\Lket{\rho(t)}$ in the \emph{various different} instantaneous eigenspaces of $W(t)$  evolve independently of each other.
Each eigenvector $\Lket{v_n(t)}$ evolves with its own 
geometric phase factor determined by a separate connection for each eigenspace $n$:
\begin{align}
  \gamma_{n}(\T) = - \oint_{\mathcal{C}} d\vec{R} A_n[\vec{R}]
  \label{eq:gamman}
  \\
  A_n[\vec{R}]:= \Lbraket{\overline{v}_n[\vec{R}] | \partial v_n[\vec{R}] }
  \label{eq:An}
	.
\end{align}
This is the most direct generalization of the Berry-Simon phase for adiabatic time evolution to open systems evolving with \Eq{eq:MarkovMeapp}.

\emph{Adiabatic-response correction.}
In \Ref{Sarandy05} the first nonadiabatic correction to the result \Eq{eq:adiabaticSolution}, here labeled ``a'' for ``adiabatic-response'' as in the main text, is shown to be
\begin{widetext}
\begin{align}	
  \Lket{\adcor{\rho}(t) } = &
  \sum_{n} \Lket{v_n(t)}
  \left\{ 	
    \sum_{m\neq n} e^{\Lambda_m(t)} e^{\gamma_m(t)} \frac{\Lbraket{\overline{v}_n(t)|\partial_t v_m(t)}}{\lambda_n(t) - \lambda_m(t)}  \rho_m(0) 
  - \sum_{m\neq n} e^{\Lambda_n(t)}e^{\gamma_n(t)} \frac{\Lbraket{\overline{v}_n(0)|\partial_t v_m(0)}}{\lambda_n(0) - \lambda_m(0)}  \rho_m(0) 
  \right. \notag
\\
  &
  \left.
 +  \sum_{m\neq n}  e^{\Lambda_n(t)} e^{\gamma_n(t)} \int_0^t d\tau   \frac{\Lbraket{\overline{v}_n(\tau)|\partial_\tau v_m(\tau)} \Lbraket{\overline{v}_m(\tau)|\partial_\tau v_n(\tau)}}{\lambda_n(\tau) - \lambda_m(\tau)}  \rho_n(0)
  \right\}
.\label{eq:adiabatic-correction}
\end{align}
\end{widetext}
Both lines of \eq{eq:adiabatic-correction} each can be understood in an intuitive picture. 
The first line describes processes where an initial state $m$ ``leaks'' into state $n$ via the coupling term 
\begin{align}
	\tfrac{\Lbraket{\overline{v}_n(\tau)|\partial_\tau v_m(\tau)}}{\lambda_n(\tau) - \lambda_m(\tau)}
\end{align}
either at final time $\tau=t$ or initial time $\tau=0$. The ``phase'' factors (exponentials) correspond to the dynamics after and before the transition.
The second term can be understood as a leakage from the state $n$ into itself: it contains the coupling terms for a transition from $n$ to $m$ and back. Although this seems to be a higher order contribution, it is in fact not since one integrates such processes over all possible times.

\emph{Gap condition.}
Reference \ref{Sarandy05} derives the gap condition ensuring the validity of the adiabatic decoupling, i.e., neglecting the correction \eq{eq:adiabatic-correction} relative to the adiabatic solution \eq{eq:adiabaticSolution},
similar to the closed system case, as
\begin{align}
  \max_{0<t<\T} \, \, 
 \left | \frac{\Lbraket{\overline{v}_n(t)|\partial_t v_m(t)}}{\lambda_n(t) - \lambda_m(t)} \right |
  \, \ll \, ||\bar{v}_n|| \,  || v_m|| \label{eq:StateGapCondition}
	.
\end{align}
for all $n$,$m$\footnote{Indeed condition \eq{eq:StateGapCondition} for the contribution \eq{eq:adiabatic-correction} to be negligible is sufficient but not necessary. 
Due to the decay in an open system some contributions are exponentially suppressed even without the condition \eq{eq:StateGapCondition}.
This condition is thus only required for the nondecaying terms.
}.
For a system driven at frequency $\Omega$ with dimensionless amplitudes $\delta\vec{R}$ and eigenvalues with relevant scale $\Gamma$ the condition \eq{eq:StateGapCondition}
gives:
\begin{align}
  \Big|  \frac{\delta \vec{R} \Omega}{\Gamma}  \Big| \ll 1 .
  \label{eq:condition-ase}
\end{align}
This is identical to \Eq{eq:condition-ar} of the main text. It is crucial to note the distinct ways in which this gap condition is used:

(a) In both the \ase and the \fcs approaches, this is the condition to \emph{neglect} the adiabatic response $\Lket{\rho^{a}}$ or $\Lket{\adcor{\rho}^{\chi}}$.

(b) In the \ar approach, at the focus of the main text,
this condition is used to justify [\App{app:iteration}] \emph{keeping} the adiabatic-response,
yet \emph{neglecting higher} adiabatic corrections (nonlinear in $\Omega$), which scale with $(\delta \vec{R} {\Omega} / {\Gamma} )^2$.

We stress this point here and in in the main text since it easily leads to confusion.
For example, in \Sec{sec:fcsAdiabatic} we show that in the \fcs the \emph{kept}  ``adiabatic'' / ''instantaneous'' contribution, through the ``$\chi$-bookkeeping-device'' of the \fcs,
in fact leads back to case (b):
one effectively uses \Eq{eq:condition-ase} to motivate \emph{keeping} a \emph{physically nonadiabatic} (!) part.

\emph{Gauge freedom in adiabatic mixed-state evolution.}
Similar to the closed-system case,
the non-degenerate instantaneous eigenvectors of the kernel $W(t)$ are only determined up to a nonzero complex time-dependent factor,
which is, however, complex:
a gauge transformation with $c_n(t) \in \mathds{C}/\{0\}$, preserving the biorthonormality,
\begin{subequations}\begin{align}
  \Lket{v_n(t)} & \rightarrow \,  c_n(t) \, \Lket{v_n(t)} \\
  \Lbra{\overline {v}_n(t)} & \rightarrow  c^{-1}_n(t) \Lbra{\overline {v}_n(t)}
\end{align}\label{eq:ase-gauge}\end{subequations}
clearly leaves the kernel $W(t)$ invariant and thus also the solution of the \mE \eq{eq:MarkovMeapp} remains unchanged.
In agreement with this, the density operator contributions \eq{eq:adiabaticSolution} and \eq{eq:adiabatic-correction} are invariant under gauge transformations:
although, the connections \eq{eq:An} are gauge dependent, transforming as $A_n \to A_n + c_n^{-1} \delR c_n$,
these enter only through the invariant geometric phases \eq{eq:gamman}.

In our discussion of the Landsberg (\Sec{sec:geometry}) and the \fcs (\Sec{sec:fcs}) approaches we made use of the following important result 
which was not discussed in \Refs{Sarandy05,Sarandy06}:
the gauge freedom \eq{eq:ase-gauge} still contains a mathematical redundancy
which can be eliminated by taking into account two physical restrictions on the evolution of the mixed-state $\rho(t)$.

(a) First, the preservation of the trace-normalization of $\rho(t)$, i.e.,
$0 = \partial_t \tr{ {\rho} } = \Lbraket{\mathds{1}|W|\rho}$, 
requires that $\Lbra{\overline{v}_0}=\Lbra{\mathds{1}} = \tr$ is always a left eigenvector of $W$ to eigenvalue $\lambda_0=0$.
The corresponding right eigenvector $\Lket{v_0(t)}$ is by our nondegeneracy assumption the only eigenvector corresponding to a nonzero trace operator:
by biorthonormality it has $\Lbraket{\unit | v_0(t) } = \tr v_0(t) = 1$ for all $t$.
This fixes
left- and right eigenvector of the $\lambda_0=0$ eigenspace and no (nonzero) geometric phase can appear here\footnote{See \Ref{Sinitsyn09} (p. 8) for a related observation in the \fcs approach.}.
Thus, the gauge freedom for the stationary eigenspace ($\lambda_0=0$) is effectively quenched. This was mentioned in \Sec{sec:LackOfPhaseState} of the main text.

(b) Similarly, preservation of Hermiticity of $\rho(t)$ can be used to reduce the gauge freedom to real-valued, nonzero eigenvalues
positive numbers, i.e., $c_n(t) \in \mathds{R}^{+}$ for each eigenspace $n\neq0$.
This corresponds to working with an eigenbasis of Hermitian operators.

\emph{Quenching of adiabatic Berry phase in the steady-state.}
Finally, we discuss how the above \ase results
simplify in the steady-state limit relevant for pumping.
This was also not part of \Refs{Sarandy05,Sarandy06}.
In the steady-state limit, i.e., for times\footnote{This is consistent with the adiabatic limit:
to describe at least one driving cycle one needs $t \geq \Omega^{-1}$ to be compatible with the gap condition \eq{eq:condition-ase} $\Omega \delta{R} \ll \Gamma$ which is fine for $t \gg \Gamma^{-1}$.}
 $t  \gg \text{min}_{n \neq 0,\vec{R} \in \mathcal{C}} \{ \lambda_n^{-1} \} \sim \Gamma^{-1}$
all non-steady-state exponentials in \Eq{eq:adiabaticSolution} and \eq{eq:adiabatic-correction} have decayed to zero and reduce to, respectively,
\begin{subequations}\begin{align}
  \Lket{\inst{\rho}(t)} =& \Lket{v_0(t)}
  \label{eq:quench}
  ,
 \\
  \Lket{\adcor{\rho}(t)} =& \sum_{n\neq 0} \Lket{v_n(t)} \frac{\Lbraket{\overline{v}_n(t)| \partial_t |v_0(t)}}{\lambda_n(t)}
                              =   \frac{1}{W} \partial_t \Lket{ \inst{\rho}(t) }
  \label{eq:quencha}
\end{align}\end{subequations}
with the pseudo inverse of $W$ [cf. \Eq{eq:Wspectral}].
These are exactly \Eq{eq:instDensityOp} and \eq{eq:adcorDensityOp} that we obtained in a much simpler way in the ``naive'' \ar approach in \Sec{sec:naive}.
However, this comparison shows that in the adiabatic evolution \eq{eq:quench} only the zero eigenvector ($\lambda_0=0$) remains, whose geometric phase is always zero
\begin{align}
  \gamma_{0}=0
  \label{eq:gamma0zero} 
\end{align}
due to the quenching of the gauge freedom.

For the \emph{nonadiabatic} correction \eq{eq:adiabatic-correction},
leading to \eq{eq:quencha},  \Refs{Sarandy05,Sarandy06} do not discuss a possible geometric-phase.
In \App{app:iteration} we show that ``iterative'' geometric phases are associated with correction \eq{eq:adiabatic-correction} and even higher ones respectively, which however are all quenched in the steady-state limit.
\section{Nonadiabatic geometric phase for mixed states
- extending Berry's adiabatic iteration\label{app:iteration}}

Equation \eq{eq:rhocont} in \Sec{sec:LackOfPhaseState} expressed the key point, that the steady-state mixed state is continuous even when accounting for higher orders of the driving frequency, assuming the Born-Markov equation~\eq{eq:MarkovMe} with nondegenerate states eigenvectors as in \App{app:ase}
In the adiabatic limit this statement follows from taking the steady-state limit of the result of Sarandy and Lidar~\cite{Sarandy05} as described in \Eq{eq:gamma0zero}.
Here we show that the nonadiabatic corrections do not break this result and
explain in which sense these corrections relate to \emph{nonadiabatic} geometric phases.

As pointed out by Berry~\cite{Berry87} for unitary closed-system evolution, nonadiabatic corrections can be obtained as a geometric phase effect \emph{relative} to an adiabatic solution
 by performing an iterative adiabatic approximation. 
Here we extend this to the nonunitary open-system evolution
given by the Born-Markov \Eq{eq:MarkovMe}:
In \App{app:ase} we have seen  by the \ase approach
that the adiabatic time-evolution superoperator,
defined by $\Lket{\inst{\rho}(t)} = \Pi^0(t) \Lket{\inst{\rho}(0)}$,
reads as [\Eq{eq:adiabaticSolution}]:
\begin{align}
  \Pi^0(t) = \sum_n e^{\phi_n(t)} \Lket{v_n(t)} \Lbra{\overline{v}_n(0)}, \label{eq:BerryIterativeU0}
\end{align}
where $\phi_n(t) = \Lambda_n(t) + \gamma_n(t)$,
 $\Lambda_n(t)=\int_0^t \lambda_n(\tau) d\tau$,
and $ \gamma_n(t):= - \int_0^t \Lbraket{\bar{v}_n(\tau)| \partial_{\tau}v_n(\tau) }  d\tau$.
The exact Born-Markov time evolution $\Pi(t)= T e^{\int_0^t W(\tau) d\tau}$ can now be expressed as a product of the adiabatic time evolution and a correction factor $\Pi^{1}(t)$:
\begin{align}
  \Pi(t) = \Pi^{0}(t) \Pi^{1}(t),
\end{align}
where $\Pi^{1}(t)$ corresponds to the time evolution in the rotating frame of the adiabatic decoupling. (This is similar to going to an interaction picture, but note that the evolutions $\Pi^1$ and $\Pi^0$ are nonunitary.)
Its evolution is generated by a new kernel $W^{1}(t)$:
\begin{align}
  \frac{d}{dt} \Pi^{1}(t) &= W^{1}(t) \Pi^{1}(t), \label{eq:IterativeMasterEquation} \\
  W^{1}(t) &=  \Pi^{0}(t)^{-1} \left [ W(t) -\partial_t \right ] \Pi^{0}(t) \label{eq:IterativeKernel}
	.
\end{align}
We again solve \Eq{eq:IterativeMasterEquation} by an adiabatic decoupling:
\begin{align}
  \Pi^{1}(t) \approx \sum_n e^{\phi_n^{1}(t)}
  \Lket{{v}^{1}_n(t)}\Lbra{\overline{v}^{1}_n(t)} 
	,
\end{align}
where now
$\phi_n^1(t) = \Lambda_n^1(t) + \gamma_n^1(t)$
with
$\Lambda^{1}_n(t)=\int_0^t \lambda_n^1(\tau) d\tau$
and
 $ \gamma_n^1(t):= - \int_0^t \Lbraket{\bar{v}_n^1(\tau)| \partial_{\tau} v_n^1(\tau) }  d\tau$
 are given in terms of the left- and right eigenvectors $\Lbra{\overline{v}^{1}_n(t)}$ and $\Lket{v^{1}_{n}(t)}$ to eigenvalues $\lambda^{1}_n(t)$ of the new kernel $W^{1}(t)$ given by \Eq{eq:IterativeKernel}.

Like the adiabatic phase in $\Pi^0$,
also this relative geometric phase is quenched when we take the steady-state limit:
in the rotating frame $\Lket{\rho^\prime(t)} = \Pi^{1}(t) \Lket{\rho(0)}$ also remains trace-normalized:
\begin{align}
  1=& \Lbra{\unit} \Pi(t) \Lket{\rho(0)} = \Lbra{\unit} \Pi^{0}(t) \Pi^{1}(t) \Lket{\rho(0)} \notag \\
  =& \Lbra{\unit} \Pi^{1}(t) \Lket{\rho(0)} =  \Lbraket{\unit | \rho^{\prime}(t)}
	.
\end{align}
In the second line we have used that $\Lbra{\unit}= \tr$ is a left eigenvector of $W^{0}$ to eigenvalue zero and therefore $\Lbra{\unit}\Pi^0 = \Lbra{\unit}$. 
Hence we know that $\Lbra{\overline{v}^{1}_{0}(t)} = \Lbra{\unit}$ is a left eigenvector of $W^{1}$ to eigenvalue zero.
Thus, the continuity \Eq{eq:rhocont} is enforced by probability normalization, even when including the adiabatic-response, $\rho(t) \approx \inst{\rho} + \adcor{\rho}$.

Clearly, we can continue this argument to find, that in each iteration step $k$ this is a left eigenvector of the corresponding generator $W^{k}$.
One then finds that the geometric and dynamic phases $\gamma_0^{k}$ and $\Lambda_0^k$, which remain in the steady-state limit, are zero
assuming\footnote{If this assumption breaks down at any iteration it would imply that the solution we seek has no unique steady state which is our working assumption.} the steady state at each iteration to be unique:
for the dynamical term this results trivially from the zero-eigenvalue $\lambda_0^{k}=0$.
For the geometric part this follows from the biorthonormality
 of $\Lbraket{\overline{v}_0^{k} | v^{k}_0(t)}=1$ for all $t$
with $t$-independent $\Lbra{\overline{v}_0^{k}} = \tr$:
\begin{subequations}\begin{align}
  \gamma_0^{k}(t) & =
-
\int_0^t d\tau
\Lbraket{\overline{v}_0^{k}(\tau) | \partial_\tau| v_0^{k}(\tau)}
\\
 & =  \int_0^t d \tau \Lbraket{\partial_t \overline{v}_0^{k}(\tau)| v_0^{k}(\tau)} =0
\end{align}\end{subequations}

Thus the geometric phase relative to the previous iteration is quenched and we can say that the nonadiabatic steady-state density operator ``exhibits no geometric phase''. 
Hence, a unique mixed steady-state will always return to itself after one driving period, i.e., it is continuous in the parameters [\Eq{eq:rhocont}].

\emph{Reduction to the steady state.}
Finally, we now show explicitly that to order $O(\dot{\vec{R}} / \Gamma)$ the steady-state solution after one iteration
coincides with the result of the more practical \emph{sum} expansion of the \ar approach in the main text.
The first iterative approximation amounts to finding the instantaneous stationary state equation \emph{in the first adiabatic frame}, i.e.,
solving $W^1 \Lket{v^1_0}=0$.
This can be done using the ansatz $\Lket{v^1_0(t)}=\Lket{v^0_0(0)}+\Lket{\delta v^1_0(t)}$ using that
$\Pi^0 \Lket{v^0_0(0)} = \Lket{v^0_0(t)}$ and $W(t) \Lket{v^0_0(t)} = 0$.
Transformed to the original frame, one obtains the approximate time-dependent steady-state solution:
\begin{align}
&\rho(t) = \Pi^0(t) \Lket{v^1_0(t)}
\\
&
\approx 
\Lket{v_0(t)} + \frac{1}{W(t) - [\partial_t \Pi^0(t)] \Pi^0(t)^{-1}} \partial_t \Lket{v_0(t)}
\notag
\end{align}
Assuming $W(t) \sim \Gamma \gg |\dot{\vec{R}}| = |\delta \vec{R}| \Omega \sim [\partial_t \Pi^0(t)] \Pi^0(t)^{-1}$
and restoring the notation $\Lket{v_0(t)}=\Lket{\inst{\rho}(t)}$
this reduces to the \ar result $\rho(t) \approx \inst{\rho} + \adcor{\rho}$ [\Eqs{eq:adiabatic-response} and \eq{eq:bothDensityOp}] of the main text:
\begin{align}
\rho(t)
\approx 
\Lket{\inst{\rho}(t)} + \frac{1}{W(t)} \partial_t \Lket{\inst{\rho}(t)}
\end{align}
However, what the above makes clear is that the first \emph{nonadiabatic} correction to the dynamics \emph{also} corresponds to a geometric phase in the quantum state,
the first-iteration phase $\gamma^1_0(\T)$, which is however zero in the steady state on general grounds.
This is the starting point for Landsberg's phase to emerge from the observable.

\section{Full counting statistics - explicit kernel\label{app:fcs}}

In this appendix we illustrate the relation of the \fcs kernel $W^\chi$ to the time-evolution kernel $W$ [\Eq{eq:CountingExpansionDensityZero}]
and to the observable current kernel $\currentKernel{X^r}$ [\eq{eq:Lambda}].
In particular, we show that the redundant $\Theta$ [\Eq{eq:Lambda}],
which obscures the relation of the $\chi$-linear part to the current kernels, is nonzero
even in a simple application.

The explicit \fcs kernel for the generating operator is taken from taken from \Refs{Nakajima15,Yoshii13} [Eq. (68) and Eq. (14), respectively] and describes the  example we discussed in \App{app:example} using the \ar approach:
\begin{align}
W^\chi= 
	-\scalemath{0.8}{
			\begin{pmatrix}  
					-W_{\uparrow,0}-W_{\downarrow,0} & W^{-\chi}_{0,\uparrow} & W^{-\chi}_{0,\downarrow} & 0 \\
					 W^{+\chi}_{\uparrow,0} & -W_{0,\uparrow}-W_{2,\uparrow} & 0 & W^{-\chi}_{\uparrow,2} \\
					 W^{+\chi}_{\downarrow,0} & 0 & -W_{0,\downarrow}-W_{2,\downarrow} & W^{-\chi}_{\downarrow,2} \\
					 0 & W^{+\chi}_{2,\uparrow} &  W^{+\chi}_{2,\downarrow} & -W_{\uparrow,2}-W_{\downarrow,2}
			\end{pmatrix}
			} \label{eq:AndersonCounting}
\end{align}
with counting-field dependent ``rates''
\begin{align}
	W_{i,j}^{\pm \chi}	= W^\text{R}_{i,j} + e^{\pm i \chi} W^\text{L}_{i,j}
  .
\end{align}
Clearly, $W^\chi |_{\chi=0}$ is exactly the quantum-state time-evolution kernel \eq{eq:AndersonMaster} in agreement with \Eq{eq:CountingExpansionDensityZero}.
However, the $\chi$-linear part of \Eq{eq:AndersonCounting} reads
\begin{align}
\left. \partial_{i\chi} W^\chi \right |_{\chi=0}= 
	-\scalemath{0.8}{
			\begin{pmatrix}  
				  0 & - W^\text{L}_{0,\uparrow} & -W^\text{L}_{0,\downarrow} & 0 \\
					 +W^\text{L}_{\uparrow,0} & 0  & 0 & -W^\text{L}_{\uparrow,2} \\
					 +W^\text{L}_{\downarrow,0} & 0 & 0 & -W^\text{L}_{\downarrow,2} \\
					 0 & +W^\text{L}_{2,\uparrow} &  +W^\text{L}_{2,\downarrow} & 0
			\end{pmatrix}
			} \label{eq:AndersonCountingLinearOrder} .
\end{align}
In contrast, computing the current kernel \eq{eq:CurrentKernelFormula} gives
\begin{align}
&	 \currentKernel{N^r}= \label{eq:AndersonCurrentKernel}
  \\ -\tfrac{1}{2} 
&	\scalemath{0.8}{
			\begin{pmatrix}  
				W^\text{L}_{0,\uparrow}+W^\text{L}_{0,\downarrow} & - W^\text{L}_{0,\uparrow} & -W^\text{L}_{0,\downarrow} & 0 \\
					 +W^\text{L}_{\uparrow,0} & W^\text{L}_{2,\uparrow}-W^\text{L}_{0,\uparrow} & 0 & -W^\text{L}_{\uparrow,2} \\
					 +W^\text{L}_{\downarrow,0} & 0 & W^\text{L}_{2,\downarrow}-W^\text{L}_{0,\downarrow} & -W^\text{L}_{\downarrow,2} \\
					 0 & +W^\text{L}_{2,\uparrow} &  + W^\text{L}_{2,\downarrow} & -W^\text{L}_{\uparrow,2}-W^\text{L}_{\downarrow,2}
			\end{pmatrix}
			}
\notag
.
\end{align}
in agreement with \Eq{eq:conservation}.
The difference between \Eq{eq:AndersonCountingLinearOrder} and \eq{eq:AndersonCurrentKernel}
\begin{align}
  &\Theta :=
    \left. \partial_{i\chi} W^\chi \right |_{\chi=0}
    -\currentKernel{N^r}
    =  \\
  &
    \tfrac{1}{2} \scalemath{0.8}{           
\begin{pmatrix}  
				W^\text{L}_{0,\uparrow}+W^\text{L}_{0,\downarrow} & + W^\text{L}_{0,\uparrow} & +W^\text{L}_{0,\downarrow} & 0 \\
					 -W^\text{L}_{\uparrow,0} & W^\text{L}_{2,\uparrow}-W^\text{L}_{0,\uparrow} & 0 & +W^\text{L}_{\uparrow,2} \\
					 -W^\text{L}_{\downarrow,0} & 0 & W^\text{L}_{2,\downarrow}-W^\text{L}_{0,\downarrow} & +W^\text{L}_{\downarrow,2} \\
					 0 & -W^\text{L}_{2,\uparrow} &  - W^\text{L}_{2,\downarrow} & -W^\text{L}_{\uparrow,2}-W^\text{L}_{\downarrow,2}
\end{pmatrix}
}
\notag
\end{align}
indeed represents a redundant superoperator with $\tr \Theta \bullet= 0$ since the sum of rows is zero, cf. \Eq{eq:AndersonMaster}:
we stress that the nonzero redundancy $\Theta$ in this relation enters \emph{both} through the choice of the current kernel used in the \ar approach [\Eq{eq:redundancy-ar} ff.] as well as in the \fcs approach [\Eq{eq:redundancyrhochi},\eq{eq:Lambda} ff.].


\begin{thebibliography}{209}%
\makeatletter
\providecommand \@ifxundefined [1]{%
 \@ifx{#1\undefined}
}%
\providecommand \@ifnum [1]{%
 \ifnum #1\expandafter \@firstoftwo
 \else \expandafter \@secondoftwo
 \fi
}%
\providecommand \@ifx [1]{%
 \ifx #1\expandafter \@firstoftwo
 \else \expandafter \@secondoftwo
 \fi
}%
\providecommand \natexlab [1]{#1}%
\providecommand \enquote  [1]{``#1''}%
\providecommand \bibnamefont  [1]{#1}%
\providecommand \bibfnamefont [1]{#1}%
\providecommand \citenamefont [1]{#1}%
\providecommand \href@noop [0]{\@secondoftwo}%
\providecommand \href [0]{\begingroup \@sanitize@url \@href}%
\providecommand \@href[1]{\@@startlink{#1}\@@href}%
\providecommand \@@href[1]{\endgroup#1\@@endlink}%
\providecommand \@sanitize@url [0]{\catcode `\\12\catcode `\$12\catcode
  `\&12\catcode `\#12\catcode `\^12\catcode `\_12\catcode `\%12\relax}%
\providecommand \@@startlink[1]{}%
\providecommand \@@endlink[0]{}%
\providecommand \url  [0]{\begingroup\@sanitize@url \@url }%
\providecommand \@url [1]{\endgroup\@href {#1}{\urlprefix }}%
\providecommand \urlprefix  [0]{URL }%
\providecommand \Eprint [0]{\href }%
\providecommand \doibase [0]{http://dx.doi.org/}%
\providecommand \selectlanguage [0]{\@gobble}%
\providecommand \bibinfo  [0]{\@secondoftwo}%
\providecommand \bibfield  [0]{\@secondoftwo}%
\providecommand \translation [1]{[#1]}%
\providecommand \BibitemOpen [0]{}%
\providecommand \bibitemStop [0]{}%
\providecommand \bibitemNoStop [0]{.\EOS\space}%
\providecommand \EOS [0]{\spacefactor3000\relax}%
\providecommand \BibitemShut  [1]{\csname bibitem#1\endcsname}%
\let\auto@bib@innerbib\@empty
\bibitem [{\citenamefont {Altland}\ and\ \citenamefont
  {Zirnbauer}(1997)}]{Altland97}%
  \BibitemOpen
  \bibfield  {author} {\bibinfo {author} {\bibfnamefont {A.}~\bibnamefont
  {Altland}}\ and\ \bibinfo {author} {\bibfnamefont {M.~R.}\ \bibnamefont
  {Zirnbauer}},\ }\href@noop {} {\bibfield  {journal} {\bibinfo  {journal}
  {Phys. Rev. B}\ }\textbf {\bibinfo {volume} {55}},\ \bibinfo {pages} {1142}
  (\bibinfo {year} {1997})}\BibitemShut {NoStop}%
\bibitem [{\citenamefont {Read}\ and\ \citenamefont {Green}(2000)}]{Read00}%
  \BibitemOpen
  \bibfield  {author} {\bibinfo {author} {\bibfnamefont {N.}~\bibnamefont
  {Read}}\ and\ \bibinfo {author} {\bibfnamefont {D.}~\bibnamefont {Green}},\
  }\href@noop {} {\bibfield  {journal} {\bibinfo  {journal} {Phys. Rev. B}\
  }\textbf {\bibinfo {volume} {61}},\ \bibinfo {pages} {10267} (\bibinfo {year}
  {2000})}\BibitemShut {NoStop}%
\bibitem [{\citenamefont {Snyder}\ and\ \citenamefont
  {Toberer}(2008)}]{Snyder08}%
  \BibitemOpen
  \bibfield  {author} {\bibinfo {author} {\bibfnamefont {G.~J.}\ \bibnamefont
  {Snyder}}\ and\ \bibinfo {author} {\bibfnamefont {E.~S.}\ \bibnamefont
  {Toberer}},\ }\href@noop {} {\bibfield  {journal} {\bibinfo  {journal} {Nat.
  Mater.}\ }\textbf {\bibinfo {volume} {7}},\ \bibinfo {pages} {105} (\bibinfo
  {year} {2008})}\BibitemShut {NoStop}%
\bibitem [{\citenamefont {Nayak}\ \emph
  {et~al.}(2008{\natexlab{a}})\citenamefont {Nayak}, \citenamefont {Simon},
  \citenamefont {Stern}, \citenamefont {Freedman},\ and\ \citenamefont
  {Das~Sarma}}]{Nayak08}%
  \BibitemOpen
  \bibfield  {author} {\bibinfo {author} {\bibfnamefont {C.}~\bibnamefont
  {Nayak}}, \bibinfo {author} {\bibfnamefont {S.~H.}\ \bibnamefont {Simon}},
  \bibinfo {author} {\bibfnamefont {A.}~\bibnamefont {Stern}}, \bibinfo
  {author} {\bibfnamefont {M.}~\bibnamefont {Freedman}}, \ and\ \bibinfo
  {author} {\bibfnamefont {S.}~\bibnamefont {Das~Sarma}},\ }\href@noop {}
  {\bibfield  {journal} {\bibinfo  {journal} {Rev. Mod. Phys.}\ }\textbf
  {\bibinfo {volume} {80}},\ \bibinfo {pages} {1083} (\bibinfo {year}
  {2008}{\natexlab{a}})}\BibitemShut {NoStop}%
\bibitem [{\citenamefont {Kitaev}(2009)}]{Kitaev09Conference}%
  \BibitemOpen
  \bibfield  {author} {\bibinfo {author} {\bibfnamefont {A.}~\bibnamefont
  {Kitaev}},\ }\href@noop {} {\bibfield  {journal} {\bibinfo  {journal} {AIP
  Conf. Proc.}\ }\textbf {\bibinfo {volume} {1134}},\ \bibinfo {pages} {22}
  (\bibinfo {year} {2009})}\BibitemShut {NoStop}%
\bibitem [{\citenamefont {Ryu}\ \emph {et~al.}(2010)\citenamefont {Ryu},
  \citenamefont {Schnyder}, \citenamefont {Furusaki},\ and\ \citenamefont
  {Ludwig}}]{Ryu10}%
  \BibitemOpen
  \bibfield  {author} {\bibinfo {author} {\bibfnamefont {S.}~\bibnamefont
  {Ryu}}, \bibinfo {author} {\bibfnamefont {A.}~\bibnamefont {Schnyder}},
  \bibinfo {author} {\bibfnamefont {A.}~\bibnamefont {Furusaki}}, \ and\
  \bibinfo {author} {\bibfnamefont {A.}~\bibnamefont {Ludwig}},\ }\href@noop {}
  {\bibfield  {journal} {\bibinfo  {journal} {New. J. Phys.}\ }\textbf
  {\bibinfo {volume} {12}},\ \bibinfo {pages} {065010} (\bibinfo {year}
  {2010})}\BibitemShut {NoStop}%
\bibitem [{\citenamefont {Hasan}\ and\ \citenamefont {Kane}(2010)}]{Hasan10}%
  \BibitemOpen
  \bibfield  {author} {\bibinfo {author} {\bibfnamefont {M.~Z.}\ \bibnamefont
  {Hasan}}\ and\ \bibinfo {author} {\bibfnamefont {C.~L.}\ \bibnamefont
  {Kane}},\ }\href@noop {} {\bibfield  {journal} {\bibinfo  {journal} {Rev.
  Mod. Phys.}\ }\textbf {\bibinfo {volume} {82}},\ \bibinfo {pages} {3045}
  (\bibinfo {year} {2010})}\BibitemShut {NoStop}%
\bibitem [{\citenamefont {Qi}\ and\ \citenamefont {Zhang}(2011)}]{Qi11}%
  \BibitemOpen
  \bibfield  {author} {\bibinfo {author} {\bibfnamefont {X.-L.}\ \bibnamefont
  {Qi}}\ and\ \bibinfo {author} {\bibfnamefont {S.-C.}\ \bibnamefont {Zhang}},\
  }\href@noop {} {\bibfield  {journal} {\bibinfo  {journal} {Rev. Mod. Phys.}\
  }\textbf {\bibinfo {volume} {83}},\ \bibinfo {pages} {1057} (\bibinfo {year}
  {2011})}\BibitemShut {NoStop}%
\bibitem [{\citenamefont {Budich}\ \emph {et~al.}(2015)\citenamefont {Budich},
  \citenamefont {Zoller},\ and\ \citenamefont {Diehl}}]{Budich15a}%
  \BibitemOpen
  \bibfield  {author} {\bibinfo {author} {\bibfnamefont {J.~C.}\ \bibnamefont
  {Budich}}, \bibinfo {author} {\bibfnamefont {P.}~\bibnamefont {Zoller}}, \
  and\ \bibinfo {author} {\bibfnamefont {S.}~\bibnamefont {Diehl}},\
  }\href@noop {} {\bibfield  {journal} {\bibinfo  {journal} {Phys. Rev. A}\
  }\textbf {\bibinfo {volume} {91}},\ \bibinfo {pages} {042117} (\bibinfo
  {year} {2015})}\BibitemShut {NoStop}%
\bibitem [{\citenamefont {Chiu}\ \emph {et~al.}(2016)\citenamefont {Chiu},
  \citenamefont {Teo}, \citenamefont {Schnyder},\ and\ \citenamefont
  {Ryu}}]{Chiu15}%
  \BibitemOpen
  \bibfield  {author} {\bibinfo {author} {\bibfnamefont {C.-K.}\ \bibnamefont
  {Chiu}}, \bibinfo {author} {\bibfnamefont {J.~C.~Y.}\ \bibnamefont {Teo}},
  \bibinfo {author} {\bibfnamefont {A.~P.}\ \bibnamefont {Schnyder}}, \ and\
  \bibinfo {author} {\bibfnamefont {S.}~\bibnamefont {Ryu}},\ }\href@noop {}
  {\bibfield  {journal} {\bibinfo  {journal} {Rev. Mod. Phys.}\ }\textbf
  {\bibinfo {volume} {88}},\ \bibinfo {pages} {035005} (\bibinfo {year}
  {2016})}\BibitemShut {NoStop}%
\bibitem [{\citenamefont {Kennedy}\ and\ \citenamefont
  {Zirnbauer}(2016)}]{Kennedy16}%
  \BibitemOpen
  \bibfield  {author} {\bibinfo {author} {\bibfnamefont {R.}~\bibnamefont
  {Kennedy}}\ and\ \bibinfo {author} {\bibfnamefont {M.~R.}\ \bibnamefont
  {Zirnbauer}},\ }\href@noop {} {\bibfield  {journal} {\bibinfo  {journal}
  {Commun. Math. Phys.}\ }\textbf {\bibinfo {volume} {342}},\ \bibinfo {pages}
  {909} (\bibinfo {year} {2016})}\BibitemShut {NoStop}%
\bibitem [{\citenamefont {Diehl}\ \emph {et~al.}(2011)\citenamefont {Diehl},
  \citenamefont {Rico}, \citenamefont {Baranov},\ and\ \citenamefont
  {Zoller}}]{Diehl11a}%
  \BibitemOpen
  \bibfield  {author} {\bibinfo {author} {\bibfnamefont {S.}~\bibnamefont
  {Diehl}}, \bibinfo {author} {\bibfnamefont {E.}~\bibnamefont {Rico}},
  \bibinfo {author} {\bibfnamefont {M.~A.}\ \bibnamefont {Baranov}}, \ and\
  \bibinfo {author} {\bibfnamefont {P.}~\bibnamefont {Zoller}},\ }\href@noop {}
  {\bibfield  {journal} {\bibinfo  {journal} {Nature Phys.}\ }\textbf {\bibinfo
  {volume} {7}},\ \bibinfo {pages} {971} (\bibinfo {year} {2011})}\BibitemShut
  {NoStop}%
\bibitem [{\citenamefont {Bardyn}\ \emph {et~al.}(2013)\citenamefont {Bardyn},
  \citenamefont {Baranov}, \citenamefont {Kraus}, \citenamefont {Rico},
  \citenamefont {İmamoğlu}, \citenamefont {Zoller},\ and\ \citenamefont
  {Diehl}}]{Bardyn13}%
  \BibitemOpen
  \bibfield  {author} {\bibinfo {author} {\bibfnamefont {C.-E.}\ \bibnamefont
  {Bardyn}}, \bibinfo {author} {\bibfnamefont {M.~A.}\ \bibnamefont {Baranov}},
  \bibinfo {author} {\bibfnamefont {C.~V.}\ \bibnamefont {Kraus}}, \bibinfo
  {author} {\bibfnamefont {E.}~\bibnamefont {Rico}}, \bibinfo {author}
  {\bibfnamefont {A.}~\bibnamefont {İmamoğlu}}, \bibinfo {author}
  {\bibfnamefont {P.}~\bibnamefont {Zoller}}, \ and\ \bibinfo {author}
  {\bibfnamefont {S.}~\bibnamefont {Diehl}},\ }\href@noop {} {\bibfield
  {journal} {\bibinfo  {journal} {New. J. Phys.}\ }\textbf {\bibinfo {volume}
  {15}},\ \bibinfo {pages} {085001} (\bibinfo {year} {2013})}\BibitemShut
  {NoStop}%
\bibitem [{\citenamefont {Iemini}\ \emph {et~al.}(2016)\citenamefont {Iemini},
  \citenamefont {Rossini}, \citenamefont {Fazio}, \citenamefont {Diehl},\ and\
  \citenamefont {Mazza}}]{Iemini16}%
  \BibitemOpen
  \bibfield  {author} {\bibinfo {author} {\bibfnamefont {F.}~\bibnamefont
  {Iemini}}, \bibinfo {author} {\bibfnamefont {D.}~\bibnamefont {Rossini}},
  \bibinfo {author} {\bibfnamefont {R.}~\bibnamefont {Fazio}}, \bibinfo
  {author} {\bibfnamefont {S.}~\bibnamefont {Diehl}}, \ and\ \bibinfo {author}
  {\bibfnamefont {L.}~\bibnamefont {Mazza}},\ }\href@noop {} {\bibfield
  {journal} {\bibinfo  {journal} {Phys. Rev. B}\ }\textbf {\bibinfo {volume}
  {93}},\ \bibinfo {pages} {115113} (\bibinfo {year} {2016})}\BibitemShut
  {NoStop}%
\bibitem [{\citenamefont {Huang}\ and\ \citenamefont {Arovas}(2014)}]{Huang14}%
  \BibitemOpen
  \bibfield  {author} {\bibinfo {author} {\bibfnamefont {Z.}~\bibnamefont
  {Huang}}\ and\ \bibinfo {author} {\bibfnamefont {D.~P.}\ \bibnamefont
  {Arovas}},\ }\href@noop {} {\bibfield  {journal} {\bibinfo  {journal} {Phys.
  Rev. Lett.}\ }\textbf {\bibinfo {volume} {113}},\ \bibinfo {pages} {076407}
  (\bibinfo {year} {2014})}\BibitemShut {NoStop}%
\bibitem [{\citenamefont {Viyuela}\ \emph {et~al.}(2015)\citenamefont
  {Viyuela}, \citenamefont {Rivas},\ and\ \citenamefont
  {Martin-Delgado}}]{Viyuela15}%
  \BibitemOpen
  \bibfield  {author} {\bibinfo {author} {\bibfnamefont {O.}~\bibnamefont
  {Viyuela}}, \bibinfo {author} {\bibfnamefont {A.}~\bibnamefont {Rivas}}, \
  and\ \bibinfo {author} {\bibfnamefont {M.~A.}\ \bibnamefont
  {Martin-Delgado}},\ }\href@noop {} {\bibfield  {journal} {\bibinfo  {journal}
  {2D Mater.}\ }\textbf {\bibinfo {volume} {2}},\ \bibinfo {pages} {034006}
  (\bibinfo {year} {2015})}\BibitemShut {NoStop}%
\bibitem [{\citenamefont {Budich}\ and\ \citenamefont
  {Diehl}(2015)}]{Budich15b}%
  \BibitemOpen
  \bibfield  {author} {\bibinfo {author} {\bibfnamefont {J.~C.}\ \bibnamefont
  {Budich}}\ and\ \bibinfo {author} {\bibfnamefont {S.}~\bibnamefont {Diehl}},\
  }\href@noop {} {\bibfield  {journal} {\bibinfo  {journal} {Phys. Rev. B}\
  }\textbf {\bibinfo {volume} {91}},\ \bibinfo {pages} {165140} (\bibinfo
  {year} {2015})}\BibitemShut {NoStop}%
\bibitem [{\citenamefont {Uhlmann}(1986)}]{Uhlmann86}%
  \BibitemOpen
  \bibfield  {author} {\bibinfo {author} {\bibfnamefont {A.}~\bibnamefont
  {Uhlmann}},\ }\href@noop {} {\bibfield  {journal} {\bibinfo  {journal} {Rep.
  Math. Phys.}\ }\textbf {\bibinfo {volume} {24}},\ \bibinfo {pages} {229 }
  (\bibinfo {year} {1986})}\BibitemShut {NoStop}%
\bibitem [{\citenamefont {Zanardi}\ and\ \citenamefont
  {Rasetti}(1999)}]{Zanardi99}%
  \BibitemOpen
  \bibfield  {author} {\bibinfo {author} {\bibfnamefont {P.}~\bibnamefont
  {Zanardi}}\ and\ \bibinfo {author} {\bibfnamefont {M.}~\bibnamefont
  {Rasetti}},\ }\href@noop {} {\bibfield  {journal} {\bibinfo  {journal} {Phys.
  Lett. A}\ }\textbf {\bibinfo {volume} {264}},\ \bibinfo {pages} {94 }
  (\bibinfo {year} {1999})}\BibitemShut {NoStop}%
\bibitem [{\citenamefont {Dennis}\ \emph {et~al.}(2002)\citenamefont {Dennis},
  \citenamefont {Kitaev}, \citenamefont {Landahl},\ and\ \citenamefont
  {Preskill}}]{DKP02}%
  \BibitemOpen
  \bibfield  {author} {\bibinfo {author} {\bibfnamefont {E.}~\bibnamefont
  {Dennis}}, \bibinfo {author} {\bibfnamefont {A.}~\bibnamefont {Kitaev}},
  \bibinfo {author} {\bibfnamefont {A.}~\bibnamefont {Landahl}}, \ and\
  \bibinfo {author} {\bibfnamefont {J.}~\bibnamefont {Preskill}},\ }\href@noop
  {} {\bibfield  {journal} {\bibinfo  {journal} {J. Math. Phys.}\ }\textbf
  {\bibinfo {volume} {43}},\ \bibinfo {pages} {4452} (\bibinfo {year}
  {2002})}\BibitemShut {NoStop}%
\bibitem [{\citenamefont {Nayak}\ \emph
  {et~al.}(2008{\natexlab{b}})\citenamefont {Nayak}, \citenamefont {Simon},
  \citenamefont {Stern}, \citenamefont {Freedman},\ and\ \citenamefont {{Das
  Sarma}}}]{Nayak08rev}%
  \BibitemOpen
  \bibfield  {author} {\bibinfo {author} {\bibfnamefont {C.}~\bibnamefont
  {Nayak}}, \bibinfo {author} {\bibfnamefont {S.~H.}\ \bibnamefont {Simon}},
  \bibinfo {author} {\bibfnamefont {A.}~\bibnamefont {Stern}}, \bibinfo
  {author} {\bibfnamefont {M.}~\bibnamefont {Freedman}}, \ and\ \bibinfo
  {author} {\bibfnamefont {S.}~\bibnamefont {{Das Sarma}}},\ }\href@noop {}
  {\bibfield  {journal} {\bibinfo  {journal} {Rep. Math. Phys.}\ }\textbf
  {\bibinfo {volume} {80}},\ \bibinfo {pages} {1083} (\bibinfo {year}
  {2008}{\natexlab{b}})}\BibitemShut {NoStop}%
\bibitem [{\citenamefont {Marsden}\ \emph {et~al.}(1990)\citenamefont
  {Marsden}, \citenamefont {Montgomery},\ and\ \citenamefont
  {Rațiu}}]{Marsden90}%
  \BibitemOpen
  \bibfield  {author} {\bibinfo {author} {\bibfnamefont {J.}~\bibnamefont
  {Marsden}}, \bibinfo {author} {\bibfnamefont {R.}~\bibnamefont {Montgomery}},
  \ and\ \bibinfo {author} {\bibfnamefont {T.}~\bibnamefont {Rațiu}},\
  }\href@noop {} {\emph {\bibinfo {title} {Reduction, Symmetry, and Phases in
  Mechanics}}},\ Memoirs of the American Mathematical Society\ (\bibinfo
  {publisher} {American Mathematical Society},\ \bibinfo {year}
  {1990})\BibitemShut {NoStop}%
\bibitem [{\citenamefont {Marsden}\ and\ \citenamefont
  {Ostrowski}(1998)}]{Marsden98}%
  \BibitemOpen
  \bibfield  {author} {\bibinfo {author} {\bibfnamefont {J.~E.}\ \bibnamefont
  {Marsden}}\ and\ \bibinfo {author} {\bibfnamefont {J.}~\bibnamefont
  {Ostrowski}},\ }\enquote {\bibinfo {title} {Control, and geometry:
  Proceedings of a symposium},}\ \ (\bibinfo  {publisher} {National Academies
  Press},\ \bibinfo {year} {1998})\ Chap.\ \bibinfo {chapter} {Symmetries in
  Motion: Geometric Foundations of Motion Control. In: Motion}, p.~\bibinfo
  {pages} {3}\BibitemShut {NoStop}%
\bibitem [{\citenamefont {Cendra}\ \emph {et~al.}(2001)\citenamefont {Cendra},
  \citenamefont {Marsden},\ and\ \citenamefont {Ratiu}}]{Cendra01}%
  \BibitemOpen
  \bibfield  {author} {\bibinfo {author} {\bibfnamefont {H.}~\bibnamefont
  {Cendra}}, \bibinfo {author} {\bibfnamefont {J.~E.}\ \bibnamefont {Marsden}},
  \ and\ \bibinfo {author} {\bibfnamefont {T.~S.}\ \bibnamefont {Ratiu}},\
  }\enquote {\bibinfo {title} {Mathematics unlimited - 2001 and beyond},}\ \
  (\bibinfo  {publisher} {Springer-Verlag, New York},\ \bibinfo {year} {2001})\
  Chap.\ \bibinfo {chapter} {Geometric mechanics, Langrangian reduction, and
  nonholonomic systems}, p.\ \bibinfo {pages} {221}\BibitemShut {NoStop}%
\bibitem [{\citenamefont {Bloch}(2003)}]{Bloch03}%
  \BibitemOpen
  \bibfield  {author} {\bibinfo {author} {\bibfnamefont {A.~M.}\ \bibnamefont
  {Bloch}},\ }\href@noop {} {\emph {\bibinfo {title} {Nonholonomic mechanics
  and control}}},\ Interdisciplinary Applied Mathematics\ (\bibinfo
  {publisher} {Springer-Verlag, New York},\ \bibinfo {year} {2003})\BibitemShut
  {NoStop}%
\bibitem [{\citenamefont {Andersson}(2003{\natexlab{a}})}]{Andersson03thesis}%
  \BibitemOpen
  \bibfield  {author} {\bibinfo {author} {\bibfnamefont {S.~B.}\ \bibnamefont
  {Andersson}},\ }\emph {\bibinfo {title} {Geometric phases in sensing and
  control}},\ \href@noop {} {Ph.D. thesis},\ \bibinfo  {school} {University of
  Maryland}, \bibinfo {address} {College Park} (\bibinfo {year}
  {2003}{\natexlab{a}})\BibitemShut {NoStop}%
\bibitem [{\citenamefont {Dabrowski}\ and\ \citenamefont
  {Grosse}(1990)}]{Dabrowski90}%
  \BibitemOpen
  \bibfield  {author} {\bibinfo {author} {\bibfnamefont {L.}~\bibnamefont
  {Dabrowski}}\ and\ \bibinfo {author} {\bibfnamefont {H.}~\bibnamefont
  {Grosse}},\ }\href@noop {} {\bibfield  {journal} {\bibinfo  {journal} {Lett.
  Math. Phys.}\ }\textbf {\bibinfo {volume} {19}},\ \bibinfo {pages} {205}
  (\bibinfo {year} {1990})}\BibitemShut {NoStop}%
\bibitem [{\citenamefont {Sj\"oqvist}\ \emph {et~al.}(2000)\citenamefont
  {Sj\"oqvist}, \citenamefont {Pati}, \citenamefont {Ekert}, \citenamefont
  {Anandan}, \citenamefont {Ericsson}, \citenamefont {Oi},\ and\ \citenamefont
  {Vedral}}]{Sjoeqvist00}%
  \BibitemOpen
  \bibfield  {author} {\bibinfo {author} {\bibfnamefont {E.}~\bibnamefont
  {Sj\"oqvist}}, \bibinfo {author} {\bibfnamefont {A.~K.}\ \bibnamefont
  {Pati}}, \bibinfo {author} {\bibfnamefont {A.}~\bibnamefont {Ekert}},
  \bibinfo {author} {\bibfnamefont {J.~S.}\ \bibnamefont {Anandan}}, \bibinfo
  {author} {\bibfnamefont {M.}~\bibnamefont {Ericsson}}, \bibinfo {author}
  {\bibfnamefont {D.~K.~L.}\ \bibnamefont {Oi}}, \ and\ \bibinfo {author}
  {\bibfnamefont {V.}~\bibnamefont {Vedral}},\ }\href@noop {} {\bibfield
  {journal} {\bibinfo  {journal} {Phys. Rev. Lett.}\ }\textbf {\bibinfo
  {volume} {85}},\ \bibinfo {pages} {2845} (\bibinfo {year}
  {2000})}\BibitemShut {NoStop}%
\bibitem [{\citenamefont {Zhou}\ \emph {et~al.}(2003)\citenamefont {Zhou},
  \citenamefont {Cho},\ and\ \citenamefont {McKenzie}}]{Zhou03}%
  \BibitemOpen
  \bibfield  {author} {\bibinfo {author} {\bibfnamefont {H.-Q.}\ \bibnamefont
  {Zhou}}, \bibinfo {author} {\bibfnamefont {S.~Y.}\ \bibnamefont {Cho}}, \
  and\ \bibinfo {author} {\bibfnamefont {R.~H.}\ \bibnamefont {McKenzie}},\
  }\href@noop {} {\bibfield  {journal} {\bibinfo  {journal} {Phys. Rev. Lett.}\
  }\textbf {\bibinfo {volume} {91}},\ \bibinfo {pages} {186803} (\bibinfo
  {year} {2003})}\BibitemShut {NoStop}%
\bibitem [{\citenamefont {Ericsson}\ \emph {et~al.}(2003)\citenamefont
  {Ericsson}, \citenamefont {Pati}, \citenamefont {Sj\"oqvist}, \citenamefont
  {Br\"annlund},\ and\ \citenamefont {Oi}}]{Ericsson03}%
  \BibitemOpen
  \bibfield  {author} {\bibinfo {author} {\bibfnamefont {M.}~\bibnamefont
  {Ericsson}}, \bibinfo {author} {\bibfnamefont {A.~K.}\ \bibnamefont {Pati}},
  \bibinfo {author} {\bibfnamefont {E.}~\bibnamefont {Sj\"oqvist}}, \bibinfo
  {author} {\bibfnamefont {J.}~\bibnamefont {Br\"annlund}}, \ and\ \bibinfo
  {author} {\bibfnamefont {D.~K.~L.}\ \bibnamefont {Oi}},\ }\href@noop {}
  {\bibfield  {journal} {\bibinfo  {journal} {Phys. Rev. Lett.}\ }\textbf
  {\bibinfo {volume} {91}},\ \bibinfo {pages} {090405} (\bibinfo {year}
  {2003})}\BibitemShut {NoStop}%
\bibitem [{\citenamefont {Cohen}(2003)}]{Cohen03}%
  \BibitemOpen
  \bibfield  {author} {\bibinfo {author} {\bibfnamefont {D.}~\bibnamefont
  {Cohen}},\ }\href@noop {} {\bibfield  {journal} {\bibinfo  {journal} {Phys.
  Rev. B}\ }\textbf {\bibinfo {volume} {68}},\ \bibinfo {pages} {201303}
  (\bibinfo {year} {2003})}\BibitemShut {NoStop}%
\bibitem [{\citenamefont {Whitney}\ and\ \citenamefont
  {Gefen}(2003)}]{Whitney03}%
  \BibitemOpen
  \bibfield  {author} {\bibinfo {author} {\bibfnamefont {R.~S.}\ \bibnamefont
  {Whitney}}\ and\ \bibinfo {author} {\bibfnamefont {Y.}~\bibnamefont
  {Gefen}},\ }\href@noop {} {\bibfield  {journal} {\bibinfo  {journal} {Phys.
  Rev. Lett.}\ }\textbf {\bibinfo {volume} {90}},\ \bibinfo {pages} {190402}
  (\bibinfo {year} {2003})}\BibitemShut {NoStop}%
\bibitem [{\citenamefont {Whitney}\ \emph {et~al.}(2005)\citenamefont
  {Whitney}, \citenamefont {Makhlin}, \citenamefont {Shnirman},\ and\
  \citenamefont {Gefen}}]{Whitney04}%
  \BibitemOpen
  \bibfield  {author} {\bibinfo {author} {\bibfnamefont {R.~S.}\ \bibnamefont
  {Whitney}}, \bibinfo {author} {\bibfnamefont {Y.}~\bibnamefont {Makhlin}},
  \bibinfo {author} {\bibfnamefont {A.}~\bibnamefont {Shnirman}}, \ and\
  \bibinfo {author} {\bibfnamefont {Y.}~\bibnamefont {Gefen}},\ }\href@noop {}
  {\bibfield  {journal} {\bibinfo  {journal} {Phys. Rev. Lett.}\ }\textbf
  {\bibinfo {volume} {94}},\ \bibinfo {pages} {070407} (\bibinfo {year}
  {2005})}\BibitemShut {NoStop}%
\bibitem [{\citenamefont {Chu}\ and\ \citenamefont {Telnov}(2004)}]{Chu04}%
  \BibitemOpen
  \bibfield  {author} {\bibinfo {author} {\bibfnamefont {S.-I.}\ \bibnamefont
  {Chu}}\ and\ \bibinfo {author} {\bibfnamefont {D.~A.}\ \bibnamefont
  {Telnov}},\ }\href@noop {} {\bibfield  {journal} {\bibinfo  {journal} {Phys.
  Rep.}\ }\textbf {\bibinfo {volume} {390}} (\bibinfo {year}
  {2004})}\BibitemShut {NoStop}%
\bibitem [{\citenamefont {Tong}\ \emph {et~al.}(2004)\citenamefont {Tong},
  \citenamefont {Sj\"oqvist}, \citenamefont {Kwek},\ and\ \citenamefont
  {Oh}}]{Tong04}%
  \BibitemOpen
  \bibfield  {author} {\bibinfo {author} {\bibfnamefont {D.~M.}\ \bibnamefont
  {Tong}}, \bibinfo {author} {\bibfnamefont {E.}~\bibnamefont {Sj\"oqvist}},
  \bibinfo {author} {\bibfnamefont {L.~C.}\ \bibnamefont {Kwek}}, \ and\
  \bibinfo {author} {\bibfnamefont {C.~H.}\ \bibnamefont {Oh}},\ }\href@noop {}
  {\bibfield  {journal} {\bibinfo  {journal} {Phys. Rev. Lett.}\ }\textbf
  {\bibinfo {volume} {93}},\ \bibinfo {pages} {080405} (\bibinfo {year}
  {2004})}\BibitemShut {NoStop}%
\bibitem [{\citenamefont {Chaturvedi}\ \emph {et~al.}(2004)\citenamefont
  {Chaturvedi}, \citenamefont {Ercolessi}, \citenamefont {Marmo}, \citenamefont
  {Morandi}, \citenamefont {Mukunda},\ and\ \citenamefont
  {Simon}}]{Chaturvedi04}%
  \BibitemOpen
  \bibfield  {author} {\bibinfo {author} {\bibfnamefont {S.}~\bibnamefont
  {Chaturvedi}}, \bibinfo {author} {\bibfnamefont {E.}~\bibnamefont
  {Ercolessi}}, \bibinfo {author} {\bibfnamefont {G.}~\bibnamefont {Marmo}},
  \bibinfo {author} {\bibfnamefont {G.}~\bibnamefont {Morandi}}, \bibinfo
  {author} {\bibfnamefont {N.}~\bibnamefont {Mukunda}}, \ and\ \bibinfo
  {author} {\bibfnamefont {R.}~\bibnamefont {Simon}},\ }\href@noop {}
  {\bibfield  {journal} {\bibinfo  {journal} {Eur. Phys. J. B}\ }\textbf
  {\bibinfo {volume} {35}},\ \bibinfo {pages} {413} (\bibinfo {year}
  {2004})}\BibitemShut {NoStop}%
\bibitem [{\citenamefont {Sarandy}\ and\ \citenamefont
  {Lidar}(2005)}]{Sarandy05}%
  \BibitemOpen
  \bibfield  {author} {\bibinfo {author} {\bibfnamefont {M.~S.}\ \bibnamefont
  {Sarandy}}\ and\ \bibinfo {author} {\bibfnamefont {D.~A.}\ \bibnamefont
  {Lidar}},\ }\href@noop {} {\bibfield  {journal} {\bibinfo  {journal} {Phys.
  Rev. A}\ }\textbf {\bibinfo {volume} {71}},\ \bibinfo {pages} {012331}
  (\bibinfo {year} {2005})}\BibitemShut {NoStop}%
\bibitem [{\citenamefont {Sarandy}\ and\ \citenamefont
  {Lidar}(2006)}]{Sarandy06}%
  \BibitemOpen
  \bibfield  {author} {\bibinfo {author} {\bibfnamefont {M.~S.}\ \bibnamefont
  {Sarandy}}\ and\ \bibinfo {author} {\bibfnamefont {D.~A.}\ \bibnamefont
  {Lidar}},\ }\href@noop {} {\bibfield  {journal} {\bibinfo  {journal} {Phys.
  Rev. A}\ }\textbf {\bibinfo {volume} {73}},\ \bibinfo {pages} {062101}
  (\bibinfo {year} {2006})}\BibitemShut {NoStop}%
\bibitem [{\citenamefont {Fujikawa}(2007)}]{Fujikawa06}%
  \BibitemOpen
  \bibfield  {author} {\bibinfo {author} {\bibfnamefont {K.}~\bibnamefont
  {Fujikawa}},\ }\href@noop {} {\bibfield  {journal} {\bibinfo  {journal} {Ann.
  Phys.}\ }\textbf {\bibinfo {volume} {322}},\ \bibinfo {pages} {1500 }
  (\bibinfo {year} {2007})}\BibitemShut {NoStop}%
\bibitem [{\citenamefont {Sinitsyn}\ and\ \citenamefont
  {Nemenman}(2007{\natexlab{a}})}]{Sinitsyn07EPL}%
  \BibitemOpen
  \bibfield  {author} {\bibinfo {author} {\bibfnamefont {N.}~\bibnamefont
  {Sinitsyn}}\ and\ \bibinfo {author} {\bibfnamefont {I.}~\bibnamefont
  {Nemenman}},\ }\href@noop {} {\bibfield  {journal} {\bibinfo  {journal} {Eur.
  Phys. Lett.}\ }\textbf {\bibinfo {volume} {77}},\ \bibinfo {pages} {58001}
  (\bibinfo {year} {2007}{\natexlab{a}})}\BibitemShut {NoStop}%
\bibitem [{\citenamefont {Goto}\ and\ \citenamefont {Ichimura}(2007)}]{Goto07}%
  \BibitemOpen
  \bibfield  {author} {\bibinfo {author} {\bibfnamefont {H.}~\bibnamefont
  {Goto}}\ and\ \bibinfo {author} {\bibfnamefont {K.}~\bibnamefont
  {Ichimura}},\ }\href@noop {} {\bibfield  {journal} {\bibinfo  {journal}
  {Phys. Rev. A}\ }\textbf {\bibinfo {volume} {76}},\ \bibinfo {pages} {012120}
  (\bibinfo {year} {2007})}\BibitemShut {NoStop}%
\bibitem [{\citenamefont {Whitney}(2010)}]{Whitney10}%
  \BibitemOpen
  \bibfield  {author} {\bibinfo {author} {\bibfnamefont {R.~S.}\ \bibnamefont
  {Whitney}},\ }\href@noop {} {\bibfield  {journal} {\bibinfo  {journal} {Phys.
  Rev. A}\ }\textbf {\bibinfo {volume} {81}},\ \bibinfo {pages} {032108}
  (\bibinfo {year} {2010})}\BibitemShut {NoStop}%
\bibitem [{\citenamefont {Avron}\ \emph
  {et~al.}(2012{\natexlab{a}})\citenamefont {Avron}, \citenamefont {Fraas},\
  and\ \citenamefont {Graf}}]{Avron12}%
  \BibitemOpen
  \bibfield  {author} {\bibinfo {author} {\bibfnamefont {J.}~\bibnamefont
  {Avron}}, \bibinfo {author} {\bibfnamefont {M.}~\bibnamefont {Fraas}}, \ and\
  \bibinfo {author} {\bibfnamefont {G.}~\bibnamefont {Graf}},\ }\href@noop {}
  {\bibfield  {journal} {\bibinfo  {journal} {J. Stat. Phys.}\ }\textbf
  {\bibinfo {volume} {148}},\ \bibinfo {pages} {800} (\bibinfo {year}
  {2012}{\natexlab{a}})}\BibitemShut {NoStop}%
\bibitem [{\citenamefont {Sinitsyn}(2009)}]{Sinitsyn09}%
  \BibitemOpen
  \bibfield  {author} {\bibinfo {author} {\bibfnamefont {N.~A.}\ \bibnamefont
  {Sinitsyn}},\ }\href@noop {} {\bibfield  {journal} {\bibinfo  {journal} {J.
  Phys. A}\ }\textbf {\bibinfo {volume} {42}},\ \bibinfo {pages} {193001}
  (\bibinfo {year} {2009})}\BibitemShut {NoStop}%
\bibitem [{\citenamefont {Grifoni}\ and\ \citenamefont
  {H\"anggi}(1998)}]{Grifoni98}%
  \BibitemOpen
  \bibfield  {author} {\bibinfo {author} {\bibfnamefont {M.}~\bibnamefont
  {Grifoni}}\ and\ \bibinfo {author} {\bibfnamefont {P.}~\bibnamefont
  {H\"anggi}},\ }\href@noop {} {\bibfield  {journal} {\bibinfo  {journal}
  {Phys. Rep.}\ }\textbf {\bibinfo {volume} {304}},\ \bibinfo {pages} {229}
  (\bibinfo {year} {1998})}\BibitemShut {NoStop}%
\bibitem [{\citenamefont {Kaestner}\ and\ \citenamefont
  {Kashcheyevs}(2015)}]{Kaestner15}%
  \BibitemOpen
  \bibfield  {author} {\bibinfo {author} {\bibfnamefont {B.}~\bibnamefont
  {Kaestner}}\ and\ \bibinfo {author} {\bibfnamefont {V.}~\bibnamefont
  {Kashcheyevs}},\ }\href@noop {} {\bibfield  {journal} {\bibinfo  {journal}
  {Rep. Prog. Phys.}\ }\textbf {\bibinfo {volume} {78}},\ \bibinfo {pages}
  {103901} (\bibinfo {year} {2015})}\BibitemShut {NoStop}%
\bibitem [{\citenamefont {Brouwer}(2001)}]{Brouwer01}%
  \BibitemOpen
  \bibfield  {author} {\bibinfo {author} {\bibfnamefont {P.~W.}\ \bibnamefont
  {Brouwer}},\ }\href@noop {} {\bibfield  {journal} {\bibinfo  {journal} {Phys.
  Rev. B}\ }\textbf {\bibinfo {volume} {63}},\ \bibinfo {pages} {R121303}
  (\bibinfo {year} {2001})}\BibitemShut {NoStop}%
\bibitem [{\citenamefont {Watanabe}\ and\ \citenamefont
  {Hayakawa}(2014)}]{Watanabe14}%
  \BibitemOpen
  \bibfield  {author} {\bibinfo {author} {\bibfnamefont {K.~L.}\ \bibnamefont
  {Watanabe}}\ and\ \bibinfo {author} {\bibfnamefont {H.}~\bibnamefont
  {Hayakawa}},\ }\href@noop {} {\bibfield  {journal} {\bibinfo  {journal}
  {Progr. Theor. Exp. Phys.}\ }\textbf {\bibinfo {volume} {2014}} (\bibinfo
  {year} {2014})}\BibitemShut {NoStop}%
\bibitem [{\citenamefont {Boeuf}\ \emph {et~al.}(2003)\citenamefont {Boeuf},
  \citenamefont {Jehl}, \citenamefont {Sanquer},\ and\ \citenamefont
  {Skotnicki}}]{Jehl03}%
  \BibitemOpen
  \bibfield  {author} {\bibinfo {author} {\bibfnamefont {F.}~\bibnamefont
  {Boeuf}}, \bibinfo {author} {\bibfnamefont {X.}~\bibnamefont {Jehl}},
  \bibinfo {author} {\bibfnamefont {M.}~\bibnamefont {Sanquer}}, \ and\
  \bibinfo {author} {\bibfnamefont {T.}~\bibnamefont {Skotnicki}},\ }\href@noop
  {} {\bibfield  {journal} {\bibinfo  {journal} {IEEE Trans. Nanotechnol}\
  }\textbf {\bibinfo {volume} {2}},\ \bibinfo {pages} {144} (\bibinfo {year}
  {2003})}\BibitemShut {NoStop}%
\bibitem [{\citenamefont {Chernyak}\ and\ \citenamefont
  {Sinitsyn}(2009)}]{Chernyak09}%
  \BibitemOpen
  \bibfield  {author} {\bibinfo {author} {\bibfnamefont {V.~Y.}\ \bibnamefont
  {Chernyak}}\ and\ \bibinfo {author} {\bibfnamefont {N.~A.}\ \bibnamefont
  {Sinitsyn}},\ }\href@noop {} {\bibfield  {journal} {\bibinfo  {journal} {J.
  Chem. Phys.}\ }\textbf {\bibinfo {volume} {131}},\ \bibinfo {pages} {181101}
  (\bibinfo {year} {2009})}\BibitemShut {NoStop}%
\bibitem [{\citenamefont {Napitu}\ and\ \citenamefont
  {Thijssen}(2015)}]{Napitu15}%
  \BibitemOpen
  \bibfield  {author} {\bibinfo {author} {\bibfnamefont {B.~D.}\ \bibnamefont
  {Napitu}}\ and\ \bibinfo {author} {\bibfnamefont {J.~M.}\ \bibnamefont
  {Thijssen}},\ }\href@noop {} {\bibfield  {journal} {\bibinfo  {journal} {J.
  Phys. Condens. Matter}\ }\textbf {\bibinfo {volume} {27}},\ \bibinfo {pages}
  {275301} (\bibinfo {year} {2015})}\BibitemShut {NoStop}%
\bibitem [{\citenamefont {B\"uttiker}\ \emph {et~al.}(1985)\citenamefont
  {B\"uttiker}, \citenamefont {Imry}, \citenamefont {Landauer},\ and\
  \citenamefont {Pinhas}}]{Buettiker85}%
  \BibitemOpen
  \bibfield  {author} {\bibinfo {author} {\bibfnamefont {M.}~\bibnamefont
  {B\"uttiker}}, \bibinfo {author} {\bibfnamefont {Y.}~\bibnamefont {Imry}},
  \bibinfo {author} {\bibfnamefont {R.}~\bibnamefont {Landauer}}, \ and\
  \bibinfo {author} {\bibfnamefont {S.}~\bibnamefont {Pinhas}},\ }\href@noop {}
  {\bibfield  {journal} {\bibinfo  {journal} {Phys. Rev. B}\ }\textbf {\bibinfo
  {volume} {31}},\ \bibinfo {pages} {6207} (\bibinfo {year}
  {1985})}\BibitemShut {NoStop}%
\bibitem [{\citenamefont {Brouwer}(1998)}]{Brouwer98}%
  \BibitemOpen
  \bibfield  {author} {\bibinfo {author} {\bibfnamefont {P.~W.}\ \bibnamefont
  {Brouwer}},\ }\href@noop {} {\bibfield  {journal} {\bibinfo  {journal} {Phys.
  Rev. B}\ }\textbf {\bibinfo {volume} {58}},\ \bibinfo {pages} {R10135}
  (\bibinfo {year} {1998})}\BibitemShut {NoStop}%
\bibitem [{\citenamefont {Switkes}\ \emph {et~al.}(1999)\citenamefont
  {Switkes}, \citenamefont {Marcus}, \citenamefont {Campman},\ and\
  \citenamefont {Gossard}}]{Switkes99}%
  \BibitemOpen
  \bibfield  {author} {\bibinfo {author} {\bibfnamefont {M.}~\bibnamefont
  {Switkes}}, \bibinfo {author} {\bibfnamefont {C.~M.}\ \bibnamefont {Marcus}},
  \bibinfo {author} {\bibfnamefont {K.}~\bibnamefont {Campman}}, \ and\
  \bibinfo {author} {\bibfnamefont {A.~C.}\ \bibnamefont {Gossard}},\
  }\href@noop {} {\bibfield  {journal} {\bibinfo  {journal} {Science}\ }\textbf
  {\bibinfo {volume} {283}},\ \bibinfo {pages} {1905} (\bibinfo {year}
  {1999})}\BibitemShut {NoStop}%
\bibitem [{\citenamefont {Splettstoesser}\ \emph {et~al.}(2006)\citenamefont
  {Splettstoesser}, \citenamefont {Governale}, \citenamefont {K\"{o}nig},\ and\
  \citenamefont {Fazio}}]{Splettstoesser06}%
  \BibitemOpen
  \bibfield  {author} {\bibinfo {author} {\bibfnamefont {J.}~\bibnamefont
  {Splettstoesser}}, \bibinfo {author} {\bibfnamefont {M.}~\bibnamefont
  {Governale}}, \bibinfo {author} {\bibfnamefont {J.}~\bibnamefont
  {K\"{o}nig}}, \ and\ \bibinfo {author} {\bibfnamefont {R.}~\bibnamefont
  {Fazio}},\ }\href@noop {} {\bibfield  {journal} {\bibinfo  {journal} {Phys.
  Rev. B}\ }\textbf {\bibinfo {volume} {74}},\ \bibinfo {pages} {085305}
  (\bibinfo {year} {2006})}\BibitemShut {NoStop}%
\bibitem [{\citenamefont {Resta}(2000)}]{Resta00}%
  \BibitemOpen
  \bibfield  {author} {\bibinfo {author} {\bibfnamefont {R.}~\bibnamefont
  {Resta}},\ }\href@noop {} {\bibfield  {journal} {\bibinfo  {journal} {J.
  Phys. Condens. Matter}\ }\textbf {\bibinfo {volume} {12}},\ \bibinfo {pages}
  {R107} (\bibinfo {year} {2000})}\BibitemShut {NoStop}%
\bibitem [{\citenamefont {Xiao}\ \emph {et~al.}(2010)\citenamefont {Xiao},
  \citenamefont {Chang},\ and\ \citenamefont {Niu}}]{Xiao10}%
  \BibitemOpen
  \bibfield  {author} {\bibinfo {author} {\bibfnamefont {D.}~\bibnamefont
  {Xiao}}, \bibinfo {author} {\bibfnamefont {M.-C.}\ \bibnamefont {Chang}}, \
  and\ \bibinfo {author} {\bibfnamefont {Q.}~\bibnamefont {Niu}},\ }\href@noop
  {} {\bibfield  {journal} {\bibinfo  {journal} {Rev. Mod. Phys.}\ }\textbf
  {\bibinfo {volume} {82}},\ \bibinfo {pages} {1959} (\bibinfo {year}
  {2010})}\BibitemShut {NoStop}%
\bibitem [{\citenamefont {Bohm}\ \emph {et~al.}(2003)\citenamefont {Bohm},
  \citenamefont {Mostafazadeh}, \citenamefont {Koizumi}, \citenamefont {Niu},\
  and\ \citenamefont {Zwanziger}}]{Bohm}%
  \BibitemOpen
  \bibfield  {author} {\bibinfo {author} {\bibfnamefont {A.}~\bibnamefont
  {Bohm}}, \bibinfo {author} {\bibfnamefont {A.}~\bibnamefont {Mostafazadeh}},
  \bibinfo {author} {\bibfnamefont {H.}~\bibnamefont {Koizumi}}, \bibinfo
  {author} {\bibfnamefont {Q.}~\bibnamefont {Niu}}, \ and\ \bibinfo {author}
  {\bibfnamefont {J.}~\bibnamefont {Zwanziger}},\ }\href@noop {} {\emph
  {\bibinfo {title} {The Geometric Phase in Quantum Systems}}}\ (\bibinfo
  {publisher} {Springer},\ \bibinfo {address} {Berlin Heidelberg},\ \bibinfo
  {year} {2003})\BibitemShut {NoStop}%
\bibitem [{\citenamefont {Nakahara}(2003)}]{Nakahara03}%
  \BibitemOpen
  \bibfield  {author} {\bibinfo {author} {\bibfnamefont {M.}~\bibnamefont
  {Nakahara}},\ }\href@noop {} {\emph {\bibinfo {title} {Geometry, topology,
  and physics}}},\ Graduate student series in physics\ (\bibinfo  {publisher}
  {Institute of Physics Publishing},\ \bibinfo {address} {Bristol,
  Philadelphia},\ \bibinfo {year} {2003})\BibitemShut {NoStop}%
\bibitem [{\citenamefont {Berry}\ and\ \citenamefont
  {Robbins}(1993)}]{Berry93}%
  \BibitemOpen
  \bibfield  {author} {\bibinfo {author} {\bibfnamefont {M.~V.}\ \bibnamefont
  {Berry}}\ and\ \bibinfo {author} {\bibfnamefont {J.~M.}\ \bibnamefont
  {Robbins}},\ }\href@noop {} {\bibfield  {journal} {\bibinfo  {journal} {Proc.
  Roy. Soc.}\ }\textbf {\bibinfo {volume} {442}},\ \bibinfo {pages} {659}
  (\bibinfo {year} {1993})}\BibitemShut {NoStop}%
\bibitem [{\citenamefont {Berry}(1984)}]{Berry84}%
  \BibitemOpen
  \bibfield  {author} {\bibinfo {author} {\bibfnamefont {M.~V.}\ \bibnamefont
  {Berry}},\ }\href@noop {} {\bibfield  {journal} {\bibinfo  {journal} {Proc.
  Roy. Soc.}\ }\textbf {\bibinfo {volume} {392}},\ \bibinfo {pages} {45}
  (\bibinfo {year} {1984})}\BibitemShut {NoStop}%
\bibitem [{\citenamefont {Pekola}\ \emph {et~al.}(2013)\citenamefont {Pekola},
  \citenamefont {Saira}, \citenamefont {Maisi}, \citenamefont {Kemppinen},
  \citenamefont {M\"ott\"onen}, \citenamefont {Pashkin},\ and\ \citenamefont
  {Averin}}]{Pekola13rev}%
  \BibitemOpen
  \bibfield  {author} {\bibinfo {author} {\bibfnamefont {J.~P.}\ \bibnamefont
  {Pekola}}, \bibinfo {author} {\bibfnamefont {O.-P.}\ \bibnamefont {Saira}},
  \bibinfo {author} {\bibfnamefont {V.~F.}\ \bibnamefont {Maisi}}, \bibinfo
  {author} {\bibfnamefont {A.}~\bibnamefont {Kemppinen}}, \bibinfo {author}
  {\bibfnamefont {M.}~\bibnamefont {M\"ott\"onen}}, \bibinfo {author}
  {\bibfnamefont {Y.~A.}\ \bibnamefont {Pashkin}}, \ and\ \bibinfo {author}
  {\bibfnamefont {D.~V.}\ \bibnamefont {Averin}},\ }\href@noop {} {\bibfield
  {journal} {\bibinfo  {journal} {Rev. Mod. Phys.}\ }\textbf {\bibinfo {volume}
  {85}},\ \bibinfo {pages} {1421} (\bibinfo {year} {2013})}\BibitemShut
  {NoStop}%
\bibitem [{\citenamefont {Roche}\ \emph {et~al.}(2013)\citenamefont {Roche},
  \citenamefont {Riwar}, \citenamefont {Voisin}, \citenamefont
  {Dupont-Ferrier}, \citenamefont {Wacquez}, \citenamefont {Vinet},
  \citenamefont {Sanquer}, \citenamefont {Splettstoesser},\ and\ \citenamefont
  {Jehl}}]{Roche13}%
  \BibitemOpen
  \bibfield  {author} {\bibinfo {author} {\bibfnamefont {B.}~\bibnamefont
  {Roche}}, \bibinfo {author} {\bibfnamefont {R.-P.}\ \bibnamefont {Riwar}},
  \bibinfo {author} {\bibfnamefont {B.}~\bibnamefont {Voisin}}, \bibinfo
  {author} {\bibfnamefont {E.}~\bibnamefont {Dupont-Ferrier}}, \bibinfo
  {author} {\bibfnamefont {R.}~\bibnamefont {Wacquez}}, \bibinfo {author}
  {\bibfnamefont {M.}~\bibnamefont {Vinet}}, \bibinfo {author} {\bibfnamefont
  {M.}~\bibnamefont {Sanquer}}, \bibinfo {author} {\bibfnamefont
  {J.}~\bibnamefont {Splettstoesser}}, \ and\ \bibinfo {author} {\bibfnamefont
  {X.}~\bibnamefont {Jehl}},\ }\href@noop {} {\bibfield  {journal} {\bibinfo
  {journal} {Nat. Commun.}\ }\textbf {\bibinfo {volume} {4}},\ \bibinfo {pages}
  {1581} (\bibinfo {year} {2013})}\BibitemShut {NoStop}%
\bibitem [{Note1()}]{Note1}%
  \BibitemOpen
  \bibinfo {note} {Pumping has also been studied intensively in superconducting
  systems. There, the pumping of Cooper-pairs can be effectively expressed as a
  closed-system geometric phase~\cite
  {Fazio03,Governale05,Brosco08,Mottonen08,Gibertini13}, i.e., a Berry-Simon
  phase picked up by a superconducting state vector during the cyclic
  evolution. Here we study situations for which a pure-state description is not
  possible.}\BibitemShut {Stop}%
\bibitem [{\citenamefont {Pothier}\ \emph {et~al.}(1992)\citenamefont
  {Pothier}, \citenamefont {Lafarge}, \citenamefont {Urbina}, \citenamefont
  {Esteve},\ and\ \citenamefont {Devoret}}]{Pothier92}%
  \BibitemOpen
  \bibfield  {author} {\bibinfo {author} {\bibfnamefont {H.}~\bibnamefont
  {Pothier}}, \bibinfo {author} {\bibfnamefont {P.}~\bibnamefont {Lafarge}},
  \bibinfo {author} {\bibfnamefont {C.}~\bibnamefont {Urbina}}, \bibinfo
  {author} {\bibfnamefont {D.}~\bibnamefont {Esteve}}, \ and\ \bibinfo {author}
  {\bibfnamefont {M.~H.}\ \bibnamefont {Devoret}},\ }\href@noop {} {\bibfield
  {journal} {\bibinfo  {journal} {Eur. Phys. Lett.}\ }\textbf {\bibinfo
  {volume} {17}},\ \bibinfo {pages} {249} (\bibinfo {year} {1992})}\BibitemShut
  {NoStop}%
\bibitem [{\citenamefont {F\`eve}\ \emph {et~al.}(2007)\citenamefont {F\`eve},
  \citenamefont {Mah\'e}, \citenamefont {Berroir}, \citenamefont {Kontos},
  \citenamefont {Pla\c{c}ais}, \citenamefont {Glattli}, \citenamefont
  {Cavanna}, \citenamefont {Etienne},\ and\ \citenamefont {Jin}}]{Feve07}%
  \BibitemOpen
  \bibfield  {author} {\bibinfo {author} {\bibfnamefont {G.}~\bibnamefont
  {F\`eve}}, \bibinfo {author} {\bibfnamefont {A.}~\bibnamefont {Mah\'e}},
  \bibinfo {author} {\bibfnamefont {J.-M.}\ \bibnamefont {Berroir}}, \bibinfo
  {author} {\bibfnamefont {T.}~\bibnamefont {Kontos}}, \bibinfo {author}
  {\bibfnamefont {B.}~\bibnamefont {Pla\c{c}ais}}, \bibinfo {author}
  {\bibfnamefont {D.~C.}\ \bibnamefont {Glattli}}, \bibinfo {author}
  {\bibfnamefont {A.}~\bibnamefont {Cavanna}}, \bibinfo {author} {\bibfnamefont
  {B.}~\bibnamefont {Etienne}}, \ and\ \bibinfo {author} {\bibfnamefont
  {Y.}~\bibnamefont {Jin}},\ }\href@noop {} {\bibfield  {journal} {\bibinfo
  {journal} {Science}\ }\textbf {\bibinfo {volume} {316}},\ \bibinfo {pages}
  {1169} (\bibinfo {year} {2007})}\BibitemShut {NoStop}%
\bibitem [{\citenamefont {Xing}\ \emph {et~al.}(2014)\citenamefont {Xing},
  \citenamefont {Yang}, \citenamefont {Sun},\ and\ \citenamefont
  {Wang}}]{Xing14}%
  \BibitemOpen
  \bibfield  {author} {\bibinfo {author} {\bibfnamefont {Y.}~\bibnamefont
  {Xing}}, \bibinfo {author} {\bibfnamefont {Z.-l.}\ \bibnamefont {Yang}},
  \bibinfo {author} {\bibfnamefont {Q.-f.}\ \bibnamefont {Sun}}, \ and\
  \bibinfo {author} {\bibfnamefont {J.}~\bibnamefont {Wang}},\ }\href@noop {}
  {\bibfield  {journal} {\bibinfo  {journal} {Phys. Rev. B}\ }\textbf {\bibinfo
  {volume} {90}},\ \bibinfo {pages} {075435} (\bibinfo {year}
  {2014})}\BibitemShut {NoStop}%
\bibitem [{\citenamefont {Dittmann}\ \emph {et~al.}(2016)\citenamefont
  {Dittmann}, \citenamefont {Splettstoesser},\ and\ \citenamefont
  {Giazotto}}]{Dittmann16}%
  \BibitemOpen
  \bibfield  {author} {\bibinfo {author} {\bibfnamefont {N.}~\bibnamefont
  {Dittmann}}, \bibinfo {author} {\bibfnamefont {J.}~\bibnamefont
  {Splettstoesser}}, \ and\ \bibinfo {author} {\bibfnamefont {F.}~\bibnamefont
  {Giazotto}},\ }\href@noop {} {\bibfield  {journal} {\bibinfo  {journal} {New.
  J. Phys.}\ }\textbf {\bibinfo {volume} {18}},\ \bibinfo {pages} {083019}
  (\bibinfo {year} {2016})}\BibitemShut {NoStop}%
\bibitem [{\citenamefont {Reckermann}\ \emph {et~al.}(2010)\citenamefont
  {Reckermann}, \citenamefont {Splettstoesser},\ and\ \citenamefont
  {Wegewijs}}]{Reckermann10a}%
  \BibitemOpen
  \bibfield  {author} {\bibinfo {author} {\bibfnamefont {F.}~\bibnamefont
  {Reckermann}}, \bibinfo {author} {\bibfnamefont {J.}~\bibnamefont
  {Splettstoesser}}, \ and\ \bibinfo {author} {\bibfnamefont {M.~R.}\
  \bibnamefont {Wegewijs}},\ }\href@noop {} {\bibfield  {journal} {\bibinfo
  {journal} {Phys. Rev. Lett.}\ }\textbf {\bibinfo {volume} {104}},\ \bibinfo
  {pages} {226803} (\bibinfo {year} {2010})}\BibitemShut {NoStop}%
\bibitem [{\citenamefont {Calvo}\ \emph {et~al.}(2012)\citenamefont {Calvo},
  \citenamefont {Classen}, \citenamefont {Splettstoesser},\ and\ \citenamefont
  {Wegewijs}}]{Calvo12a}%
  \BibitemOpen
  \bibfield  {author} {\bibinfo {author} {\bibfnamefont {H.~L.}\ \bibnamefont
  {Calvo}}, \bibinfo {author} {\bibfnamefont {L.}~\bibnamefont {Classen}},
  \bibinfo {author} {\bibfnamefont {J.}~\bibnamefont {Splettstoesser}}, \ and\
  \bibinfo {author} {\bibfnamefont {M.~R.}\ \bibnamefont {Wegewijs}},\
  }\href@noop {} {\bibfield  {journal} {\bibinfo  {journal} {Phys. Rev. B}\
  }\textbf {\bibinfo {volume} {86}},\ \bibinfo {pages} {245308} (\bibinfo
  {year} {2012})}\BibitemShut {NoStop}%
\bibitem [{\citenamefont {Riwar}\ \emph {et~al.}(2013)\citenamefont {Riwar},
  \citenamefont {Splettstoesser},\ and\ \citenamefont {K\"onig}}]{Riwar13}%
  \BibitemOpen
  \bibfield  {author} {\bibinfo {author} {\bibfnamefont {R.-P.}\ \bibnamefont
  {Riwar}}, \bibinfo {author} {\bibfnamefont {J.}~\bibnamefont
  {Splettstoesser}}, \ and\ \bibinfo {author} {\bibfnamefont {J.}~\bibnamefont
  {K\"onig}},\ }\href@noop {} {\bibfield  {journal} {\bibinfo  {journal} {Phys.
  Rev. B}\ }\textbf {\bibinfo {volume} {87}},\ \bibinfo {pages} {195407}
  (\bibinfo {year} {2013})}\BibitemShut {NoStop}%
\bibitem [{\citenamefont {Chernyak}\ \emph
  {et~al.}(2012{\natexlab{a}})\citenamefont {Chernyak}, \citenamefont {Klein},\
  and\ \citenamefont {Sinitsyn}}]{Chernyak12a}%
  \BibitemOpen
  \bibfield  {author} {\bibinfo {author} {\bibfnamefont {V.~Y.}\ \bibnamefont
  {Chernyak}}, \bibinfo {author} {\bibfnamefont {J.~R.}\ \bibnamefont {Klein}},
  \ and\ \bibinfo {author} {\bibfnamefont {N.~A.}\ \bibnamefont {Sinitsyn}},\
  }\href@noop {} {\bibfield  {journal} {\bibinfo  {journal} {J. Chem. Phys.}\
  }\textbf {\bibinfo {volume} {136}},\ \bibinfo {eid} {154107} (\bibinfo {year}
  {2012}{\natexlab{a}})}\BibitemShut {NoStop}%
\bibitem [{\citenamefont {Chernyak}\ \emph
  {et~al.}(2012{\natexlab{b}})\citenamefont {Chernyak}, \citenamefont {Klein},\
  and\ \citenamefont {Sinitsyn}}]{Chernyak12b}%
  \BibitemOpen
  \bibfield  {author} {\bibinfo {author} {\bibfnamefont {V.~Y.}\ \bibnamefont
  {Chernyak}}, \bibinfo {author} {\bibfnamefont {J.~R.}\ \bibnamefont {Klein}},
  \ and\ \bibinfo {author} {\bibfnamefont {N.~A.}\ \bibnamefont {Sinitsyn}},\
  }\href@noop {} {\bibfield  {journal} {\bibinfo  {journal} {J. Chem. Phys.}\
  }\textbf {\bibinfo {volume} {136}},\ \bibinfo {eid} {154108} (\bibinfo {year}
  {2012}{\natexlab{b}})}\BibitemShut {NoStop}%
\bibitem [{\citenamefont {Chernyak}\ and\ \citenamefont
  {Sinitsyn}(2010)}]{Chernyak10}%
  \BibitemOpen
  \bibfield  {author} {\bibinfo {author} {\bibfnamefont {Y.}~\bibnamefont
  {Chernyak}, \bibfnamefont {Vladimir}}\ and\ \bibinfo {author} {\bibfnamefont
  {A.}~\bibnamefont {Sinitsyn}, \bibfnamefont {N}},\ }\href@noop {} {\bibfield
  {journal} {\bibinfo  {journal} {J. Stat. Mech. Theor. Exp.}\ ,\ \bibinfo
  {pages} {L07001}} (\bibinfo {year} {2010})}\BibitemShut {NoStop}%
\bibitem [{\citenamefont {Ivanov}\ and\ \citenamefont
  {Abanov}(2010)}]{Ivanov10}%
  \BibitemOpen
  \bibfield  {author} {\bibinfo {author} {\bibfnamefont {D.~A.}\ \bibnamefont
  {Ivanov}}\ and\ \bibinfo {author} {\bibfnamefont {A.~G.}\ \bibnamefont
  {Abanov}},\ }\href@noop {} {\bibfield  {journal} {\bibinfo  {journal} {Eur.
  Phys. Lett.}\ }\textbf {\bibinfo {volume} {92}},\ \bibinfo {pages} {37008}
  (\bibinfo {year} {2010})}\BibitemShut {NoStop}%
\bibitem [{\citenamefont {Nakajima}\ \emph {et~al.}(2015)\citenamefont
  {Nakajima}, \citenamefont {Taguchi}, \citenamefont {Kubo},\ and\
  \citenamefont {Tokura}}]{Nakajima15}%
  \BibitemOpen
  \bibfield  {author} {\bibinfo {author} {\bibfnamefont {S.}~\bibnamefont
  {Nakajima}}, \bibinfo {author} {\bibfnamefont {M.}~\bibnamefont {Taguchi}},
  \bibinfo {author} {\bibfnamefont {T.}~\bibnamefont {Kubo}}, \ and\ \bibinfo
  {author} {\bibfnamefont {Y.}~\bibnamefont {Tokura}},\ }\href@noop {}
  {\bibfield  {journal} {\bibinfo  {journal} {Phys. Rev. B}\ }\textbf {\bibinfo
  {volume} {92}},\ \bibinfo {pages} {195420} (\bibinfo {year}
  {2015})}\BibitemShut {NoStop}%
\bibitem [{\citenamefont {B\"uttiker}\ \emph {et~al.}(1993)\citenamefont
  {B\"uttiker}, \citenamefont {Pr\^etre},\ and\ \citenamefont
  {Thomas}}]{Buettiker93}%
  \BibitemOpen
  \bibfield  {author} {\bibinfo {author} {\bibfnamefont {M.}~\bibnamefont
  {B\"uttiker}}, \bibinfo {author} {\bibfnamefont {A.}~\bibnamefont
  {Pr\^etre}}, \ and\ \bibinfo {author} {\bibfnamefont {H.}~\bibnamefont
  {Thomas}},\ }\href@noop {} {\bibfield  {journal} {\bibinfo  {journal} {Phys.
  Rev. Lett.}\ }\textbf {\bibinfo {volume} {70}},\ \bibinfo {pages} {4114}
  (\bibinfo {year} {1993})}\BibitemShut {NoStop}%
\bibitem [{\citenamefont {B\"uttiker}\ \emph {et~al.}(1994)\citenamefont
  {B\"uttiker}, \citenamefont {Thomas},\ and\ \citenamefont
  {Pr\^etre}}]{Buettiker94}%
  \BibitemOpen
  \bibfield  {author} {\bibinfo {author} {\bibfnamefont {M.}~\bibnamefont
  {B\"uttiker}}, \bibinfo {author} {\bibfnamefont {H.}~\bibnamefont {Thomas}},
  \ and\ \bibinfo {author} {\bibfnamefont {A.}~\bibnamefont {Pr\^etre}},\
  }\href@noop {} {\bibfield  {journal} {\bibinfo  {journal} {Z. Phys. B Con.
  Mat.}\ }\textbf {\bibinfo {volume} {94}},\ \bibinfo {pages} {133} (\bibinfo
  {year} {1994})}\BibitemShut {NoStop}%
\bibitem [{\citenamefont {Altshuler}\ and\ \citenamefont
  {Glazman}(1999)}]{Altshuler99}%
  \BibitemOpen
  \bibfield  {author} {\bibinfo {author} {\bibfnamefont {B.~L.}\ \bibnamefont
  {Altshuler}}\ and\ \bibinfo {author} {\bibfnamefont {L.~I.}\ \bibnamefont
  {Glazman}},\ }\href@noop {} {\bibfield  {journal} {\bibinfo  {journal}
  {Science}\ }\textbf {\bibinfo {volume} {283}},\ \bibinfo {pages} {1864}
  (\bibinfo {year} {1999})}\BibitemShut {NoStop}%
\bibitem [{\citenamefont {Dundas}\ \emph {et~al.}(2009)\citenamefont {Dundas},
  \citenamefont {McEniry},\ and\ \citenamefont {Todorov}}]{Dundas09}%
  \BibitemOpen
  \bibfield  {author} {\bibinfo {author} {\bibfnamefont {D.}~\bibnamefont
  {Dundas}}, \bibinfo {author} {\bibfnamefont {E.~J.}\ \bibnamefont {McEniry}},
  \ and\ \bibinfo {author} {\bibfnamefont {T.~N.}\ \bibnamefont {Todorov}},\
  }\href@noop {} {\bibfield  {journal} {\bibinfo  {journal} {Nat.
  Nanotechnol.}\ }\textbf {\bibinfo {volume} {4}},\ \bibinfo {pages} {99}
  (\bibinfo {year} {2009})}\BibitemShut {NoStop}%
\bibitem [{\citenamefont {Bode}\ \emph {et~al.}(2011)\citenamefont {Bode},
  \citenamefont {Kusminskiy}, \citenamefont {Egger},\ and\ \citenamefont {von
  Oppen}}]{Bode11}%
  \BibitemOpen
  \bibfield  {author} {\bibinfo {author} {\bibfnamefont {N.}~\bibnamefont
  {Bode}}, \bibinfo {author} {\bibfnamefont {S.~V.}\ \bibnamefont
  {Kusminskiy}}, \bibinfo {author} {\bibfnamefont {R.}~\bibnamefont {Egger}}, \
  and\ \bibinfo {author} {\bibfnamefont {F.}~\bibnamefont {von Oppen}},\
  }\href@noop {} {\bibfield  {journal} {\bibinfo  {journal} {Phys. Rev. Lett.}\
  }\textbf {\bibinfo {volume} {107}},\ \bibinfo {pages} {036804} (\bibinfo
  {year} {2011})}\BibitemShut {NoStop}%
\bibitem [{\citenamefont {Thomas}\ \emph {et~al.}(2012)\citenamefont {Thomas},
  \citenamefont {Karzig}, \citenamefont {Kusminskiy}, \citenamefont
  {Zar\'and},\ and\ \citenamefont {von Oppen}}]{Thomas12b}%
  \BibitemOpen
  \bibfield  {author} {\bibinfo {author} {\bibfnamefont {M.}~\bibnamefont
  {Thomas}}, \bibinfo {author} {\bibfnamefont {T.}~\bibnamefont {Karzig}},
  \bibinfo {author} {\bibfnamefont {S.~V.}\ \bibnamefont {Kusminskiy}},
  \bibinfo {author} {\bibfnamefont {G.}~\bibnamefont {Zar\'and}}, \ and\
  \bibinfo {author} {\bibfnamefont {F.}~\bibnamefont {von Oppen}},\ }\href@noop
  {} {\bibfield  {journal} {\bibinfo  {journal} {Phys. Rev. B}\ }\textbf
  {\bibinfo {volume} {86}},\ \bibinfo {pages} {195419} (\bibinfo {year}
  {2012})}\BibitemShut {NoStop}%
\bibitem [{\citenamefont {Todorov}\ \emph {et~al.}(2014)\citenamefont
  {Todorov}, \citenamefont {Dundas}, \citenamefont {L{\"u}}, \citenamefont
  {Brandbyge},\ and\ \citenamefont {Hedegård}}]{Todorov14}%
  \BibitemOpen
  \bibfield  {author} {\bibinfo {author} {\bibfnamefont {T.~N.}\ \bibnamefont
  {Todorov}}, \bibinfo {author} {\bibfnamefont {D.}~\bibnamefont {Dundas}},
  \bibinfo {author} {\bibfnamefont {J.-T.}\ \bibnamefont {L{\"u}}}, \bibinfo
  {author} {\bibfnamefont {M.}~\bibnamefont {Brandbyge}}, \ and\ \bibinfo
  {author} {\bibfnamefont {P.}~\bibnamefont {Hedegård}},\ }\href@noop {}
  {\bibfield  {journal} {\bibinfo  {journal} {Eur. J. Phys.}\ }\textbf
  {\bibinfo {volume} {35}},\ \bibinfo {pages} {065004} (\bibinfo {year}
  {2014})}\BibitemShut {NoStop}%
\bibitem [{\citenamefont {L\"u}\ \emph {et~al.}(2015)\citenamefont {L\"u},
  \citenamefont {Christensen}, \citenamefont {Wang}, \citenamefont
  {Hedeg\aa{}rd},\ and\ \citenamefont {Brandbyge}}]{Lue15}%
  \BibitemOpen
  \bibfield  {author} {\bibinfo {author} {\bibfnamefont {J.-T.}\ \bibnamefont
  {L\"u}}, \bibinfo {author} {\bibfnamefont {R.~B.}\ \bibnamefont
  {Christensen}}, \bibinfo {author} {\bibfnamefont {J.-S.}\ \bibnamefont
  {Wang}}, \bibinfo {author} {\bibfnamefont {P.}~\bibnamefont {Hedeg\aa{}rd}},
  \ and\ \bibinfo {author} {\bibfnamefont {M.}~\bibnamefont {Brandbyge}},\
  }\href@noop {} {\bibfield  {journal} {\bibinfo  {journal} {Phys. Rev. Lett.}\
  }\textbf {\bibinfo {volume} {114}},\ \bibinfo {pages} {096801} (\bibinfo
  {year} {2015})}\BibitemShut {NoStop}%
\bibitem [{\citenamefont {H\"anggi}\ and\ \citenamefont
  {Marchesoni}(2009)}]{Haenggi09}%
  \BibitemOpen
  \bibfield  {author} {\bibinfo {author} {\bibfnamefont {P.}~\bibnamefont
  {H\"anggi}}\ and\ \bibinfo {author} {\bibfnamefont {F.}~\bibnamefont
  {Marchesoni}},\ }\href@noop {} {\bibfield  {journal} {\bibinfo  {journal}
  {Rev. Mod. Phys.}\ }\textbf {\bibinfo {volume} {81}},\ \bibinfo {pages} {387}
  (\bibinfo {year} {2009})}\BibitemShut {NoStop}%
\bibitem [{\citenamefont {Seldenthuis}\ \emph {et~al.}(2010)\citenamefont
  {Seldenthuis}, \citenamefont {Prins}, \citenamefont {Thijssen},\ and\
  \citenamefont {van~der Zant}}]{Seldenthuis10}%
  \BibitemOpen
  \bibfield  {author} {\bibinfo {author} {\bibfnamefont {J.~S.}\ \bibnamefont
  {Seldenthuis}}, \bibinfo {author} {\bibfnamefont {F.}~\bibnamefont {Prins}},
  \bibinfo {author} {\bibfnamefont {J.~M.}\ \bibnamefont {Thijssen}}, \ and\
  \bibinfo {author} {\bibfnamefont {H.~S.~J.}\ \bibnamefont {van~der Zant}},\
  }\href@noop {} {\bibfield  {journal} {\bibinfo  {journal} {ACS Nano}\
  }\textbf {\bibinfo {volume} {4}},\ \bibinfo {pages} {6681} (\bibinfo {year}
  {2010})}\BibitemShut {NoStop}%
\bibitem [{\citenamefont {Bustos-Mar\'un}\ \emph {et~al.}(2013)\citenamefont
  {Bustos-Mar\'un}, \citenamefont {Refael},\ and\ \citenamefont {von
  Oppen}}]{Bustos13}%
  \BibitemOpen
  \bibfield  {author} {\bibinfo {author} {\bibfnamefont {R.}~\bibnamefont
  {Bustos-Mar\'un}}, \bibinfo {author} {\bibfnamefont {G.}~\bibnamefont
  {Refael}}, \ and\ \bibinfo {author} {\bibfnamefont {F.}~\bibnamefont {von
  Oppen}},\ }\href@noop {} {\bibfield  {journal} {\bibinfo  {journal} {Phys.
  Rev. Lett.}\ }\textbf {\bibinfo {volume} {111}},\ \bibinfo {pages} {060802}
  (\bibinfo {year} {2013})}\BibitemShut {NoStop}%
\bibitem [{\citenamefont {Arrachea}\ and\ \citenamefont {von
  Oppen}(2016)}]{Arrachea16}%
  \BibitemOpen
  \bibfield  {author} {\bibinfo {author} {\bibfnamefont {L.}~\bibnamefont
  {Arrachea}}\ and\ \bibinfo {author} {\bibfnamefont {F.}~\bibnamefont {von
  Oppen}},\ }\href@noop {} {\bibfield  {journal} {\bibinfo  {journal} {Physica
  E}\ ,\ } (\bibinfo {year} {2016})}\BibitemShut {NoStop}%
\bibitem [{\citenamefont {Splettstoesser}\ \emph {et~al.}(2005)\citenamefont
  {Splettstoesser}, \citenamefont {Governale}, \citenamefont {K\"{o}nig},\ and\
  \citenamefont {Fazio}}]{Splettstoesser05}%
  \BibitemOpen
  \bibfield  {author} {\bibinfo {author} {\bibfnamefont {J.}~\bibnamefont
  {Splettstoesser}}, \bibinfo {author} {\bibfnamefont {M.}~\bibnamefont
  {Governale}}, \bibinfo {author} {\bibfnamefont {J.}~\bibnamefont
  {K\"{o}nig}}, \ and\ \bibinfo {author} {\bibfnamefont {R.}~\bibnamefont
  {Fazio}},\ }\href@noop {} {\bibfield  {journal} {\bibinfo  {journal} {Phys.
  Rev. Lett.}\ }\textbf {\bibinfo {volume} {95}},\ \bibinfo {pages} {246803}
  (\bibinfo {year} {2005})}\BibitemShut {NoStop}%
\bibitem [{\citenamefont {Sela}\ and\ \citenamefont {Oreg}(2006)}]{Sela06}%
  \BibitemOpen
  \bibfield  {author} {\bibinfo {author} {\bibfnamefont {E.}~\bibnamefont
  {Sela}}\ and\ \bibinfo {author} {\bibfnamefont {Y.}~\bibnamefont {Oreg}},\
  }\href@noop {} {\bibfield  {journal} {\bibinfo  {journal} {Phys. Rev. Lett.}\
  }\textbf {\bibinfo {volume} {96}},\ \bibinfo {pages} {166802} (\bibinfo
  {year} {2006})}\BibitemShut {NoStop}%
\bibitem [{\citenamefont {Fioretto}\ and\ \citenamefont
  {Silva}(2008)}]{Fioretto08}%
  \BibitemOpen
  \bibfield  {author} {\bibinfo {author} {\bibfnamefont {D.}~\bibnamefont
  {Fioretto}}\ and\ \bibinfo {author} {\bibfnamefont {A.}~\bibnamefont
  {Silva}},\ }\href@noop {} {\bibfield  {journal} {\bibinfo  {journal} {Phys.
  Rev. Lett.}\ }\textbf {\bibinfo {volume} {100}},\ \bibinfo {pages} {236803}
  (\bibinfo {year} {2008})}\BibitemShut {NoStop}%
\bibitem [{\citenamefont {Pluecker}\ \emph {et~al.}()\citenamefont {Pluecker},
  \citenamefont {Calvo}, \citenamefont {Wegewijs},\ and\ \citenamefont
  {Splettstoesser}}]{Pluecker16b}%
  \BibitemOpen
  \bibfield  {author} {\bibinfo {author} {\bibfnamefont {T.}~\bibnamefont
  {Pluecker}}, \bibinfo {author} {\bibfnamefont {H.}~\bibnamefont {Calvo}},
  \bibinfo {author} {\bibfnamefont {M.~R.}\ \bibnamefont {Wegewijs}}, \ and\
  \bibinfo {author} {\bibfnamefont {J.}~\bibnamefont {Splettstoesser}},\
  }\href@noop {} {}\BibitemShut {NoStop}%
\bibitem [{\citenamefont {Sinitsyn}\ and\ \citenamefont
  {Nemenman}(2007{\natexlab{b}})}]{Sinitsyn07PRL}%
  \BibitemOpen
  \bibfield  {author} {\bibinfo {author} {\bibfnamefont {N.~A.}\ \bibnamefont
  {Sinitsyn}}\ and\ \bibinfo {author} {\bibfnamefont {I.}~\bibnamefont
  {Nemenman}},\ }\href@noop {} {\bibfield  {journal} {\bibinfo  {journal}
  {Phys. Rev. Lett.}\ }\textbf {\bibinfo {volume} {99}},\ \bibinfo {pages}
  {220408} (\bibinfo {year} {2007}{\natexlab{b}})}\BibitemShut {NoStop}%
\bibitem [{\citenamefont {Ren}\ \emph {et~al.}(2010)\citenamefont {Ren},
  \citenamefont {H\"anggi},\ and\ \citenamefont {Li}}]{Ren10}%
  \BibitemOpen
  \bibfield  {author} {\bibinfo {author} {\bibfnamefont {J.}~\bibnamefont
  {Ren}}, \bibinfo {author} {\bibfnamefont {P.}~\bibnamefont {H\"anggi}}, \
  and\ \bibinfo {author} {\bibfnamefont {B.}~\bibnamefont {Li}},\ }\href@noop
  {} {\bibfield  {journal} {\bibinfo  {journal} {Phys. Rev. Lett.}\ }\textbf
  {\bibinfo {volume} {104}},\ \bibinfo {pages} {170601} (\bibinfo {year}
  {2010})}\BibitemShut {NoStop}%
\bibitem [{\citenamefont {Liu}\ \emph {et~al.}(2013)\citenamefont {Liu},
  \citenamefont {Agarwalla}, \citenamefont {Wang},\ and\ \citenamefont
  {Li}}]{Liu13}%
  \BibitemOpen
  \bibfield  {author} {\bibinfo {author} {\bibfnamefont {S.}~\bibnamefont
  {Liu}}, \bibinfo {author} {\bibfnamefont {B.~K.}\ \bibnamefont {Agarwalla}},
  \bibinfo {author} {\bibfnamefont {J.-S.}\ \bibnamefont {Wang}}, \ and\
  \bibinfo {author} {\bibfnamefont {B.}~\bibnamefont {Li}},\ }\href@noop {}
  {\bibfield  {journal} {\bibinfo  {journal} {Phys. Rev. E}\ }\textbf {\bibinfo
  {volume} {87}},\ \bibinfo {pages} {022122} (\bibinfo {year}
  {2013})}\BibitemShut {NoStop}%
\bibitem [{\citenamefont {Bagrets}\ and\ \citenamefont
  {Nazarov}(2003)}]{Bagrets03}%
  \BibitemOpen
  \bibfield  {author} {\bibinfo {author} {\bibfnamefont {D.~A.}\ \bibnamefont
  {Bagrets}}\ and\ \bibinfo {author} {\bibfnamefont {Y.~V.}\ \bibnamefont
  {Nazarov}},\ }\href@noop {} {\bibfield  {journal} {\bibinfo  {journal} {Phys.
  Rev. B}\ }\textbf {\bibinfo {volume} {67}},\ \bibinfo {pages} {085316}
  (\bibinfo {year} {2003})}\BibitemShut {NoStop}%
\bibitem [{\citenamefont {Utsumi}(2007)}]{Utsumi07}%
  \BibitemOpen
  \bibfield  {author} {\bibinfo {author} {\bibfnamefont {Y.}~\bibnamefont
  {Utsumi}},\ }\href@noop {} {\bibfield  {journal} {\bibinfo  {journal} {Phys.
  Rev. B}\ }\textbf {\bibinfo {volume} {75}},\ \bibinfo {pages} {035333}
  (\bibinfo {year} {2007})}\BibitemShut {NoStop}%
\bibitem [{\citenamefont {Yuge}\ \emph {et~al.}(2012)\citenamefont {Yuge},
  \citenamefont {Sagawa}, \citenamefont {Sugita},\ and\ \citenamefont
  {Hayakawa}}]{Yuge12}%
  \BibitemOpen
  \bibfield  {author} {\bibinfo {author} {\bibfnamefont {T.}~\bibnamefont
  {Yuge}}, \bibinfo {author} {\bibfnamefont {T.}~\bibnamefont {Sagawa}},
  \bibinfo {author} {\bibfnamefont {A.}~\bibnamefont {Sugita}}, \ and\ \bibinfo
  {author} {\bibfnamefont {H.}~\bibnamefont {Hayakawa}},\ }\href@noop {}
  {\bibfield  {journal} {\bibinfo  {journal} {Phys. Rev. B}\ }\textbf {\bibinfo
  {volume} {86}},\ \bibinfo {pages} {235308} (\bibinfo {year}
  {2012})}\BibitemShut {NoStop}%
\bibitem [{\citenamefont {Yoshii}\ and\ \citenamefont
  {Hayakawa}(2013)}]{Yoshii13}%
  \BibitemOpen
  \bibfield  {author} {\bibinfo {author} {\bibfnamefont {R.}~\bibnamefont
  {Yoshii}}\ and\ \bibinfo {author} {\bibfnamefont {H.}~\bibnamefont
  {Hayakawa}},\ }\href@noop {} {\enquote {\bibinfo {title} {Analytical
  expression of geometrical pumping for a quantum dot based on quantum master
  equation},}\ } (\bibinfo {year} {2013}),\ \bibinfo {note}
  {arXiv:1312.3772}\BibitemShut {NoStop}%
\bibitem [{\citenamefont {Yuge}\ \emph {et~al.}(2013)\citenamefont {Yuge},
  \citenamefont {Sagawa}, \citenamefont {Sugita},\ and\ \citenamefont
  {Hayakawa}}]{Yuge13}%
  \BibitemOpen
  \bibfield  {author} {\bibinfo {author} {\bibfnamefont {T.}~\bibnamefont
  {Yuge}}, \bibinfo {author} {\bibfnamefont {T.}~\bibnamefont {Sagawa}},
  \bibinfo {author} {\bibfnamefont {A.}~\bibnamefont {Sugita}}, \ and\ \bibinfo
  {author} {\bibfnamefont {H.}~\bibnamefont {Hayakawa}},\ }\href@noop {}
  {\bibfield  {journal} {\bibinfo  {journal} {J. Stat. Phys.}\ }\textbf
  {\bibinfo {volume} {153}},\ \bibinfo {pages} {412} (\bibinfo {year}
  {2013})}\BibitemShut {NoStop}%
\bibitem [{\citenamefont {Robbins}\ and\ \citenamefont
  {Berry}(1992)}]{Robbins92}%
  \BibitemOpen
  \bibfield  {author} {\bibinfo {author} {\bibfnamefont {J.~M.}\ \bibnamefont
  {Robbins}}\ and\ \bibinfo {author} {\bibfnamefont {M.~V.}\ \bibnamefont
  {Berry}},\ }\href@noop {} {\bibfield  {journal} {\bibinfo  {journal} {J.
  Phys. A}\ }\textbf {\bibinfo {volume} {25}},\ \bibinfo {pages} {L961}
  (\bibinfo {year} {1992})}\BibitemShut {NoStop}%
\bibitem [{\citenamefont {Berry}\ and\ \citenamefont
  {Sinclair}(1997)}]{Berry97}%
  \BibitemOpen
  \bibfield  {author} {\bibinfo {author} {\bibfnamefont {M.~V.}\ \bibnamefont
  {Berry}}\ and\ \bibinfo {author} {\bibfnamefont {E.~C.}\ \bibnamefont
  {Sinclair}},\ }\href@noop {} {\bibfield  {journal} {\bibinfo  {journal} {J.
  Phys. A}\ }\textbf {\bibinfo {volume} {30}},\ \bibinfo {pages} {2853}
  (\bibinfo {year} {1997})}\BibitemShut {NoStop}%
\bibitem [{\citenamefont {Bellissard}(2002)}]{Bellissard02}%
  \BibitemOpen
  \bibfield  {author} {\bibinfo {author} {\bibfnamefont {J.}~\bibnamefont
  {Bellissard}},\ }\enquote {\bibinfo {title} {Dynamics of dissipation},}\ \
  (\bibinfo {year} {2002})\ Chap.\ \bibinfo {chapter} {Coherent and Dissipative
  Transport in Aperiodic Solids: An Overview}\BibitemShut {NoStop}%
\bibitem [{\citenamefont {Gebauer}\ and\ \citenamefont
  {Car}(2004)}]{Gebauer04}%
  \BibitemOpen
  \bibfield  {author} {\bibinfo {author} {\bibfnamefont {R.}~\bibnamefont
  {Gebauer}}\ and\ \bibinfo {author} {\bibfnamefont {R.}~\bibnamefont {Car}},\
  }\href@noop {} {\bibfield  {journal} {\bibinfo  {journal} {Phys. Rev. Lett.}\
  }\textbf {\bibinfo {volume} {93}},\ \bibinfo {pages} {160404} (\bibinfo
  {year} {2004})}\BibitemShut {NoStop}%
\bibitem [{\citenamefont {Bodor}\ and\ \citenamefont
  {Di\'osi}(2006)}]{Bodor06}%
  \BibitemOpen
  \bibfield  {author} {\bibinfo {author} {\bibfnamefont {A.}~\bibnamefont
  {Bodor}}\ and\ \bibinfo {author} {\bibfnamefont {L.}~\bibnamefont
  {Di\'osi}},\ }\href@noop {} {\bibfield  {journal} {\bibinfo  {journal} {Phys.
  Rev. A}\ }\textbf {\bibinfo {volume} {73}},\ \bibinfo {pages} {064101}
  (\bibinfo {year} {2006})}\BibitemShut {NoStop}%
\bibitem [{\citenamefont {Salmilehto}\ \emph {et~al.}(2012)\citenamefont
  {Salmilehto}, \citenamefont {Solinas},\ and\ \citenamefont
  {M\"ott\"onen}}]{Salmilehto12}%
  \BibitemOpen
  \bibfield  {author} {\bibinfo {author} {\bibfnamefont {J.}~\bibnamefont
  {Salmilehto}}, \bibinfo {author} {\bibfnamefont {P.}~\bibnamefont {Solinas}},
  \ and\ \bibinfo {author} {\bibfnamefont {M.}~\bibnamefont {M\"ott\"onen}},\
  }\href@noop {} {\bibfield  {journal} {\bibinfo  {journal} {Phys. Rev. A}\
  }\textbf {\bibinfo {volume} {85}},\ \bibinfo {pages} {032110} (\bibinfo
  {year} {2012})}\BibitemShut {NoStop}%
\bibitem [{\citenamefont {Landsberg}(1992)}]{Landsberg92}%
  \BibitemOpen
  \bibfield  {author} {\bibinfo {author} {\bibfnamefont {A.~S.}\ \bibnamefont
  {Landsberg}},\ }\href@noop {} {\bibfield  {journal} {\bibinfo  {journal}
  {Phys. Rev. Lett.}\ }\textbf {\bibinfo {volume} {69}},\ \bibinfo {pages}
  {865} (\bibinfo {year} {1992})}\BibitemShut {NoStop}%
\bibitem [{\citenamefont {Andersson}(2005)}]{Andersson05}%
  \BibitemOpen
  \bibfield  {author} {\bibinfo {author} {\bibfnamefont {S.~B.}\ \bibnamefont
  {Andersson}},\ }in\ \href@noop {} {\emph {\bibinfo {booktitle} {IFAC
  Symposium on Nonlinear Control Systems}}}\ (\bibinfo {year}
  {2005})\BibitemShut {NoStop}%
\bibitem [{\citenamefont {Kepler}\ and\ \citenamefont
  {Kagan}(1991)}]{Kepler91}%
  \BibitemOpen
  \bibfield  {author} {\bibinfo {author} {\bibfnamefont {T.~B.}\ \bibnamefont
  {Kepler}}\ and\ \bibinfo {author} {\bibfnamefont {M.~L.}\ \bibnamefont
  {Kagan}},\ }\href@noop {} {\bibfield  {journal} {\bibinfo  {journal} {Phys.
  Rev. Lett.}\ }\textbf {\bibinfo {volume} {66}},\ \bibinfo {pages} {847}
  (\bibinfo {year} {1991})}\BibitemShut {NoStop}%
\bibitem [{\citenamefont {Ning}\ and\ \citenamefont {Haken}(1992)}]{Ning92}%
  \BibitemOpen
  \bibfield  {author} {\bibinfo {author} {\bibfnamefont {C.~Z.}\ \bibnamefont
  {Ning}}\ and\ \bibinfo {author} {\bibfnamefont {H.}~\bibnamefont {Haken}},\
  }\href@noop {} {\bibfield  {journal} {\bibinfo  {journal} {Phys. Rev. Lett.}\
  }\textbf {\bibinfo {volume} {68}},\ \bibinfo {pages} {2109} (\bibinfo {year}
  {1992})}\BibitemShut {NoStop}%
\bibitem [{\citenamefont {K{\"o}nig}\ \emph {et~al.}(1995)\citenamefont
  {K{\"o}nig}, \citenamefont {Schoeller},\ and\ \citenamefont
  {Sch{\"o}n}}]{Koenig95}%
  \BibitemOpen
  \bibfield  {author} {\bibinfo {author} {\bibfnamefont {J.}~\bibnamefont
  {K{\"o}nig}}, \bibinfo {author} {\bibfnamefont {H.}~\bibnamefont
  {Schoeller}}, \ and\ \bibinfo {author} {\bibfnamefont {G.}~\bibnamefont
  {Sch{\"o}n}},\ }\href@noop {} {\bibfield  {journal} {\bibinfo  {journal}
  {Eur. Phys. Lett.}\ }\textbf {\bibinfo {volume} {31}},\ \bibinfo {pages} {31}
  (\bibinfo {year} {1995})}\BibitemShut {NoStop}%
\bibitem [{\citenamefont {Schoeller}(2009)}]{Schoeller09a}%
  \BibitemOpen
  \bibfield  {author} {\bibinfo {author} {\bibfnamefont {H.}~\bibnamefont
  {Schoeller}},\ }\href@noop {} {\bibfield  {journal} {\bibinfo  {journal}
  {Eur. Phys. Journ. Special Topics}\ }\textbf {\bibinfo {volume} {168}},\
  \bibinfo {pages} {179} (\bibinfo {year} {2009})}\BibitemShut {NoStop}%
\bibitem [{\citenamefont {Aghassi}\ \emph {et~al.}(2006)\citenamefont
  {Aghassi}, \citenamefont {Thielmann}, \citenamefont {Hettler},\ and\
  \citenamefont {Sch\"{o}n}}]{Aghassi06}%
  \BibitemOpen
  \bibfield  {author} {\bibinfo {author} {\bibfnamefont {J.}~\bibnamefont
  {Aghassi}}, \bibinfo {author} {\bibfnamefont {A.}~\bibnamefont {Thielmann}},
  \bibinfo {author} {\bibfnamefont {M.~H.}\ \bibnamefont {Hettler}}, \ and\
  \bibinfo {author} {\bibfnamefont {G.}~\bibnamefont {Sch\"{o}n}},\ }\href@noop
  {} {\bibfield  {journal} {\bibinfo  {journal} {Phys. Rev. B}\ }\textbf
  {\bibinfo {volume} {73}},\ \bibinfo {pages} {195323} (\bibinfo {year}
  {2006})}\BibitemShut {NoStop}%
\bibitem [{\citenamefont {Schuricht}\ and\ \citenamefont
  {Schoeller}(2009)}]{Schuricht09}%
  \BibitemOpen
  \bibfield  {author} {\bibinfo {author} {\bibfnamefont {D.}~\bibnamefont
  {Schuricht}}\ and\ \bibinfo {author} {\bibfnamefont {H.}~\bibnamefont
  {Schoeller}},\ }\href@noop {} {\bibfield  {journal} {\bibinfo  {journal}
  {Phys. Rev. B}\ }\textbf {\bibinfo {volume} {80}},\ \bibinfo {pages} {075120}
  (\bibinfo {year} {2009})}\BibitemShut {NoStop}%
\bibitem [{\citenamefont {Schmidt}\ \emph {et~al.}(2010)\citenamefont
  {Schmidt}, \citenamefont {Hettler},\ and\ \citenamefont
  {Sch\"on}}]{BSchmidt10}%
  \BibitemOpen
  \bibfield  {author} {\bibinfo {author} {\bibfnamefont {B.~B.}\ \bibnamefont
  {Schmidt}}, \bibinfo {author} {\bibfnamefont {M.~H.}\ \bibnamefont
  {Hettler}}, \ and\ \bibinfo {author} {\bibfnamefont {G.}~\bibnamefont
  {Sch\"on}},\ }\href@noop {} {\bibfield  {journal} {\bibinfo  {journal} {Phys.
  Rev. B}\ }\textbf {\bibinfo {volume} {82}},\ \bibinfo {pages} {155113}
  (\bibinfo {year} {2010})}\BibitemShut {NoStop}%
\bibitem [{Note2()}]{Note2}%
  \BibitemOpen
  \bibinfo {note} {Sinitsyn's master equation~\Eq
  {eq:GeneralizedMasterEquation} for the generating-operator of the \fcs can be
  derived in both, the ``real-time'' and the ``Nakayima-Zwanzig' approach. This
  underlines that neither label is a meaningful labels for distinguishing
  different approaches to pumping.}\BibitemShut {Stop}%
\bibitem [{\citenamefont {Braggio}\ \emph {et~al.}(2006)\citenamefont
  {Braggio}, \citenamefont {K\"onig},\ and\ \citenamefont {Fazio}}]{Braggio05}%
  \BibitemOpen
  \bibfield  {author} {\bibinfo {author} {\bibfnamefont {A.}~\bibnamefont
  {Braggio}}, \bibinfo {author} {\bibfnamefont {J.}~\bibnamefont {K\"onig}}, \
  and\ \bibinfo {author} {\bibfnamefont {R.}~\bibnamefont {Fazio}},\
  }\href@noop {} {\bibfield  {journal} {\bibinfo  {journal} {Phys. Rev. Lett.}\
  }\textbf {\bibinfo {volume} {96}},\ \bibinfo {pages} {026805} (\bibinfo
  {year} {2006})}\BibitemShut {NoStop}%
\bibitem [{\citenamefont {Eckel}\ \emph {et~al.}(2010)\citenamefont {Eckel},
  \citenamefont {Heidrich-Meisner}, \citenamefont {Jakobs}, \citenamefont
  {Thorwart}, \citenamefont {Pletyukhov},\ and\ \citenamefont
  {Egger}}]{Eckel10}%
  \BibitemOpen
  \bibfield  {author} {\bibinfo {author} {\bibfnamefont {J.}~\bibnamefont
  {Eckel}}, \bibinfo {author} {\bibfnamefont {F.}~\bibnamefont
  {Heidrich-Meisner}}, \bibinfo {author} {\bibfnamefont {S.}~\bibnamefont
  {Jakobs}}, \bibinfo {author} {\bibfnamefont {M.}~\bibnamefont {Thorwart}},
  \bibinfo {author} {\bibfnamefont {M.}~\bibnamefont {Pletyukhov}}, \ and\
  \bibinfo {author} {\bibfnamefont {R.}~\bibnamefont {Egger}},\ }\href@noop {}
  {\bibfield  {journal} {\bibinfo  {journal} {New. J. Phys.}\ }\textbf
  {\bibinfo {volume} {12}},\ \bibinfo {pages} {043042} (\bibinfo {year}
  {2010})}\BibitemShut {NoStop}%
\bibitem [{\citenamefont {Pletyukhov}\ \emph {et~al.}(2010)\citenamefont
  {Pletyukhov}, \citenamefont {Schuricht},\ and\ \citenamefont
  {Schoeller}}]{Pletyukhov10}%
  \BibitemOpen
  \bibfield  {author} {\bibinfo {author} {\bibfnamefont {M.}~\bibnamefont
  {Pletyukhov}}, \bibinfo {author} {\bibfnamefont {D.}~\bibnamefont
  {Schuricht}}, \ and\ \bibinfo {author} {\bibfnamefont {H.}~\bibnamefont
  {Schoeller}},\ }\href@noop {} {\bibfield  {journal} {\bibinfo  {journal}
  {Phys. Rev. Lett.}\ }\textbf {\bibinfo {volume} {104}},\ \bibinfo {pages}
  {106801} (\bibinfo {year} {2010})}\BibitemShut {NoStop}%
\bibitem [{\citenamefont {Andergassen}\ \emph {et~al.}(2011)\citenamefont
  {Andergassen}, \citenamefont {Pletyukhov}, \citenamefont {Schuricht},
  \citenamefont {Schoeller},\ and\ \citenamefont {Borda}}]{Andergassen11a}%
  \BibitemOpen
  \bibfield  {author} {\bibinfo {author} {\bibfnamefont {S.}~\bibnamefont
  {Andergassen}}, \bibinfo {author} {\bibfnamefont {M.}~\bibnamefont
  {Pletyukhov}}, \bibinfo {author} {\bibfnamefont {D.}~\bibnamefont
  {Schuricht}}, \bibinfo {author} {\bibfnamefont {H.}~\bibnamefont
  {Schoeller}}, \ and\ \bibinfo {author} {\bibfnamefont {L.}~\bibnamefont
  {Borda}},\ }\href@noop {} {\bibfield  {journal} {\bibinfo  {journal} {Phys.
  Rev. B}\ }\textbf {\bibinfo {volume} {83}},\ \bibinfo {pages} {205103}
  (\bibinfo {year} {2011})}\BibitemShut {NoStop}%
\bibitem [{\citenamefont {Pletyukhov}\ and\ \citenamefont
  {Schoeller}(2012)}]{Pletyukhov12a}%
  \BibitemOpen
  \bibfield  {author} {\bibinfo {author} {\bibfnamefont {M.}~\bibnamefont
  {Pletyukhov}}\ and\ \bibinfo {author} {\bibfnamefont {H.}~\bibnamefont
  {Schoeller}},\ }\href@noop {} {\bibfield  {journal} {\bibinfo  {journal}
  {Phys. Rev. Lett.}\ }\textbf {\bibinfo {volume} {108}},\ \bibinfo {pages}
  {260601} (\bibinfo {year} {2012})}\BibitemShut {NoStop}%
\bibitem [{\citenamefont {Saptsov}\ and\ \citenamefont
  {Wegewijs}(2012)}]{Saptsov12a}%
  \BibitemOpen
  \bibfield  {author} {\bibinfo {author} {\bibfnamefont {R.~B.}\ \bibnamefont
  {Saptsov}}\ and\ \bibinfo {author} {\bibfnamefont {M.~R.}\ \bibnamefont
  {Wegewijs}},\ }\href@noop {} {\bibfield  {journal} {\bibinfo  {journal}
  {Phys. Rev. B}\ }\textbf {\bibinfo {volume} {86}},\ \bibinfo {pages} {235432}
  (\bibinfo {year} {2012})}\BibitemShut {NoStop}%
\bibitem [{\citenamefont {Klochan}\ \emph {et~al.}(2013)\citenamefont
  {Klochan}, \citenamefont {Micolich}, \citenamefont {Hamilton}, \citenamefont
  {Reuter}, \citenamefont {Wieck}, \citenamefont {Reininghaus}, \citenamefont
  {Pletyukhov},\ and\ \citenamefont {Schoeller}}]{Klochan13}%
  \BibitemOpen
  \bibfield  {author} {\bibinfo {author} {\bibfnamefont {O.}~\bibnamefont
  {Klochan}}, \bibinfo {author} {\bibfnamefont {A.~P.}\ \bibnamefont
  {Micolich}}, \bibinfo {author} {\bibfnamefont {A.~R.}\ \bibnamefont
  {Hamilton}}, \bibinfo {author} {\bibfnamefont {D.}~\bibnamefont {Reuter}},
  \bibinfo {author} {\bibfnamefont {A.~D.}\ \bibnamefont {Wieck}}, \bibinfo
  {author} {\bibfnamefont {F.}~\bibnamefont {Reininghaus}}, \bibinfo {author}
  {\bibfnamefont {M.}~\bibnamefont {Pletyukhov}}, \ and\ \bibinfo {author}
  {\bibfnamefont {H.}~\bibnamefont {Schoeller}},\ }\href@noop {} {\bibfield
  {journal} {\bibinfo  {journal} {Phys. Rev. B}\ }\textbf {\bibinfo {volume}
  {87}},\ \bibinfo {pages} {201104} (\bibinfo {year} {2013})}\BibitemShut
  {NoStop}%
\bibitem [{\citenamefont {Kashuba}\ \emph {et~al.}(2012)\citenamefont
  {Kashuba}, \citenamefont {Schoeller},\ and\ \citenamefont
  {Splettstoesser}}]{Kashuba12}%
  \BibitemOpen
  \bibfield  {author} {\bibinfo {author} {\bibfnamefont {O.}~\bibnamefont
  {Kashuba}}, \bibinfo {author} {\bibfnamefont {H.}~\bibnamefont {Schoeller}},
  \ and\ \bibinfo {author} {\bibfnamefont {J.}~\bibnamefont {Splettstoesser}},\
  }\href@noop {} {\bibfield  {journal} {\bibinfo  {journal} {Eur. Phys. Lett.}\
  }\textbf {\bibinfo {volume} {98}},\ \bibinfo {pages} {57003} (\bibinfo {year}
  {2012})}\BibitemShut {NoStop}%
\bibitem [{\citenamefont {Mora}\ and\ \citenamefont {LeHur}(2010)}]{Mora10}%
  \BibitemOpen
  \bibfield  {author} {\bibinfo {author} {\bibfnamefont {C.}~\bibnamefont
  {Mora}}\ and\ \bibinfo {author} {\bibfnamefont {K.}~\bibnamefont {LeHur}},\
  }\href@noop {} {\bibfield  {journal} {\bibinfo  {journal} {Nat. Phys.}\
  }\textbf {\bibinfo {volume} {6}},\ \bibinfo {pages} {697} (\bibinfo {year}
  {2010})}\BibitemShut {NoStop}%
\bibitem [{\citenamefont {Hamamoto}\ \emph {et~al.}(2010)\citenamefont
  {Hamamoto}, \citenamefont {Jonckheere}, \citenamefont {Kato},\ and\
  \citenamefont {Martin}}]{Hamamoto10}%
  \BibitemOpen
  \bibfield  {author} {\bibinfo {author} {\bibfnamefont {Y.}~\bibnamefont
  {Hamamoto}}, \bibinfo {author} {\bibfnamefont {T.}~\bibnamefont
  {Jonckheere}}, \bibinfo {author} {\bibfnamefont {T.}~\bibnamefont {Kato}}, \
  and\ \bibinfo {author} {\bibfnamefont {T.}~\bibnamefont {Martin}},\
  }\href@noop {} {\bibfield  {journal} {\bibinfo  {journal} {Phys. Rev. B}\
  }\textbf {\bibinfo {volume} {81}},\ \bibinfo {pages} {153305} (\bibinfo
  {year} {2010})}\BibitemShut {NoStop}%
\bibitem [{\citenamefont {Lee}\ \emph {et~al.}(2011)\citenamefont {Lee},
  \citenamefont {L\'opez}, \citenamefont {Choi}, \citenamefont {Jonckheere},\
  and\ \citenamefont {Martin}}]{Lee11}%
  \BibitemOpen
  \bibfield  {author} {\bibinfo {author} {\bibfnamefont {M.}~\bibnamefont
  {Lee}}, \bibinfo {author} {\bibfnamefont {R.}~\bibnamefont {L\'opez}},
  \bibinfo {author} {\bibfnamefont {M.-S.}\ \bibnamefont {Choi}}, \bibinfo
  {author} {\bibfnamefont {T.}~\bibnamefont {Jonckheere}}, \ and\ \bibinfo
  {author} {\bibfnamefont {T.}~\bibnamefont {Martin}},\ }\href@noop {}
  {\bibfield  {journal} {\bibinfo  {journal} {Phys. Rev. B}\ }\textbf {\bibinfo
  {volume} {83}},\ \bibinfo {pages} {201304} (\bibinfo {year}
  {2011})}\BibitemShut {NoStop}%
\bibitem [{\citenamefont {Wegewijs}\ \emph {et~al.}()\citenamefont {Wegewijs},
  \citenamefont {Pluecker},\ and\ \citenamefont
  {Splettstoesser}}]{Pluecker17a}%
  \BibitemOpen
  \bibfield  {author} {\bibinfo {author} {\bibfnamefont {M.~R.}\ \bibnamefont
  {Wegewijs}}, \bibinfo {author} {\bibfnamefont {T.}~\bibnamefont {Pluecker}},
  \ and\ \bibinfo {author} {\bibfnamefont {J.}~\bibnamefont {Splettstoesser}},\
  }\href@noop {} {}\BibitemShut {NoStop}%
\bibitem [{\citenamefont {Vanevi\ifmmode~\acute{c}\else \'{c}\fi{}}\ \emph
  {et~al.}(2016)\citenamefont {Vanevi\ifmmode~\acute{c}\else \'{c}\fi{}},
  \citenamefont {Gabelli}, \citenamefont {Belzig},\ and\ \citenamefont
  {Reulet}}]{Vanevic16}%
  \BibitemOpen
  \bibfield  {author} {\bibinfo {author} {\bibfnamefont {M.}~\bibnamefont
  {Vanevi\ifmmode~\acute{c}\else \'{c}\fi{}}}, \bibinfo {author} {\bibfnamefont
  {J.}~\bibnamefont {Gabelli}}, \bibinfo {author} {\bibfnamefont
  {W.}~\bibnamefont {Belzig}}, \ and\ \bibinfo {author} {\bibfnamefont
  {B.}~\bibnamefont {Reulet}},\ }\href@noop {} {\bibfield  {journal} {\bibinfo
  {journal} {Phys. Rev. B}\ }\textbf {\bibinfo {volume} {93}},\ \bibinfo
  {pages} {041416} (\bibinfo {year} {2016})}\BibitemShut {NoStop}%
\bibitem [{Note3()}]{Note3}%
  \BibitemOpen
  \bibinfo {note} {For experiments one should keep in mind that driving gate
  voltages defining a quantum dot changes the screening properties~\cite
  {Kaasbjerg08} and thus the effective interaction. This may well contribute to
  pumping and can be accounted for in our approach.}\BibitemShut {Stop}%
\bibitem [{\citenamefont {Pedersen}\ and\ \citenamefont
  {B\"uttiker}(1998)}]{Pedersen98}%
  \BibitemOpen
  \bibfield  {author} {\bibinfo {author} {\bibfnamefont {M.~H.}\ \bibnamefont
  {Pedersen}}\ and\ \bibinfo {author} {\bibfnamefont {M.}~\bibnamefont
  {B\"uttiker}},\ }\href@noop {} {\bibfield  {journal} {\bibinfo  {journal}
  {Phys. Rev. B}\ }\textbf {\bibinfo {volume} {58}},\ \bibinfo {pages} {12993}
  (\bibinfo {year} {1998})}\BibitemShut {NoStop}%
\bibitem [{\citenamefont {Battista}\ \emph {et~al.}(2014)\citenamefont
  {Battista}, \citenamefont {Haupt},\ and\ \citenamefont
  {Splettstoesser}}]{Battista14a}%
  \BibitemOpen
  \bibfield  {author} {\bibinfo {author} {\bibfnamefont {F.}~\bibnamefont
  {Battista}}, \bibinfo {author} {\bibfnamefont {F.}~\bibnamefont {Haupt}}, \
  and\ \bibinfo {author} {\bibfnamefont {J.}~\bibnamefont {Splettstoesser}},\
  }\href@noop {} {\bibfield  {journal} {\bibinfo  {journal} {Phys. Rev. B}\
  }\textbf {\bibinfo {volume} {90}},\ \bibinfo {pages} {085418} (\bibinfo
  {year} {2014})}\BibitemShut {NoStop}%
\bibitem [{Note4()}]{Note4}%
  \BibitemOpen
  \bibinfo {note} {The seemingly inconvenient cancellation of time-dependences
  in \Eq {eq:rhoresb} in the time-constant $\env {\rho }$ is actually
  advantageous.}\BibitemShut {Stop}%
\bibitem [{Note5()}]{Note5}%
  \BibitemOpen
  \bibinfo {note} {We use a hat ($\protect \mathaccentV {hat}05E{\protect
  \tmspace +\thinmuskip {.1667em} }$) only when operators may be confused with
  their expectation values.}\BibitemShut {Stop}%
\bibitem [{\citenamefont {Esposito}\ \emph
  {et~al.}(2009{\natexlab{a}})\citenamefont {Esposito}, \citenamefont
  {Harbola},\ and\ \citenamefont {Mukamel}}]{Esposito09rev}%
  \BibitemOpen
  \bibfield  {author} {\bibinfo {author} {\bibfnamefont {M.}~\bibnamefont
  {Esposito}}, \bibinfo {author} {\bibfnamefont {U.}~\bibnamefont {Harbola}}, \
  and\ \bibinfo {author} {\bibfnamefont {S.}~\bibnamefont {Mukamel}},\
  }\href@noop {} {\bibfield  {journal} {\bibinfo  {journal} {Rev. Mod. Phys.}\
  }\textbf {\bibinfo {volume} {81}},\ \bibinfo {pages} {1665} (\bibinfo {year}
  {2009}{\natexlab{a}})}\BibitemShut {NoStop}%
\bibitem [{Note6()}]{Note6}%
  \BibitemOpen
  \bibinfo {note} {The “Wangsness-Bloch” approach used here to obtain \Eqs
  {eq:nonLocalMasterEquation} and \eq {eq:DynamicsKernel} runs into problems
  when going beyond the weak coupling approximation, see~\Ref {Koller10} for a
  discussion. The real-time approach allows for a systematic derivation of
  corrections \cite {Splettstoesser06} to \Eq {eq:nonLocalMasterEquation} and
  \eq {eq:DynamicsKernel} including higher-order coupling effects as well as
  non-Markovian effects. As a result of these corrections to the
  frozen-parameter approximation, the kernel's time dependence will in general
  not be mediated solely by the parameters as in \Eq
  {eq:MarkovMe}.}\BibitemShut {Stop}%
\bibitem [{\citenamefont {K\"onig}\ \emph {et~al.}(1996)\citenamefont
  {K\"onig}, \citenamefont {Schmid}, \citenamefont {Schoeller},\ and\
  \citenamefont {Sch\"on}}]{Koenig96b}%
  \BibitemOpen
  \bibfield  {author} {\bibinfo {author} {\bibfnamefont {J.}~\bibnamefont
  {K\"onig}}, \bibinfo {author} {\bibfnamefont {J.}~\bibnamefont {Schmid}},
  \bibinfo {author} {\bibfnamefont {H.}~\bibnamefont {Schoeller}}, \ and\
  \bibinfo {author} {\bibfnamefont {G.}~\bibnamefont {Sch\"on}},\ }\href@noop
  {} {\bibfield  {journal} {\bibinfo  {journal} {Phys. Rev. B}\ }\textbf
  {\bibinfo {volume} {54}},\ \bibinfo {pages} {16820} (\bibinfo {year}
  {1996})}\BibitemShut {NoStop}%
\bibitem [{\citenamefont {Schoeller}\ and\ \citenamefont
  {Reininghaus}(2009)}]{Schoeller09b}%
  \BibitemOpen
  \bibfield  {author} {\bibinfo {author} {\bibfnamefont {H.}~\bibnamefont
  {Schoeller}}\ and\ \bibinfo {author} {\bibfnamefont {F.}~\bibnamefont
  {Reininghaus}},\ }\href@noop {} {\bibfield  {journal} {\bibinfo  {journal}
  {Phys. Rev. B}\ }\textbf {\bibinfo {volume} {80}},\ \bibinfo {pages} {045117}
  (\bibinfo {year} {2009})}\BibitemShut {NoStop}%
\bibitem [{Note7()}]{Note7}%
  \BibitemOpen
  \bibinfo {note} {``Partial'' distinguishes it from the usual operation of
  normal ordering that ensures that \protect \emph {any single} Wick
  contraction of an operator expression is zero instead of just the
  average.}\BibitemShut {Stop}%
\bibitem [{Note8()}]{Note8}%
  \BibitemOpen
  \bibinfo {note} {This implies that the gauge transformations $X^r \to X^r +
  g(t) \unit $ with \protect \emph {arbitrary} time-dependent functions $g(t$),
  introduced in \Sec {sec:geometry}, do not break the validity of the Markov
  approximations.}\BibitemShut {Stop}%
\bibitem [{\citenamefont {Splettstoesser}\ \emph {et~al.}(2008)\citenamefont
  {Splettstoesser}, \citenamefont {Governale},\ and\ \citenamefont
  {K\"onig}}]{Splettstoesser08a}%
  \BibitemOpen
  \bibfield  {author} {\bibinfo {author} {\bibfnamefont {J.}~\bibnamefont
  {Splettstoesser}}, \bibinfo {author} {\bibfnamefont {M.}~\bibnamefont
  {Governale}}, \ and\ \bibinfo {author} {\bibfnamefont {J.}~\bibnamefont
  {K\"onig}},\ }\href@noop {} {\bibfield  {journal} {\bibinfo  {journal} {Phys.
  Rev. B}\ }\textbf {\bibinfo {volume} {77}},\ \bibinfo {pages} {195320}
  (\bibinfo {year} {2008})}\BibitemShut {NoStop}%
\bibitem [{\citenamefont {Winkler}\ \emph {et~al.}(2009)\citenamefont
  {Winkler}, \citenamefont {Governale},\ and\ \citenamefont
  {K\"onig}}]{Winkler09}%
  \BibitemOpen
  \bibfield  {author} {\bibinfo {author} {\bibfnamefont {N.}~\bibnamefont
  {Winkler}}, \bibinfo {author} {\bibfnamefont {M.}~\bibnamefont {Governale}},
  \ and\ \bibinfo {author} {\bibfnamefont {J.}~\bibnamefont {K\"onig}},\
  }\href@noop {} {\bibfield  {journal} {\bibinfo  {journal} {Phys. Rev. B}\
  }\textbf {\bibinfo {volume} {79}},\ \bibinfo {pages} {235309} (\bibinfo
  {year} {2009})}\BibitemShut {NoStop}%
\bibitem [{\citenamefont {Haupt}\ \emph {et~al.}(2013)\citenamefont {Haupt},
  \citenamefont {Leijnse}, \citenamefont {Calvo}, \citenamefont {Classen},
  \citenamefont {Splettstoesser},\ and\ \citenamefont {Wegewijs}}]{Haupt13}%
  \BibitemOpen
  \bibfield  {author} {\bibinfo {author} {\bibfnamefont {F.}~\bibnamefont
  {Haupt}}, \bibinfo {author} {\bibfnamefont {M.}~\bibnamefont {Leijnse}},
  \bibinfo {author} {\bibfnamefont {H.~L.}\ \bibnamefont {Calvo}}, \bibinfo
  {author} {\bibfnamefont {L.}~\bibnamefont {Classen}}, \bibinfo {author}
  {\bibfnamefont {J.}~\bibnamefont {Splettstoesser}}, \ and\ \bibinfo {author}
  {\bibfnamefont {M.~R.}\ \bibnamefont {Wegewijs}},\ }\href@noop {} {\bibfield
  {journal} {\bibinfo  {journal} {Phys. Stat. Solidi B}\ }\textbf {\bibinfo
  {volume} {250}},\ \bibinfo {pages} {2315} (\bibinfo {year}
  {2013})}\BibitemShut {NoStop}%
\bibitem [{\citenamefont {Winkler}\ \emph {et~al.}(2013)\citenamefont
  {Winkler}, \citenamefont {Governale},\ and\ \citenamefont
  {K\"onig}}]{Winkler13}%
  \BibitemOpen
  \bibfield  {author} {\bibinfo {author} {\bibfnamefont {N.}~\bibnamefont
  {Winkler}}, \bibinfo {author} {\bibfnamefont {M.}~\bibnamefont {Governale}},
  \ and\ \bibinfo {author} {\bibfnamefont {J.}~\bibnamefont {K\"onig}},\
  }\href@noop {} {\bibfield  {journal} {\bibinfo  {journal} {Phys. Rev. B}\
  }\textbf {\bibinfo {volume} {87}},\ \bibinfo {pages} {155428} (\bibinfo
  {year} {2013})}\BibitemShut {NoStop}%
\bibitem [{\citenamefont {Rojek}\ \emph {et~al.}(2014)\citenamefont {Rojek},
  \citenamefont {Governale},\ and\ \citenamefont {K{\"o}nig}}]{Rojek14}%
  \BibitemOpen
  \bibfield  {author} {\bibinfo {author} {\bibfnamefont {S.}~\bibnamefont
  {Rojek}}, \bibinfo {author} {\bibfnamefont {M.}~\bibnamefont {Governale}}, \
  and\ \bibinfo {author} {\bibfnamefont {J.}~\bibnamefont {K{\"o}nig}},\
  }\href@noop {} {\bibfield  {journal} {\bibinfo  {journal} {Phys. Stat. Solidi
  B}\ }\textbf {\bibinfo {volume} {251}},\ \bibinfo {pages} {1912} (\bibinfo
  {year} {2014})}\BibitemShut {NoStop}%
\bibitem [{Note9()}]{Note9}%
  \BibitemOpen
  \bibinfo {note} {Note the difference between ``lag'' (Markovian,
  non-adiabatic) that we keep and ``memory'' (non-Markovian) that we neglect:
  Since thermal fluctuations are much faster than both coupling and driving, $T
  \gg \Gamma , |\delta \vec {R} | \Omega $, we can neglect the ``memory'' in
  the kernel. This results in Markovian dynamics of $\rho (t)$ [\Eq
  {eq:MarkovMe}] on time scale $\Gamma ^{-1}$. For driving velocities slower
  than this, i.e., $|\delta \vec {R} | \Omega \ll \Gamma $, the solution $\rho
  (t)$ of \Eq {eq:MarkovMe} develops a small ``lag'' responsible for pumping
  that we do take into account.}\BibitemShut {Stop}%
\bibitem [{\citenamefont {Avron}\ \emph
  {et~al.}(2012{\natexlab{b}})\citenamefont {Avron}, \citenamefont {Fraas},
  \citenamefont {Graf},\ and\ \citenamefont {Grech}}]{Avron12a}%
  \BibitemOpen
  \bibfield  {author} {\bibinfo {author} {\bibfnamefont {J.~E.}\ \bibnamefont
  {Avron}}, \bibinfo {author} {\bibfnamefont {M.}~\bibnamefont {Fraas}},
  \bibinfo {author} {\bibfnamefont {G.~M.}\ \bibnamefont {Graf}}, \ and\
  \bibinfo {author} {\bibfnamefont {P.}~\bibnamefont {Grech}},\ }\href@noop {}
  {\bibfield  {journal} {\bibinfo  {journal} {Communications in Mathematical
  Physics}\ }\textbf {\bibinfo {volume} {314}},\ \bibinfo {pages} {163}
  (\bibinfo {year} {2012}{\natexlab{b}})}\BibitemShut {NoStop}%
\bibitem [{Note10()}]{Note10}%
  \BibitemOpen
  \bibinfo {note} {In \Eq {eq:Obs} $W_{\protect \mathaccentV {hat}05E{I}_{X^r}}
  \propto V^2 \propto \Gamma $ by \Eq {eq:ObservableKernel} and \eq
  {eq:WIXnaive}.}\BibitemShut {Stop}%
\bibitem [{Note11()}]{Note11}%
  \BibitemOpen
  \bibinfo {note} {\Sec {sec:application} shows that via Stoke's theorem the
  pumped charge can be expressed as an area integral which for small driving
  cycles scales as $|\delta \vec {R}|^2$.}\BibitemShut {Stop}%
\bibitem [{Note12()}]{Note12}%
  \BibitemOpen
  \bibinfo {note} {Often the quoted condition $\Omega \ll \Gamma $ for pumping
  implicitly assumes $|\delta \vec {R}| \sim 1$.}\BibitemShut {Stop}%
\bibitem [{\citenamefont {Juergens}\ \emph {et~al.}(2013)\citenamefont
  {Juergens}, \citenamefont {Haupt}, \citenamefont {Moskalets},\ and\
  \citenamefont {Splettstoesser}}]{Juergens13}%
  \BibitemOpen
  \bibfield  {author} {\bibinfo {author} {\bibfnamefont {S.}~\bibnamefont
  {Juergens}}, \bibinfo {author} {\bibfnamefont {F.}~\bibnamefont {Haupt}},
  \bibinfo {author} {\bibfnamefont {M.}~\bibnamefont {Moskalets}}, \ and\
  \bibinfo {author} {\bibfnamefont {J.}~\bibnamefont {Splettstoesser}},\
  }\href@noop {} {\bibfield  {journal} {\bibinfo  {journal} {Phys. Rev. B}\
  }\textbf {\bibinfo {volume} {87}},\ \bibinfo {pages} {245423} (\bibinfo
  {year} {2013})}\BibitemShut {NoStop}%
\bibitem [{\citenamefont {Splettstoesser}\ \emph {et~al.}(2010)\citenamefont
  {Splettstoesser}, \citenamefont {Governale}, \citenamefont {K\"onig},\ and\
  \citenamefont {B\"uttiker}}]{Splettstoesser10}%
  \BibitemOpen
  \bibfield  {author} {\bibinfo {author} {\bibfnamefont {J.}~\bibnamefont
  {Splettstoesser}}, \bibinfo {author} {\bibfnamefont {M.}~\bibnamefont
  {Governale}}, \bibinfo {author} {\bibfnamefont {J.}~\bibnamefont {K\"onig}},
  \ and\ \bibinfo {author} {\bibfnamefont {M.}~\bibnamefont {B\"uttiker}},\
  }\href@noop {} {\bibfield  {journal} {\bibinfo  {journal} {Phys. Rev. B}\
  }\textbf {\bibinfo {volume} {81}},\ \bibinfo {pages} {165318} (\bibinfo
  {year} {2010})}\BibitemShut {NoStop}%
\bibitem [{\citenamefont {Landsberg}(1993)}]{Landsberg93}%
  \BibitemOpen
  \bibfield  {author} {\bibinfo {author} {\bibfnamefont {A.~S.}\ \bibnamefont
  {Landsberg}},\ }\href@noop {} {\bibfield  {journal} {\bibinfo  {journal}
  {Modern Physics Letters B}\ }\textbf {\bibinfo {volume} {07}},\ \bibinfo
  {pages} {71} (\bibinfo {year} {1993})}\BibitemShut {NoStop}%
\bibitem [{\citenamefont {Thouless}(1983)}]{Thouless83}%
  \BibitemOpen
  \bibfield  {author} {\bibinfo {author} {\bibfnamefont {D.~J.}\ \bibnamefont
  {Thouless}},\ }\href@noop {} {\bibfield  {journal} {\bibinfo  {journal}
  {Phys. Rev. B}\ }\textbf {\bibinfo {volume} {27}},\ \bibinfo {pages} {6083}
  (\bibinfo {year} {1983})}\BibitemShut {NoStop}%
\bibitem [{\citenamefont {Avron}\ \emph {et~al.}(2011)\citenamefont {Avron},
  \citenamefont {Fraas}, \citenamefont {Graf},\ and\ \citenamefont
  {Kenneth}}]{Avron11}%
  \BibitemOpen
  \bibfield  {author} {\bibinfo {author} {\bibfnamefont {J.}~\bibnamefont
  {Avron}}, \bibinfo {author} {\bibfnamefont {M.}~\bibnamefont {Fraas}},
  \bibinfo {author} {\bibfnamefont {G.}~\bibnamefont {Graf}}, \ and\ \bibinfo
  {author} {\bibfnamefont {O.}~\bibnamefont {Kenneth}},\ }\href@noop {}
  {\bibfield  {journal} {\bibinfo  {journal} {New. J. Phys.}\ }\textbf
  {\bibinfo {volume} {13}},\ \bibinfo {pages} {053042} (\bibinfo {year}
  {2011})}\BibitemShut {NoStop}%
\bibitem [{Note13()}]{Note13}%
  \BibitemOpen
  \bibinfo {note} {See \cite {Andersson03thesis,Andersson05} for a detailed
  exposition and generalization to the non-Abelian case and the review \Ref
  {Sinitsyn09} for related references.}\BibitemShut {Stop}%
\bibitem [{\citenamefont {Berry}(1987)}]{Berry87}%
  \BibitemOpen
  \bibfield  {author} {\bibinfo {author} {\bibfnamefont {M.~V.}\ \bibnamefont
  {Berry}},\ }\href@noop {} {\bibfield  {journal} {\bibinfo  {journal} {Proc.
  Roy. Soc.}\ }\textbf {\bibinfo {volume} {414}},\ \bibinfo {pages} {31}
  (\bibinfo {year} {1987})}\BibitemShut {NoStop}%
\bibitem [{Note14()}]{Note14}%
  \BibitemOpen
  \bibinfo {note} {Whereas often gauge-invariance (dependence) is a test for
  ``(un)physicality'' of computed quantities, this is not the case here: Only a
  \protect \emph {change} of an expectation value is gauge invariant. One-point
  measurements, e.g., of the current $\expec {I_{X}}(\tau )$ in the interval
  $\tau \in [0,\T ]$ are gauge dependent which is perfectly physical: changing
  the meter gauge changes the measured current.}\BibitemShut {Stop}%
\bibitem [{\citenamefont {Bruder}\ and\ \citenamefont
  {Schoeller}(1994)}]{Bruder94}%
  \BibitemOpen
  \bibfield  {author} {\bibinfo {author} {\bibfnamefont {C.}~\bibnamefont
  {Bruder}}\ and\ \bibinfo {author} {\bibfnamefont {H.}~\bibnamefont
  {Schoeller}},\ }\href@noop {} {\bibfield  {journal} {\bibinfo  {journal}
  {Phys. Rev. Lett.}\ }\textbf {\bibinfo {volume} {72}},\ \bibinfo {pages}
  {1076} (\bibinfo {year} {1994})}\BibitemShut {NoStop}%
\bibitem [{Note15()}]{Note15}%
  \BibitemOpen
  \bibinfo {note} {This split-up is relative to $\expec { \protect \mathaccentV
  {hat}05E{X}^{r}}(0)$ since it can only be defined via the corresponding
  split-up of the current. The latter exploits the gauge freedom \Eq
  {eq:Xrgauge} that emerges only for periodic driving.}\BibitemShut {Stop}%
\bibitem [{Note16()}]{Note16}%
  \BibitemOpen
  \bibinfo {note} {The rewriting of the result \Eq {eq:ObservableKernel} in
  \App {app:kernels} involves adding a commutator $[\protect \mathaccentV
  {hat}05E{X}^r,\bullet ]$ to the expression, which is invariant under the
  physical gauge transformations \Eq {eq:Xrgauge}. This means one can do such
  rewriting at \protect \emph {any} stage of the calculation.}\BibitemShut
  {Stop}%
\bibitem [{Note17()}]{Note17}%
  \BibitemOpen
  \bibinfo {note} {The freedom in the assignment of a current operator to an
  observable was discussed in \Ref {Avron12} for geometric pumping of \protect
  \emph {system} observables connected to a single reservoir, motivated by
  other works~\cite {Bellissard02,Gebauer04,Bodor06,Salmilehto12}. Here we
  consider more general \protect \emph {nonsystem} observables and multiple
  reservoirs, requiring consideration of current kernels. This allows
  steady-state transport \protect \emph {through} the system to be discussed.
  See further \App {app:pumpsys}.}\BibitemShut {Stop}%
\bibitem [{Note18()}]{Note18}%
  \BibitemOpen
  \bibinfo {note} {Throughout the paper we assume that the ``bare'', ungauged
  observable $\protect \mathaccentV {hat}05E{X}^r$ does not depend on the
  parameters: it is the ``probe'' used to detect a response of the driving
  ($\vec {R}$) and should be independent of the stimulus. However, when
  observable ${\protect \mathaccentV {hat}05E{X}^r+\protect \mathaccentV
  {hat}05E{X}}$ is conserved the corresponding system observable $\protect
  \mathaccentV {hat}05E{X}$ may well dependent on $\vec {R}$ [see discussion
  after \Eq {eq:Xconservation}].}\BibitemShut {Stop}%
\bibitem [{Note19()}]{Note19}%
  \BibitemOpen
  \bibinfo {note} {For simplicity we consider the parameters $\vec {R}$ to form
  a vector space. For general nontrivial parameter manifolds one should instead
  consider the exterior 2-form $B_{X^r}[\vec {R}] = \Lbra {\unit } [ \protect
  \text {d} ( W_{\protect \mathaccentV {hat}05E{I}_{X^r}} {W}^{-1} ) \wedge
  \Lket { \protect \text {d} \inst {\rho } }$ on the tangent space of the
  manifold.}\BibitemShut {Stop}%
\bibitem [{Note20()}]{Note20}%
  \BibitemOpen
  \bibinfo {note} {For bilinearly coupled quantum dots, the energy current is
  conserved only to the first order in the coupling, which is considered here.
  See \Ref {Gergs17a} for a detailed study of the corrections.}\BibitemShut
  {Stop}%
\bibitem [{Note21()}]{Note21}%
  \BibitemOpen
  \bibinfo {note} {We assume $[X^r,H^r]=0$ as in Eq. (15) of \Ref
  {Esposito09}.}\BibitemShut {Stop}%
\bibitem [{\citenamefont {Flindt}\ \emph {et~al.}(2008)\citenamefont {Flindt},
  \citenamefont {Novotn\'y}, \citenamefont {Braggio}, \citenamefont
  {Sassetti},\ and\ \citenamefont {Jauho}}]{Flindt08}%
  \BibitemOpen
  \bibfield  {author} {\bibinfo {author} {\bibfnamefont {C.}~\bibnamefont
  {Flindt}}, \bibinfo {author} {\bibfnamefont {T.}~\bibnamefont {Novotn\'y}},
  \bibinfo {author} {\bibfnamefont {A.}~\bibnamefont {Braggio}}, \bibinfo
  {author} {\bibfnamefont {M.}~\bibnamefont {Sassetti}}, \ and\ \bibinfo
  {author} {\bibfnamefont {A.-P.}\ \bibnamefont {Jauho}},\ }\href@noop {}
  {\bibfield  {journal} {\bibinfo  {journal} {Phys. Rev. Lett.}\ }\textbf
  {\bibinfo {volume} {100}},\ \bibinfo {pages} {150601} (\bibinfo {year}
  {2008})}\BibitemShut {NoStop}%
\bibitem [{Note22()}]{Note22}%
  \BibitemOpen
  \bibinfo {note} {Beyond the first moment, non-Markovian, higher order effects
  and initial correlations are important as shown in \Ref
  {Braggio05}.}\BibitemShut {Stop}%
\bibitem [{Note23()}]{Note23}%
  \BibitemOpen
  \bibinfo {note} {$K$ is the antilinear superoperator that effects Hermitian
  conjugation of an operator $x$, see \Ref {Saptsov12a}, App. G.}\BibitemShut
  {Stop}%
\bibitem [{Note24()}]{Note24}%
  \BibitemOpen
  \bibinfo {note} {As in the \ar approach, the kernel's eigenvalue with largest
  real part is assumed to be nondegenerate in the \fcs approach.}\BibitemShut
  {Stop}%
\bibitem [{Note25()}]{Note25}%
  \BibitemOpen
  \bibinfo {note} {See Eq. (19) ff. in \Ref {Sinitsyn07EPL} and p. 25. of \Ref
  {Sinitsyn09}.}\BibitemShut {Stop}%
\bibitem [{Note26()}]{Note26}%
  \BibitemOpen
  \bibinfo {note} {This is possible for driving cycles that can be covered by
  one single coordinate patch of the parameter manifold. For other cycles one
  can glue the solutions together in the standard way using the gauge
  invariance of the curvature, see, e.g., \Ref {Bohm}.}\BibitemShut {Stop}%
\bibitem [{Note27()}]{Note27}%
  \BibitemOpen
  \bibinfo {note} {The general relation reads $ v^\chi _0[\vec {R}] = \beta
  ^\chi [v^{-\chi }_0[\vec {R}]]^\protect \dag $ where $\beta ^\chi $ is a
  nonvanishing complex function which is restricted as $ \beta ^{\chi *} \beta
  ^{-\chi } =1 \label {eq:alpha} $ by consistency when taking the adjoint of
  the relation and setting $\chi \to -\chi $. By taking $\beta ^\chi =1$ for
  all $\chi $ we obtain \Eq {eq:hermgauge}.}\BibitemShut {Stop}%
\bibitem [{Note28()}]{Note28}%
  \BibitemOpen
  \bibinfo {note} {This requires the transformation \eq {eq:Achitransform2} of
  the \fcs connection.}\BibitemShut {Stop}%
\bibitem [{Note29()}]{Note29}%
  \BibitemOpen
  \bibinfo {note} {Since we focus on the first moment, cumulants need not be
  introduced here.}\BibitemShut {Stop}%
\bibitem [{Note30()}]{Note30}%
  \BibitemOpen
  \bibinfo {note} {Compare with a similar derivation given in \Ref {Nakajima15}
  see Eq. (22) there.}\BibitemShut {Stop}%
\bibitem [{Note31()}]{Note31}%
  \BibitemOpen
  \bibinfo {note} {See Eq. (20) ff. in \Ref {Sinitsyn07EPL}.}\BibitemShut
  {Stop}%
\bibitem [{\citenamefont {Ren}\ and\ \citenamefont {Sinitsyn}(2013)}]{Ren13}%
  \BibitemOpen
  \bibfield  {author} {\bibinfo {author} {\bibfnamefont {J.}~\bibnamefont
  {Ren}}\ and\ \bibinfo {author} {\bibfnamefont {N.~A.}\ \bibnamefont
  {Sinitsyn}},\ }\href@noop {} {\bibfield  {journal} {\bibinfo  {journal}
  {Phys. Rev. E}\ }\textbf {\bibinfo {volume} {87}},\ \bibinfo {pages} {050101}
  (\bibinfo {year} {2013})}\BibitemShut {NoStop}%
\bibitem [{\citenamefont {Li}\ \emph {et~al.}(2014)\citenamefont {Li},
  \citenamefont {Ren},\ and\ \citenamefont {Sinitsyn}}]{Li14}%
  \BibitemOpen
  \bibfield  {author} {\bibinfo {author} {\bibfnamefont {F.}~\bibnamefont
  {Li}}, \bibinfo {author} {\bibfnamefont {J.}~\bibnamefont {Ren}}, \ and\
  \bibinfo {author} {\bibfnamefont {N.~A.}\ \bibnamefont {Sinitsyn}},\
  }\href@noop {} {\bibfield  {journal} {\bibinfo  {journal} {Eur. Phys. Lett.}\
  }\textbf {\bibinfo {volume} {105}},\ \bibinfo {pages} {27001} (\bibinfo
  {year} {2014})}\BibitemShut {NoStop}%
\bibitem [{\citenamefont {Levitov}\ and\ \citenamefont
  {Lesovik}(1993)}]{Levitov93}%
  \BibitemOpen
  \bibfield  {author} {\bibinfo {author} {\bibfnamefont {L.~S.}\ \bibnamefont
  {Levitov}}\ and\ \bibinfo {author} {\bibfnamefont {G.~B.}\ \bibnamefont
  {Lesovik}},\ }\href@noop {} {\bibfield  {journal} {\bibinfo  {journal} {JETP
  Lett.}\ }\textbf {\bibinfo {volume} {58}},\ \bibinfo {pages} {230} (\bibinfo
  {year} {1993})}\BibitemShut {NoStop}%
\bibitem [{\citenamefont {Schaller}\ \emph {et~al.}(2009)\citenamefont
  {Schaller}, \citenamefont {Kie\ss{}lich},\ and\ \citenamefont
  {Brandes}}]{Schaller09}%
  \BibitemOpen
  \bibfield  {author} {\bibinfo {author} {\bibfnamefont {G.}~\bibnamefont
  {Schaller}}, \bibinfo {author} {\bibfnamefont {G.}~\bibnamefont
  {Kie\ss{}lich}}, \ and\ \bibinfo {author} {\bibfnamefont {T.}~\bibnamefont
  {Brandes}},\ }\href@noop {} {\bibfield  {journal} {\bibinfo  {journal} {Phys.
  Rev. B}\ }\textbf {\bibinfo {volume} {80}},\ \bibinfo {pages} {245107}
  (\bibinfo {year} {2009})}\BibitemShut {NoStop}%
\bibitem [{Note32()}]{Note32}%
  \BibitemOpen
  \bibinfo {note} {This requires consideration of observable operators that are
  sufficiently general, i.e., explicitly time-dependent and not partially
  normal-ordered with respect to the reservoirs.}\BibitemShut {Stop}%
\bibitem [{\citenamefont {Andersson}(2003{\natexlab{b}})}]{AnderssonThesis}%
  \BibitemOpen
  \bibfield  {author} {\bibinfo {author} {\bibfnamefont {S.}~\bibnamefont
  {Andersson}},\ }\emph {\bibinfo {title} {Geometric Phases in Sensing and
  Control}},\ \href@noop {} {Ph.D. thesis},\ \bibinfo  {school} {University of
  Maryland}, \bibinfo {address} {Mailand} (\bibinfo {year}
  {2003}{\natexlab{b}})\BibitemShut {NoStop}%
\bibitem [{Note33()}]{Note33}%
  \BibitemOpen
  \bibinfo {note} {The further \protect \emph {Markov} approximation is
  discussed in detail in the main text. Consistent with the Born-Markov
  approximation the kernels \Eq {eq:AppKernelFormula} and \eq
  {eq:CurrentKernelFormula} can be calculated for \protect \emph {frozen}
  parameters as discussed in the main text.}\BibitemShut {Stop}%
\bibitem [{\citenamefont {Koller}\ \emph {et~al.}(2010)\citenamefont {Koller},
  \citenamefont {Grifoni}, \citenamefont {Leijnse},\ and\ \citenamefont
  {Wegewijs}}]{Koller10}%
  \BibitemOpen
  \bibfield  {author} {\bibinfo {author} {\bibfnamefont {S.}~\bibnamefont
  {Koller}}, \bibinfo {author} {\bibfnamefont {M.}~\bibnamefont {Grifoni}},
  \bibinfo {author} {\bibfnamefont {M.}~\bibnamefont {Leijnse}}, \ and\
  \bibinfo {author} {\bibfnamefont {M.~R.}\ \bibnamefont {Wegewijs}},\
  }\href@noop {} {\bibfield  {journal} {\bibinfo  {journal} {Phys. Rev. B}\
  }\textbf {\bibinfo {volume} {82}},\ \bibinfo {pages} {235307} (\bibinfo
  {year} {2010})}\BibitemShut {NoStop}%
\bibitem [{Note34()}]{Note34}%
  \BibitemOpen
  \bibinfo {note} {The system Hamiltonian may then depend on reservoir
  parameters (e.g., temperature, electrochemical potentials).}\BibitemShut
  {Stop}%
\bibitem [{\citenamefont {Schoeller}(1997)}]{Schoeller97hab}%
  \BibitemOpen
  \bibfield  {author} {\bibinfo {author} {\bibfnamefont {H.}~\bibnamefont
  {Schoeller}},\ }\enquote {\bibinfo {title} {Mesoscopic electron transport},}\
  \ (\bibinfo  {publisher} {Kluwer},\ \bibinfo {year} {1997})\ Chap.\ \bibinfo
  {chapter} {Transport through interacting quantum dots}, p.\ \bibinfo {pages}
  {291}\BibitemShut {NoStop}%
\bibitem [{Note35()}]{Note35}%
  \BibitemOpen
  \bibinfo {note} {This can be done of course computationally if one has first
  measured the instantaneous currents at all parameter values accessed by the
  driving curve, but this seems less accurate as measurement errors for
  different times may accumulate.}\BibitemShut {Stop}%
\bibitem [{Note36()}]{Note36}%
  \BibitemOpen
  \bibinfo {note} {Such a procedure is usually used to rectify the
  instantaneous current by suppressing it to values $\ll 1$ for parameter
  values $\vec {R}$ where the current flows in an undesired direction and
  setting it to $1$ otherwise.}\BibitemShut {Stop}%
\bibitem [{\citenamefont {Sinitsyn}(2007)}]{Sinitsyn07PRB}%
  \BibitemOpen
  \bibfield  {author} {\bibinfo {author} {\bibfnamefont {N.~A.}\ \bibnamefont
  {Sinitsyn}},\ }\href@noop {} {\bibfield  {journal} {\bibinfo  {journal}
  {Phys. Rev. B}\ }\textbf {\bibinfo {volume} {76}},\ \bibinfo {pages} {153314}
  (\bibinfo {year} {2007})}\BibitemShut {NoStop}%
\bibitem [{\citenamefont {Jehl}\ \emph {et~al.}(2013)\citenamefont {Jehl},
  \citenamefont {Voisin}, \citenamefont {Charron}, \citenamefont {Clapera},
  \citenamefont {Ray}, \citenamefont {Roche}, \citenamefont {Sanquer},
  \citenamefont {Djordjevic}, \citenamefont {Devoille}, \citenamefont
  {Wacquez},\ and\ \citenamefont {Vinet}}]{Jehl13}%
  \BibitemOpen
  \bibfield  {author} {\bibinfo {author} {\bibfnamefont {X.}~\bibnamefont
  {Jehl}}, \bibinfo {author} {\bibfnamefont {B.}~\bibnamefont {Voisin}},
  \bibinfo {author} {\bibfnamefont {T.}~\bibnamefont {Charron}}, \bibinfo
  {author} {\bibfnamefont {P.}~\bibnamefont {Clapera}}, \bibinfo {author}
  {\bibfnamefont {S.}~\bibnamefont {Ray}}, \bibinfo {author} {\bibfnamefont
  {B.}~\bibnamefont {Roche}}, \bibinfo {author} {\bibfnamefont
  {M.}~\bibnamefont {Sanquer}}, \bibinfo {author} {\bibfnamefont
  {S.}~\bibnamefont {Djordjevic}}, \bibinfo {author} {\bibfnamefont
  {L.}~\bibnamefont {Devoille}}, \bibinfo {author} {\bibfnamefont
  {R.}~\bibnamefont {Wacquez}}, \ and\ \bibinfo {author} {\bibfnamefont
  {M.}~\bibnamefont {Vinet}},\ }\href@noop {} {\bibfield  {journal} {\bibinfo
  {journal} {Phys. Rev. X}\ }\textbf {\bibinfo {volume} {3}},\ \bibinfo {pages}
  {021012} (\bibinfo {year} {2013})}\BibitemShut {NoStop}%
\bibitem [{\citenamefont {Schulenborg}\ \emph {et~al.}(2016)\citenamefont
  {Schulenborg}, \citenamefont {Saptsov}, \citenamefont {Haupt}, \citenamefont
  {Splettstoesser},\ and\ \citenamefont {Wegewijs}}]{Schulenborg16a}%
  \BibitemOpen
  \bibfield  {author} {\bibinfo {author} {\bibfnamefont {J.}~\bibnamefont
  {Schulenborg}}, \bibinfo {author} {\bibfnamefont {R.~B.}\ \bibnamefont
  {Saptsov}}, \bibinfo {author} {\bibfnamefont {F.}~\bibnamefont {Haupt}},
  \bibinfo {author} {\bibfnamefont {J.}~\bibnamefont {Splettstoesser}}, \ and\
  \bibinfo {author} {\bibfnamefont {M.~R.}\ \bibnamefont {Wegewijs}},\
  }\href@noop {} {\bibfield  {journal} {\bibinfo  {journal} {Phys. Rev. B}\
  }\textbf {\bibinfo {volume} {93}},\ \bibinfo {pages} {081411} (\bibinfo
  {year} {2016})}\BibitemShut {NoStop}%
\bibitem [{Note37()}]{Note37}%
  \BibitemOpen
  \bibinfo {note} {Choosing two tunnel rates $\Gamma ^\protect \text {L} /
  \protect \mathaccentV {bar}016{\Gamma }$ and $ \Gamma ^\protect \text {R} /
  \protect \mathaccentV {bar}016{\Gamma }$ as driving parameters does not lead
  to pumping. For nonzero rates the reason for this is that the pumping
  contribution depends only on the ratio of tunnel rates. Therefore driving of
  $\Gamma ^\protect \text {L} / \protect \mathaccentV {bar}016{\Gamma }$ and $
  \Gamma ^\protect \text {R} / \protect \mathaccentV {bar}016{\Gamma }$
  effectively amounts to single parameter driving.}\BibitemShut {Stop}%
\bibitem [{\citenamefont {Contreras-Pulido}\ \emph {et~al.}(2012)\citenamefont
  {Contreras-Pulido}, \citenamefont {Splettstoesser}, \citenamefont
  {Governale}, \citenamefont {K\"onig},\ and\ \citenamefont
  {B\"uttiker}}]{Contreras12}%
  \BibitemOpen
  \bibfield  {author} {\bibinfo {author} {\bibfnamefont {L.~D.}\ \bibnamefont
  {Contreras-Pulido}}, \bibinfo {author} {\bibfnamefont {J.}~\bibnamefont
  {Splettstoesser}}, \bibinfo {author} {\bibfnamefont {M.}~\bibnamefont
  {Governale}}, \bibinfo {author} {\bibfnamefont {J.}~\bibnamefont {K\"onig}},
  \ and\ \bibinfo {author} {\bibfnamefont {M.}~\bibnamefont {B\"uttiker}},\
  }\href@noop {} {\bibfield  {journal} {\bibinfo  {journal} {Phys. Rev. B}\
  }\textbf {\bibinfo {volume} {85}},\ \bibinfo {pages} {075301} (\bibinfo
  {year} {2012})}\BibitemShut {NoStop}%
\bibitem [{\citenamefont {Saptsov}\ and\ \citenamefont
  {Wegewijs}(2014)}]{Saptsov14a}%
  \BibitemOpen
  \bibfield  {author} {\bibinfo {author} {\bibfnamefont {R.~B.}\ \bibnamefont
  {Saptsov}}\ and\ \bibinfo {author} {\bibfnamefont {M.~R.}\ \bibnamefont
  {Wegewijs}},\ }\href@noop {} {\bibfield  {journal} {\bibinfo  {journal}
  {Phys. Rev. B}\ }\textbf {\bibinfo {volume} {90}},\ \bibinfo {pages} {045407}
  (\bibinfo {year} {2014})}\BibitemShut {NoStop}%
\bibitem [{Note38()}]{Note38}%
  \BibitemOpen
  \bibinfo {note} {The explicit pumping curvature result \Eq {eq:brouwer2}
  corresponds to expressing in the intermediate calculation, for example, the
  \ar charge current as $\adcor {I}_{N^r}(t) = \Lbraket {\Phi _{N^r}|N} \cdot
  \protect \frac {d}{dt}\inst {\expec {\protect \mathaccentV {hat}05E{N}}}$
  [\Ref {Splettstoesser06}] and $\adcor {I}_{N^r}(t) = \Lbraket {\Phi _{N^r}|N}
  \cdot \protect \frac {d}{dt}\inst {\expec {\protect \mathaccentV
  {hat}05E{N}}} + \Lbraket {\Phi _{N^r}|S_z} \cdot \protect \frac {d}{dt}\inst
  {\expec {\protect \mathaccentV {hat}05E{S}_z}}$ [\Refs
  {Reckermann10a}].}\BibitemShut {Stop}%
\bibitem [{Note39()}]{Note39}%
  \BibitemOpen
  \bibinfo {note} {The adiabatic solution is the instantaneous solution with
  the right ``phase'' factors.}\BibitemShut {Stop}%
\bibitem [{Note40()}]{Note40}%
  \BibitemOpen
  \bibinfo {note} {Indeed condition \eq {eq:StateGapCondition} for the
  contribution \eq {eq:adiabatic-correction} to be negligible is sufficient but
  not necessary. Due to the decay in an open system some contributions are
  exponentially suppressed even without the condition \eq
  {eq:StateGapCondition}. This condition is thus only required for the
  nondecaying terms.}\BibitemShut {Stop}%
\bibitem [{Note41()}]{Note41}%
  \BibitemOpen
  \bibinfo {note} {See \Ref {Sinitsyn09} (p. 8) for a related observation in
  the \fcs approach.}\BibitemShut {Stop}%
\bibitem [{Note42()}]{Note42}%
  \BibitemOpen
  \bibinfo {note} {This is consistent with the adiabatic limit: to describe at
  least one driving cycle one needs $t \geq \Omega ^{-1}$ to be compatible with
  the gap condition \eq {eq:condition-ase} $\Omega \delta {R} \ll \Gamma $
  which is fine for $t \gg \Gamma ^{-1}$.}\BibitemShut {Stop}%
\bibitem [{Note43()}]{Note43}%
  \BibitemOpen
  \bibinfo {note} {If this assumption breaks down at any iteration it would
  imply that the solution we seek has no unique steady state which is our
  working assumption.}\BibitemShut {Stop}%
\bibitem [{\citenamefont {Fazio}\ \emph {et~al.}(2003)\citenamefont {Fazio},
  \citenamefont {Hekking},\ and\ \citenamefont {Pekola}}]{Fazio03}%
  \BibitemOpen
  \bibfield  {author} {\bibinfo {author} {\bibfnamefont {R.}~\bibnamefont
  {Fazio}}, \bibinfo {author} {\bibfnamefont {F.~W.~J.}\ \bibnamefont
  {Hekking}}, \ and\ \bibinfo {author} {\bibfnamefont {J.~P.}\ \bibnamefont
  {Pekola}},\ }\href@noop {} {\bibfield  {journal} {\bibinfo  {journal} {Phys.
  Rev. B}\ }\textbf {\bibinfo {volume} {68}},\ \bibinfo {pages} {054510}
  (\bibinfo {year} {2003})}\BibitemShut {NoStop}%
\bibitem [{\citenamefont {Governale}\ \emph {et~al.}(2005)\citenamefont
  {Governale}, \citenamefont {Taddei}, \citenamefont {Fazio},\ and\
  \citenamefont {Hekking}}]{Governale05}%
  \BibitemOpen
  \bibfield  {author} {\bibinfo {author} {\bibfnamefont {M.}~\bibnamefont
  {Governale}}, \bibinfo {author} {\bibfnamefont {F.}~\bibnamefont {Taddei}},
  \bibinfo {author} {\bibfnamefont {R.}~\bibnamefont {Fazio}}, \ and\ \bibinfo
  {author} {\bibfnamefont {F.~W.~J.}\ \bibnamefont {Hekking}},\ }\href@noop {}
  {\bibfield  {journal} {\bibinfo  {journal} {Phys. Rev. Lett.}\ }\textbf
  {\bibinfo {volume} {95}},\ \bibinfo {pages} {256801} (\bibinfo {year}
  {2005})}\BibitemShut {NoStop}%
\bibitem [{\citenamefont {Brosco}\ \emph {et~al.}(2008)\citenamefont {Brosco},
  \citenamefont {Fazio}, \citenamefont {Hekking},\ and\ \citenamefont
  {Joye}}]{Brosco08}%
  \BibitemOpen
  \bibfield  {author} {\bibinfo {author} {\bibfnamefont {V.}~\bibnamefont
  {Brosco}}, \bibinfo {author} {\bibfnamefont {R.}~\bibnamefont {Fazio}},
  \bibinfo {author} {\bibfnamefont {F.~W.~J.}\ \bibnamefont {Hekking}}, \ and\
  \bibinfo {author} {\bibfnamefont {A.}~\bibnamefont {Joye}},\ }\href@noop {}
  {\bibfield  {journal} {\bibinfo  {journal} {Phys. Rev. Lett.}\ }\textbf
  {\bibinfo {volume} {100}},\ \bibinfo {pages} {027002} (\bibinfo {year}
  {2008})}\BibitemShut {NoStop}%
\bibitem [{\citenamefont {M\"ott\"onen}\ \emph {et~al.}(2008)\citenamefont
  {M\"ott\"onen}, \citenamefont {Vartiainen},\ and\ \citenamefont
  {Pekola}}]{Mottonen08}%
  \BibitemOpen
  \bibfield  {author} {\bibinfo {author} {\bibfnamefont {M.}~\bibnamefont
  {M\"ott\"onen}}, \bibinfo {author} {\bibfnamefont {J.~J.}\ \bibnamefont
  {Vartiainen}}, \ and\ \bibinfo {author} {\bibfnamefont {J.~P.}\ \bibnamefont
  {Pekola}},\ }\href@noop {} {\bibfield  {journal} {\bibinfo  {journal} {Phys.
  Rev. Lett.}\ }\textbf {\bibinfo {volume} {100}},\ \bibinfo {pages} {177201}
  (\bibinfo {year} {2008})}\BibitemShut {NoStop}%
\bibitem [{\citenamefont {Gibertini}\ \emph {et~al.}(2013)\citenamefont
  {Gibertini}, \citenamefont {Fazio}, \citenamefont {Polini},\ and\
  \citenamefont {Taddei}}]{Gibertini13}%
  \BibitemOpen
  \bibfield  {author} {\bibinfo {author} {\bibfnamefont {M.}~\bibnamefont
  {Gibertini}}, \bibinfo {author} {\bibfnamefont {R.}~\bibnamefont {Fazio}},
  \bibinfo {author} {\bibfnamefont {M.}~\bibnamefont {Polini}}, \ and\ \bibinfo
  {author} {\bibfnamefont {F.}~\bibnamefont {Taddei}},\ }\href@noop {}
  {\bibfield  {journal} {\bibinfo  {journal} {Phys. Rev. B}\ }\textbf {\bibinfo
  {volume} {88}},\ \bibinfo {pages} {140508} (\bibinfo {year}
  {2013})}\BibitemShut {NoStop}%
\bibitem [{\citenamefont {Kaasbjerg}\ and\ \citenamefont
  {Flensberg}(2008)}]{Kaasbjerg08}%
  \BibitemOpen
  \bibfield  {author} {\bibinfo {author} {\bibfnamefont {K.}~\bibnamefont
  {Kaasbjerg}}\ and\ \bibinfo {author} {\bibfnamefont {K.}~\bibnamefont
  {Flensberg}},\ }\href@noop {} {\bibfield  {journal} {\bibinfo  {journal}
  {Nano Lett.}\ }\textbf {\bibinfo {volume} {8}},\ \bibinfo {pages} {3809}
  (\bibinfo {year} {2008})}\BibitemShut {NoStop}%
\bibitem [{\citenamefont {Gergs}\ \emph {et~al.}()\citenamefont {Gergs},
  \citenamefont {Saptsov}, \citenamefont {Schuricht},\ and\ \citenamefont
  {Wegewijs}}]{Gergs17a}%
  \BibitemOpen
  \bibfield  {author} {\bibinfo {author} {\bibfnamefont {N.~M.}\ \bibnamefont
  {Gergs}}, \bibinfo {author} {\bibfnamefont {R.~B.}\ \bibnamefont {Saptsov}},
  \bibinfo {author} {\bibfnamefont {D.}~\bibnamefont {Schuricht}}, \ and\
  \bibinfo {author} {\bibfnamefont {M.~R.}\ \bibnamefont {Wegewijs}},\
  }\href@noop {} {}\BibitemShut {NoStop}%
\bibitem [{\citenamefont {Esposito}\ \emph
  {et~al.}(2009{\natexlab{b}})\citenamefont {Esposito}, \citenamefont
  {Lindenberg},\ and\ \citenamefont {{van den Broeck}}}]{Esposito09}%
  \BibitemOpen
  \bibfield  {author} {\bibinfo {author} {\bibfnamefont {M.}~\bibnamefont
  {Esposito}}, \bibinfo {author} {\bibfnamefont {K.}~\bibnamefont
  {Lindenberg}}, \ and\ \bibinfo {author} {\bibfnamefont {C.}~\bibnamefont
  {{van den Broeck}}},\ }\href@noop {} {\bibfield  {journal} {\bibinfo
  {journal} {Eur. Phys. Lett.}\ }\textbf {\bibinfo {volume} {85}},\ \bibinfo
  {pages} {60010} (\bibinfo {year} {2009}{\natexlab{b}})}\BibitemShut {NoStop}%
\end{thebibliography}
\end{document}